\documentclass[mnsc,nonblindrev]{informs3_hide}

\OneAndAHalfSpacedXI % current default line spacing
\usepackage{bm}
\usepackage[normalem]{ulem}
\usepackage[dvipsnames]{xcolor}
\usepackage{booktabs,tabularx}
\usepackage{multirow}
\usepackage{makecell}
\usepackage{longtable}
\usepackage[ruled]{algorithm2e} % For algorithms

\usepackage{algorithmic}
\usepackage{xcolor}
\usepackage{hyperref}
\hypersetup{hidelinks}
\usepackage{bbm}
\usepackage{subfig}

\newenvironment{assumptionp}[1]{
  \renewcommand\theassumption{#1}
  \assumption
}{\endassumption}
\newenvironment{definitionp}[1]{
  \renewcommand\thedefinition{#1}
  \definition
}{\enddefinition}
\newenvironment{theoremp}[1]{
  \renewcommand\thetheorem{#1}
  \theorem
}{\endtheorem}

\usepackage[normalem]{ulem}

\newcommand{\bI}{\mathbbm{1}}
\newcommand{\independent}{\perp \!\!\! \perp}
\newcommand{\Authornote}[3]{{\textcolor{#2}{\sf$<${  #1: #3}$>$}}}
\newcommand{\jnote}[1]{\Authornote{Jinglong Comments}{red}{ }}
\newcommand{\yuan}[1]{\Authornote{Yuan Comments}{blue}{ }}
\newcommand{\shan}[1]{\Authornote{Shan Comments}{orange}{ }}
\newcommand{\bX}{\mathbb{X}}
\newcommand{\bY}{\mathbb{Y}}

\newcommand{\bS}{\mathbb{S}}
\newcommand{\bR}{\mathbb{R}}
\newcommand{\bT}{\mathbb{T}}
\newcommand{\bE}{\mathbb{E}}
\newcommand{\cF}{\mathcal{F}}
\newcommand{\cS}{\mathcal{S}}

\newcommand{\bN}{\mathbb{N}}

\usepackage{natbib}
 \bibpunct[, ]{(}{)}{,}{a}{}{,}%
 \def\bibfont{\small}%
\TheoremsNumberedThrough     % Preferred (Theorem 1, Lemma 1, Theorem 2)
\ECRepeatTheorems

\EquationsNumberedThrough    % Default: (1), (2), ...
\MANUSCRIPTNO{}

\begin{document}
%%%%%%%%%%%%%%%%

% Outcomment only when entries are known. Otherwise leave as is and
%   default values will be used.
%\setcounter{page}{1}
%\VOLUME{00}%
%\NO{0}%
%\MONTH{Xxxxx}% (month or a similar seasonal id)
%\YEAR{0000}% e.g., 2005
%\FIRSTPAGE{000}%
%\LASTPAGE{000}%
%\SHORTYEAR{00}% shortened year (two-digit)
%\ISSUE{0000} %
%\LONGFIRSTPAGE{0001} %
%\DOI{10.1287/xxxx.0000.0000}%

% Author's names for the running heads
% Sample depending on the number of authors;
% \RUNAUTHOR{Jones}
% \RUNAUTHOR{Jones and Wilson}
% \RUNAUTHOR{Jones, Miller, and Wilson}
% \RUNAUTHOR{Jones et al.} % for four or more authors
% Enter authors following the given pattern:
\RUNAUTHOR{Huang et al.}

% Title or shortened title suitable for running heads. Sample:
% \RUNTITLE{Bundling Information Goods of Decreasing Value}
% Enter the (shortened) title:
\RUNTITLE{Long-Term Treatments}

\TITLE{Estimating Effects of Long-Term Treatments}

\begingroup\renewcommand\thefootnote{*}
\footnotetext{Authors are listed in alphabetical order.}
\endgroup
\begingroup\renewcommand\thefootnote{†}
\footnotetext{To whom correspondence should be addressed.}
\endgroup

\ARTICLEAUTHORS{%
\AUTHOR{Shan Huang\textsuperscript{*}\textsuperscript{†}}
\AFF{The University of Hong Kong, \EMAIL{shanhh@hku.hk}} %, \URL{}}
\AUTHOR{Chen Wang\textsuperscript{*}}
\AFF{The University of Hong Kong, \EMAIL{annacwang@connect.hku.hk}} %, \URL{}}
\AUTHOR{Yuan Yuan\textsuperscript{*}}
\AFF{University of California, Davis, \EMAIL{yuyuan@ucdavis.edu}}
\AUTHOR{Jinglong Zhao\textsuperscript{*}}
\AFF{Boston University, Questrom School of Business, \EMAIL{jinglong@bu.edu}} %, \URL{}}
\AUTHOR{Brocco (Jingjing) Zhang}
\AFF{Tencent, Inc., \EMAIL{broccozhang@tencent.com}} %, \URL{}}
} % end of the block

\ABSTRACT{Estimating the effects of long-term treatments through A/B testing is challenging. Treatments, such as updates to product functionalities, user interface designs, and recommendation algorithms, are intended to persist within the system for a long duration of time after their initial launches. However, due to the constraints of conducting long-term experiments, practitioners often rely on short-term experimental results to make product launch decisions. It remains open how to accurately estimate the effects of long-term treatments using short-term experimental data. To address this question, we introduce a longitudinal surrogate framework that decomposes the long-term effects into functions based on user attributes, short-term metrics, and treatment assignments. We outline identification assumptions, estimation strategies, inferential techniques, and validation methods under this framework. Empirically, we demonstrate that our approach outperforms existing solutions by using data from two real-world experiments, each involving more than a million users on WeChat, one of the world's largest social networking platforms. 
}


\maketitle

\section{Introduction}
\label{sec:Intro}

Online controlled experiments, often referred to as A/B tests, have become the gold standard for evaluating the impact of product updates for technology companies. These updates can include the introduction of new product functions, user interface designs, and recommendation algorithms \citep{bakshy2014designing, bojinov2022online, kohavi2013online, Larsen2022, xu2015infrastructure, ye2023cold}. By randomly assigning experimental units (e.g., users) to different groups and exposing them to different product versions, A/B tests can measure the effects of the product update and guide business decisions. Modern technology companies deploy thousands of experiments daily to enable rapid iterations in their product development \citep{hohnhold2015focusing, kohavi2013online, leng2021calibration, ye2023deep}.

Estimating the effects of product updates presents a challenge in A/B testing \citep{gupta2019top, kohavi2020trustworthy}. When companies deploy a product update, it is usually intended to remain in the system for a long duration, typically spanning several months or over a year. Ideally, companies need to conduct long-term experiments to ensure that these updates have a lasting positive impact on user satisfaction and improve key product metrics. However, in practice, A/B tests are often short-term, typically lasting only several days or weeks. This is due to the considerable costs associated with long-term experiments, such as occupying substantial user traffic for an extended period and causing potential delays in the product iteration process \citep{kohavi2020trustworthy, bojinov2023design}. Short-term A/B tests offer the benefits of rapid feedback and lower costs, allowing companies to economize resources and maintain their agility in a competitive market.

The treatment effects derived from these short-term experiments can substantially differ from the actual effects of long-term product updates~\citep{hohnhold2015focusing, kohavi2012trustworthy, munro2021treatment}. A notable phenomenon here is the ``novelty effect": users may show higher levels of interest or response to a new or unfamiliar feature, resulting in stronger short-term outcomes in the treatment group. However, as users become more acquainted with this feature, this effect often diminishes over time~\citep{xu2015infrastructure}. Similarly, the ``primacy effect" arises when the benefits of a new feature only become evident after users have had sufficient time to become familiar with it, leading to a gradual increase in treatment effects over time~\citep{kohavi2020trustworthy}. Moreover, the introduction of new product changes in online marketplaces can cause disturbances in the product ecosystem, which could take a long duration to stabilize~\citep{bright2022reducing, farias2022markovian, glynn2020adaptive, hu2022switchback, johari2022experimental, wager2021experimenting}. Although practitioners often rely on the treatment effects in short-term experiments to represent the impact of long-term product changes in decision-making, the above scenarios underscore that this practice can mislead their decisions.

To address the above challenge, we introduce the ``longitudinal surrogate framework'' in this paper. Our theoretical results and empirical evidence suggest the feasibility of making trustworthy estimation of the effects of long-term treatments using data collected from short-term experiments. 
Our framework proposes to use ``longitudinal surrogates,'' which are the intermediate outcomes that saturate the causal links between historical treatments and future outcomes. We iteratively make use of these longitudinal surrogates and define the ``longitudinal surrogate index'' and ``pivot index'' functions. 
These index functions enable us to extrapolate the longitudinal surrogates from the short-term experimental periods to the long-term future periods.
Within this framework, we explain the underlying identification assumptions, the estimation strategies, inferential techniques, and strategies for validating our assumptions.

Empirically, we collaborated with WeChat, one of the world's largest social networking platforms, to validate the effectiveness of our framework through two large-scale, long-term experiments, each involving over a million users.
To leverage the long-term nature of these experiments, we partition the horizon into an ``experimental period'' and a ``future period.'' 
At the end of the experimental period, we apply our approach to estimate the treatment effects in the future period and compare our estimates with the true treatment effects observed in those periods. 
We show that our approach consistently outperforms two baseline approaches --- the Constant Extrapolation and the Vector Autoregressive Model \citep{stock2001vector} --- as well as several related existing solutions.
Compared to the baseline approaches, our approach reduces the estimation bias across different experimental periods by $59.8\%$, averaged across both experiments in our study, 
without increasing mean squared errors (MSE). 
Additionally, we conduct synthetic experiments to supplement our real-world experiments. 
We also conduct tests for the assumptions made under our framework, and discuss the practical guidelines to facilitate the applications of our methods in real-world settings.

Our longitudinal surrogate framework builds on the literature on proxies and surrogates~\citep{weir2006statistical,joffe2009related,prentice1989surrogate,athey2019surrogate, yang2020targeting, anderer2022adaptive, imbens2022long}. 
Yet our work differs from these previous studies in both the problem it addresses and the solutions it offers. 
Previous studies often employ surrogates to estimate the ``long-term effects of short-term treatments,'' as seen in applications such as job training programs \citep{athey2019surrogate} and marketing campaigns \citep{yang2020targeting}. In contrast, our framework is designed to estimate the ``long-term effects of long-term treatments,'' where subjects receive continuous treatments over extended periods. This context necessitates the estimation of the combined effects of both past and ongoing treatments, requiring a novel approach. For a comprehensive comparison of our work with that of \cite{athey2019surrogate}, please refer to Appendix~\ref{sec:appendix:Athey}. Similarly, \cite{battocchi2021estimating} address treatment effect estimation in long-term time series using surrogates. Their research focuses on a dynamic treatment setting, where treatment decisions in each period are influenced by previous treatments and outcomes, differing from the question in our study where the same treatment is employed over a long-term period.

Prior works, such as \cite{hohnhold2015focusing, munro2023causal} from online advertising applications, take a different approach when estimating the long-term effects. They model user learning behavior over time using parametric models with stronger assumptions, and combine such parametric models with non-trivial (i.e., Cookie-Cookie-Day) experiments. In contrast, our approach focuses on traditional randomized experiments, and conducts non-trivial post-experiment analysis. Our approach is designed to integrate with the conventional A/B testing pipelines at modern technology companies, avoiding the additional conceptual or implementation cost associated with executing non-trivial experiments. 

More broadly, our work is also related to panel data experiments.
In panel data experiments, subjects are not only repeatedly measured over time, but the treatment itself is also flexibly introduced, modified, or removed at different points in time \citep[e.g.,][]{abadie2021synthetic, athey2021matrix, basse2019minimax, chen2021learning, doudchenko2019designing, doudchenko2021synthetic, ni2023design, xiong2023optimal, xiong2023data}.
The major difference is that our approach only uses data collected from short-term experiments with standard A/B testing procedures, instead of using the entire panel.

\section{The Longitudinal Surrogate Framework}
\label{sec:Model}

\subsection{Problem Setup}
\label{sec:Setup}
Consider an A/B testing problem that an experimenter faces on an online platform. 
The platform conducts an A/B test to evaluate the effects of introducing a new product update.
To do so, the platform includes a total of $N$ experimental subjects, denoted by set $[N] = \{1,2,\ldots,N\}$.
Each subject is typically an active user. 
Each subject $i \in [N]$ is endowed with some $R$-dimensional covariates $\bm{X}_i \in \bX \subseteq \bR^R$, which we refer to as the pre-treatment variables.
For example, the pre-treatment variables $\bm{X}_i$ are typically user demographics at online platforms.
In this paper, we only consider the setting where the pre-treatment variables are low-dimensional, that is, the dimension of $\bm{X}_i$ is much smaller than the number of experimental subjects $N$. 

\begin{figure}[]
\centering
\includegraphics[width=0.6\textwidth]{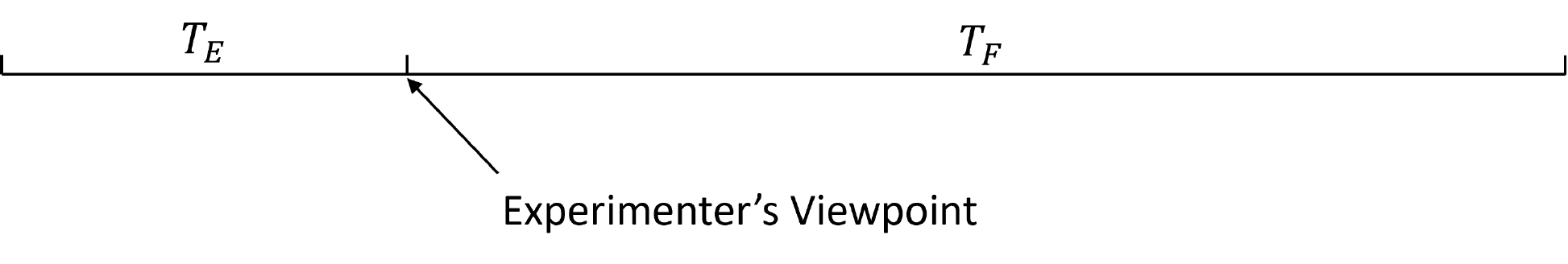}
\vspace{-3mm}
\caption{An illustrator of experimental periods, future periods, and the experimenter's viewpoint}
\vspace{-3mm}
\label{fig:TimePeriods}
\end{figure}

The experimenter is interested in understanding the effects of a long-term treatment yet they can only run the experiment for a shorter duration. 
We explain the horizon as follows.
Let there be a discrete, finite time horizon consisting of $T = T_E + T_F$ time periods in chronological order.
Out of these $T$ time periods, the first $T_E$ time periods are referred to as the \textit{experimental} periods, and the last $T_F$ time periods are referred to as the \textit{future} periods.
After conducting the experiment until the end of the experimental periods $T_E$, the experimenter has access to data collected from periods $1$ to $T_E$, and is interested in some causal effects that will not be directly observed until the end of period $T$.
In our running example, the experimenter could run the experiment for a few weeks, and then use the experimental data to estimate what would happen if the intervention continues to last for additional weeks. 
See Figure~\ref{fig:TimePeriods} for an illustration.

We consider two versions of treatments although our approach can easily extend to multiple treatments. One version is the control condition (or, simply, ``control''), which represents the status-quo of the product; the other version is the active treatment (or, simply, ``treatment''), which represents the product with the new feature.
Let $W_{i,t}$ be the random treatment assignment that subject $i \in [N]$ receives in  time period $t \in [T]$.
$W_{i,t}$ takes values from $\{0,1\}$, where $0$ stands for control and $1$ stands for treatment.
For each subject, we use $\bm{W}_{i, 1:t}$ to stand for the treatment assignments that subject $i \in [N]$ receives during periods $1$ to $t$.
Following convention, we use $\bm{W}_{i,1:t}$ to stand for a random treatment assignment and $\bm{w}_{i,1:t}$ to stand for one realization.
When the subscript $i$ is clear from the context, we sometimes drop it for brevity, and write $\bm{W}_{1:t}$ instead.

We conduct a randomized experiment wherein once a subject is assigned into either the treatment or control group, it stays in that group during the entire horizon.
If subject $i$ is assigned into the treatment group, then $\bm{W}_{i,1:T} = \bm{1}_T$; if subject $i$ is assigned into the control group, then $\bm{W}_{i,1:T} = \bm{0}_T$, where we use $\bm{1}_{t}$ and $\bm{0}_t$ to stand for a length-$t$ vector of ones and zeros, respectively.
As we stand at the end of period $T_E$, we have only conducted the experiment during the first $T_E$ experimental periods, and not yet in the last $T_F$ future periods.

We do not consider other types of treatment patterns that change the treatment assignment in the middle of the horizon, such as a step-wedge design (i.e., a staggered adoption pattern, \citet{brown2006stepped, hussey2007design, hemming2015stepped, li2018optimal, xiong2023optimal}) or a switchback design \citep{cochran1941double, glynn2020adaptive, bojinov2023design, hu2022switchback, xiong2023data}.
This implies that, for simplicity, we could just use a single binary variable to indicate if a subject is assigned to the treatment or control group.
But for clarity, we would rather carry the treatment assignment vector.
While the treatment assignments remain the same over time, the treatment probabilities across different subjects can be different.
Our framework allows treatment assignments to be dependent on $\bm{X}_i$ (i.e., stratified randomization), although we have only conducted complete randomization in our empirical execution.

During the $T_E$ experimental periods, the experimenter observes several quantities of interest.
For each subject $i\in[N]$ and at each time period $t\in[T_E]$, the experimenter observes a primary outcome $Y_{it}$ that takes values from $\bY \subseteq \bR$ and $D$ intermediate outcomes $\bm{S}_{it}$ that take values from $\bS \subseteq \bR^{D}$.
In our running example, the primary outcome could be the click through rate and the intermediate outcomes could include a number of user activity metrics such as log-in frequency, average usage duration, number of total searches, and the numbers of searches in each category.

Following the potential outcomes framework \citep{neyman1923application} and under the Stable Unit Treatment Value Assumption \citep{rubin1974estimating, holland1986statistics, imbens2015causal}, each subject $i\in[N]$ at each time period $t\in[T_E]$ has a set of potential outcomes $Y_{it}(\bm{W}_{i,1:t})$ and $\bm{S}_{it}(\bm{W}_{i,1:t})$.
Each observed outcome, either the primary outcome or the intermediate outcome, is related to its respective potential outcomes as follows,
\begin{align*}
Y_{it} = Y_{it}(\bm{w}_{1:t}), \quad \bm{S}_{it} = \bm{S}_{it}(\bm{w}_{1:t}), && \text{if} \ \bm{W}_{i,1:t}=\bm{w}_{1:t}.
\end{align*}
During the future periods $\{T_E+1,...,T\}$, we could also define the same quantities as above, although the observed outcomes have not been observed by the experimenter.
See Table~\ref{tbl:Notation} for an illustration of our problem setup and summary of notations.

\begin{table}[!tb]
\TABLE{Illustration of our problem setup and summary of notations.
\label{tbl:Notation}}
{\small
\begin{tabular}{|>{\centering}p{3.2cm}|>{\centering}p{5.8cm}|>{\centering}p{5cm}|c}
\cline{1-3}
\multirow{2}{*}{}             & Experimental   periods                                                                            & Future periods                 & \\
                              & $t \in \{1,2,…,T_E\}$                                                                             & $t \in \{T_E+1, T_E+2, …, T\}$ & \\ \cline{1-3}
Treatment group               & $W_{it} = 1$, observe $\left(   Y_{it}(\bm{1}_t), \bm{S}_{it}(\bm{1}_t) \right)$ & missing       & \\ \cline{1-3}
% $\bm{X}_i$    &                                                                                                   &                                & \\ \cline{1-3}
Control group                 & $W_{it} = 0$, observe $\left(   Y_{it}(\bm{0}_t), \bm{S}_{it}(\bm{0}_t) \right)$ & missing       & \\ \cline{1-3}
% $\bm{X}_i$    &                                                                                                   &                                & \\ \cline{1-3}
\end{tabular}
}
{\footnotesize \textit{Note}: The treatment assignments $W_{it}$, primary outcomes $Y_{it}$, and surrogate outcomes $\bm{S}_{it}$ are all missing from the future periods, as our viewpoint is at the end of the experimental periods. }
\vspace{-3mm}
\end{table}

In addition, let $\bm{S}_{i0}$ be some pre-treatment intermediate outcomes at time $0$, which may reflect subject-level heterogeneity before the experiment.
For notational convenience, we collect $\bm{Y}_i = \{Y_{it}(\bm{w}_{1:t})\}_{t \in [T],\bm{w}_{1:t}}$ and $\bm{S}_i = \{\bm{S}_{i0}, \bm{S}_{it}(\bm{w}_{1:t})\}_{t \in [T],\bm{w}_{1:t}}$ to be all the potential outcomes.
Further, we introduce a short-hand notation to emphasize the most recent treatment assignments.
For any $i \in [N]$ and any $t < t' \in [T]$, if $\bm{W}_{i,1:t} = \bm{0}_{1:t}$, then we write $Y_{it'}(\bm{W}_{i,t+1:t'}) := Y_{it'}(\bm{W}_{i,1:t'})$.
Note that this is only a short-hand notation, and does not impose any assumptions.

In this paper, we postulate a super-population that each subject is sampled from with replacement, so that each subject $i \in [N]$ is identically and independently distributed. 
For each $i \in [N]$, let $\cF$ be the joint probability distribution that $(\bm{X}_i, \bm{Y}_i, \bm{S}_i)$ is sampled from.
There are two sources of randomness in our experiment: one comes from the randomized experiment, i.e., the treatment assignments are random; the other comes from the sampling from a super-population, i.e., the pre-treatment variables and all the potential outcomes are random.

The experimenter is interested in understanding the average effect of long-term treatments on the primary outcome,
\begin{align}
\tau_T = \bE_{\cF}\bigg[ Y_{iT}(\bm{1}_{T}) - Y_{iT}(\bm{0}_{T}) \bigg]. \label{eqn:Estimand}
\end{align}
Such causal effects often emerge when experimenters aim to permanently launch a new product. In our running example, this relates to click-through rates over weeks or months.

\subsection{Conventional Wisdom and New Challenges}
\label{sec:Challenges}

In this paper, the duration of treatments spans the entire horizon, which we refer to as long-term treatments. 
To estimate the effects of long-term treatments, the ideal approach is to conduct experiments for an extended duration of time in the future periods $\{T_E+1, ..., T\}$ and directly estimate $\tau_T$ from such an ideal experiment. 
However, as discussed in Section~\ref{sec:Intro}, the experimenter is often unable to assign treatments for a long-term duration, and there is no observation from the future periods at the moment of estimation.
The fundamental challenges associated with this problem are two-fold:
\begin{enumerate}
\item \textbf{(Missing treatments)} At the moment of estimation, the experimenter has not conducted any treatment in the future periods.
\item \textbf{(Missing observations)} At the moment of estimation, the experimenter has not observed any outcome in the future periods.
\end{enumerate}

The presence of the above two challenges requires a new method that explicitly considers the longitudinal nature of the treatments, where the existing surrogate approach \citep{athey2019surrogate, joffe2009related, prentice1989surrogate, yang2020targeting, weir2006statistical} does not directly apply.
For example, \citet{athey2019surrogate} and 
\citet{yang2020targeting} examine the treatment effects, where the duration of treatments is relatively short compared to the length of future periods and the treatments never occurred during the future periods. 
We thus refer to the effect they studied as the long-term effects of short-term treatments; in other words, they focus on  estimating the long-term ``carryover effects,'' i.e.,
\begin{align*}
\bE_{\cF}\bigg[ Y_{iT}(\bm{1}_{T_E}, \bm{0}_{T_F}) - Y_{iT}(\bm{0}_{T}) \bigg].
\end{align*}
Therefore, the existing surrogate approach addresses the second challenge only and  establishes a surrogate predictor using the historical data, which is used to extrapolate from the short-term observations.
Unless the treatments in the future periods have no direct effects, i.e., $\bE_{\cF}[ Y_{iT}(\bm{1}_T)] = \bE_{\cF}[ Y_{iT}(\bm{1}_{T_E}, \bm{0}_{T_F})]$, the existing surrogate approach will lead to biased estimation of $\tau_T$ the average effect of long-term treatments.

To address the above two challenges, we propose a framework to extend the existing surrogate approaches to the longitudinal setting discussed above.
Below we introduce a few identification assumptions that we make in the longitudinal surrogate framework.

\subsection{Identification Assumptions}
\label{sec:Assumptions}
Below we first introduce the \textit{longitudinal surrogate model} and the two required identification assumptions. 
These two identification assumptions are what we refer to as the first level of assumptions.
Since the longitudinal surrogate model may suffer from the potentially limited sample size (see Section~\ref{sec:Estimation} for details), we introduce an additional assumption to the first level of assumptions, leading to the \textit{linear surrogate model}.
\footnote{In addition to the longitudinal surrogate model and the linear surrogate model, we also introduce the \textit{linear additive model}, which requires a different additional assumption to the first level of assumptions. Although the additional assumption is intuitive, it does not seem to hold in many real-world applications. Our empirical estimation shows that its performance is often unsatisfactory. We present more details in Appendix~\ref{sec:appendix:additive}.}

\subsubsection{Longitudinal surrogate model.}
We start with the basic assumptions that lay out the foundations of estimating the causal effect.
There are two such basic assumptions.

\begin{assumption}[Longitudinal Surrogacy]
\label{asp:Surrogacy}
The treatment assignment at an earlier period is independent of the primary and intermediate outcomes at a later period, conditional on the intermediate outcomes at a middle period, i.e., there exists a subset of time indices $\bT = \{t_1,t_2,...,t_K\} \subseteq [T]$, such that for any $i \in [N]$, any $t \in \bT$, and any $t' > t$,
\begin{align*}
\left(Y_{it'}, \bm{S}_{it'}\right) \independent \bm{W}_{i, 1:t} \vert \bm{S}_{it}, \bm{X}_i.
\end{align*}
Moreover, we refer to the intermediate outcomes at the time periods $t \in \bT$ as surrogate outcomes, or, simply, surrogates.
\end{assumption}

\begin{figure}[!tb]
\centering
\includegraphics[width=0.7\textwidth]{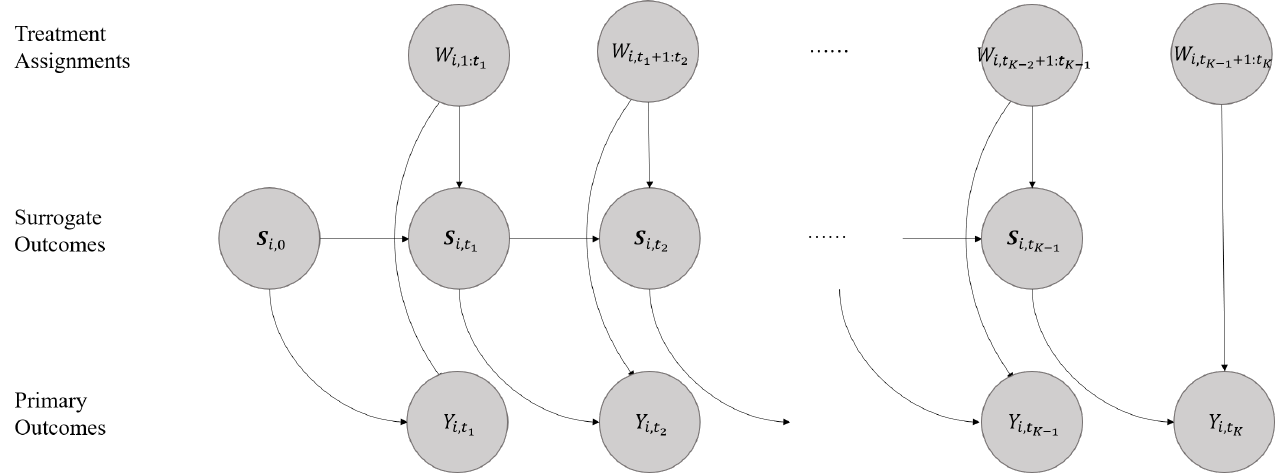}
\caption{An illustrator of the Longitudinal Surrogacy assumption using directed acyclic graph representation.\protect\footnotemark}
\vspace{-3mm}
\label{fig:Surrogacy}
\end{figure}

\footnotetext{In this illustrator, each solid line represents a causal path. Each treatment assignment at an earlier period impacts the surrogate outcomes and the primary outcome at a later period; each surrogate outcome and the primary outcome at an earlier period impacts the primary outcome at a later period. Each treatment assignment at an earlier period does not directly impact the primary and surrogate outcomes at a later period without going through the surrogate outcomes and the primary outcome at the middle period. For simplicity, pre-treatment variables are not explicitly included in this figure. However, the subscript $i$ in the surrogate and primary outcomes implicitly suggests that we could incorporate pre-treatment variables.}

Assumption~\ref{asp:Surrogacy} is the longitudinal extension of the surrogacy assumption in the literature \citep{athey2019surrogate, joffe2009related, prentice1989surrogate, yang2020targeting, weir2006statistical}.
Intuitively, Assumption~\ref{asp:Surrogacy} implies that the surrogate outcomes at a middle period fully saturate the causal link between the treatment assignment at an earlier period and the primary and intermediate outcomes at a later period.
In other words, there is no effect of the treatment assignment at an earlier period on the primary and intermediate outcomes at a later period that does not pass through the surrogate outcomes at the middle period.
See Figure~\ref{fig:Surrogacy} for an illustration using the directed acyclic graph representation \citep{pearl1995causal}.
We discuss practical guidelines for choosing surrogates in Section~\ref{sec:guidelines}.

There are two direct implications of Assumption~\ref{asp:Surrogacy}.
The first implication is that, if Assumption~\ref{asp:Surrogacy} holds for some $\bT$, it also holds for any subset of $\bT$, i.e., for any $\bT' \subseteq \bT$, Assumption~\ref{asp:Surrogacy} also holds for $\bT'$.
The second implication is that, for any $i \in [N]$, any $t \in \bT$, and any $t' > t \geq t''$,
\begin{align*}
\left(Y_{it'}, \bm{S}_{it'}\right) \independent \bm{S}_{i t''} \vert \bm{S}_{it}, \bm{X}_i.
\end{align*}
This is because, if $\left(Y_{it'}, \bm{S}_{it'}\right)$ and $\bm{S}_{it''}$ are not independent, then $\bm{W}_{i, 1:t''}$ and $\bm{S}_{i t''}$ will not be independent, violating Assumption~\ref{asp:Surrogacy}.

In the longitudinal surrogate model, the surrogate outcomes serve as critical links in the causal diagram in two ways.
First, conditional on the surrogate outcomes, we extrapolate to the primary outcomes in the future periods using what we refer to as the longitudinal surrogate index, which we define below in Definition~\ref{defn:SurrogateIndex}.
Second, conditional on the surrogate outcomes at an earlier period, we build our understanding of the future surrogate outcomes using what we refer to as the pivot index, which we define below in Definition~\ref{defn:PivotIndex}.

\begin{definition}[Longitudinal Surrogate Index]
\label{defn:SurrogateIndex}
For any $t \in [T], \bm{s} \in \bS, \bm{x} \in \bX, \bm{w}_{1:t} \in \{\bm{0}_t, \bm{1}_t\}$, the surrogate index is the conditional expectation of the primary outcome at time $t$, given the surrogate outcomes at time $0$, the pre-treatment variables, and the treatment assignments, i.e.,
\begin{align*}
h_t(\bm{s}, \bm{x}, \bm{w}_{1:t}) = \bE_{\cF} \left[ Y_{it} \vert \bm{S}_{i0} = \bm{s}, \bm{X}_i = \bm{x}, \bm{W}_{i,1:t} = \bm{w}_{1:t}\right],
\end{align*}
where the expectation is taken over $Y_{it}$.
\end{definition}

Intuitively, the longitudinal surrogate index serves as a prediction of future primary outcomes using the current intermediate outcomes, the pre-treatment variables, and the treatment assignments. 
This index has a time-dependent subscript, which reflects the longitudinal nature of our setup, and is different from the surrogate index as originally defined in \citet{athey2019surrogate}.

In addition to the longitudinal surrogate index, we introduce the pivot index as defined below.\footnote{For notational convenience, if two random variables $X'$ and $X''$ have the same distribution, we write $X' \sim X''$.}

\begin{definition}[Pivot Index]
\label{defn:PivotIndex}
For any $t \in [T]$, $\bm{s} \in \bS$, $\bm{x} \in \bX$, $\bm{w}_{1:t} \in \{\bm{0}_t, \bm{1}_t\}$, the pivot index is a vector of the conditional expectations of the surrogate outcomes at time $t$, given the surrogate outcomes at time $0$, the pre-treatment variables, and the treatment assignments, i.e.,
\begin{align*}
\bm{g}_t(\bm{s}, \bm{x}, \bm{w}_{1:t}) = \bE_{\cF} \left[ \bm{S}_{it} \vert \bm{S}_{i0} = \bm{s}, \bm{X}_i = \bm{x}, \bm{W}_{i,1:t} = \bm{w}_{1:t}\right],
\end{align*}
where the expectation is taken over $\bm{S}_{it}$.
Moreover, we denote the conditional surrogate outcomes at time $t$, given the surrogate outcomes at time $0$, the pre-treatment variables, and the treatment assignments to be,
\begin{align*}
\bm{G}_t(\bm{s}, \bm{x}, \bm{w}_{1:t}) \sim \bm{S}_{it} \vert \bm{S}_{i0} = \bm{s}, \bm{X}_i = \bm{x}, \bm{W}_{i,1:t} = \bm{w}_{1:t}.
\end{align*}
\end{definition}

The pivot indices (or the conditional surrogate outcomes, depending on which identification strategy to use) are the key idea behind our longitudinal surrogate framework. ``$\sim$'' indicates following the same distributions. 
Intuitively, they bridge the surrogates at the earlier periods and the surrogates at the later periods. 
The use of pivot indices is necessary in our model because the experimental duration is short, and what we learn from the experimental data needs the pivot indices (or the conditional surrogate outcomes) to iterate and extrapolate to the future periods.
Note that the definition of pivot indices replaces the primary outcomes as defined in Definition~\ref{defn:SurrogateIndex} by the surrogate outcomes.

\begin{assumption}[Comparability]
\label{asp:Comparability}
The primary and intermediate outcomes across different periods share the same support. 
The distribution of the primary and intermediate outcomes at a later period, conditional on the intermediate outcomes at an earlier period, on the treatment assignments during the earlier and later periods, and on the pre-treatment variables, is the same across different time periods, i.e., for any $t,t' \in [T]$, and any positive integer $\delta \in \bN^+$, 
\begin{align*}
(Y_{it}, \bm{S}_{it}) \vert \bm{S}_{i(t-\delta)}, \bm{W}_{i,t-\delta+1:t}, \bm{X}_i \ \sim \ (Y_{it'}, \bm{S}_{it'}) \vert \bm{S}_{i(t'-\delta)}, \bm{W}_{i,t'-\delta+1:t'}, \bm{X}_i.
\end{align*}
\end{assumption}

Intuitively, Assumption~\ref{asp:Comparability} implies that the relationship between the primary and intermediate outcomes at a later period and the intermediate outcomes at an earlier period is the same at other time periods.
So we could use data collected from the experimental periods to learn the relationship and apply it to future periods.
Note that Assumption~\ref{asp:Comparability} does not necessarily assume the primary outcomes  or the surrogate outcomes are time-homogeneous; 
instead, Assumption~\ref{asp:Comparability} assumes the \textit{functions} of the surrogate index and the pivot indices to be time-homogeneous.

Assumptions~\ref{asp:Surrogacy} --~\ref{asp:Comparability} are the most basic level of assumptions.
Under Assumptions~\ref{asp:Surrogacy} --~\ref{asp:Comparability}, and using the succinct notations from Definitions~\ref{defn:SurrogateIndex} --~\ref{defn:PivotIndex}, we present the first identification result as follows.

We first introduce a special case to illustrate the key idea behind our main theorem.
\begin{lemma}
\label{lem:SpecialCaseL1}
Consider the special case when $T_E = T_F$.
Under Assumptions~\ref{asp:Surrogacy} --~\ref{asp:Comparability}, where Assumption~\ref{asp:Surrogacy} holds for $\bT = \{T_E\}$, the average effect of long-term treatments on the primary outcome is equal to the following expression, 
\begin{align*}
\tau_T = \bE_{\cF} \left[ h_{T_E}\left( \bm{G}_{T_E}(\bm{S}_{i0}, \bm{X}_i, \bm{1}_{T_E}), \bm{X}_i, \bm{1}_{T_E} \right) \right] - \bE_{\cF} \left[ h_{T_E}\left( \bm{G}_{T_E}(\bm{S}_{i0}, \bm{X}_i, \bm{0}_{T_E}), \bm{X}_i, \bm{0}_{T_E} \right) \right].
\end{align*}
\end{lemma}

Lemma~\ref{lem:SpecialCaseL1} consists of two components: the surrogate index component $h_{T_E}(\cdot, \cdot, \cdot)$ that predicts the primary outcomes using the pivots, and a conditional surrogate outcomes component $\bm{G}_{T_E}(\cdot, \cdot, \cdot)$ that re-weighs the distributions of the random surrogate outcomes using the pre-treatment surrogate outcomes.
Lemma~\ref{lem:SpecialCaseL1} illustrates how the surrogate outcomes at $T_E$ as the outputs of the inner loop re-weighting are used as the input of the outer loop surrogate index.
The surrogate outcomes at $T_E$ effectively serve as the link between the two components.

In the more general setting when $T_F > T_E$, we need to have more surrogate outcomes to serve as the links.
We split the horizon of $T$ periods into several intervals, each length of which is no larger than the length of the experimental periods.
Mathematically, denote $\Delta t_k := t_k - t_{k-1}$.
The above condition suggests that $T_E \geq \max_{k\in[K+1]} \Delta t_{k}$. We write $t_{K+1}=T$ and $t_0=0$ as the end and start of all periods.
Then, we apply the same method as in Lemma~\ref{lem:SpecialCaseL1} on each interval and update the surrogate outcomes iteratively.
We formalize the above intuition as follows.

\begin{theorem}[Longitudinal Surrogate Model]
\label{thm:IdentificationL1}
Under Assumptions~\ref{asp:Surrogacy} and~\ref{asp:Comparability}, where Assumption~\ref{asp:Surrogacy} holds for $\bT = \{t_1,t_2,...,t_K\}$, the average effect of long-term treatments on the primary outcome is equal to the following expression, 
\begin{multline*}
\tau_T = \bE_{\cF} \left[ h_{\Delta t_{K+1}}\left( \bm{G}_{\Delta t_K}( ... \bm{G}_{\Delta t_1}(\bm{S}_{i0}, \bm{X}_i, \bm{1}_{\Delta t_1}) ... ,\bm{X}_i, \bm{1}_{\Delta t_K}), \bm{X}_i, \bm{1}_{\Delta t_{K+1}} \right) \right] \\
- \bE_{\cF} \left[ h_{\Delta t_{K+1}}\left( \bm{G}_{\Delta t_K}( ... \bm{G}_{\Delta t_1}(\bm{S}_{i0}, \bm{X}_i, \bm{0}_{\Delta t_1}) ... ,\bm{X}_i, \bm{0}_{\Delta t_K}), \bm{X}_i, \bm{0}_{\Delta t_{K+1}} \right) \right],
\end{multline*}
where the expectation is taken over $\bm{S}_{i0}, \bm{X}_i$, as well as the conditional surrogate outcomes $\bm{G}_{\Delta t_1}, ..., \bm{G}_{\Delta t_K}$.
\end{theorem}
Theorem~\ref{thm:IdentificationL1} consists of a sequence of iterative components.
There is one surrogate index component $h_{\Delta t_{K+1}}(\cdot, \cdot, \cdot)$ that predicts the primary outcomes during the last interval, using the conditional surrogate outcomes re-weighted from the second last interval.
There is a sequence of conditional surrogate outcomes $\bm{G}_{\cdot}(\cdot, \cdot, \cdot)$ that re-weighs the distributions using the conditional surrogate outcomes re-weighted from the previous interval.
Both components (i.e., the surrogate index and the conditional surrogate outcomes) can be estimated from the data during the experimental periods.

\subsubsection{Linear surrogate model.} 

Although general, the first identification strategy as suggested by Lemma~\ref{lem:SpecialCaseL1} and Theorem~\ref{thm:IdentificationL1} suffers from a major challenge resulting from the random nature of conditional surrogate outcomes and potentially limited sample sizes.
We will revisit this challenge in greater details in Section~\ref{sec:Estimation}.
To address this, we introduce an additional assumption to the two basic assumptions.
This set of three assumptions is the second level of assumptions.

\begin{assumption}[Linearity of Surrogates]
\label{asp:Linearity}
\begin{enumerate}
\item The surrogate index function is linear with respect to the surrogates, i.e., there exists $\alpha_d(\bm{x}, \bm{w}_{1:t})$, $\forall d\in\{0,1,...,D\}, \bm{x}\in\bX, \bm{w}_{1:t}\in\{\bm{0}_t, \bm{1}_t\}$, 
such that
\begin{align}
h_t(\bm{s}, \bm{x}, \bm{w}_{1:t}) = \alpha_0(\bm{x}, \bm{w}_{1:t}) + \sum_{d=1}^D s_d \cdot \alpha_d(\bm{x}, \bm{w}_{1:t}). \label{eqn:Linearity:SurrogateIndex}
\end{align}
\item The pivot index function is linear with respect to the surrogates, i.e., there exists $\beta_{d,d'}(\bm{x}, \bm{w}_{1:t})$, $\forall d\in[D], d'\in\{0,1,...,D\}, \bm{x}\in\bX, \bm{w}_{1:t}\in\{\bm{0}_t, \bm{1}_t\}$, 
such that for each $d\in[D]$,
\begin{align}
g_{t,d}(\bm{s}, \bm{x}, \bm{w}_{1:t}) = \beta_{d,0}(\bm{x}, \bm{w}_{1:t}) + \sum_{d'=1}^D s_d \cdot \beta_{d,d'}(\bm{x}, \bm{w}_{1:t}), \label{eqn:Linearity:PivotIndex}
\end{align}
where $g_{t,d}(\bm{s}, \bm{x}, \bm{w}_{1:t})$ stands for the $d$-th component of $\bm{g}_t(\bm{s}, \bm{x}, \bm{w}_{1:t})$ the pivot index.
\end{enumerate}
\end{assumption}

Assumption~\ref{asp:Linearity} specifies a linear functional form to the surrogate index and the pivot index.
It is worth mentioning that Assumption~\ref{asp:Linearity} assumes both the surrogate index and the pivot index to be linear with respect to the surrogates, but not necessarily with respect to the pre-treatment variables.
Under this additional Assumption~\ref{asp:Linearity}, we simplify Theorem~\ref{thm:IdentificationL1} and introduce the second identification result as follows.

\begin{theorem}[Linear Surrogate Model]
\label{thm:IdentificationL2}
Under Assumptions~\ref{asp:Surrogacy},~\ref{asp:Comparability}, and~\ref{asp:Linearity}, where Assumption~\ref{asp:Surrogacy} holds for $\bT = \{t_1,t_2,...,t_K\}$, the average effect of long-term treatments on the primary outcome is equal to the following expression, 
\begin{multline*}
\tau_T = \bE_{\cF} \Big[ h_{\Delta t_{K+1}} \big( \bm{g}_{\Delta t_K}( ... \bm{g}_{\Delta t_1}(\bm{S}_{i0}, \bm{X}_i, \bm{1}_{\Delta t_1}) ... ,\bm{X}_i, \bm{1}_{\Delta t_K}), \bm{X}_i, \bm{1}_{\Delta t_{K+1}} \big) \Big] \\
- \bE_{\cF} \Big[ h_{\Delta t_{K+1}}\left( \bm{g}_{\Delta t_K}( ... \bm{g}_{\Delta t_1}(\bm{S}_{i0}, \bm{X}_i, \bm{0}_{\Delta t_1}) ... ,\bm{X}_i, \bm{0}_{\Delta t_K}), \bm{X}_i, \bm{0}_{\Delta t_{K+1}} \right) \Big],
\end{multline*}
where the expectation is taken over $\bm{S}_{i0}, \bm{X}_i$.
\end{theorem}

Theorem~\ref{thm:IdentificationL2} involves both the surrogate index and the pivot index.
The input of an outer iteration is the output of an inner iteration, which, under the linearity assumption, is simply the pivot index in the inner iteration.
With this linear model, the identification strategy as suggested by Theorem~\ref{thm:IdentificationL2} properly mitigates the issues of large sample sizes as required by the longitudinal surrogate model, and thus estimate the future treatment effects with reasonable sample sizes.

\section{Estimation and Inference}
\label{sec:EstimationInference}

In this section, we discuss the estimation strategies, inference strategies, and model validation strategies for the models discussed above. 
We focus on conventional randomized experiments where subjects are randomly assigned into the treatment or the control groups under (covariate-independent) complete randomization. 
Let $N_1$ and $N_0$ be the number of users in the treatment and the control group, respectively, which are fixed quantities under complete randomization. 
Our approach readily applies to more general randomization schemes, which we omit in this paper. 

\subsection{Estimation Strategies}
\label{sec:Estimation}

Recall that in Section~\ref{sec:Assumptions} we introduce two levels of identification assumptions.
Below we introduce two estimation strategies, each requiring one level of assumptions discussed in Section~\ref{sec:Assumptions}.

\subsubsection{Estimators for the longitudinal surrogate model.}
\label{sec:LongitudinalEstimator}
Given estimators of the surrogate index and estimators of the conditional surrogate outcomes, we follow Theorem~\ref{thm:IdentificationL1} and obtain the following plug-in estimator,
{\small
\vspace{1mm}
\begin{align}
\widehat{\tau}_T = & \ \frac{1}{N_1} \sum_{i\in[N]} \bI\{\bm{W}_{i,1:T_E} = \bm{1}_{T_E}\} \bE_{\widehat{\bm{G}}_{\Delta t_1}, ..., \widehat{\bm{G}}_{\Delta t_K}}\left[ \widehat{h}_{\Delta t_{K+1}}\left( \widehat{\bm{G}}_{\Delta t_K}( ... \widehat{\bm{G}}_{\Delta t_1}(\bm{S}_{i0}, \bm{X}_i, \bm{1}_{\Delta t_1}) ... ,\bm{X}_i, \bm{1}_{\Delta t_K}), \bm{X}_i, \bm{1}_{\Delta t_{K+1}} \right) \right] \nonumber \\
& - \frac{1}{N_0}\sum_{i\in[N]} \bI\{\bm{W}_{i,1:T_E} = \bm{0}_{T_E}\} \bE_{\widehat{\bm{G}}_{\Delta t_1}, ..., \widehat{\bm{G}}_{\Delta t_K}}\left[ \widehat{h}_{\Delta t_{K+1}}\left( \widehat{\bm{G}}_{\Delta t_K}( ... \widehat{\bm{G}}_{\Delta t_1}(\bm{S}_{i0}, \bm{X}_i, \bm{0}_{\Delta t_1}) ... ,\bm{X}_i, \bm{0}_{\Delta t_K}), \bm{X}_i, \bm{0}_{\Delta t_{K+1}} \right) \right]. \label{eqn:EstimationL1}
\end{align}
}

We explain how to estimate the surrogate index functions in \eqref{eqn:EstimationL1}.
For any $t \in [T_E]$, $\bm{x} \in \bX$, $\bm{s} \in \bS$, 
one naive estimator of the surrogate index under consecutive controls is given by
\begin{align*}
\widehat{h}_T(\bm{s}, \bm{x}, \bm{0}_t) = \frac{\sum_{i \in [N]}Y_{it} \bI\{\bm{X}_i = \bm{x}, \bm{S}_{i0} = \bm{s}, \bm{W}_{i,1:t} = \bm{0}_t\}}{\sum_{i \in [N]} \bI\{\bm{X}_i = \bm{x}, \bm{S}_{i0} = \bm{s}, \bm{W}_{i,1:t} = \bm{0}_t\}}. 
\end{align*}
Under complete randomization, such an estimator is unbiased for the surrogate index function.
Similarly, for any $t \in [T_E]$, $\bm{x} \in \bX$, $\bm{s} \in \bS$, one naive estimator of the surrogate index under consecutive treatments is given by
\begin{align*}
\widehat{h}_t(\bm{s}, \bm{x}, \bm{1}_t) = \frac{\sum_{i \in [N]}Y_{it} \bI\{\bm{X}_i = \bm{x}, \bm{S}_{i0} = \bm{s}, \bm{W}_{i,1:t} = \bm{1}_t\}}{\sum_{i \in [N]} \bI\{\bm{X}_i = \bm{x}, \bm{S}_{i0} = \bm{s}, \bm{W}_{i,1:t} = \bm{1}_t\}}.
\end{align*}
Under complete randomization, such an estimator is unbiased for the surrogate index function.
Yet given the oftentimes multi-dimensional nature of $\bm{s}$ and $\bm{x}$, and the limited number of treatment subjects in the experimental periods, the above two estimators are not always well-behaved.
For each combination of $\bm{s}$ and $\bm{x}$, we need a sufficiently large number of samples in the experimental periods to have reasonably accurate estimation, which is often challenging in practice.

\subsubsection{Estimators for the linear surrogate model.}
Due to the limitations of the longitudinal surrogate model, we introduce the linear surrogate model, which requires the additional Assumption~\ref{asp:Linearity}. 
Given the surrogate and pivot index estimators, we follow Theorem~\ref{thm:IdentificationL2} and obtain the following plug-in estimator,
\begin{multline}
\widehat{\tau}_T = \frac{1}{N_1} \sum_{i\in[N]} \bI\{\bm{W}_{i,1:T_E} = \bm{1}_{T_E}\} \widehat{h}_{\Delta t_{K+1}}\left( \widehat{\bm{g}}_{\Delta t_K}( ... \widehat{\bm{g}}_{\Delta t_1}(\bm{S}_{i0}, \bm{X}_i, \bm{1}_{\Delta t_1}) ... ,\bm{X}_i, \bm{1}_{\Delta t_K}), \bm{X}_i, \bm{1}_{\Delta t_{K+1}} \right) \\
- \frac{1}{N_0} \sum_{i\in[N]} \bI\{\bm{W}_{i,1:T_E} = \bm{0}_{T_E}\} \widehat{h}_{\Delta t_{K+1}}\left( \widehat{\bm{g}}_{\Delta t_K}( ... \widehat{\bm{g}}_{\Delta t_1}(\bm{S}_{i0}, \bm{X}_i, \bm{0}_{\Delta t_1}) ... ,\bm{X}_i, \bm{0}_{\Delta t_K}), \bm{X}_i, \bm{0}_{\Delta t_{K+1}} \right)
\end{multline}
Note that since $S_{i\Delta t_1}$ is directly observable, we can use the observed $S_{i\Delta t_1}$ to replace $\widehat{\bm{g}}_{\Delta t_1}(\bm{S}_{i0}, \bm{X}_i, \bm{1}_{\Delta t_1})$ in the first (inner) plug-in. 
We use the following plug-in estimator in empirical estimation. 
\begin{multline}
\widehat{\tau}_T = \frac{1}{N_1} \sum_{i\in[N]} \bI\{\bm{W}_{i,1:T_E} = \bm{1}_{T_E}\} \widehat{h}_{\Delta t_{K+1}}\left( \widehat{\bm{g}}_{\Delta t_K}( ... \widehat{\bm{g}}_{\Delta t_2}(\bm{S}_{i\Delta t_1}, \bm{X}_i, \bm{1}_{\Delta t_2}) ... ,\bm{X}_i, \bm{1}_{\Delta t_K}), \bm{X}_i, \bm{1}_{\Delta t_{K+1}} \right) \\
- \frac{1}{N_0} \sum_{i\in[N]} \bI\{\bm{W}_{i,1:T_E} = \bm{0}_{T_E}\} \widehat{h}_{\Delta t_{K+1}}\left( \widehat{\bm{g}}_{\Delta t_K}( ... \widehat{\bm{g}}_{\Delta t_2}(\bm{S}_{i\Delta t_1}, \bm{X}_i, \bm{0}_{\Delta t_2}) ... ,\bm{X}_i, \bm{0}_{\Delta t_K}), \bm{X}_i, \bm{0}_{\Delta t_{K+1}} \right) \label{eqn:EstimationL2}
\end{multline}

We explain how to estimate the surrogate and pivot index functions in \eqref{eqn:EstimationL2}.
We first consider a proper discretization of the pre-treatment variables $\bm{x}$.
Then, for each $\bm{x}$ and under homoscedasticity, a naive estimator of the coefficients of the surrogate index function is given by
\begin{align*}
\left( \widehat{\alpha}_0(\bm{x},\bm{1}_t), ..., \widehat{\alpha}_D(\bm{x},\bm{1}_t) \right) = \arg\min_{\alpha_0,...,\alpha_D} \sum_{i\in[N]} \left( Y_{it} - \alpha_0 - \sum_{d=1}^D S_{i0,d} \alpha_d \right)^2 \bI\{\bm{X}_i = \bm{x}, \bm{W}_{i,1:t}=\bm{1}_t\},
\end{align*}
and for each $d \in [D]$, the pivot index function is given by
\begin{align*}
\left( \widehat{\beta}_{d,0}(\bm{x},\bm{1}_t), ..., \widehat{\beta}_{d,D}(\bm{x},\bm{1}_t) \right) = \arg\min_{\beta_{d,0},...,\beta_{d,D}} \sum_{i\in[N]} \left( S_{it,d} - \beta_{d,0} - \sum_{d'=1}^D S_{i0,d'} \beta_{d,d'} \right)^2 \bI\{\bm{X}_i = \bm{x}, \bm{W}_{i,1:t}=\bm{1}_t\},
\end{align*}
where $S_{i0,d}$ and $S_{it,d}$ stand for the $d$-th dimension of surrogate outcomes $\bm{S}_{i0}$ and $\bm{S}_{it}$, respectively.
The estimators of the surrogate and pivot index functions are obtained by replacing the coefficients in \eqref{eqn:Linearity:SurrogateIndex} and \eqref{eqn:Linearity:PivotIndex} with their estimated counterparts.
Under complete randomization, both estimators are unbiased for the linear coefficients in \eqref{eqn:Linearity:SurrogateIndex} and \eqref{eqn:Linearity:PivotIndex}.
The second part in \eqref{eqn:EstimationL2} can be estimated similarly.
See Lemma~\ref{lem:LinearConsistency} in the Appendix~\ref{sec:appendix:lemma2}. 

The above two least squares estimators find the coefficients for any $\bm{x} \in \bX$.
This is suitable when the pre-treatment variables are low-dimensional and discrete.
Given the multi-dimensional nature of $\bm{x}$, and especially when $\bm{x}$ is continuous, the least squares estimators are not always well-behaved.
To address the above concern, we could include the pre-treatment variables $\bm{X}_i$ in the least square term instead of conditioning on them. 
Instead of estimating $\widehat{\alpha}_d(\bm{x},\bm{1}_t)$ and $\widehat{\beta}_{d,d'}(\bm{x},\bm{1}_t)$, we pool the data and run the following linear regression to estimate $\widehat{\alpha}_d(\bm{1}_t)$ and $\widehat{\beta}_{d,d'}(\bm{1}_t)$, as well as $\widehat{\phi}_r(\bm{1}_t)$ and $\widehat{\psi}_{d,r}(\bm{1}_t)$.
\vspace{-3mm}
\begin{multline*}
\left( \widehat{\alpha}_0(\bm{1}_t), ..., \widehat{\alpha}_D(\bm{1}_t), \widehat{\phi}_1(\bm{1}_t), ..., \widehat{\phi}_R(\bm{1}_t) \right) = \\ \argmin_{\substack{\alpha_0,...,\alpha_D, \\ \phi_1,...,\phi_R}} \sum_{i\in[N]} \left( Y_{it} - \alpha_0 - \sum_{d=1}^D S_{i0,d} \alpha_d - \sum_{r=1}^R X_{i,r} \phi_r \right)^2 \bI\{\bm{W}_{i,1:t}=\bm{1}_t\},
\end{multline*}
and for each $d \in [D]$, 
\vspace{-3mm}
\begin{multline*}
\left( \widehat{\beta}_{d,0}(\bm{1}_t), ..., \widehat{\beta}_{d,D}(\bm{1}_t), \widehat{\psi}_{d,1}(\bm{1}_t), ..., \widehat{\psi}_{d,R}(\bm{1}_t) \right) = \\
\argmin_{\substack{\beta_{d,0},...,\beta_{d,D}, \\ \psi_{d,1}, ..., \psi_{d,R}}} \sum_{i\in[N]} \left( S_{it,d} - \beta_{d,0} - \sum_{d'=1}^D S_{i0,d'} \beta_{d,d'} - \sum_{r=1}^R X_{i,r} \psi_{d,r} \right)^2 \bI\{\bm{W}_{i,1:t}=\bm{1}_t\}.
\end{multline*}
The second part in \eqref{eqn:EstimationL2} can be estimated similarly.
The above expressions find the best linear unbiased estimator for the coefficients of the pre-treatment variables. 
They mitigate the issue of requiring a large sample size in the longitudinal surrogate model. 

\subsection{Inference and Testing}
\label{sec:Inference}
Our estimator leverages an additional layer of randomness from the random treatment assignments.
Here we propose a Fisher's exact test to draw inference from the collected data.
We consider the following sharp null hypothesis of no treatment effect at any time period for any subject:
\begin{align}
~\label{eqn:null}
H_{0}: \ (Y_{it}(\bm{1}_t), \bm{S}_{it}(\bm{1}_t)) = (Y_{it}(\bm{0}_t), \bm{S}_{it}(\bm{0}_t)), \ \forall t \in [T], i \in [N].
\end{align}

We can conduct exact tests by leveraging the completely randomized experiment to simulate new treatment assignments; see Algorithm~\ref{alg:Exact} in the Online Appendix.
To obtain a confidence interval, we propose inverting a sequence of exact hypothesis tests to identify the region outside of which \eqref{eqn:null} is violated at the pre-specified nominal level \citep[Chapter~5]{imbens2015causal}. 
Alternatively, one could also use bootstrap to obtain a confidence interval.
The source of randomness comes from our random treatment assignments;
see Algorithm~\ref{alg:bootstrap} in the Online Appendix. In later empirical sections, we mainly report the results using the bootstrap method. 

Our work is also related to forecasting methods in the time series analysis and the macroeconometrics literature, such as autoregressive models, Vector Autoregression (VAR), and Autoregressive Integrated Moving Average (ARIMA) \citep{stock2001vector, stock2020introduction, hamilton2020time, fuller2009introduction, andersen2003modeling}. 
The macroeconometrics literature has also provided ways to construct confidence intervals by leveraging the randomness of the joint probability distribution that $(\bm{X}_i, \bm{Y}_i, \bm{S}_i)$ is sampled from. 
Such confidence intervals are generally recognized to have more power than Fisher's exact test, which relies on the randomness of the random treatment assignments. 
For simplicity, we adopted the simpler approach of the Fisher's exact test and the bootstrap method.

\subsection{Validation of Assumptions}
\label{sec:validation}

As the longitudinal surrogacy assumption (Assumption~\ref{asp:Surrogacy}) and the comparability assumption (Assumption~\ref{asp:Comparability}) play a critical role in determining the validity of our method in practice, we explore approaches to validate whether these assumptions are satisfied in this section.\footnote{Intuitively, we validate whether the dynamics of the carryover effects satisfy certain patterns.
Assumption~\ref{asp:Surrogacy} restricts that the carryover effects should be fully mediated by the selected surrogate variables. 
This is essentially the Markovian assumption in modeling the surrogate outcomes.
Assumption~\ref{asp:Comparability} can be relaxed into Assumption~\ref{asp:ParallelTrends:appendix} when combined with the linearity assumption. 
Intuitively, our method allows for distributional shifts in the primary outcomes, as long as the difference in the primary outcomes between the treatment group and the control group (i.e., the dynamics of carryover effects) remains stable over time.}

\subsubsection{Validation of Assumption~\ref{asp:Surrogacy}.}

Similar to the tests on the validity of instrumental variables, Assumption~\ref{asp:Surrogacy} cannot be directly tested. 
Instead, we propose conducting a sensitivity analysis to determine how sensitive the treatment effect estimation is when Assumption~\ref{asp:Surrogacy} is violated. 
Our approach is inspired by the literature on sensitivity analysis of instrumental variables \citep{baiocchi2014instrumental}. 
Arguably, the most common violation of Assumption~\ref{asp:Surrogacy} occurs when there are \textit{omitted surrogates}. Figure~\ref{fig:SurrogacyViolation} in Appendix illustrates such a scenario: Assumption~\ref{asp:Surrogacy} is violated because the treatment assignment during the experimental periods $1:T_E$ affects the primary outcome through both variables $\bm{S}_{T_E}$ and $\bm{U}_{T_E}$. Here only $\bm{S}_{T_E}$ are considered as the surrogate variables, while $\bm{U}_{T_E}$ represent the omitted surrogates that remain unidentified or uncollected.

First, a straightforward approach for sensitivity analysis on Assumption~\ref{asp:Surrogacy} is to assess the fluctuation in estimation given that only a \textit{subset} of surrogate outcomes are applied as surrogates.
This analysis reveals how the estimation is impacted by the exclusion of certain already collected surrogates. We demonstrate that as more surrogates are removed, the estimation performance deteriorates, aligning with our intuition. Overall, our estimation approach is relatively robust across different subsets of surrogates. Detailed analysis of this approach is provided in Appendix~\ref{sec:appendix:surrogateSubset}. 

Second, we design an approach to test the sensitivity of omitted surrogates, focusing on assessing the model's sensitivity to surrogates that were never observed. This approach can be particularly valuable in real-world experiments where some of the surrogates can be potentially unobservable and missing from our estimation. 
Our method can be seen as an adaptation of the sensitivity analysis for assessing the \textit{Exclusion Restriction} assumption for instrumental variables~\citep{baiocchi2014instrumental}.

Suppose, for any $i \in [N]$, $t \in \bT$, and $\bm{w}_{1:t} \in \{\bm{0}_{t}, \bm{1}_{t}\}$, the treatment assignment affects the primary outcome not only through the identified surrogates, but also via a missing variable $\zeta_{it}$.
We create this variable $\zeta_{it}$ following a normal distribution with mean zero, and variance equal to the average variance of the $Y$ during the experimental periods. 
We manually introduce an additional causal path between the treatment assignment and the primary outcome through variable $\zeta_{it}$:
\begin{align*}
\tilde{Y}_{it}(\bm{w}_{1:t}) = Y_{it}(\bm{w}_{1:t}) + \theta\cdot \zeta_{it}\cdot \mathbbm{1}[\bm{w}_{1:t}=\bm{1}_{t}],
\end{align*}
where $\theta$ is a parameter that we generate to vary the degree of omitted surrogates
and $\mathbbm{1}[\cdot]$ is the indicator function.
In this sensitivity analysis, we treat $\tilde{Y_{it}}$ instead of $Y_{it}$ as the primary outcome and consider only the observed surrogates $S_{it}$, as if the omitted surrogate $\zeta_{it}$ was neither observed nor collected. Clearly, Assumption~\ref{asp:Surrogacy} is violated due to the omitted surrogate $\zeta_{it}$, and a larger $\theta$ indicates a greater violation of Assumption~\ref{asp:Surrogacy}. We then follow the same procedure to estimate the average effect of long-term treatments.
Finally, we compare these estimates with the ones obtained using $Y_{it}$ as the true primary outcome variable, where Assumption~\ref{asp:Surrogacy} is not violated. This approach allows us to examine the sensitivity of our estimation results to varying degrees of violation of the surrogacy assumption. A detailed demonstration of this sensitivity analysis, along with empirical experiments, is provided in Appendix~\ref{sec:appendix:missingSurrogates}. The results show that the bias and RMSE remain stable when $\theta$ is relatively small, demonstrating the robustness of the estimation.

\subsubsection{Validation of Assumption~\ref{asp:Comparability}.}
We begin by introducing a straightforward test directly for Assumption~\ref{asp:Comparability} (the comparability assumption). Moreover, we discuss that even when Assumption~\ref{asp:Comparability} does not hold, we can still apply our longitudinal surrogate framework, by leveraging a relaxation of Assumption~\ref{asp:Comparability}, which we refer to as the \textit{Parallel Trends} assumption (Assumption~\ref{asp:ParallelTrends:appendix}). We also provide a test for this parallel trends assumption.

\textit{Direct Test for Assumption~\ref{asp:Comparability}.} 
The objective of this test is to identify matched observations across two distinct time periods, \( t \) and \( t' \), based on exact matching criteria involving the surrogates $\bm{S}_i$, the treatment assignments $\bm{W}_i$, and the pre-treatment variables $\bm{X}_i$. 
More specifically, we begin by specifying the two time periods of interest, \( t \) and \( t' \), and the lag parameter \( \delta \). 
For each unit \( i \) at time \( t \), we collect the following information: $\bm{S}_{i,t-\delta}, \bm{W}_{i,t-\delta+1:t}$, and $\bm{X}_i$.
Next, 
we search for any unit $i'$ at time \( t' \) that satisfies the following conditions,
\[
\bm{S}_{i',t' - \delta} = \bm{S}_{i,t - \delta}, \quad \bm{W}_{i',t' - \delta+1:t'} = \bm{W}_{i,t - \delta+1:t}, \quad \bm{X}_{i'} = \bm{X}_i.
\]
All pairs of observations \( (i, i') \) that meet the above conditions are included in the analysis pool, which results in two groups of observations from each of the two time periods \( t \) and \( t' \), with the corresponding outcomes \( (Y_{it}, Y_{i't'}) \).
If no observations meet the requirement at time \( t \), the test for that specific condition is excluded from further analysis. 
For each possible combination of \(\bm{s},\bm{w},\bm{x}\), we perform statistical tests to examine the difference between \( Y_{it} \) and \( Y_{i't'} \) and report $p$-values.

\textit{Parallel Trends Test.}
To make our longitudinal surrogate framework more useful to practitioners, we relax Assumption~\ref{asp:Comparability} to Assumption~\ref{asp:ParallelTrends:appendix}, which we call the \textit{Parallel Trends Assumption}. When combined with the linearity assumption and under certain conditions, this new assumption still guarantees that Theorem~\ref{thm:IdentificationL2} holds. The detailed theory of Assumption \ref{asp:ParallelTrends:appendix} is presented in Appendix~\ref{sec:appendix:comparability}. Assumption~\ref{asp:ParallelTrends:appendix} can be more robust to real-world settings.

Below, we introduce a statistical test to evaluate whether the parallel trends assumption holds by focusing on two distinct time periods, denoted as \( t \) and \( t' \), along with a specified positive integer \( \delta \). 
The first step is a \textit{matching procedure}.
For each unit \( i \) in the treatment group characterized by the pre-period surrogates \( \bm{S}_{i,t-\delta} \) and pre-treatment covariates \( \bm{X}_i \) at time \( t \), where the treatment assignment satisfies \( \bm{W}_{i,t-\delta+1:t} = \bm{1}_{\delta} \), we identify an exact match in the time period \( t' \). The matching criteria require that the matched unit \( i' \) satisfies:
\[
\bm{S}_{i',t'-\delta} = \bm{S}_{i,t-\delta}, \quad \bm{X}_{i'} = \bm{X}_i, \quad  \bm{W}_{i',t'-\delta+1:t'} = \bm{1}_{\delta}.
\]
Upon locating an exact match, one observation from period \( t' \) is randomly selected to form a matched pair \((Y_{it}, Y_{i't'})\) within the treatment group. Observations without an exact match are excluded from the evaluation. This matching process is similarly applied to the control group, where the treatment assignment condition is \( \bm{W}_{i,t-\delta+1:t} = \bm{0}_{\delta} \), resulting in matched pairs \((Y_{it}, Y_{i't'})\) within the control group.

The second step is a \textit{regression analysis}.
This exact matching ensures that the paired observations in both the treatment and control groups are conditioned on identical distributions of pre-period surrogates and pre-treatment covariates. The regression model is specified as follows for the\textit{ matched pairs only}:
\[
Y_{i\cdot} = \beta_0 + \beta_1 \cdot \mathbbm{1}[\bm{W}_{i,\delta} = \bm{1}_{\delta}] + \beta_2 \cdot \mathbbm{1}[\text{period} = t] + \beta_3 \cdot \mathbbm{1}[\bm{W}_{i,\delta} = \bm{1}_{\delta} \text{ and } \text{period} = t] + \epsilon_i,
\]

We estimate the parameters of this regression model and conduct a t-test for the null hypothesis \( H_0: \beta_3 = 0 .\) Failure to reject \( H_0 \) suggests that the parallel trends assumption may not be violated.
Note that a comprehensive discussion on the validation of the comparability assumption and parallel trends assumption, including theorem, related proof, and the statistical testing results derived from empirical experiments, is provided in Appendix~\ref{sec:appendix:comparability}.

\section{Empirical Validation}
\label{sec:Empirical}
We collaborated with WeChat and analyzed two real-world long-term experiments on WeChat Search to validate the effectiveness of our proposed approach.\footnote{These two experiments were the only ones conducted to examine single treatments and over a long-term at WeChat Search during our observational period, due to the high costs and infrequency of long-term experiments.} 
WeChat Search serves as a function within WeChat, enabling users to search for information both internally and externally to the WeChat platform.\footnote{Network interference is not a major concern in these two experiments, as user engagement with the Search function is largely driven by their individual experiences with the features, rather than interactions between users.} 
These experiments offer valuable data, enabling us to observe the ground truth of treatment effects in the future periods and compare them with our estimates made at the end of the experimental period.\footnote{In this section, all data were gathered with user consent through the contract between users and the platform and have been obfuscated to ensure user privacy.} 
Sections \ref{sec:experiment1} and \ref{sec:experiment2} offer detailed descriptions of the experiment background and our empirical strategy and results. 

After analyzing the experimental results, we further validate the effectiveness of our approach using multiple synthetic experiments, detailed in Section \ref{sec:synthetic}. These synthetic experiments discuss scenarios not necessarily represented in the two real-world experiments, offering a thorough examination of our proposed method. In Section~\ref{sec:robustcheck}, we provide additional robustness analyses of our real-world experiments.

\subsection{Experiment 1: Mini-programs in  Search History} \label{sec:experiment1}
\subsubsection{Experiment background.}
Similar to many other social media platforms, WeChat provides a search box that allows users to search for a variety of embedded WeChat features, such as chat history, news articles, and mini-programs (embedded third-party apps). In Experiment 1, practitioners aimed to test whether displaying recently searched mini-programs as part of the search query history in the search box would affect user activity on WeChat Search.

\begin{figure}
\centering\makebox[\linewidth]{
\includegraphics[width=1.2\linewidth]{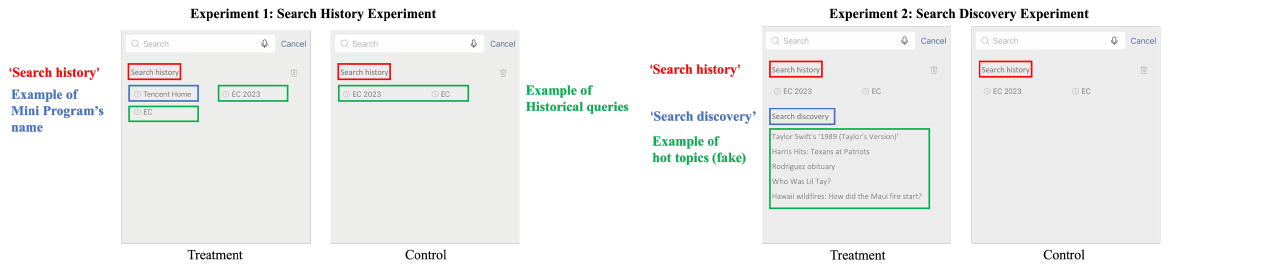}}
\caption{Illustration of user interfaces of the treatment and control groups in two empirical experiments}
\vspace{-3mm}
\label{fig:treatmentIllustrator}
\end{figure}

As presented in Figure~\ref{fig:treatmentIllustrator}, the ``search history" panel provides a shortcut for users to quickly access the search results of keywords they previously searched.
In the treatment condition, the experiment extended the functionality of the ``search history" panel by providing additional shortcuts to access mini-programs that users had recently used. The control condition did not show this new function and remained as the status quo.
The experimenters hypothesized that with this new feature, users would be more likely to visit their frequently used mini-programs through the shortcuts provided by WeChat Search, rather than swiping down on WeChat and scrolling to find the target mini-programs. The business objective of this treatment was to encourage users to engage more with WeChat Search, thereby increasing its user engagement. Figure~\ref{fig:treatmentIllustrator} illustrates the user interfaces for both the treatment and control groups. 

In the experiment, about 1.3 million users were randomly assigned to treatment or control groups. The treatment group consists of 667,206 users, while the control group consists of 665,830 users. The primary outcome of interest is weekly \textit{search\_uv}, the average number of days that a user has searched in a week.\footnote{\textit{search\_uv} is the key metric for WeChat Search to evaluate their product performance. We aggregated it at the week level to remove the impact of strong weekly periodicity on the outcome and average treatment effect. This enhances the satisfaction of the comparability assumption and allows for a more accurate analysis of the treatment effect.} During this 7-week experiment, the results showed a positive treatment effect with a sharp increasing trend in the short term (the first two weeks), setting high initial expectations for the new feature's potential. However, the positive treatment effect becomes stable, albeit slightly diminished, in the long term (see the trends for the ground truth in Figure~\ref{fig:searchHistoryNesting}). As a result, the treatment was launched to all users after the experiment.

For randomization checks, we performed two tests. First, we conducted the sample ratio mismatch (SRM) test \citep{fabijan2019diagnosing}, which uses chi-squared tests to examine whether the sample sizes of the two groups are not significantly different, as 50\% of the number of experiment participants were assigned to each group. The experiment passed the chi-squared test, indicating no sample ratio mismatch problem. Second, we observed that there were no significant differences in the pre-treatment variables between the two groups before the experiment. We performed $t$-tests for mean comparisons, where all the $p$ values are larger than 0.1, suggesting the insignificant differences and the validity of our randomization process. See details in Appendix~\ref{sec:appendix:randomization}.

\subsubsection{Empirical strategy.}
In our analysis, we divided the seven-week experimental period into two phases: the experimental periods $1$ to $T_E$ and the future periods $T_E+1$ to $T$.
During the experimental periods, we collected data and observed the effects of the treatment. 
After the experimental periods end, our goal is to predict the treatment effects for each week in the future periods, starting from week $T_E+1$ and continuing through the last period.
While making these predictions, we do not use data from the future periods, as they have not been observed yet at time $T_E$. 
We use our model to estimate the treatment effects during the future periods. Finally, we compare these estimated effects with the actual treatment effects observed during the future period. These observed effects in the long-term experiment serve as the ``ground truth'' to evaluate the accuracy of our approach.

We consider variables that capture various aspects of user behavior during the search process as our surrogates. Detailed descriptions of all surrogate and primary outcomes are provided in Table~\ref{tb:searchHistorydef}. These surrogates are not only responsive to the treatments but also reflect the diverse aspects of user behavior that lead to variations in primary outcomes over time~\citep{duan2021online,deng2013improving}. Note that we include past primary outcomes as a subset of the surrogate variables, as they are shown to be useful in modeling the future primary outcomes~\citep{deng2013improving}.
This is a little different from the causal diagram shown in Figure~\ref{fig:Surrogacy}, yet this still satisfies Assumption~\ref{asp:Surrogacy}.
To see this, consider the following simplest example with only two periods $t_k$ and $t_{k-1}$ for any $k \in \{2,...,K\}$. 
Let the surrogates consist of two parts $\bm{S}_{i,t_k} = (Y_{i,t_k}, Y_{i,t_{k-1}}, \tilde{\bm{S}}_{i,t_k})$ where $Y_{i,t_k}$ and $Y_{i,t_{k-1}}$ are primary outcomes, and $\tilde{\bm{S}}_{i,t_k}$ are the other surrogate outcomes.
We still have 
\begin{align*}
(Y_{i,t_k}, Y_{i,t_{k-1}}, \tilde{\bm{S}}_{i,t_k}) \independent \bm{W}_{i, 1:t_{k-1}} \vert (Y_{i,t_{k-1}}, Y_{i,t_{k-2}}, \tilde{\bm{S}}_{i,t_{k-1}}, \bm{X}_i),
\end{align*} 
because $Y_{i,t_{k-1}}$ is a constant when conditional on $Y_{i,t_{k-1}}$, and it is conditionally independent of $\bm{W}_{i, 1:t_{k-1}}$.
This could enlarge the surrogate space and potentially better satisfy the longitudinal surrogacy assumption. We provide detailed practical guidelines for choosing surrogates in Section~\ref{sec:guidelines:surrogates}.

Note that we not only use surrogates from the immediate preceding time period $t-1$ but also incorporate surrogates including primary outcomes from earlier periods — $t-2,\ t-3, \cdots$ up to $t-T_E+1$ — into our model. For example, \textit{search\_uv} in period $t-2$ can be seen as the ``\textit{search\_uv} two weeks ago," which is then used as a surrogate in period $t$. Therefore, to establish the models, we use the surrogates (including the primary outcomes) from week $1$ to week $T_E-1$ in total to be our training features, and the surrogates and primary outcome of the week $T_E$ serve as the training outcomes. As we have five surrogate variables, our prediction model has $5\times (T_E-1)$ training features.\footnote{In reality, 
companies would typically have broader access to their internal user behavior data than us as external researchers, enabling companies to curate a more extensive set of surrogates, which ensures a better alignment with the longitudinal surrogacy assumption.} By employing this approach, we effectively broaden the surrogate space, thereby enhancing the precision of our predictions.

After establishing the models for the primary outcome and surrogates, we iteratively use each model to estimate surrogate and primary outcome values for each week during the future period, i.e., $T_E+1, T_E+2, \cdots, T$. Note that the prediction model is not supposed to have access to the actual values of any surrogates or primary outcomes post $T_E$. Consequently, the input features for each model are based on both the observed surrogate and primary outcome values during the $T_E$ experimental periods and their predicted values post $T_E$. For example, we employ observed surrogates from weeks 2 to $T_E$ to project those in $T_E+1$, 
and then we utilize the surrogates observed from weeks 3 to $T_E$ as well as the surrogates previously predicted for time $T_E+1$ to estimate those in $T_E+2$ (and so on). With this iteration, we are able to predict both primary outcomes and surrogates until time $T$.

We focus on presenting the results from our main model, the linear surrogate model\footnote{We construct an additional linear surrogate model that includes both surrogate and pre-treatment variables. See details in Appendix~\ref{sec:appendix:estimation:covariate}.}. We construct confidence intervals using the bootstrapping technique~\citep{efron1987better,efron1994introduction}.
We use a bootstrapping approach to estimate the confidence intervals for the long-term treatment effects. We resample 50\% of the users with replacement to create each replica, selecting half of the original sample to form a new subsample.\footnote{We adopt such subsampling approach for straightforward implementation in our analysis. For comprehensive validation, we supplement this method by resampling all users with replacement, detailed in Appendix~\ref{sec:appendix:bootstrap}.} For each replica, we build a separate prediction model using only this subsample. Based on this model, we then estimate the long-term treatment effects for each replica. This process is repeated 100 times to determine a 95\% confidence interval for the true treatment effect. This method allows us to account for variability from both the random assignment of subjects and the model itself.

\subsubsection{Baselines.} 
We employ two different baselines with confidence intervals using the same bootstrap technique described above.

\begin{itemize}
\item \textit{Constant Extrapolation Baseline} (CEB): We use the average treatment effects observed during the first $T_E$ weeks of the experiment to predict the treatment effects for the future period. Although obviously this approach cannot capture any increasing or decreasing trends in the treatment effects, this serves as a common industry practice.

\item \textit{Vector Autoregressive} (VAR) Model: 
We employ a Vector Autoregressive (VAR) model with lag order \( p = T_E - 2 \) on the initial \( T_E \) weeks of the multivariate time series, using the average values of four surrogates and one primary outcome variable as input candidates. This allows the VAR model to forecast future outcomes based on past values of all included variables, though VAR is traditionally used for forecasting rather than causal inference \citep{stock2001vector}.\footnote{The forecasted effect is calculated by taking the difference between the predicted average primary outcomes of the treatment and control groups at each future time point. The lag order \( p \) is selected as the largest feasible term to maintain model performance. When \( p < 5 \), we include the primary outcome and randomly select \( p-1 \) of the surrogates; otherwise, all five variables are included. For the edge case ($T_E = 2$), the result from constant extrapolation is used instead. We choose $p=T_E-2$ because this is the the largest possible term that can be selected, ensuring the VAR model's performance.}
\end{itemize}

\subsubsection{Results.} 
\begin{figure}
\centering\subfloat[\scriptsize Experiment 1\label{fig:searchHistoryNesting}]{\includegraphics[width=.51\textwidth]{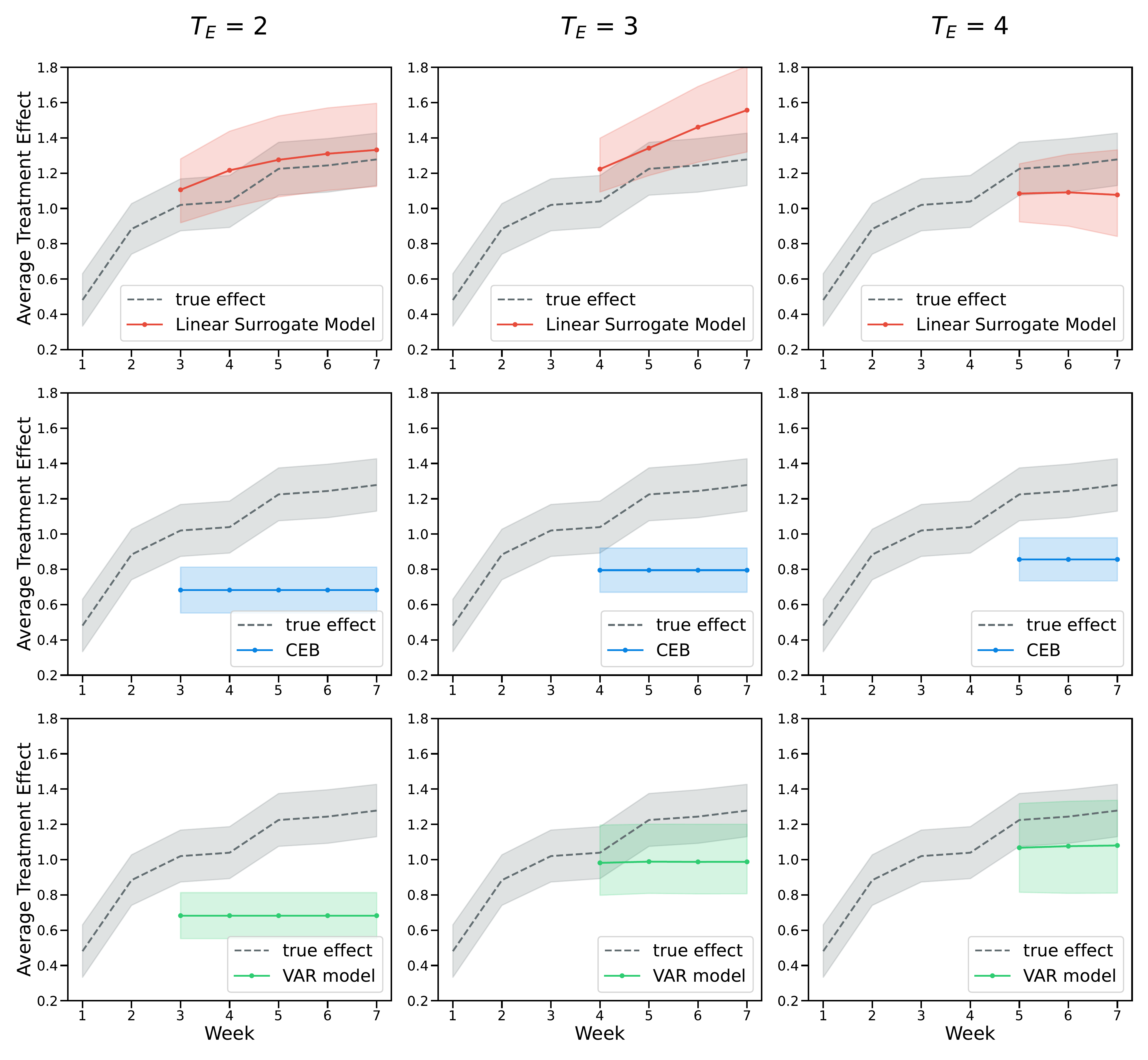}}
\subfloat[\scriptsize Experiment 2\label{fig:searchDiscoveryNesting}]{\includegraphics[width=.51\textwidth]{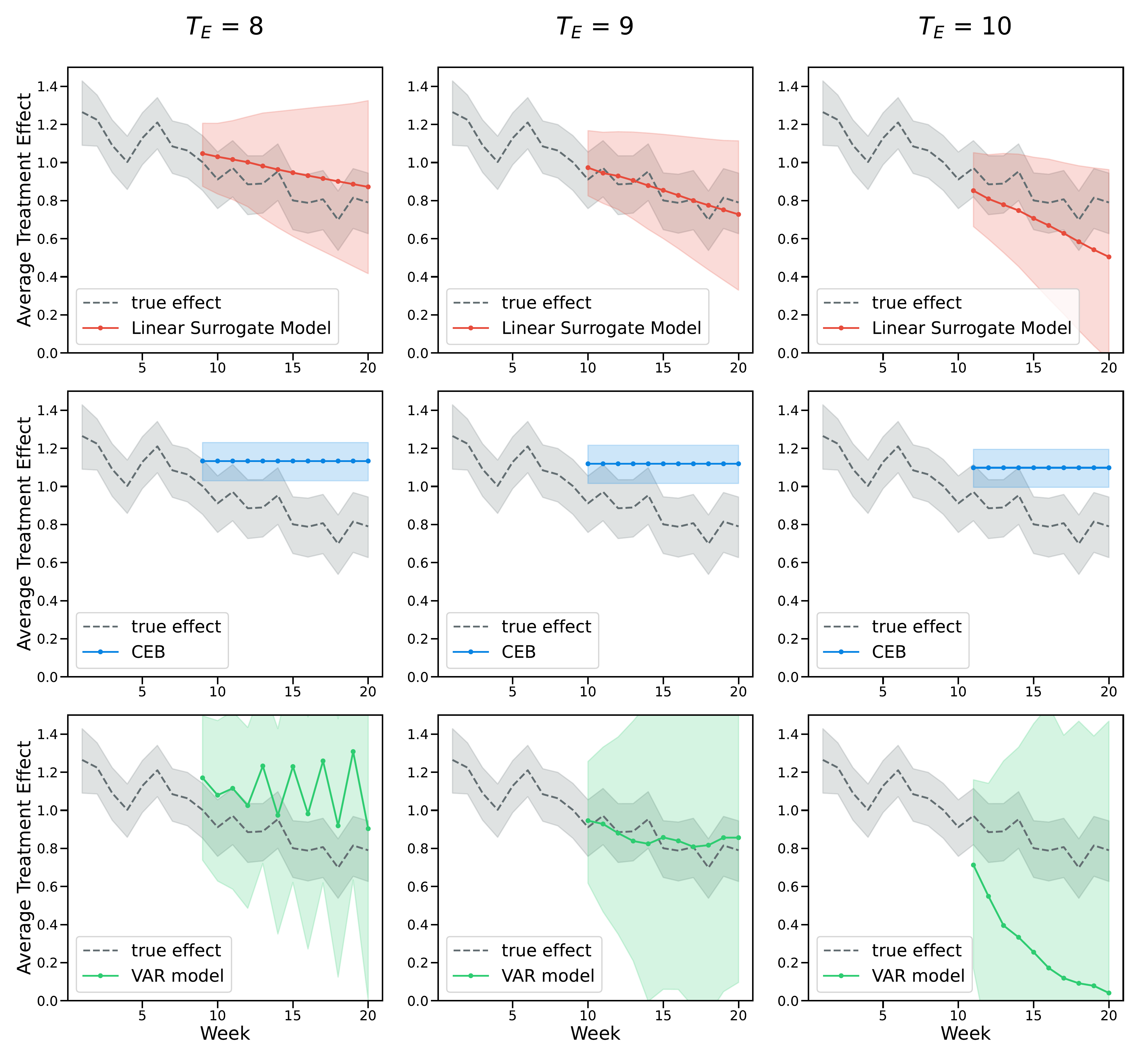}}
\caption{Estimated Effects of Long-term Treatment using Linear Surrogate Model, CEB, and VAR model \protect\footnotemark}
\vspace{-3mm}
\end{figure}

We present the estimates of the linear surrogate model, the baselines, and true effects in Figure~\ref{fig:searchHistoryNesting}. We vary the value of $T_E$ from 2 to 4 to ensure that $T_E$ is meaningfully short compared to the entire duration, constituting around half or less of the entire horizon. We observe that the CEB  consistently underestimates the effects of long-term treatments. The vector auto-regressive models perform slightly better than CEB, especially when $T_E$ is larger. However, these baseline models cannot predict the long-term increasing trend of the treatment effect. 

By contrast, our estimation, indicated by red curves, can successfully capture an increasing trend in the treatment effect regardless of the choice of $T_E$. 
For instance, our estimation successfully predicts both a long-term increasing trend at $T_E=2$ and a stable trend at $T_E=4$, which other baseline models fail to do. In practice, successfully predicting the trend of treatment effects over time is critical for making product decisions. In addition, using the first two weeks only, our estimation of the effect of long-term treatment in week 7 is 1.347, which is less biased compared to the true effect (1.278) compared to baselines.

\footnotetext{Grey dashed curves represent the true average treatment effect on \textit{search\_uv} from week 1 to week 7 for Experiment 1 (from week 1 to week 20 for Experiment 2). Solid red curves represent the estimated effects with the linear surrogate model. Solid blue curves are the Constant Extrapolation Baseline that uses short-term effect to extrapolate. Solid green curves are the estimated effects with the VAR model. Shadows indicate 95\% confidence intervals. The three panels represent the scenarios when we use the first $T_E$ weeks as the experimental period and the last $T_F$ weeks as the future period. For $T_E=2$ in  VAR, a constant extrapolation is used due to the limited length of the time series.}

Further, we compare the bias and mean squared errors (MSE) between our model and the baselines. Specifically, we present their averages over all weeks during future periods and present the results in Table~\ref{tb:searchHistoryMSE}. Overall, considering the three choices of $T_E$, our model consistently outperforms the baseline model in terms of bias, and outperforms the baseline model in terms of MSE in the majority of cases. 
These results further underscore the effectiveness of our approach.\footnote{As highlighted earlier, \cite{athey2019surrogate} addresses a fundamentally different problem, entailing assumptions and methodologies that are not directly applicable to our context. While their model is not suited for our setting, we offer an estimation derived from their approach. The results shown in Appendix~\ref{sec:appendix:Athey} confirm our argument about the difference in the problem setup.}

As \( T_E \) increases, an estimation model is generally expected to have higher bias and MSE, as predictions are made further into the future and are less anchored by current observations. However, temporal fluctuations—such as seasonal effects, holidays, and rare events—can introduce additional complexity that disrupts this trend. Such events can make certain short- or mid-term periods more challenging to predict accurately than other periods further in the future. As a result, while a general increase in bias and MSE with forecast horizon length may hold, this trend is not strictly monotonic and can vary based on the occurrence of these less predictable events.

\begin{table}[tb]
\centering
\caption{Comparison result between different methods in terms of Bias and MSE for Experiment 1}
\label{tb:searchHistoryMSE}
\scriptsize
  \begin{tabular}{c ccc ccc}
    \toprule
    \multirow{2}{*}{\textbf{Method}}&\multicolumn{3}{c}{\textbf{Bias}}&\multicolumn{3}{c}{\textbf{MSE}}\\
      & {$\mathbf{T_E=2}$} & {$\mathbf{T_E=3}$} & {$\mathbf{T_E=4}$} & {$\mathbf{T_E=2}$} & {$\mathbf{T_E=3}$} & {$\mathbf{T_E=4}$} \\
      \midrule
    Linear Surrogate Model & 0.087 & 0.199 & 0.165 & 0.327 & 0.324 & 0.314 \\
    CEB                    & 0.479 & 0.401 & 0.393 & 0.342 & 0.263 & 0.243 \\
    VAR Model              & 0.479 & 0.210 & 0.174 & 0.342 & 0.328 & 0.435 \\
    \bottomrule
    \vspace{-7mm}
  \end{tabular}
\end{table}

We also examine whether the surrogacy and comparability assumptions hold for the experiment using the methods proposed in Section~\ref{sec:validation}. First, we conduct a sensitivity analysis for the surrogacy assumptions, presenting the results in Appendix~\ref{sec:appendix:surrogateSubset} and Appendix~\ref{sec:appendix:missingSurrogates}, to demonstrate the robustness of our estimation to the potential of omitted surrogates. Moreover, we perform both the tests for comparability and parallel trends assumptions. Detailed results are presented in Appendix~\ref{sec:appendix:comparabilityDirect} and Appendix~\ref{sec:appendix:comparabilityParallel}.

\subsection{Experiment 2: Search Discovery} \label{sec:experiment2}
\subsubsection{Experiment background.}
Similar to Experiment 1, Experiment 2 also involves a change in WeChat Search. Instead of adding shortcuts to mini-programs in the ``search history,'' practitioners aimed to test whether displaying hot topics as part of the search discovery'' in the search box would affect user activity on WeChat Search. The experimenters hypothesized that with this new feature, users would be more likely to read and engage with these new shortcuts to trendy topics. The business objective of this treatment was to encourage users to engage more with WeChat Search, thereby increasing its user engagement. Figure~\ref{fig:treatmentIllustrator} illustrates the user interfaces for both the treatment and control groups. In the treatment condition, the users were offered this new feature, while the users were not in the control condition.
However, the long-term effect of this treatment remains uncertain and critical since including this new panel of hot topics might also crowd out users' intention to search. Different from Experiment 1, where the new feature mainly assists in searching for the mini-programs based on individuals' search history, the new feature for Experiment 2 is to provide shortcuts that help users to explore hot topics, which might affect their initial search intention. Thus, WeChat launched this experiment for a total of 20 weeks.

This 20-week experiment involves 3.6 million WeChat users. Among them, 1,807,335 users were randomly assigned to the treatment group, while 1,803,675 users were randomly assigned to the control group. Again, the primary outcome of interest is \textit{search\_uv}. Since both experiments focus on WeChat Search and share the same primary outcome, the same set of surrogates described in Table~\ref{tb:searchHistorydef} is used for Experiment 2. In this experiment, the goal is to predict the treatment effects over a long period until period $T$ (the 20th week) using the data available at the end of period $T_E$.

To examine the validity of randomization, we also performed the SRM test and mean comparisons on the pre-treatment variables between the treatment and control groups, similar to Experiment 1. The results confirm the validity of our randomization process, showing that there is no statistically significant difference in the sample sizes and no statistically significant differences in the pre-treatment variables between the two groups. More details are discussed in Section \ref{sec:appendix:randomization} of the Appendix.
In addition, we present the summary statistics in Table~\ref{tb:summaryStatistics}. 

The average treatment effect shows continuous fluctuations without an apparent downward trend signal in the first seven weeks, while followed by a continuous decline after eight weeks of treatment over time. It is suspected that there is a long-lasting novelty effect for this treatment, and the effect is likely to decay over time. As a result, this new product change (treatment) was not eventually adopted or launched to all users. Nevertheless, the valuable insights gained from this experiment have inspired the development of other significant product strategies.

\subsubsection{Empirical strategy, baselines and results.} 
Both experiments conducted on WeChat Search have the same primary outcomes and surrogates, so we use the same empirical strategy as in Experiment 1. The potential consistency of surrogates among different experiments can enable easy scalability of our approach in practice. Since this experiment is longer (20 weeks), we employ the linear surrogate model and showcase results for ($T_E = 8, 9, 10$) in the main text, while presenting results with different choices of ($T_E$) in Appendix~\ref{sec:exp2:more}. Additionally, we use the same baselines for validation in Experiment 2 as those in Experiment 1 for consistency.

The estimation results are presented in Figure~\ref{fig:searchDiscoveryNesting}. We observe that our approach effectively captures the decreasing trend of the average treatment effect in the long run. By contrast, the CEB model consistently overestimates the treatment effects during the future period $T_F$, as it fails to capture the decreasing trend of the treatment effect. The VAR model exhibits fluctuating estimates over time and unstable prediction trends across different experimental periods $T_E$, due to the volatility of the primary outcome ($Y$), \textit{search\_uv}, over time in both the treatment and control groups. The VAR model appears to be unable to handle this scenario well.

Table~\ref{tb:searchDiscoveryMSE} reports the average bias and mean squared error (MSE) over the \( T_F \) future periods for each \( T_E \). Consistent with the results from Experiment~1, our method outperforms both baseline models (CEB and VAR) in terms of bias across all values of \( T_E \). As the forecast horizon extends beyond the experimental period, our model tends to exhibit increased estimation variance for more distant future periods. This phenomenon occurs because errors in near-term predictions can propagate and amplify when used as inputs for subsequent, longer-term forecasts. A similar issue arises with the VAR baseline model, which also relies on near-future periods' information for extended predictions. Despite both our method and the VAR model exhibiting higher variance compared to the trivial constant extrapolation, we consider this an acceptable bias-variance tradeoff.
Similar to Experiment 1, we conduct analyses proposed in Section~\ref{sec:validation} to examine surrogacy and comparability assumptions. We demonstrate our estimation's robustness to both assumptions and present the results in Appendices~\ref{sec:appendix:surrogateSubset}, \ref{sec:appendix:missingSurrogates}, \ref{sec:appendix:comparabilityDirect}, and \ref{sec:appendix:comparabilityParallel}.

\begin{table}[tb]
\centering
\caption{Comparison result between different methods in terms of Bias and MSE for Experiment 2}
\scriptsize
  \begin{tabular}{c ccc ccc}
    \toprule
    \multirow{2}{*}{\textbf{Method}}&\multicolumn{3}{c}{\textbf{Bias}}&\multicolumn{3}{c}{\textbf{MSE}}\\
      & {$\mathbf{T_E=8}$} & {$\mathbf{T_E=9}$} & {$\mathbf{T_E=10}$} & {$\mathbf{T_E=8}$} & {$\mathbf{T_E=9}$} & {$\mathbf{T_E=10}$} \\
      \midrule
    Linear Surrogate Model & 0.098 & 0.048 & 0.158 & 0.233 & 0.136 & 0.201 \\
    CEB                    & 0.274 & 0.272 & 0.258 & 0.106 & 0.103 & 0.096 \\
    VAR model              & 0.240 & 0.054 & 0.565 & 0.520 & 0.645 & 2.552 \\
    \bottomrule
    \vspace{-6mm}
  \end{tabular}
\label{tb:searchDiscoveryMSE}
\end{table}

\subsection{Simulations Using Synthetic Data} \label{sec:synthetic}
In addition, to encompass scenarios not necessarily represented in the real-world experiments, we undertake synthetic experiments for a more thorough evaluation of our approach.

\subsubsection{Stabilized treatment effect.}\label{sec:Stabilized}
In our synthetic experiments, the first scenario we investigate is when the effects of long-term treatments plateau or stabilize over time. To illustrate this, we set up the following synthetic experiment: The simulation presupposes four surrogates, \(\bm{S}_{it}\), for each time period $t$ and unit $i$. For each dimension $d$, each of its corresponding surrogates draws from a normal distribution, 
\(S_{it,d} \sim \mathcal{N}(\mu_d,\sigma^2_d)\). Surrogates in different dimensions are independent from each other. In this synthetic experiment, subjects assigned to the control group experience no deviation from the status quo; 
as a result, the surrogates' distribution remains unchanged. In contrast, for those in the treatment group, there is a time-dependent decay in the four surrogates, governed by decay factors  \(\bm{\gamma}=(0.8,0.6,0.4,0.2)\) respectively (e.g., $S_{it,d+1}=\gamma_d \cdot S_{it,d}$). In order to comply with both the surrogacy and linearity assumptions, the primary outcome, \(Y_{it}\), is designed as a linear combination of these four surrogates.

In the first synthetic experiment, we set the parameter \(\mu_d \sim \mathcal{N}(2,1)\) and \(\sigma_d \sim \mathcal{N}(2,1)\), and the primary outcome \(Y\) in period $t$ is formulated as \(Y_{it} = -(0.1S_{it,1} + 0.1S_{it,2} + 0.4S_{it,3} + 0.4S_{it,4})\). In this setup, the effect of long-term treatments on \(Y_{it}\) initially increases and then stabilizes, showcasing a characteristic ``level off" pattern. Using experimental data spanning \(T_E = 2, 3, \text{and } 4\) periods, we compare our approach's estimates with the true future effects. The first row of Figure~\ref{fig:sim1Baseline} in Appendix demonstrates a precise estimation of the effects of long-term treatments.

The second simulation shares the settings with the first one except for the parameter \(\mu_d \sim \mathcal{N}(1.5,1)\) and \(\sigma_d \sim \mathcal{N}(1,1)\), and the primary outcome being formulated as \(Y_{i,t+1} = 0.1S_{it,1} + 0.1S_{it,2} + 0.4S_{it,3} + 0.4S_{it,4}\).\footnote{Another subtle difference is that we draw surrogates in the control group from the distribution \(S_{it,d} \sim \mathcal{N}(\mu_d-2,\sigma^2_d)\) in order to overall shift the treatment effect into positive values. This change does not affect our conclusion.}
This configuration leads to the effect of the long-term treatment on \(Y\) initially declining and then stabilizing, exemplifying another typical ``level off" trend.   The first row of Figure~\ref{fig:sim2Baseline} presents the estimation results, demonstrating that our approach can accurately capture the future treatment effects. 

In both synthetic experiments, our estimates closely align with the true effects of long-term treatment, demonstrating our approach's capability to account for scenarios where treatment effects stabilize over time. Figures~\ref{fig:sim1Baseline} and~\ref{fig:sim2Baseline} showcase the graphical comparison between our approach and all the other baseline models, including the CEB model and the VAR model in two synthetic experiments. Moreover, numerical comparison between our approach and multiple baselines in terms of bias and MSE, is provided in Tables~\ref{tb:synthetic1MSE} and~\ref{tb:synthetic2MSE}. 
Our approach surpasses all of the baseline models in both synthetic experiments regarding bias and MSE. Collectively, these analyses further show the validity and generalization of our approach to various empirical settings. 

Further, we performed a sensitivity analysis for the surrogacy assumption (Assumption~\ref{asp:Surrogacy}) on both of the two synthetic experiments to demonstrate the relationship between the degree of violation of Assumption~\ref{asp:Surrogacy} and estimation accuracy. 
The results in Appendix~\ref{sec:appendix:surrogateSubset} show that performance worsens with more severe violations of Assumption~\ref{asp:Surrogacy}, but a longer observational experimental period can mitigate this deterioration to some extent.

\subsubsection{Additional synthetic experiments.}
To complement our real-world experiments, we conduct additional synthetic experiments that challenge certain assumptions or alter the behavior of long-term treatment effects. We explore two scenarios:

\paragraph{Violation of Comparability:} In this experiment, we create synthetic contexts where the comparability assumption may not hold and test whether our framework can detect these violations, as well as observe how its performance changes accordingly.
We simulate scenarios with varying degrees of comparability assumption violations. In this simulation, the primary outcome for users in the treatment group is defined as 
$Y_{it} = -\gamma \times (0.1S_{it,1} + 0.4S_{it,2})$.
When \( t = 2 \) and \( i \) is in the treatment group, we vary \(\gamma\) over the values \([1, 1.5, 2, 2.5, 3]\) to control the extent of the comparability violation. 
For all other time periods for the treatment group and for all time periods in the control group (including \( t = 2 \)), we set \(\gamma = 1\). We demonstrate that both the comparability and parallel trends assumption tests we proposed can effectively  detect this violation. Moreover, as the degree of violation increases (i.e., as \(\gamma\) becomes larger), estimation bias increases accordingly. Please refer to Appendix~\ref{sec:appendix:additionalNonComparable} for more details.

\paragraph{Non-linear Outcome Function:} Although our main results rely on the linearity assumption in the linear surrogate model, we also create synthetic contexts where this assumption may not hold and test how our method's performance may be affected.
We evaluate our method under a non-linear outcome function \(Y_{it}\) by introducing two surrogates \(S_{it,1}\) and \(S_{it,2}\). 
The primary outcome is
$Y_{i,t+1} = -\left(S_{it,1} + \theta e^{S_{it,2}}\right)$,
where \(\theta\) adjusts the degree of non-linearity. 
Note that to create the treatment effect, we allow the surrogates in the treatment group to decay over \( t \), while the surrogates for users in the control group do not exhibit this decay; their difference is the treatment effect.
Our method yields accurate long-term estimates when linearity is not severely violated, demonstrating the robustness of our approach to linearity to some extent.
The detailed setups and results are provided in Appendix~\ref{sec:appendix:additionalNonlinear}.

\paragraph{No Long-Term Treatment Effect:} Here, the long-term treatment effect diminishes over time, with surrogates following the same distribution across the treatment and control groups. The outcome for the treatment group includes a diminishing term $\frac{(-1)^t}{(t+2)^3}$. That is, $Y_{i,t+1} = -(0.1S_{it,1} + 0.1S_{it,2} + 0.4S_{it,3} + 0.4S_{it,4}) + \frac{(-1)^t}{(t+2)^3}$.
By contrast, the outcome for the control group is the same but excludes the term \(\frac{(-1)^t}{(t+2)^3}\). The difference between the treatment and control groups reveals a treatment effect that gradually fades, converging to zero as \( t \) increases.
Our empirical results show that our method effectively predicts this decline using short-term data, showing its capability with transient effects. The detailed setups, results, and analyses are provided in Appendix~\ref{sec:appendix:additionalNoeffect}. 

Overall, the findings from these experiments demonstrate that our approach remains effective even when some  assumptions are moderately relaxed or when the treatment effects exhibit different temporal patterns, demonstrating its applicability in a variety of real-world settings.

\subsection{Robustness Checks}\label{sec:robustcheck}
The following analyses demonstrate the robustness of our methods further. First, instead of using the full sample, we focused on each heterogeneous user group in the two WeChat experiments. Detailed implementations are illustrated in Appendix~\ref{sec:appendix:robust:hetero}. Figures~\ref{fig:searchHistoryHTE} and~\ref{fig:searchDiscoveryHTE} illustrate the estimated long-term treatment effects for each group in the two separate experiments. We also present the biases and MSEs for each subgroup in Tables~\ref{tb:synthetic2MSE1} and \ref{tb:synthetic2MSE2}. The results show a close alignment of our estimation with the true effects across various heterogeneous groups.

Second, to address the challenge of the curse of dimensionality in surrogates, we implemented a linear surrogate model with elastic net regularization to mitigate potential overfitting issues. The details of the methodology and the empirical validations are presented in Appendix~\ref{sec:appendix:robust:regularization}. The effectiveness of this approach is confirmed by the consistency in long-term effect estimation shown in Figures~\ref{fig:searchHistoryRegularization} and~\ref{fig:searchDiscoveryRegularization}, compared to prior models, underscoring the robustness and predictive accuracy of our linear surrogate model with regularization.

\section{Conclusions and Future Research}
\label{sec:Conclusion}
In this paper, we propose a longitudinal surrogate framework to estimate the long-term effects of long-term treatments using data collected from short-term experiments, which has remained an open challenge in the existing literature. We used two real-world long-term experiments conducted on WeChat to validate the effectiveness of our proposed framework. 
Our framework emphasizes the practical relevance of applying our method in real-world A/B testing scenarios, allowing practitioners to evaluate the effects of long-term product updates without incurring high costs and an extended waiting period. We discuss the limitations of our model in Section~\ref{sec:appendix:limitation}, by providing examples when our modeling assumptions do not hold.
This serves as a cautionary note on when to apply our method in practice.

We outline several future research directions. One such direction is the integration of our concept of estimating future experimental effects with the existing literature on optimal stopping in A/B testing~\citep{deng2016continuous,xiong2023optimal,berman2022false}. Specifically, a valuable direction would be developing a method to optimally determine the parameter $T_E$, the experimental period duration. This approach would allow practitioners to conclude the experiment earlier, thereby directing towards the most beneficial treatment arm more efficiently.
Second, it would be interesting to combine structural information, such as user behavior modeling, with estimating the effects of long-term treatments. In our current empirical study, we recognize that certain outcome variables, such as retention rates and subscription fees, may not show significant changes in the short term, due to factors such as data scarcity. Leveraging structural information may potentially improve the performance when the data sample is limited.

\renewcommand*{\bibfont}{\footnotesize}
{\SingleSpacedXI

\bibliographystyle{informs2014} 
\bibliography{bibliography}
}
\clearpage

\clearpage

\begin{APPENDICES}
% \appendix
\renewcommand\thefigure{\thesection\arabic{figure}}    
\renewcommand\thetable{\thesection\arabic{table}}    
\setcounter{figure}{0}    
\setcounter{table}{0}    
\centerline{\large \textbf{Online Appendix}}
\vspace{0.5em}

\setcounter{page}{1}
\section{Linear Additive Model}
\label{sec:appendix:additive}

In addition to the linear surrogate model described in the main text, we introduce a \textit{linear additive model}. 
This model requires slightly stronger assumptions. 
We first outline the theoretical framework and then show our empirical findings. 
Our results suggest that the assumption required for this model is only applicable to a few restrictive situations. 
Therefore, we advise against using this approach in common settings.

\subsection{Model and Assumption}

We introduce one additional assumption that helps establish the linear additive model. 
Note that this additional assumption is strong and may not hold in a number of scenarios.
The linear additive model should thus be used with cautions. 
Combined with the previous three assumptions, this new set of four assumptions is the third level of assumptions.

\begin{assumption}[Additive Treatment Effects]
\label{asp:Additive}
The average treatment effect of long-term treatments is linear additive to the treatments, i.e., there exists a subset of time indices $\bT = \{t_1, t_2, ..., t_K\} \subseteq [T]$, such that for any $i \in [N]$,
\begin{align*}
\tau = \sum_{k=1}^{K+1} \left\{ \bE_{\cF}\bigg[ Y_{iT} (\bm{0}_{1:t_{k-1}}, \bm{1}_{t_{k-1}+1:t_k}, \bm{0}_{t_k+1:T}) \bigg] - \bE_{\cF}\bigg[ Y_{iT}(\bm{0}_T) \bigg] \right\}.
\end{align*}
\end{assumption}

Assumption~\ref{asp:Additive} suggests that the average effect of long-term treatments is the summation of several components. 
Each component is the effect of a subset of short-term treatments. 
Assumption~\ref{asp:Additive} holds when the effect of long-term treatments can be decomposed into the summation of a number of carryover effects. 
For example, when advertisements are sent to users regularly, each advertisement may marginally increase the average click-through rate where the effects of the earlier advertisements quickly decay with the time. 
This could be a context when Assumption~\ref{asp:Additive} holds. 
In contrast, Assumption~\ref{asp:Additive} does not hold in some other contexts, such as in estimating the ``novelty effect,'' where users click a button more frequently when its color is changed, but their clicks quickly decrease back to normal as they become familiar with the new color. 

Under Assumption~\ref{asp:Additive}, we leverage multiple surrogate indices whose subscripts are different, and re-write the average effect of long-term treatments as follows.

\begin{theorem}[Linear Additive Model]
\label{thm:IdentificationL3}
Under Assumptions~\ref{asp:Surrogacy},~\ref{asp:Comparability}, and~\ref{asp:Additive}, where Assumptions~\ref{asp:Surrogacy} and~\ref{asp:Additive} hold for $\bT=\{t_1,t_2,...,t_K\}$, the average effect of long-term treatments on the primary outcome is equal to the following expression,
\begin{multline*}
\tau_T = \sum_{k=1}^{K} \bE_{\cF} \bigg[ h_{T-t_k}\left( \bm{G}_{t_k}\left(\bm{S}_{i0},\bm{X}_i,(\bm{0}_{1:t_{k-1}}, \bm{1}_{t_{k-1}+1:t_k})\right),\bm{X}_i,\bm{0}_{T-t_k} \right) \bigg] \\
+ \bE_{\cF} \bigg[ h_{T}\left( \bm{S}_{i0},\bm{X}_i, (\bm{0}_{1:t_K},\bm{1}_{t_K+1:T}) \right) \bigg] - \sum_{k=1}^{K+1} \bE_{\cF}\left[h_{T}\left(\bm{S}_{i0},\bm{X}_i,\bm{0}_T\right)\right].
\end{multline*}
\end{theorem}

Theorem~\ref{thm:IdentificationL3} is similar to Lemma~\ref{lem:SpecialCaseL1}, in the sense that for each $k$, each potential outcome $Y_{iT} (\bm{0}_{1:t_{k-1}}, \bm{1}_{t_{k-1}+1:t_k}, \bm{0}_{t_k+1:T})$ is decomposed into two iterations using only one conditional surrogate outcomes $\bm{G}_{t_k}$.
However, different from Lemma~\ref{lem:SpecialCaseL1}, Theorem~\ref{thm:IdentificationL3} does not require the length of future periods $T_F$ to be equal to the length of experimental periods $T_E$.
Compared with Theorem~\ref{thm:IdentificationL1}, Theorem~\ref{thm:IdentificationL3} makes one additional Assumption~\ref{asp:Additive}.
This additional assumption enables Theorem~\ref{thm:IdentificationL3} to avoid iterating the pivots for multiple times and use summation instead.

To fully leverage the benefits of the linear additive model, we assume the availability of historical data, which is collected during the \textit{observational} periods without any intervention.
During the observational periods $t \in \{0,-1,-2,...\}$, we can observe both the surrogate and primary outcomes $(\bm{S}_{it}, Y_{it})$.
Compared with the number of treatment subjects in the experimental periods, there are many more subjects in the observational periods.
But no subject is treated during the observational periods, i.e, $W_{it} = 0, \forall t \in \{0,-1,-2,...\}$.
See Figure~\ref{fig:ObservationalPeriods} for an illustration.
\begin{figure}[tb]
\centering
\includegraphics[width=0.8\textwidth]{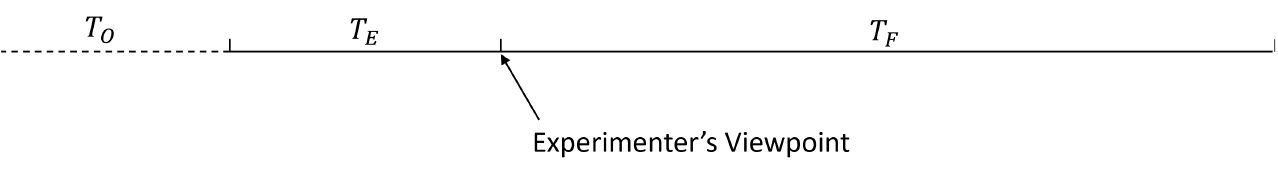}
\caption{An illustrator of the observational periods, the experimental periods and the future periods.}
\label{fig:ObservationalPeriods}
\end{figure}
The linear additive model as described in Theorem~\ref{thm:IdentificationL3} allows us to estimate the surrogate indices $h_{T-t_k}(\cdot, \cdot, \bm{0}_{T-t_k}), \forall k \in [K]$ and $h_{T}(\cdot, \cdot, \bm{0}_{T})$ all from abundant historical data.
This identification strategy does not suffer from the length of experimental periods $T_E$ being too small.

\subsection{Estimators for the Linear Additive Model}
Finally, we propose the third estimation strategy to estimate the surrogate index functions from the abundant data in the observational periods.
Given the estimators of the surrogate index and estimators of the conditional surrogate outcomes, we follow Theorem~\ref{thm:IdentificationL3} and obtain the following plug-in estimator,
\begin{multline}
\widehat{\tau}_T = \frac{1}{N_1} \sum_{k=1}^{K} \sum_{i\in[N]} \bI\{\bm{W}_{i,1:T_E} = \bm{1}_{T_E}\} \bE_{\widehat{\bm{G}}_{t_k}}\left[ \widehat{h}_{T-t_k}\left( \widehat{\bm{G}}_{t_k}\left(\bm{s}_{i0}, \bm{x}_i, (\bm{0}_{1:t_{k-1}}, \bm{1}_{t_{k-1}+1:t_k})\right), \bm{x}_i, \bm{0}_{T-t_k} \right) \right] \\
+ \frac{1}{N_1} \sum_{i\in[N]} \bI\{\bm{W}_{i,1:T_E} = \bm{1}_{T_E}\} \widehat{h}_{T}\left( \bm{s}_{i0},\bm{x}_i, (\bm{0}_{1:t_K},\bm{1}_{t_K+1:T}) \right) \\
- (K+1) \frac{1}{N_0} \sum_{i\in[N]} \bI\{\bm{W}_{i,1:T_E} = \bm{0}_{T_E}\} \widehat{h}_T(\bm{s}_{i0}, \bm{x}_i, \bm{0}_T). 
\label{eqn:EstimationL3}
\end{multline}

We explain how to estimate the surrogate index functions in \eqref{eqn:EstimationL3}.
For any $t \in [T_E] \cup \{0,-1,-2,...\}$, $\bm{x} \in \bX$, $\bm{s} \in \bS$, one naive estimator of the surrogate index under consecutive controls is given by
\begin{align*}
\widehat{h}_{T-t_k}(\bm{s}, \bm{x}, \bm{0}_{T-t_k}) = \frac{\sum_{i \in [N]}Y_{it} \bI\{\bm{X}_i = \bm{x}, \bm{S}_{i(t-T+t_k)} = \bm{s}, \bm{W}_{i,t-T+t_k+1:t} = \bm{0}_{T-t_k}\}}{\sum_{i \in [N]} \bI\{\bm{X}_i = \bm{x}, \bm{S}_{i(t-T+t_k)} = \bm{s}, \bm{W}_{i,t-T+t_k+1:t} = \bm{0}_{T-t_k}\}}.
\end{align*}
Under Assumptions~\ref{asp:Comparability}, such an estimator is unbiased for the surrogate index function.

To estimate the middle term of \eqref{eqn:EstimationL3}, for any $\bm{x} \in \bX$, $\bm{s} \in \bS$, one naive estimator of the surrogate index under $t_K$ controls followed by $\Delta t_{K+1}$ treatments is given by
\begin{align*}
\widehat{h}_{T}\left( \bm{s}, \bm{x}, (\bm{0}_{1:t_K},\bm{1}_{t_K+1:T}) \right) = \frac{\sum_{i \in [N]}Y_{i\Delta t_{K+1}} \bI\{\bm{X}_i = \bm{x}, \bm{S}_{i(-t_K)} = \bm{s}, \bm{W}_{i,-t_K+1:\Delta t_{K+1}} = (\bm{0}_{1:t_K},\bm{1}_{t_K+1:T})\}}{\sum_{i \in [N]} \bI\{\bm{X}_i = \bm{x}, \bm{S}_{i(-t_K)} = \bm{s}, \bm{W}_{i,-t_K+1:\Delta t_{K+1}} = (\bm{0}_{1:t_K},\bm{1}_{t_K+1:T})\}}.
\end{align*}
Under Assumptions~\ref{asp:Comparability}, such an estimator is unbiased for the surrogate index function.
Note that, such an estimator above uses data from the treatment subjects only, because of the part $\bm{W}_{i,-t_K+1:\Delta t_{K+1}} = (\bm{0}_{1:t_K},\bm{1}_{t_K+1:T})$ in the indicators.
In words, $\bm{W}_{i,-t_K+1:\Delta t_{K+1}} = (\bm{0}_{1:t_K},\bm{1}_{t_K+1:T})$ refers to the $t_K$ treatment assignments right before the experimental periods (which are all controls), and the next $\Delta t_{K+1}$ treatment assignments in the experimental periods (which are all treatments).
Such an estimator takes the advantage of the treatment subjects having received consecutive controls during the observational periods.

Finally, we estimate the distributions of the conditional surrogate outcomes in \eqref{eqn:EstimationL3}.
Borrowing ideas from Markov Chains, for any 
$k\in[K]$, $\bm{x} \in \bX$, $\bm{s} \in \bS$, and $\cS \subseteq \bS$, one naive estimator of the transition kernel is given by
\begin{align*}
\widehat{p}_{t_k}(\bm{s}, \cS) = \frac{\sum_{i \in [N]} \bI\{\bm{X}_i = \bm{x}, \bm{S}_{i(-t_{k-1})} = \bm{s}, \bm{W}_{i,-t_{k-1}+1:\Delta t_k} = (\bm{0}_{1:t_{k-1}}, \bm{1}_{t_{k-1}+1:t_k}), \bm{S}_{it_k} \in \cS\}}{\sum_{i \in [N]} \bI\{\bm{X}_i = \bm{x}, \bm{S}_{i(-t_{k-1})} = \bm{s}, \bm{W}_{i,-t_{k-1}+1:\Delta t_k} = (\bm{0}_{1:t_{k-1}}, \bm{1}_{t_{k-1}+1:t_k})\}}.
\end{align*}
Similar to the surrogate index estimator under $t_K$ controls followed by $\Delta t_{K+1}$ treatments, this estimator of the probability also takes the advantage of the treatment subjects having received consecutive controls during the observational periods.

\subsection{Empirical Results}

We apply the linear surrogate model in Experiment 1 to illustrate the practical performance. The result is shown in Figure~\ref{fig:searchHistoryAdding}. We observe that the linear additive model tends
to overestimate the treatment effect over extended periods, which indicates the potential violation of Assumption~\ref{asp:Additive} in this experiment. 
Therefore, we advise against using this approach in common settings.

\begin{figure}[tb]
\centering
\includegraphics[width=1\linewidth]
{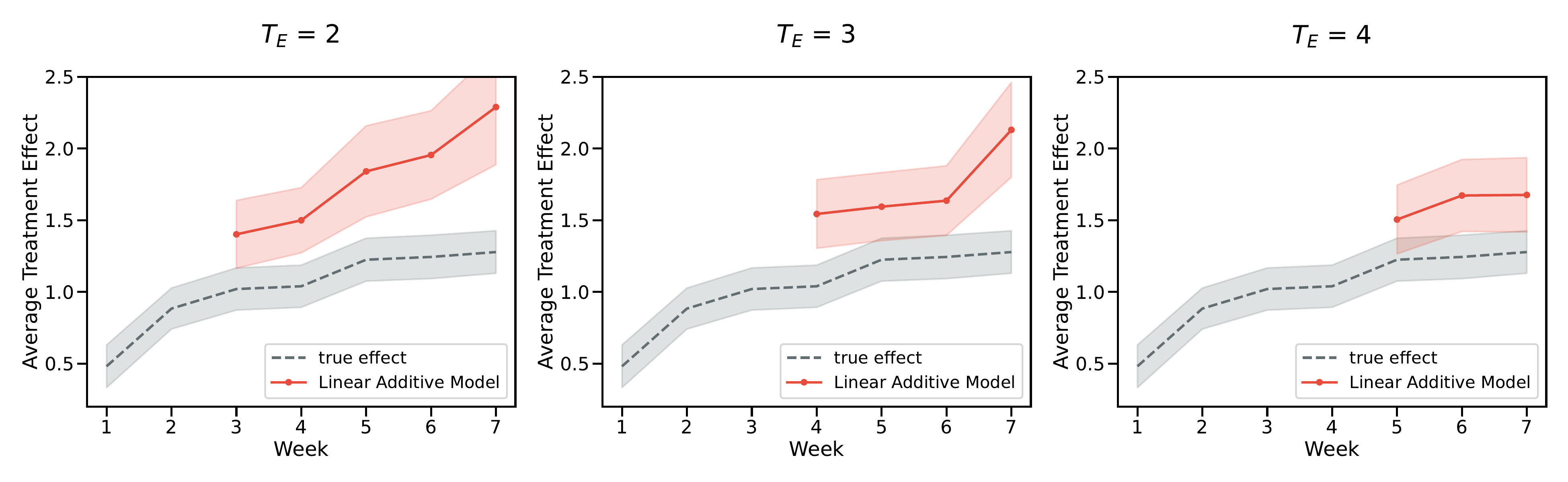}
\caption{Estimated effect for 7-week treatment with $T_E$ weeks observed data under the Linear Additive Model for Experiment 1}
{\footnotesize \textit{Note}: Grey dashed curves represent the true average treatment effect on \textit{search\_uv} from week 1 to week 7. Solid red curves represent the estimated effects with the linear additive model. Shadows indicate 95\% confidence intervals. The three panels represent the scenarios when we use the first $T_E$ weeks as the experimental period and the last $T_F$ weeks as the future period.}
\label{fig:searchHistoryAdding}
\end{figure}

\section{Missing Algorithms for Inference and Testing}

In Section~\ref{sec:Inference} we discuss procedures to  test the sharp null hypothesis in Eq.~\eqref{eqn:null} and calculate the $p$-value, and (2) calculate the variance and obtain confidence intervals.
Now we provide the detailed description of our algorithms for  conducting the permutation test, and  conducting the bootstrap.

\begin{algorithm}[!htb]
\begin{algorithmic}[1]\small
\caption{Algorithm for testing the sharp-null hypothesis Eq.~\eqref{eqn:null}}
\label{alg:Exact}
\REQUIRE Fix $M$, total number of samples drawn.
\FOR{$m$ in $1:M$}
\STATE For all subjects $i \in [N]$, sample treatment assignments $\bm{W}_{i,1:T}^{[m]}$ according to $\pi(\bm{X}_i)$.
\STATE Hold $(Y_{it}, \bm{S}_{it})$ unchanged. Compute $\widehat{\tau}_T^{[m]}$ according to one of the estimation strategies \eqref{eqn:EstimationL1}, \eqref{eqn:EstimationL2}, or \eqref{eqn:EstimationL3}.
More specifically, estimate $\widehat{h}^{[m]}_t(\cdot, \cdot, \cdot)$, $\widehat{\bm{g}}^{[m]}_t(\cdot, \cdot, \cdot)$, and $\widehat{\bm{G}}^{[m]}_t(\cdot, \cdot, \cdot)$ using $\bm{W}_{i,1:T}^{[m]}$ instead of $\bm{W}_{i,1:T}$.
\ENDFOR
\STATE Compute $\widehat{p} = M^{-1} \sum_{m=1}^M \bI\{|\widehat{\tau}_T^{[m]}| > | \widehat{\tau}_T|\}$
\RETURN $\widehat{p}$, the estimated $p$-value. For large $M$, this is exact.
\end{algorithmic}
\end{algorithm}

The permutation test depends on the fact that, under the sharp null hypothesis of no treatment effect \eqref{eqn:null}, any treatment assignment $\bm{w}^{[m]}_{i,1:T}$ leads to the same primary and surrogate outcomes, for any subject $i\in[N]$ at any time $t\in[T]$.
In particular, in Step 3, we assume the observed primary and surrogate outcomes remain unchanged.

\begin{algorithm}[!htb]
\begin{algorithmic}[1]\small
\caption{Algorithm for bootstrap}
\label{alg:bootstrap}
\REQUIRE Fix $M$, total number of samples drawn.
\FOR{$m$ in $1:M$}
\STATE For all subjects $i \in [N]$, sample treatment assignments $\bm{W}_{i,1:T}^{[m]}$ according to $\pi(\bm{X}_i)$.
\STATE Hold $(Y_{it}, \bm{S}_{it})$ unchanged. Compute $\widehat{\tau}_T^{[m]}$ according to one of the estimation strategies \eqref{eqn:EstimationL1}, \eqref{eqn:EstimationL2}, or \eqref{eqn:EstimationL3}.
More specifically, estimate $\widehat{h}^{[m]}_t(\cdot, \cdot, \cdot)$, $\widehat{\bm{g}}^{[m]}_t(\cdot, \cdot, \cdot)$, and $\widehat{\bm{G}}^{[m]}_t(\cdot, \cdot, \cdot)$ using $\bm{W}_{i,1:T}^{[m]}$ instead of $\bm{W}_{i,1:T}$.
\ENDFOR
\STATE Compute sample variance $\widehat{\mathrm{Var}}(\widehat{\tau}_T) = (M-1)^{-1} \sum_{m=1}^M \Big(\widehat{\tau}_T^{[m]} - M^{-1} \sum_{l=1}^M \widehat{\tau}_T^{[l]}\Big)^2$.
\RETURN $\widehat{\mathrm{Var}}(\widehat{\tau}_T)$, the estimated variance.
\end{algorithmic}
\end{algorithm}

The bootstrap procedure depends on that we conduct a completely randomized experiment.
We acknowledge that, in practice, obtaining confidence intervals through this approach could be computationally challenging.
One could parallelize the computation for different values of $m$ to obtain the results more efficiently.

\section{Missing Statements and Proofs}

\subsection{Proofs of Lemma~\ref{lem:SpecialCaseL1} and Theorem~\ref{thm:IdentificationL1}}

Since Lemma~\ref{lem:SpecialCaseL1} is a special case of Theorem~\ref{thm:IdentificationL1}, the proof of Theorem~\ref{thm:IdentificationL1} applies to Lemma~\ref{lem:SpecialCaseL1}.
However, we prove Lemma~\ref{lem:SpecialCaseL1} first as a separate proof for easier understanding.

\proof{Proof of Lemma~\ref{lem:SpecialCaseL1}.}
\label{prf:Lemma1}
We prove Lemma~\ref{lem:SpecialCaseL1} by definition.
From the definition of the causal effect,
\begin{align*}
\tau = \bE_{\cF}\bigg[ Y_{iT}(\bm{1}_{T}) - Y_{iT}(\bm{0}_{T}) \bigg],
\end{align*}
we start from the first part $\bE_{\cF}\left[ Y_{iT}(\bm{1}_{T})\right]$.
\begin{align}
\bE_{\cF}\left[ Y_{iT}(\bm{1}_{T}) \right] = \bE_{\cF}\left[ Y_{iT} \bigg\vert \bm{W}_{i,1:T} = \bm{1}_T \right] = \sum_{\bm{x}_i \in \bX, \bm{s}_{i0} \in \bS} \bE_{\cF}\left[ Y_{iT} \bigg\vert \bm{W}_{i,1:T} = \bm{1}_T, \bm{x}_i, \bm{s}_{i0} \right] \Pr(\bm{x}_i, \bm{s}_{i0}). \label{eqn:IPWDecomposition:lem:simpler}
\end{align}
Using the law of total expectation,
\begin{multline*}
\bE_{\cF}\left[ Y_{iT}  \bigg\vert \bm{W}_{i,1:T} = \bm{1}_{T}, \bm{x}_i, \bm{s}_{i0} \right] \\
= \sum_{\bm{s} \in \bS} \bE_{\cF}\left[ Y_{iT}  \bigg\vert  \bm{S}_{i T_E} = \bm{s}, \bm{W}_{i,1:T} = \bm{1}_{T}, \bm{x}_i, \bm{s}_{i0} \right] \cdot \Pr\left(\bm{S}_{i T_E} = \bm{s} \bigg\vert \bm{W}_{i,1:T} = \bm{1}_{T}, \bm{x}_i, \bm{s}_{i0}\right).
\end{multline*}

Note that,
\begin{align*}
\bE_{\cF}\left[ Y_{iT} \bigg\vert \bm{S}_{i T_E} = \bm{s}, \bm{W}_{i,1:T} = \bm{1}_{T}, \bm{x}_i, \bm{s}_{i0} \right] = & \bE_{\cF}\left[ Y_{iT} \bigg\vert \bm{S}_{i T_E} = \bm{s}, \bm{W}_{i,T_E+1:T} = \bm{1}_{T_F}, \bm{x}_i \right] \\
= & \bE_{\cF}\left[Y_{i T_E} \bigg\vert \bm{S}_{i 0} = \bm{s}, \bm{W}_{i,1:T_E} = \bm{1}_{T_E}, \bm{x}_i \right],
\end{align*}
where the first equality is due to Assumption~\ref{asp:Surrogacy} when $t=T_E, t'=T$;
the second equality is due to Assumption~\ref{asp:Comparability} when $t=T_E, t'=T, \delta=T_E=T_F$.

Note also that,
\begin{align*}
\Pr\left(\bm{S}_{i T_E} = \bm{s}  \bigg\vert \bm{W}_{i,1:T} = \bm{1}_{T}, \bm{x}_i, \bm{s}_{i0}\right) = \Pr\left(\bm{S}_{i T_E} = \bm{s}  \bigg\vert \bm{W}_{i,1:T_E} = \bm{1}_{T_E}, \bm{x}_i, \bm{s}_{i0}\right).
\end{align*}

Combining both parts,
\begin{multline*}
\bE_{\cF}\left[ Y_{iT} \bigg\vert \bm{W}_{i,1:T} = \bm{1}_{T}, \bm{x}_i, \bm{s}_{i0} \right] \\
= \sum_{\bm{s} \in \bS} \bE_{\cF}\left[Y_{i T_E} \bigg\vert \bm{S}_{i 0} = \bm{s}, \bm{W}_{i,1:T_E} = \bm{1}_{T_E}, \bm{x}_i \right] \cdot \Pr\left(\bm{S}_{i T_E} = \bm{s}  \bigg\vert \bm{W}_{i,1:T_E} = \bm{1}_{T_E}, \bm{x}_i, \bm{s}_{i0}\right).
\end{multline*}

Putting the above expression in \eqref{eqn:IPWDecomposition:lem:simpler},
\begin{align*}
& \ \bE_{\cF}\left[ Y_{iT}(\bm{1}_{T}) \right] \\
= & \sum_{\bm{x}_i \in \bX, \bm{s}_{i0} \in \bS} \bE_{\cF}\left[ Y_{iT} \bigg\vert \bm{W}_{i,1:T} = \bm{1}_T, \bm{x}_i, \bm{s}_{i0} \right] \Pr(\bm{x}_i, \bm{s}_{i0}) \\
= & \sum_{\bm{x}_i \in \bX, \bm{s}_{i0} \in \bS} \sum_{\bm{s} \in \bS} \bE_{\cF}\left[Y_{i T_E} \bigg\vert \bm{S}_{i 0} = \bm{s}, \bm{W}_{i,1:T_E} = \bm{1}_{T_E}, \bm{x}_i \right] \cdot \Pr\left(\bm{S}_{i T_E} = \bm{s}  \bigg\vert \bm{W}_{i,1:T_E} = \bm{1}_{T_E}, \bm{x}_i, \bm{s}_{i0}\right) \cdot \Pr(\bm{x}_i, \bm{s}_{i0}) \\
= & \sum_{\bm{x}_i \in \bX, \bm{s}_{i0} \in \bS} \sum_{\bm{s} \in \bS} h_{T_E}(\bm{s}, \bm{x}_i, \bm{1}_{T_E}) \cdot \Pr\left(\bm{S}_{i T_E} = \bm{s} \bigg\vert \bm{W}_{i,1:T_E} = \bm{1}_{T_E}, \bm{x}_i, \bm{s}_{i0}\right) \cdot \Pr(\bm{x}_i, \bm{s}_{i0}) \\
= & \sum_{\bm{x}_i \in \bX, \bm{s}_{i0} \in \bS} \bE_{\cF}\left[h_{T_E}\left(\bm{G}_{T_E}\left(\bm{s}_{i0}, \bm{x}_i, \bm{1}_{T_E}\right), \bm{x}_i, \bm{1}_{T_E}\right) \right] \cdot \Pr(\bm{x}_i, \bm{s}_{i0}) \\
= & \ \bE_{\cF} \left[ h_{T_E}\left( \bm{G}_{T_E}(\bm{S}_{i0}, \bm{X}_i, \bm{1}_{T_E}), \bm{X}_i, \bm{1}_{T_E} \right) \right]
\end{align*}

Next, we focus on the second part $\bE_{\cF}\left[ Y_{iT}(\bm{0}_{T})\right]$.
Similarly we have 
\begin{align*}
\bE_{\cF}\left[ Y_{iT}(\bm{0}_{T}) \right] = \bE_{\cF} \left[ h_{T_E}\left( \bm{G}_{T_E}(\bm{S}_{i0}, \bm{X}_i, \bm{0}_{T_E}), \bm{X}_i, \bm{0}_{T_E} \right) \right].
\end{align*}

Combining both parts,
\begin{align*}
\tau_T = & \bE_{\cF} \left[ h_{T_E}\left( \bm{G}_{T_E}(\bm{S}_{i0}, \bm{X}_i, \bm{1}_{T_E}), \bm{X}_i, \bm{1}_{T_E} \right) \right] - \bE_{\cF} \left[ h_{T_E}\left( \bm{G}_{T_E}(\bm{S}_{i0}, \bm{X}_i, \bm{0}_{T_E}), \bm{X}_i, \bm{0}_{T_E} \right) \right],
\end{align*}
which finishes the proof.
\hfill \halmos
\endproof

\proof{Proof of Theorem~\ref{thm:IdentificationL1}.}
We prove Theorem~\ref{thm:IdentificationL1} by definition.
From the definition of the causal effect,
\begin{align*}
\tau = \bE_{\cF}\bigg[ Y_{iT}(\bm{1}_{T}) - Y_{iT}(\bm{0}_{T}) \bigg],
\end{align*}
and we start from the first part $\bE_{\cF}\left[ Y_{iT}(\bm{1}_{T})\right]$.
\begin{align}
\bE_{\cF}\left[ Y_{iT}(\bm{1}_{T}) \right] = \bE_{\cF}\left[ Y_{iT} \bigg\vert \bm{W}_{i,1:T} = \bm{1}_T \right] = \sum_{\bm{x}_i \in \bX, \bm{s}_{i0} \in \bS} \bE_{\cF}\left[ Y_{iT} \bigg\vert \bm{W}_{i,1:T} = \bm{1}_T, \bm{x}_i, \bm{s}_{i0} \right] \Pr(\bm{x}_i, \bm{s}_{i0}). \label{eqn:IPWDecomposition}
\end{align}

Using the law of total expectation,
\begin{multline}
\bE_{\cF}\left[ Y_{iT}  \bigg\vert \bm{W}_{i,1:T} = \bm{1}_{T}, \bm{x}_i, \bm{s}_{i0} \right] \\
= \sum_{\bm{s}_{t_K} \in \bS} \bE_{\cF}\left[ Y_{iT}  \bigg\vert  \bm{S}_{i t_K} = \bm{s}_{t_K}, \bm{W}_{i,1:T} = \bm{1}_{T}, \bm{x}_i, \bm{s}_{i0} \right] \cdot \Pr\left(\bm{S}_{i t_K} = \bm{s}_{t_K} \bigg\vert \bm{W}_{i,1:T} = \bm{1}_{T}, \bm{x}_i, \bm{s}_{i0}\right). \label{eqn:proof:thm:LawTotalExpectations}
\end{multline}

Note that,
\begin{align}
\bE_{\cF}\left[ Y_{iT} \bigg\vert \bm{S}_{i t_K} = \bm{s}_{t_K}, \bm{W}_{i,1:T} = \bm{1}_{T}, \bm{x}_i, \bm{s}_{i0} \right] = & \bE_{\cF}\left[ Y_{iT} \bigg\vert \bm{S}_{i t_K} = \bm{s}_{t_K}, \bm{W}_{i,t_K+1:T} = \bm{1}_{\Delta t_{K+1}}, \bm{x}_i \right]\nonumber \\
= & \bE_{\cF}\left[Y_{i \Delta t_{K+1}} \bigg\vert \bm{S}_{i 0} = \bm{s}_{t_K}, \bm{W}_{i,1:\Delta t_{K+1}} = \bm{1}_{\Delta t_{K+1}}, \bm{x}_i \right], \label{eqn:proof:thm:ConditionalExpectation}
\end{align}
where the first equality is due to Assumption~\ref{asp:Surrogacy} when $t=t_K, t'=T$;
the second equality is due to Assumption~\ref{asp:Comparability} when $t=t_K, t'=T, \delta=t_K$.

Note also that,
\begin{align}
& \ \Pr\left(\bm{S}_{i t_K} = \bm{s}_{t_K} \bigg\vert \bm{W}_{i,1:T} = \bm{1}_{T}, \bm{x}_i, \bm{s}_{i0}\right) = \ \Pr\left(\bm{S}_{i t_K} = \bm{s}_{t_K} \bigg\vert \bm{W}_{i,1:t_K} = \bm{1}_{t_K}, \bm{x}_i, \bm{s}_{i0}\right) \nonumber \\
= & \sum_{\bm{s}_{t_{K-1}} \in \bS} \Pr\left(\bm{S}_{i t_K} = \bm{s}_{t_K} \bigg\vert \bm{W}_{i,1:t_K} = \bm{1}_{t_K}, \bm{S}_{i t_{K-1}} = \bm{s}_{t_{K-1}}, \bm{x}_i, \bm{s}_{i0}\right) \cdot \Pr\left(\bm{S}_{i t_{K-1}} = \bm{s}_{t_{K-1}} \bigg\vert \bm{W}_{i,1:t_{K-1}} = \bm{1}_{t_{K-1}}, \bm{x}_i, \bm{s}_{i0}\right) \nonumber \\
= & \ldots \nonumber \\
= & \sum_{\bm{s}_{t_{K-1}}, \bm{s}_{t_{K-2}}, ..., \bm{s}_{t_{2}}, \bm{s}_{t_1} \in \bS} 
\Pr\left(\bm{S}_{i t_K} = \bm{s}_{t_K} \bigg\vert \bm{W}_{i,1:t_K} = \bm{1}_{t_K}, \bm{S}_{i t_{K-1}} = \bm{s}_{t_{K-1}}, \bm{x}_i, \bm{s}_{i0}\right) \cdot \nonumber \\
& \hspace{3.5cm} \Pr\left(\bm{S}_{i t_{K-1}} = \bm{s}_{t_{K-1}} \bigg\vert \bm{W}_{i,1:t_{K-1}} = \bm{1}_{t_{K-1}}, \bm{S}_{i t_{K-2}} = \bm{s}_{t_{K-2}}, \bm{x}_i, \bm{s}_{i0}\right) \cdot \nonumber \\
& \hspace{3.2cm} \ldots \cdot \Pr\left(\bm{S}_{i t_2} = \bm{s}_{t_2} \bigg\vert \bm{W}_{i,1:t_2} = \bm{1}_{t_2}, \bm{S}_{i t_{1}} = \bm{s}_{t_{1}}, \bm{x}_i, \bm{s}_{i0}\right) \cdot \Pr\left(\bm{S}_{i t_{1}} = \bm{s}_{t_{1}} \bigg\vert \bm{W}_{i,1:t_{1}} = \bm{1}_{t_{1}}, \bm{x}_i, \bm{s}_{i0}\right) \nonumber \\
= & \sum_{\bm{s}_{t_{K-1}}, \bm{s}_{t_{K-2}}, ..., \bm{s}_{t_{2}}, \bm{s}_{t_1} \in \bS} 
\Pr\left(\bm{S}_{i t_K} = \bm{s}_{t_K} \bigg\vert \bm{W}_{i,t_{K-1}+1:t_K} = \bm{1}_{\Delta t_K}, \bm{S}_{i t_{K-1}} = \bm{s}_{t_{K-1}}, \bm{x}_i\right) \cdot \nonumber \\
& \hspace{3.5cm} \Pr\left(\bm{S}_{i t_{K-1}} = \bm{s}_{t_{K-1}} \bigg\vert \bm{W}_{i,t_{K-2}+1:t_{K-1}} = \bm{1}_{\Delta t_{K-1}}, \bm{S}_{i t_{K-2}} = \bm{s}_{t_{K-2}}, \bm{x}_i \right) \cdot \nonumber \\
& \hspace{3.2cm} \ldots \cdot \Pr\left(\bm{S}_{i t_2} = \bm{s}_{t_2} \bigg\vert \bm{W}_{i,t_1+1:t_2} = \bm{1}_{\Delta t_2}, \bm{S}_{i t_{1}} = \bm{s}_{t_{1}}, \bm{x}_i \right) \cdot \Pr\left(\bm{S}_{i t_{1}} = \bm{s}_{t_{1}} \bigg\vert \bm{W}_{i,1:t_{1}} = \bm{1}_{\Delta t_{1}}, \bm{x}_i, \bm{s}_{i0}\right), \label{eqn:proof:thm:Probabilities}
\end{align}
where the last equality is due to Assumption~\ref{asp:Surrogacy}.

Putting the above expressions \eqref{eqn:proof:thm:LawTotalExpectations}--\eqref{eqn:proof:thm:Probabilities} into \eqref{eqn:IPWDecomposition},
\begin{align*}
& \ \bE_{\cF}\left[ Y_{iT}(\bm{1}_{T}) \right] \\
= & \sum_{\bm{x}_i \in \bX, \bm{s}_{i0} \in \bS} \bE_{\cF}\left[ Y_{iT} \bigg\vert \bm{W}_{i,1:T} = \bm{1}_T, \bm{x}_i, \bm{s}_{i0} \right] \Pr(\bm{x}_i, \bm{s}_{i0}) \\
= & \sum_{\bm{x}_i \in \bX, \bm{s}_{i0} \in \bS} \sum_{\bm{s}_{t_K} \in \bS} \bE_{\cF}\left[Y_{i \Delta t_{K+1}} \bigg\vert \bm{S}_{i 0} = \bm{s}_{t_K}, \bm{W}_{i,1:\Delta t_{K+1}} = \bm{1}_{\Delta t_{K+1}}, \bm{x}_i \right] \cdot \Pr\left(\bm{S}_{i t_K} = \bm{s}_{t_K} \bigg\vert \bm{W}_{i,1:t_K} = \bm{1}_{t_K}, \bm{x}_i, \bm{s}_{i0}\right) \cdot \Pr(\bm{x}_i, \bm{s}_{i0}) \\
= & \sum_{\bm{x}_i \in \bX, \bm{s}_{i0} \in \bS} \sum_{\bm{s}_{t_K} \in \bS} h_{\Delta t_{K+1}}(\bm{s}_{t_K}, \bm{x}_i, \bm{1}_{\Delta t_{K+1}}) \cdot \Pr\left(\bm{S}_{i t_K} = \bm{s}_{t_K} \bigg\vert \bm{W}_{i,1:t_K} = \bm{1}_{t_K}, \bm{x}_i, \bm{s}_{i0}\right) \cdot \Pr(\bm{x}_i, \bm{s}_{i0}) \\
= & \sum_{\bm{x}_i \in \bX, \bm{s}_{i0} \in \bS} \sum_{\bm{s}_{t_K}, ..., \bm{s}_{t_1} \in \bS} h_{\Delta t_{K+1}}(\bm{s}_{t_K}, \bm{x}_i, \bm{1}_{\Delta t_{K+1}}) \cdot \Pr\left(\bm{S}_{i t_K} = \bm{s}_{t_K} \bigg\vert \bm{W}_{i,t_{K-1}+1:t_K} = \bm{1}_{\Delta t_K}, \bm{S}_{i t_{K-1}} = \bm{s}_{t_{K-1}}, \bm{x}_i\right) \cdot \\
& \hspace{4cm} \Pr\left(\bm{S}_{i t_{K-1}} = \bm{s}_{t_{K-1}} \bigg\vert \bm{W}_{i,t_{K-2}+1:t_{K-1}} = \bm{1}_{\Delta t_{K-1}}, \bm{S}_{i t_{K-2}} = \bm{s}_{t_{K-2}}, \bm{x}_i \right) \cdot \\
& \hspace{4.5cm} ... \cdot \Pr\left(\bm{S}_{i t_{1}} = \bm{s}_{t_{1}} \bigg\vert \bm{W}_{i,1:t_{1}} = \bm{1}_{\Delta t_{1}}, \bm{x}_i, \bm{s}_{i0}\right) \cdot \Pr(\bm{x}_i, \bm{s}_{i0}) \\
= & \sum_{\bm{x}_i \in \bX, \bm{s}_{i0} \in \bS} \bE_{\cF} \left[ h_{\Delta t_{K+1}}\left( \bm{G}_{\Delta t_K}( ... \bm{G}_{\Delta t_1}(\bm{s}_{i0}, \bm{x}_i, \bm{1}_{\Delta t_1}) ... ,\bm{x}_i, \bm{1}_{\Delta t_K}), \bm{x}_i, \bm{1}_{\Delta t_{K+1}} \right) \bigg\vert \bm{x}_i, \bm{s}_{i0} \right] \cdot \Pr(\bm{x}_i, \bm{s}_{i0}) \\ 
= & \ \bE_{\cF} \left[ h_{\Delta t_{K+1}}\left( \bm{G}_{\Delta t_K}( ... \bm{G}_{\Delta t_1}(\bm{S}_{i0}, \bm{X}_i, \bm{1}_{\Delta t_1}) ... ,\bm{X}_i, \bm{1}_{\Delta t_K}), \bm{X}_i, \bm{1}_{\Delta t_{K+1}} \right) \right],
\end{align*}
where the second equality is plugging in \eqref{eqn:proof:thm:LawTotalExpectations}--\eqref{eqn:proof:thm:ConditionalExpectation}; the third equality is using Definition~\ref{defn:SurrogateIndex}; the fourth equality is plugging in \eqref{eqn:proof:thm:Probabilities}.

Next, we focus on the second part $\bE_{\cF}\left[ Y_{iT}(\bm{0}_{T})\right]$.
Similarly, we have 
\begin{align*}
\bE_{\cF}\left[ Y_{iT}(\bm{0}_{T}) \right] = \bE_{\cF} \left[ h_{\Delta t_{K+1}}\left( \bm{G}_{\Delta t_K}( ... \bm{G}_{\Delta t_1}(\bm{S}_{i0}, \bm{X}_i, \bm{0}_{\Delta t_1}) ... ,\bm{X}_i, \bm{0}_{\Delta t_K}), \bm{X}_i, \bm{0}_{\Delta t_{K+1}} \right) \right].
\end{align*}

Combining both parts,
\begin{multline*}
\tau_T = \bE_{\cF} \left[ h_{\Delta t_{K+1}}\left( \bm{G}_{\Delta t_K}( ... \bm{G}_{\Delta t_1}(\bm{S}_{i0}, \bm{X}_i, \bm{1}_{\Delta t_1}) ... ,\bm{X}_i, \bm{1}_{\Delta t_K}), \bm{X}_i, \bm{1}_{\Delta t_{K+1}} \right) \right] \\
- \bE_{\cF} \left[ h_{\Delta t_{K+1}}\left( \bm{G}_{\Delta t_K}( ... \bm{G}_{\Delta t_1}(\bm{S}_{i0}, \bm{X}_i, \bm{0}_{\Delta t_1}) ... ,\bm{X}_i, \bm{0}_{\Delta t_K}), \bm{X}_i, \bm{0}_{\Delta t_{K+1}} \right) \right].
\end{multline*}
which finishes the proof.
\hfill \halmos
\endproof

\subsection{Proof of Theorem~\ref{thm:IdentificationL2}}
\proof{Proof of Theorem~\ref{thm:IdentificationL2}.}
\label{prf:Theorem1}
Under Assumptions~\ref{asp:Surrogacy} and~\ref{asp:Comparability}, Theorem~\ref{thm:IdentificationL1} yields
\begin{multline*}
\tau_T = \bE_{\cF} \left[ h_{\Delta t_{K+1}}\left( \bm{G}_{\Delta t_K}( ... \bm{G}_{\Delta t_1}(\bm{S}_{i0}, \bm{X}_i, \bm{1}_{\Delta t_1}) ... ,\bm{X}_i, \bm{1}_{\Delta t_K}), \bm{X}_i, \bm{1}_{\Delta t_{K+1}} \right) \right] \\
- \bE_{\cF} \left[ h_{\Delta t_{K+1}}\left( \bm{G}_{\Delta t_K}( ... \bm{G}_{\Delta t_1}(\bm{S}_{i0}, \bm{X}_i, \bm{0}_{\Delta t_1}) ... ,\bm{X}_i, \bm{0}_{\Delta t_K}), \bm{X}_i, \bm{0}_{\Delta t_{K+1}} \right) \right].
\end{multline*}

We start from the first part 
\begin{align*}
\bE_{\cF} \left[ h_{\Delta t_{K+1}}\left( \bm{G}_{\Delta t_K}( ... \bm{G}_{\Delta t_1}(\bm{S}_{i0}, \bm{X}_i, \bm{1}_{\Delta t_1}) ... ,\bm{X}_i, \bm{1}_{\Delta t_K}), \bm{X}_i, \bm{1}_{\Delta t_{K+1}} \right) \right].
\end{align*}

Denote $\bm{G}_{\Delta t_K} = (G_{\Delta t_K, 1}, ..., G_{\Delta t_K, D})$ and $\bm{g}_{\Delta t_K} = (g_{\Delta t_K, 1}, ..., g_{\Delta t_K, D})$.
\begin{align}
& \bE_{\cF} \left[h_{\Delta t_{K+1}}(\bm{G}_{\Delta t_K}( ... \bm{G}_{\Delta t_1}(\bm{S}_{i0}, \bm{X}_i, \bm{1}_{\Delta t_1}) ... ,\bm{X}_i, \bm{1}_{\Delta t_K}), \bm{X}_i, \bm{1}_{\Delta t_{K+1}})\right] \nonumber \\
= & \bE_{\cF} \left[ \alpha_0(\bm{X}_i, \bm{1}_{\Delta t_{K+1}}) + \sum_{d=1}^D G_{\Delta t_K, d}( ... \bm{G}_{\Delta t_1}(\bm{S}_{i0}, \bm{X}_i, \bm{1}_{\Delta t_1}) ... ,\bm{X}_i, \bm{1}_{\Delta t_K}) \cdot \alpha_d(\bm{X}_i, \bm{1}_{\Delta t_{K+1}})\right] \nonumber \\
= & \alpha_0(\bm{X}_i, \bm{1}_{\Delta t_{K+1}}) + \sum_{d=1}^D \bE_{\cF} \left[ G_{\Delta t_K, d}( ... \bm{G}_{\Delta t_1}(\bm{S}_{i0}, \bm{X}_i, \bm{1}_{\Delta t_1}) ... ,\bm{X}_i, \bm{1}_{\Delta t_K})\right] \cdot \alpha_d(\bm{X}_i, \bm{1}_{\Delta t_{K+1}}) \nonumber \\
= & \alpha_0(\bm{X}_i, \bm{1}_{\Delta t_{K+1}}) + \sum_{d=1}^D \bE_{\cF} \left[ g_{\Delta t_K, d}( ... \bm{G}_{\Delta t_1}(\bm{S}_{i0}, \bm{X}_i, \bm{1}_{\Delta t_1}) ... ,\bm{X}_i, \bm{1}_{\Delta t_K})\right] \cdot \alpha_d(\bm{X}_i, \bm{1}_{\Delta t_{K+1}}) \nonumber \\
= & h_{\Delta t_{K+1}}\left(\bE_{\cF} \left[\bm{g}_{\Delta t_K}( ... \bm{G}_{\Delta t_1}(\bm{S}_{i0}, \bm{X}_i, \bm{1}_{\Delta t_1}) ... ,\bm{X}_i, \bm{1}_{\Delta t_K}), \bm{X}_i, \bm{1}_{\Delta t_{K+1}}\right]\right), \label{eqn:proof:thmL2:h:iteration}
\end{align}
where the first equality is due to Assumption~\ref{asp:Linearity}; the second equality is due to linearity of expectation; in the third equality, we write the expectation of $G_{\Delta t_K, d}$ as $g_{\Delta t_K, d}$, but there is still uncertainty from $\bm{G}_{\Delta t_{K-1}, d}$, which keeps the expectation notion.

Next, denote $\bm{G}_{\Delta t_{K-1}} = (G_{\Delta t_{K-1}, 1}, ..., G_{\Delta t_{K-1}, D})$ and $\bm{g}_{\Delta t_{K-1}} = (g_{\Delta t_{K-1}, 1}, ..., g_{\Delta t_{K-1}, D})$.
For each $d \in [D]$, we focus on $g_{\Delta t_K, d}$ as follows,
\begin{align*}
& \bE_{\cF} \left[g_{\Delta t_{K},d}(\bm{G}_{\Delta t_{K-1}}( ... \bm{G}_{\Delta t_1}(\bm{S}_{i0}, \bm{X}_i, \bm{1}_{\Delta t_1}) ... ,\bm{X}_i, \bm{1}_{\Delta t_{K-1}}), \bm{X}_i, \bm{1}_{\Delta t_K})\right] \\
= & \bE_{\cF} \left[ \beta_{d,0}(\bm{X}_i, \bm{1}_{\Delta t_{K}}) + \sum_{d'=1}^D G_{\Delta t_{K-1}, d'}( ... \bm{G}_{\Delta t_1}(\bm{S}_{i0}, \bm{X}_i, \bm{1}_{\Delta t_1}) ... ,\bm{X}_i, \bm{1}_{\Delta t_{K-1}}) \cdot \beta_{d,d'}(\bm{X}_i, \bm{1}_{\Delta t_{K}})\right] \\
= & \beta_{d,0}(\bm{X}_i, \bm{1}_{\Delta t_{K}}) + \sum_{d'=1}^D \bE_{\cF} \left[G_{\Delta t_{K-1}, d'}( ... \bm{G}_{\Delta t_1}(\bm{S}_{i0}, \bm{X}_i, \bm{1}_{\Delta t_1}) ... ,\bm{X}_i, \bm{1}_{\Delta t_{K-1}})\right] \cdot \beta_{d,d'}(\bm{X}_i, \bm{1}_{\Delta t_{K}}) \\
= & \beta_{d,0}(\bm{X}_i, \bm{1}_{\Delta t_{K}}) + \sum_{d'=1}^D \bE_{\cF} \left[g_{\Delta t_{K-1}, d'}( ... \bm{G}_{\Delta t_1}(\bm{S}_{i0}, \bm{X}_i, \bm{1}_{\Delta t_1}) ... ,\bm{X}_i, \bm{1}_{\Delta t_{K-1}})\right] \cdot \beta_{d,d'}(\bm{X}_i, \bm{1}_{\Delta t_{K}}) \\
= & g_{\Delta t_{K}, d}\left(\bE_{\cF} \left[\bm{g}_{\Delta t_{K-1}}( ... \bm{G}_{\Delta t_1}(\bm{S}_{i0}, \bm{X}_i, \bm{1}_{\Delta t_1}) ... ,\bm{X}_i, \bm{1}_{\Delta t_{K-1}}), \bm{X}_i, \bm{1}_{\Delta t_{K}}\right)\right].
\end{align*}

Collecting the above equality by vector form,
\begin{multline}
\bE_{\cF} \left[\bm{g}_{\Delta t_{K}}(\bm{G}_{\Delta t_{K-1}}( ... \bm{G}_{\Delta t_1}(\bm{S}_{i0}, \bm{X}_i, \bm{1}_{\Delta t_1}) ... ,\bm{X}_i, \bm{1}_{\Delta t_{K-1}}), \bm{X}_i, \bm{1}_{\Delta t_K})\right] = \\
\bm{g}_{\Delta t_{K}}\left(\bE_{\cF} \left[\bm{g}_{\Delta t_{K-1}}( ... \bm{G}_{\Delta t_1}(\bm{S}_{i0}, \bm{X}_i, \bm{1}_{\Delta t_1}) ... ,\bm{X}_i, \bm{1}_{\Delta t_{K-1}}), \bm{X}_i, \bm{1}_{\Delta t_{K}}\right)\right]. \label{eqn:proof:thmL2:g:iteration}
\end{multline}

Iteratively applying \eqref{eqn:proof:thmL2:g:iteration} and combining with \eqref{eqn:proof:thmL2:h:iteration} finishes the proof.
\hfill \halmos
\endproof

\subsection{Proof of Theorem~\ref{thm:IdentificationL3}}
\proof{Proof of Theorem~\ref{thm:IdentificationL3}.}
For each $k \in \{1,2,...,K+1\}$, we start with the first part. 
\begin{align}
& \bE_{\cF}\bigg[ Y_{iT} (\bm{0}_{1:t_{k-1}}, \bm{1}_{t_{k-1}+1:t_k}, \bm{0}_{t_k+1:T})\bigg] \nonumber \\
= & \bE_{\cF}\left[ Y_{iT} \bigg\vert \bm{W}_{i,1:T} = (\bm{0}_{1:t_{k-1}}, \bm{1}_{t_{k-1}+1:t_k}, \bm{0}_{t_k+1:T}) \right] \nonumber \\
= & \sum_{\bm{x}_i \in \bX, \bm{s}_{i0} \in \bS} \bE_{\cF}\left[ Y_{iT} \bigg\vert \bm{W}_{i,1:T} = (\bm{0}_{1:t_{k-1}}, \bm{1}_{t_{k-1}+1:t_k}, \bm{0}_{t_k+1:T}), \bm{x}_i, \bm{s}_{i0} \right] \Pr(\bm{x}_i, \bm{s}_{i0}). \label{eqn:IPWDecomposition:lem:simpler2}
\end{align}

Using the law of total expectation,
\begin{multline*}
\bE_{\cF}\left[ Y_{iT}  \bigg\vert \bm{W}_{i,1:T} = (\bm{0}_{1:t_{k-1}}, \bm{1}_{t_{k-1}+1:t_k}, \bm{0}_{t_k+1:T}), \bm{x}_i, \bm{s}_{i0} \right] \\
= \sum_{\bm{s} \in \bS} \bE_{\cF}\left[ Y_{iT}  \bigg\vert  \bm{S}_{i T_E} = \bm{s}, \bm{W}_{i,1:T} = (\bm{0}_{1:t_{k-1}}, \bm{1}_{t_{k-1}+1:t_k}, \bm{0}_{t_k+1:T}), \bm{x}_i, \bm{s}_{i0} \right] \\
\cdot \Pr\left(\bm{S}_{i T_E} = \bm{s} \bigg\vert \bm{W}_{i,1:T} = (\bm{0}_{1:t_{k-1}}, \bm{1}_{t_{k-1}+1:t_k}, \bm{0}_{t_k+1:T}), \bm{x}_i, \bm{s}_{i0}\right).
\end{multline*}

Note that,
\begin{align*}
& \bE_{\cF}\left[ Y_{iT} \bigg\vert \bm{S}_{i T_E} = \bm{s}, \bm{W}_{i,1:T} = (\bm{0}_{1:t_{k-1}}, \bm{1}_{t_{k-1}+1:t_k}, \bm{0}_{t_k+1:T}), \bm{x}_i, \bm{s}_{i0} \right] \\
= & \bE_{\cF}\left[ Y_{iT} \bigg\vert \bm{S}_{i t_k} = \bm{s}, \bm{W}_{i,t_k+1:T} = \bm{0}_{T-t_k}, \bm{x}_i \right] \\
= & \bE_{\cF}\left[Y_{i (T-t_k)} \bigg\vert \bm{S}_{i 0} = \bm{s}, \bm{W}_{i,1:T-t_k} = \bm{0}_{T-t_k}, \bm{x}_i \right],
\end{align*}
where the first equality is due to Assumption~\ref{asp:Surrogacy} when $t=t_k, t'=T$;
the second equality is due to Assumption~\ref{asp:Comparability} when $t=t_k, t'=T, \delta=t_k$.

Also note that,
\begin{align*}
& \Pr\left(\bm{S}_{i t_k} = \bm{s}  \bigg\vert \bm{W}_{i,1:T} = (\bm{0}_{1:t_{k-1}}, \bm{1}_{t_{k-1}+1:t_k}, \bm{0}_{t_k+1:T}), \bm{x}_i, \bm{s}_{i0}\right) \\
= & \Pr\left(\bm{S}_{i t_k} = \bm{s}  \bigg\vert \bm{W}_{i,1:t_k} = (\bm{0}_{1:t_{k-1}}, \bm{1}_{t_{k-1}+1:t_k}), \bm{x}_i, \bm{s}_{i0}\right).
\end{align*}

Combining both parts,
\begin{multline*}
\bE_{\cF}\left[ Y_{iT}  \bigg\vert \bm{W}_{i,1:T} = (\bm{0}_{1:t_{k-1}}, \bm{1}_{t_{k-1}+1:t_k}, \bm{0}_{t_k+1:T}), \bm{x}_i, \bm{s}_{i0} \right] \\
= \sum_{\bm{s} \in \bS} \bE_{\cF}\left[Y_{i (T-t_k)} \bigg\vert \bm{S}_{i 0} = \bm{s}, \bm{W}_{i,1:T-t_k} = \bm{0}_{T-t_k}, \bm{x}_i \right] \cdot \Pr\left(\bm{S}_{i t_k} = \bm{s}  \bigg\vert \bm{W}_{i,1:t_k} = (\bm{0}_{1:t_{k-1}}, \bm{1}_{t_{k-1}+1:t_k}), \bm{x}_i, \bm{s}_{i0}\right).
\end{multline*}

Putting the above expression in \eqref{eqn:IPWDecomposition:lem:simpler2},
\begin{align*}
& \ \bE_{\cF}\bigg[ Y_{iT} (\bm{0}_{1:t_{k-1}}, \bm{1}_{t_{k-1}+1:t_k}, \bm{0}_{t_k+1:T})\bigg] \\
= & \sum_{\bm{x}_i \in \bX, \bm{s}_{i0} \in \bS} \bE_{\cF}\left[ Y_{iT} \bigg\vert \bm{W}_{i,1:T} = (\bm{0}_{1:t_{k-1}}, \bm{1}_{t_{k-1}+1:t_k}, \bm{0}_{t_k+1:T}), \bm{x}_i, \bm{s}_{i0} \right] \Pr(\bm{x}_i, \bm{s}_{i0}) \\
= & \sum_{\bm{x}_i \in \bX, \bm{s}_{i0} \in \bS} \sum_{\bm{s} \in \bS} \bE_{\cF}\left[Y_{i (T-t_k)} \bigg\vert \bm{S}_{i 0} = \bm{s}, \bm{W}_{i,1:T-t_k} = \bm{0}_{T-t_k}, \bm{x}_i \right] \\
& \hspace{3cm} \cdot \Pr\left(\bm{S}_{i t_k} = \bm{s}  \bigg\vert \bm{W}_{i,1:t_k} = (\bm{0}_{1:t_{k-1}}, \bm{1}_{t_{k-1}+1:t_k}), \bm{x}_i, \bm{s}_{i0}\right) \cdot \Pr(\bm{x}_i, \bm{s}_{i0}) \\
= & \sum_{\bm{x}_i \in \bX, \bm{s}_{i0} \in \bS} \sum_{\bm{s} \in \bS} h_{T-t_k}(\bm{s}, \bm{x}_i, \bm{0}_{T-t_k}) \cdot \Pr\left(\bm{S}_{i t_k} = \bm{s}  \bigg\vert \bm{W}_{i,1:t_k} = (\bm{0}_{1:t_{k-1}}, \bm{1}_{t_{k-1}+1:t_k}), \bm{x}_i, \bm{s}_{i0}\right) \cdot \Pr(\bm{x}_i, \bm{s}_{i0}) \\
= & \sum_{\bm{x}_i \in \bX, \bm{s}_{i0} \in \bS} \bE_{\cF}\left[h_{T-t_k}\left(\bm{G}_{t_k}\left(\bm{s}_{i0}, \bm{x}_i, (\bm{0}_{1:t_{k-1}}, \bm{1}_{t_{k-1}+1:t_k})\right), \bm{x}_i, \bm{0}_{T-t_k}\right) \right] \cdot \Pr(\bm{x}_i, \bm{s}_{i0}) \\
= & \ \bE_{\cF} \left[ h_{T-t_k}\left( \bm{G}_{t_k}\left(\bm{S}_{i0}, \bm{X}_i, (\bm{0}_{1:t_{k-1}}, \bm{1}_{t_{k-1}+1:t_k})\right), \bm{X}_i, \bm{0}_{T-t_k} \right) \right]
\end{align*}

Next, we focus on the second part $\bE_{\cF}\left[ Y_{iT}(\bm{0}_{T})\right]$.
\begin{multline*}
\bE_{\cF}\left[ Y_{iT}(\bm{0}_{T}) \right] = \bE_{\cF}\left[ Y_{iT} \bigg\vert \bm{W}_{i,1:T} = \bm{0}_T \right] \\ = \sum_{\bm{x}_i \in \bX, \bm{s}_{i0} \in \bS} \bE_{\cF}\left[ Y_{iT} \bigg\vert \bm{W}_{i,1:T} = \bm{0}_T, \bm{x}_i, \bm{s}_{i0} \right] \Pr(\bm{x}_i, \bm{s}_{i0}) = \bE_{\cF}\left[h_{T}\left(\bm{S}_{i0},\bm{X}_i,\bm{0}_T\right)\right].
\end{multline*}

Combining both parts we finish the proof.
\hfill \halmos
\endproof

\subsection{Statement and Proof of Lemma~\ref{lem:LinearConsistency}}\label{sec:appendix:lemma2}

\begin{lemma}
\label{lem:LinearConsistency}
Under Assumption~\ref{asp:Linearity}, assume the coefficients $\widehat{\alpha}_{d}, \forall d \in \{0,1,...,D\}$ and $\widehat{\beta}_{d,d'}, \forall d \in [D], d' \in \{0,1,...,D\}$ are consistently estimated, that is, for any $\bm{x} \in \bX, \bm{w}_{1:t} \in \{\bm{0}_t, \bm{1}_t\}$,
\begin{align*}
\Pr\Big( \lim_{N \to \infty} \widehat{\alpha}_{d}^{(N)}(\bm{x}, \bm{w}_{1:t}) = \alpha_{d}(\bm{x}, \bm{w}_{1:t}) \Big) & = 1, & \Pr\Big( \lim_{N \to \infty} \widehat{\beta}_{d,d'}^{(N)}(\bm{x}, \bm{w}_{1:t}) = \beta_{d,d'}(\bm{x}, \bm{w}_{1:t}) \Big) & = 1,
\end{align*}
where we use the superscript $(N)$ to stand for the dependence on sample size $N$.
Then, $\widehat{\tau}_T$ as defined in \eqref{eqn:EstimationL2} is a consistent estimator of $\tau_T$ as defined in \eqref{eqn:Estimand}, that is,
\begin{align*}
\Pr\Big( \lim_{N \to \infty} \widehat{\tau}_{T}^{(N)} = \tau_{T} \Big) & = 1.
\end{align*}
\end{lemma}

\proof{Proof of Lemma~\ref{lem:LinearConsistency}.}
Note that, with probability one, $\lim_{N \to \infty} \widehat{\beta}_{d,d'}^{(N)}(\bm{x}, \bm{w}_{1:t}) = \beta_{d,d'}(\bm{x}, \bm{w}_{1:t})$.
This implies that, for any $\bm{s} \in \bS, \bm{x} \in \bX, \bm{w}_{1:t} \in \{\bm{0}_t, \bm{1}_t\}$, and $d\in[D]$, with probability one,
\begin{multline*}
\lim_{N \to \infty} \widehat{g}^{(N)}_{t,d}(\bm{s}, \bm{x}, \bm{w}_{1:t}) = \lim_{N \to \infty} \Big( \widehat{\beta}^{(N)}_{d,0}(\bm{x}, \bm{w}_{1:t}) + \sum_{d'=1}^D s_d \cdot \widehat{\beta}^{(N)}_{d,d'}(\bm{x}, \bm{w}_{1:t}) \Big) \\
= \beta_{d,0}(\bm{x}, \bm{w}_{1:t}) + \sum_{d'=1}^D s_d \cdot \beta_{d,d'}(\bm{x}, \bm{w}_{1:t}) = g_{t,d}(\bm{s}, \bm{x}, \bm{w}_{1:t}). 
\end{multline*}
Furthermore, for any $t$ and $t'$, with probability one, 
\begin{multline*}
\lim_{N \to \infty} \widehat{g}^{(N)}_{t',d}(\widehat{g}^{(N)}_{t,d}(\bm{s}, \bm{x}, \bm{w}_{1:t}), \bm{x}, \bm{w}_{1:t'}) = \lim_{N \to \infty} \Big( \widehat{\beta}^{(N)}_{d,0}(\bm{x}, \bm{w}_{1:t'}) + \sum_{d'=1}^D \widehat{g}^{(N)}_{t,d}(\bm{s}, \bm{x}, \bm{w}_{1:t}) \cdot \widehat{\beta}^{(N)}_{d,d'}(\bm{x}, \bm{w}_{1:t'}) \Big) \\
= \beta_{d,0}(\bm{x}, \bm{w}_{1:t'}) + \sum_{d'=1}^D g_{t,d'}(\bm{s}, \bm{x}, \bm{w}_{1:t}) \cdot \beta_{d,d'}(\bm{x}, \bm{w}_{1:t'}) = g_{t',d}(\bm{g}_{t}(\bm{s}, \bm{x}, \bm{w}_{1:t}), \bm{x}, \bm{w}_{1:t'}). 
\end{multline*}
where the second equality is due to Slutsky's theorem.
Similarly, we can establish the above equality for surrogate index function $h_t(\cdot, \cdot, \cdot)$.
In addition, using the fact that all units are i.i.d. sampled, we know that with probability one,
\begin{align*}
& \lim_{N \to \infty} \widehat{\tau}^{(N)}_T \\
= & \frac{1}{N_1} \lim_{N \to \infty} \sum_{i\in[N]} \bI\{\bm{W}_{i,1:T_E} = \bm{1}_{T_E}\} h_{\Delta t_{K+1}}\left( \bm{g}_{\Delta t_K}( ... \bm{g}_{\Delta t_1}(\bm{S}_{i0}, \bm{X}_i, \bm{1}_{\Delta t_1}) ... ,\bm{X}_i, \bm{1}_{\Delta t_K}), \bm{X}_i, \bm{1}_{\Delta t_{K+1}} \right) \\
& - \frac{1}{N_0} \lim_{N \to \infty} \sum_{i\in[N]} \bI\{\bm{W}_{i,1:T_E} = \bm{0}_{T_E}\} h_{\Delta t_{K+1}}\left( \bm{g}_{\Delta t_K}( ... \bm{g}_{\Delta t_1}(\bm{S}_{i0}, \bm{X}_i, \bm{0}_{\Delta t_1}) ... ,\bm{X}_i, \bm{0}_{\Delta t_K}), \bm{X}_i, \bm{0}_{\Delta t_{K+1}} \right) \\
= & \bE_\cF\left[ h_{\Delta t_{K+1}}\left( \bm{g}_{\Delta t_K}( ... \bm{g}_{\Delta t_1}(\bm{S}_{i0}, \bm{X}_i, \bm{1}_{\Delta t_1}) ... ,\bm{X}_i, \bm{1}_{\Delta t_K}), \bm{X}_i, \bm{1}_{\Delta t_{K+1}} \right) \right] \\
& - \bE_\cF\left[ h_{\Delta t_{K+1}}\left( \bm{g}_{\Delta t_K}( ... \bm{g}_{\Delta t_1}(\bm{S}_{i0}, \bm{X}_i, \bm{0}_{\Delta t_1}) ... ,\bm{X}_i, \bm{0}_{\Delta t_K}), \bm{X}_i, \bm{0}_{\Delta t_{K+1}} \right) \right] \\
= & \tau_T,
\end{align*}
where the second equality is due to Theorem~\ref{thm:IdentificationL2}.
\hfill \halmos \endproof

\section{Practical Guidelines}\label{sec:guidelines}
In this section, we provide additional insights on how to apply our methodology in real-world settings—specifically, the practice of A/B testing in companies—more effectively. 

First, choosing surrogates is the critical decision to be made in applying our method. In practice, even hundreds of metrics are measured and collected for each experiment. Therefore, how to choose surrogates that better satisfy Assumption~\ref{asp:Surrogacy} --- the longitudinal surrogacy assumption --- can be a challenge. In particular, the assumption is not testable. In the selection of surrogates for longitudinal studies, it is essential to choose metrics that are highly responsive to treatments and reflect the diverse aspects of user behavior that lead to variations in primary outcomes over time~\citep{duan2021online,deng2013improving}. Surrogates should capture both up-stream behaviors (like search queries and navigation patterns) and down-stream behaviors (such as post-click actions including purchases or sharing), as these can robustly predict future primary outcomes due to their close alignment with user intentions and deeper engagement respectively. In
our real-world experiments, we mostly use down-stream behaviors as surrogates and demonstrated
their effectiveness. The current primary outcome
itself also serves as a strong predictive surrogate. 

To streamline surrogate selection in A/B testing, creating a library of surrogates based on metrics from prior experiments can considerably enhance efficiency and consistency across studies. This repository approach not only saves time and effort but also ensures the reliability of surrogate effectiveness across different contexts. However, selecting an appropriate number of surrogates is crucial; too few may not adequately capture the causal links necessary for reliable estimates, leading to potential biases, while too many can cause overfitting due to the curse of dimensionality. Techniques such as elastic net regularization are recommended to manage the number of surrogates by reducing dimensionality and focusing on the most predictive metrics, thus balancing the need for comprehensive data representation against the risks of overfitting. See detailed discussion in Appendix~\ref{sec:guidelines:surrogates}.

Additionally, in our main framework, the linear surrogate model, which extends the basic longitudinal surrogate model by incorporating a linearity assumption, generally outperforms non-linear models such as nearest neighbors, random forests, or neural networks in predicting future treatment effects. Our empirical tests, including those using the $k$-nearest neighbors (kNN) model, reveal that while non-linear models can fit within the confidence bounds of the treatment's actual long-term effect, they tend to produce unaffordably large variance and consequently, larger mean squared errors (MSE) compared to linear models. This is further evidenced by our tests with a relatively large $k$ (e.g., $k=20$) in the kNN model, which, despite its computational efficiency in finding close matches in the surrogate space, still results in higher variance and MSE. Hence, we recommend using linear models over non-linear approaches for treatment effect estimation, as they provide more reliable and precise estimates with smaller confidence intervals, as outlined in Appendix~\ref{sec:appendix:otherModelts}.

\section{More Details for Empirical Experiment Results}

Here we provide a number of additional experimental results mentioned in the main text.

\subsection{Description of Surrogates}
The detailed descriptions of surrogates and primary outcome we used in both empirical experiments is illustrated in Table~\ref{tb:searchHistorydef}.

%%%%%%%%%%%%%%%%%%%%%%%%%%%%%%%%%%%%%%%%%%%%%%%%%%%
\begin{table}[h]
\renewcommand{\arraystretch}{1.5} 
\centering\scriptsize{
\caption{The detailed descriptions of surrogates and primary outcome in Experiments 1 and 2}
\label{tb:searchHistorydef}
\begin{tabularx}{\textwidth}{@{}llX@{}}
\toprule
\multicolumn{1}{l}{\textbf{Variable}}&\multicolumn{1}{l}{\textbf{Role}}&\multicolumn{1}{l}{\textbf{Description}}\\
\midrule
\textit{search\_uv} & {Primary outcome, Surrogate} & A float number representing the average number of days that the user has searched in a week, normalized by being divided by 7\\
\textit{search\_qv} & Surrogate & A float number representing the average number of times that the user has searched in a week, normalized by being divided by 7\\
\textit{recall\_qv} & Surrogate & A float number representing the average number of times of search with results that the user has made in a week, normalized by being divided by 7.\\
\textit{expose\_qv} & Surrogate & A float number representing the average number of times of search with results that have been exposed to the user in a week, normalized by being divided by 7.\\
\textit{click\_qv} & Surrogate & A float number representing the average number of times of search with results that have been clicked by the user in a week, normalized by being divided by 7.\\
\bottomrule
\end{tabularx}}
\end{table}
%%%%%%%%%%%%%%%%%%%%%%%%%%%%%%%%%%%%%%%%%%%%%%%%%%%%%%

\subsection{Randomization Check}
\label{sec:appendix:randomization}

To illustrate the sample balance in our empirical experiments, we conducted the Sample Ratio Mismatch (SRM) \citep{fabijan2019diagnosing} test between the treatment and control groups and provided summary statistics on pre-treatment variables for both experiments.

In Experiment 1, the treatment group includes 667,206 users, while the control group includes 665,830 users. Given the expected ratio of 50\% for both groups, the experiment passes the chi-squared test with a statistic of 1.420 and a $p$-value of 0.233. This result indicates no significant difference between the observed and expected group sizes, suggesting no sample ratio mismatch problems. In Experiment 2, the treatment group includes 1,807,335 users, while the control group includes 1,803,675 users, with the expected ratio again being 50\% for both groups. The chi-squared test result for Experiment 2 shows a statistic of 3.710 and a $p$-value of 0.054, also implying no sample ratio mismatch problems.

Table~\ref{tb:summaryStatistics} presents the summary statistics for several pre-treatment variables across the treatment and control groups for both empirical experiments. These pre-treatment variables hold the same meaning as the surrogate variables used in the model but are collected before the experiment started. For Experiment 1, each variable records its average value over the 7 weeks prior to the experiment, while for Experiment 2, the average value is recorded over the 20 weeks before the experiment began. The table shows that the treatment and control groups are essentially balanced on these pre-treatment variables, indicating the randomness of both empirical experiments.

\begin{table}[htbp]
    \centering
    \caption{Summary statistics for two empirical experiments across treatment and control groups}\tiny
    \begin{tabular}{cccccccccccccc}
        \toprule
        &  & \multicolumn{5}{c}{\textbf{Treatment}} & \multicolumn{5}{c}{\textbf{Control}} & \multicolumn{2}{c}{\textbf{Compare Means}}\\
        \cmidrule(lr){3-7} \cmidrule(lr){8-12} \cmidrule(lr){13-14}
        \textbf{Expt.} & \shortstack{\textbf{Pre-treatment} \\ \textbf{Variable}} & \textbf{Count} & \textbf{Mean} & \textbf{Std.} & \textbf{Max} & \textbf{Min} & \textbf{Count} & \textbf{Mean} & \textbf{Std.} & \textbf{Max} & \textbf{Min} & \textbf{t-statistic} & \textbf{p-value}\\
        \midrule
        \multirow{5}{*}{Exp 1} & \textit{search\_uv} & \multirow{5}{*}{667,206} & 0.146 & 0.161 & 1.000 & 0.000 & \multirow{5}{*}{665,830} & 0.146 & 0.161 & 1.000 & 0.000 & 0.288 & 0.773\\
        & \textit{search\_qv} & & 0.369 & 0.936 & 127.918 & 0.000 & & 0.370 & 0.941 & 135.020 & 0.000 & 0.918 & 0.359\\
        & \textit{recall\_qv} & & 0.363 & 0.926 & 127.653 & 0.000 & & 0.364 & 0.932 & 131.612 & 0.000 & 0.954 & 0.340\\
        & \textit{expose\_qv} & & 0.347 & 0.915 & 127.306 & 0.000 & & 0.349 & 0.921 & 130.878 & 0.000 & 1.018 & 0.309\\
        & \textit{click\_qv} & & 0.226 & 0.629 & 75.612 & 0.000 & & 0.228 & 0.654 & 108.735 & 0.000 & 1.123 & 0.261\\
        \midrule
        \multirow{5}{*}{Exp 2} & \textit{search\_uv} & \multirow{5}{*}{1,807,335} & 0.117 & 0.134 & 0.979 & 0.000 & \multirow{5}{*}{1,803,675} & 0.117 & 0.135 & 0.979 & 0.000 & 0.366 & 0.714\\
        & \textit{search\_qv} & & 0.282 & 0.760 & 119.421 & 0.000 & & 0.283 & 0.766 & 162.084 & 0.000 & 0.660 & 0.509\\
        & \textit{recall\_qv} & & 0.277 & 0.749 & 119.214 & 0.000 & & 0.277 & 0.757 & 161.941 & 0.000 & 0.674 & 0.500\\
        & \textit{expose\_qv} & & 0.257 & 0.734 & 117.788 & 0.000 & & 0.258 & 0.742 & 161.427 & 0.000 & 0.737 & 0.461\\
        & \textit{click\_qv} & & 0.171 & 0.526 & 81.371 & 0.000 & & 0.172 & 0.535 & 122.720 & 0.000 & 1.025 & 0.305\\
        % \midrule
        \bottomrule
    \end{tabular}
    \label{tb:summaryStatistics}
\end{table}

\subsection{Different Choices of $T_E$ in Experiment 2}
\label{sec:exp2:more}

In the main text, for space reasons we only present the results with $T_E=8,9,10$. Here we provide additional results for the performance of the linear surrogate model given different choices for the duration of the experimental periods $T_E$.

\begin{figure}[h!]
\centering
\includegraphics[width=0.9\linewidth]
{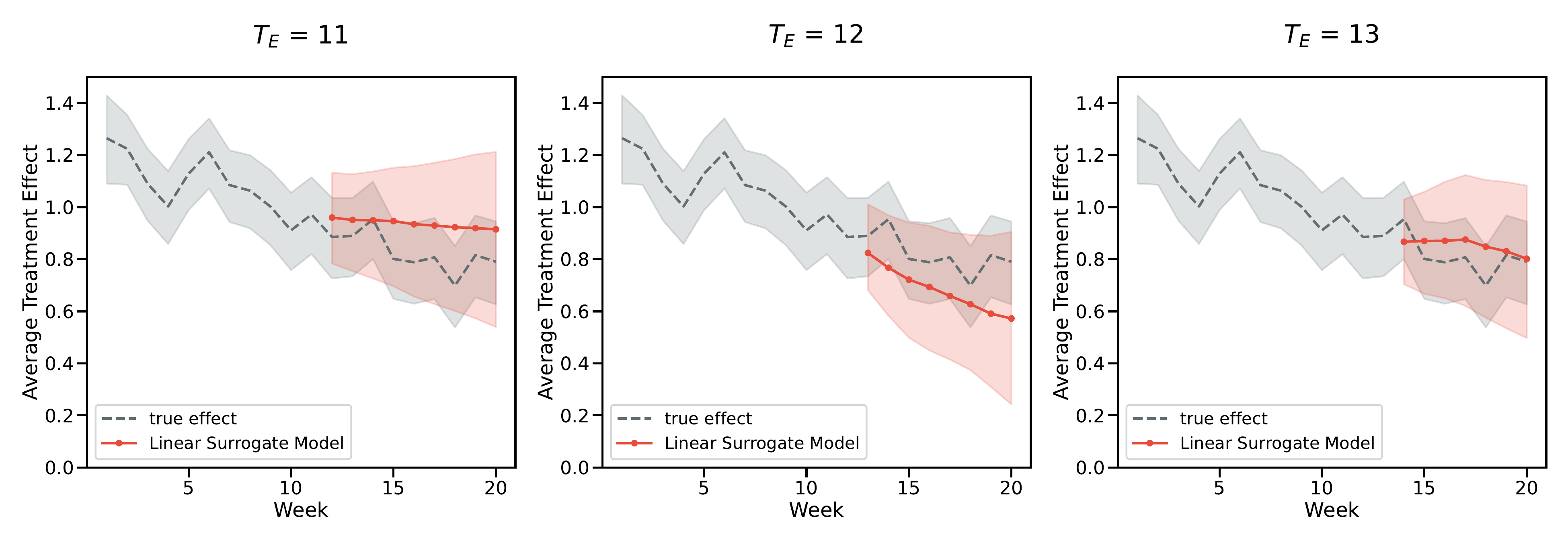}
\caption{Long-term effect estimation for Experiment 2 with $T_E=11,12,13$ for the linear surrogate model }
{\footnotesize \textit{Note}: Grey dashed curves represent the true average treatment effect on \textit{search\_uv} from week 1 to week 20. Solid red curves represent the estimated effects with the linear surrogate model. Shadows indicate 95\% confidence intervals. The three panels represent the scenarios when we use the first $T_E$ weeks as the experimental period and the last $T_F$ weeks as the future period.}
\label{fig:searchDiscoveryNesting11_13}
\end{figure}

Here, we select $T_E=11, 12, 13$ and present the corresponding  results  in Figure~\ref{fig:searchDiscoveryNesting11_13}. As shown in the figures, our estimates still consistently outperform those of the baseline and effectively capture the decreasing trend of the treatment effects.

\subsection{\cite{athey2019surrogate}}\label{sec:appendix:Athey}
As discussed in the main text, \cite{athey2019surrogate} addresses a fundamentally different problem from the one we explore. The detailed comparison is listed in Table~\ref{tab:comparison_ourwork_athey}.

\begin{table}[ht]
\centering
\scriptsize
\begin{tabular}{p{0.15\linewidth}p{0.4\linewidth}p{0.4\linewidth}}
\hline
 & \textbf{Our Work} & \textbf{Athey et al.} \\
\hline
\textbf{Focus} & Long-term effects of \textbf{long-term treatments} & Long-term effects of \textbf{short-term treatments} \\[1em]
\textbf{Motivation} & Evaluating the impact of persistent interventions over extended periods, common in A/B testing scenarios & Estimating long-term impacts when outcomes are not immediately observable \\[1em]
\textbf{Technical Framework} & Sequential longitudinal framework accounting for interventions, surrogates, and outcomes at each period & Surrogate index built on historical data without consideration of multiple time periods \\[1em]
\textbf{Estimation Approach} & Proposed three novel estimation strategies for long-term treatments & Surrogate index method using observed short-term outcomes to predict long-term effects \\[1em]
\textbf{Type of Treatment} & Continuous or ongoing treatments & One-shot or short-term treatments \\[1em]
\textbf{Applications} & Continuous healthcare management, ongoing software updates, etc. & Short-lived campaigns such as promotional events, educational workshops, marketing campaigns, policy changes \\[1em]
\hline
\end{tabular}
\caption{Comparative Analysis between Our Work and Athey et al.}
\label{tab:comparison_ourwork_athey}
\end{table}

In a nutshell, our paper focuses on the persistent long-term effects of treatments that extend beyond the observed experimental period into future periods. This involves analyzing both the ``carryover effect''—the impact of treatments during the experimental periods as measured in subsequent future periods—and the ``direct effect''—the impact of treatments during future periods measured concurrently.

In contrast, the methodology presented by \cite{athey2019surrogate} is specifically designed to capture the ``carryover effect,'' but does not account for the effects of treatments administered in future periods. This limitation arises from the method's inability to identify a set of surrogate variables that can capture the effects of future treatments from the perspective of current experimental periods. In other words, the approach described in \cite{athey2019surrogate} is not appropriate for estimating these long-term effects due to its failure to meet the necessary surrogacy assumptions. Our paper addresses this gap by establishing a longitudinal surrogate model premised on the longitudinal surrogacy assumption.

That said, we provide the estimation results of applying 
\cite{athey2019surrogate} to Experiment 1 as an illustration. 
We present the estimation results in Figure~\ref{fig:searchHistoryAthey}. 
We observe that
the estimates by \cite{athey2019surrogate} consistently  underestimate the treatment effects, yielding values much lower than the ground truth. This discrepancy can be explained by the approach's exclusion of the direct treatment effects in the future period $T_F$. In this experiment, the effects during $T_F$ are positive, and their approach, which only considers the carryover effects from the previous treatment, tends to underestimate the treatment effects in $T_F$. 

\begin{figure}[h!]
\centering
\includegraphics[width=1\linewidth]
{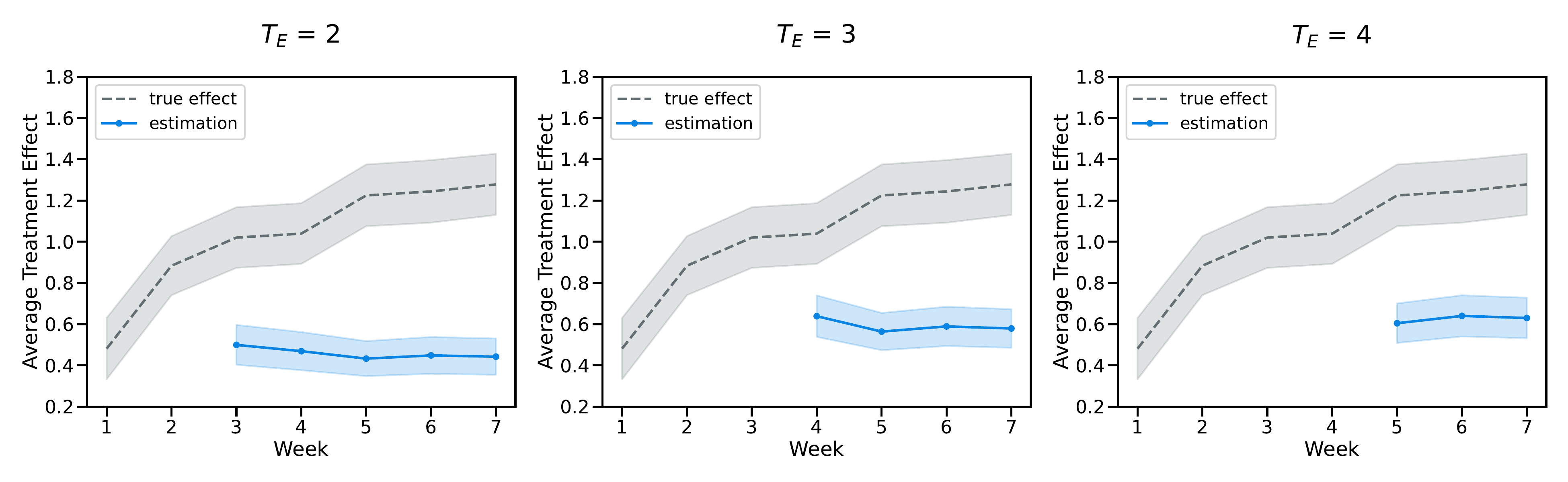}
\caption{Estimated effect for 7-week treatment with $T_E$ weeks observed data under the traditional surrogate model (\cite{athey2019surrogate}) for Experiment 1}
{\footnotesize \textit{Note}: Grey dashed curves represent the true average treatment effect on \textit{search\_uv} from week 1 to week 7. Solid blue curves represent the estimated effects with the traditional surrogate model introduced in \cite{athey2019surrogate}. Shadows indicate 95\% confidence intervals. The three panels represent the scenarios when we use the first $T_E$ weeks as the experimental period and the last $T_F$ weeks as the future period.}
\label{fig:searchHistoryAthey}
\end{figure}

We include the following toy example to further illustrate the difference between our method and Athey's method.

\begin{example}[Athey's method cannot handle consistent treatment effect]
Consider a simple setting with no covariate $\bm{X_i}$ for subject $i$ involved. The objective is to estimate the treatment effects on the primary outcome $Y_i$, where the total treatment duration spans $T = T_E + T_F$ time periods, with data collected from periods $1$ to $T_E$. Suppose that $T_E = 2\cdot T_F$, and the relationship between the primary outcome $Y_i$ and the surrogate variable $S_i$ can be functionally expressed as

\begin{equation*}
    \begin{cases}
      & Y_{iT}(\bm{1}_{T_E}, \bm{0}_{T_F}) = \beta_0 + \beta_1\cdot S_{iT_E}\\
      & Y_{iT}(\bm{0}_{T_E}, \bm{1}_{T_F}) = \alpha\\
      & Y_{iT}(\bm{1}_{T}) = \alpha + \beta_0 + \beta_1\cdot S_{iT_E}\\
      & Y_{iT}(\bm{0}_{T}) = 0
    \end{cases}       
\end{equation*}

It is evident that $S_i$ fully captures the treatment effects on $Y_i$ during periods $1$ to $T_E$. Following the instruction proposed in \cite{athey2019surrogate}, Athey’s method constructs a surrogate index $\hat{Y}_{iT}$ by leveraging historical data spanning $T$ periods based on the observed surrogate variable $S_{iT_E}$. By estimating a linear regression model using OLS, it derives
$$\tilde{Y}_{iT} = \beta_0 + \beta_1\cdot S_{iT_E} + \epsilon_i $$
and the estimate of the treatment eﬀect based on the surrogate index is then calculated as
$$\hat{\tau}_T = \frac{1}{N_1}\sum^{N}_{i=1}\tilde{Y}_{iT}\bm{W}_i - \frac{1}{N_0}\sum^{N}_{i=1}\tilde{Y}_{iT}(1-\bm{W}_i)$$
which systematically underestimates the true effect with a bias of magnitude approximately $\alpha$.

On the other hand, our method first discretize the $T$ periods into three intervals, $\Delta t_1,\Delta t_2$ and $\Delta t_3$, with each spanning $T_F$ periods. Leveraging data from the first $T_E$ ($\Delta t_1$ and $\Delta t_2$) periods, we construct a longitudinal surrogate model by fitting the following linear regression
$$\hat{Y}_{iT_E} = \alpha + \beta_0 + \beta_1\cdot S_{i\Delta t_1} + e_i $$
Plug-in the estimation of $S_{iT_E}$ (the process is omitted) the  $\hat{Y}_{iT}$ can be obtained as
$$\hat{Y}_{iT} = \alpha + \beta_0 + \beta_1\cdot \hat{S}_{iT_E} + u_i$$

Using the same estimator of the treatment eﬀect $\hat{\tau}_T$ stated above, we can see that the estimation with our longitudinal surrogate model is unbiased.

\end{example}

In summary, this result showcases that it is inappropriate to directly apply \cite{athey2019surrogate} to our distinct setting. It is necessary to employ a longitudinal perspective and develop the assumptions and theoretical conclusions applied in our study.

\subsection{Robustness Check}
\label{sec:appendix:robust}
\subsubsection{Model Performance on Heterogeneous Groups}
\label{sec:appendix:robust:hetero}

Although we showcase two experiments in our main results, 
we can partition the sample into several subgroups based on covariates and examine the performance of our approach when applied to each subgroup.
Note that due to confidentiality concerns, we cannot leverage the demographic information of users involved in the experiment. Instead, we used the value of pre-treatment variables, (i.e., \textit{search\_uv}), to categorize sample into different groups. We divided the users into five groups based on the [20, 40, 60, 80] percentiles of the average \textit{search\_uv} value from seven days before the experiment began.

Figures~\ref{fig:searchHistoryHTE} and~\ref{fig:searchDiscoveryHTE} present the estimated long-term treatment effect for each heterogeneous group in Experiment 1 and Experiment 2 respectively. The value in the last column, denoted as $X$, represents a subgroup ($X=0,1,...,4$ corresponding to 0-20\%, 20-40\%, 40-60\%, 60-80\%, and 80-100\%). A larger value signifies a higher average of \textit{search\_uv} during the pre-treatment period, indicating a more active user group. We also present the biases and MSEs for each subgroups in Tables~\ref{tb:synthetic2MSE1} and \ref{tb:synthetic2MSE2}. From these figures and tables, we observe that the estimation result closely aligns with the true effect across multiple groups, suggesting that our method exhibits considerable robustness.

\begin{figure}[tb]
\centering
\includegraphics[width=0.9\linewidth]
{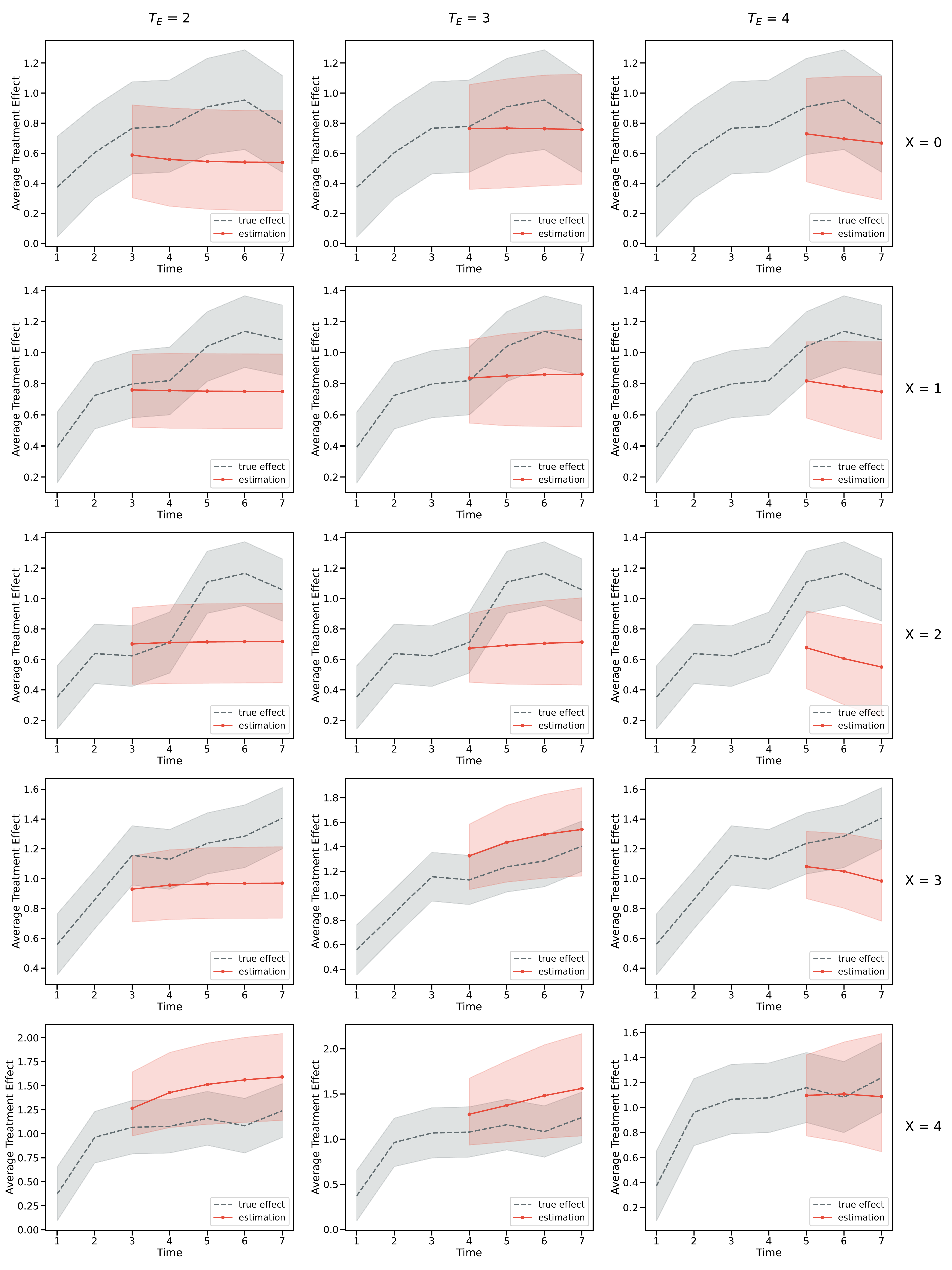}
\caption{Estimated effect using the linear surrogate model on heterogeneous groups for Experiment 1}
{\footnotesize \textit{Note}: Grey dashed curves represent the true average treatment effect on \textit{search\_uv} from week 1 to week 7. Solid red curves represent the estimated effects with the linear surrogate model. Shadows indicate 95\% confidence intervals. The five rows correspond to specific heterogeneous user groups, with a larger value indicating a more active user group. The three panels represent the scenarios when we use the first $T_E$ weeks as the experimental period and the last $T_F$ weeks as the future period.
}
\label{fig:searchHistoryHTE}
\end{figure}

\begin{figure}[tb]
\centering
\includegraphics[width=0.9\linewidth]
{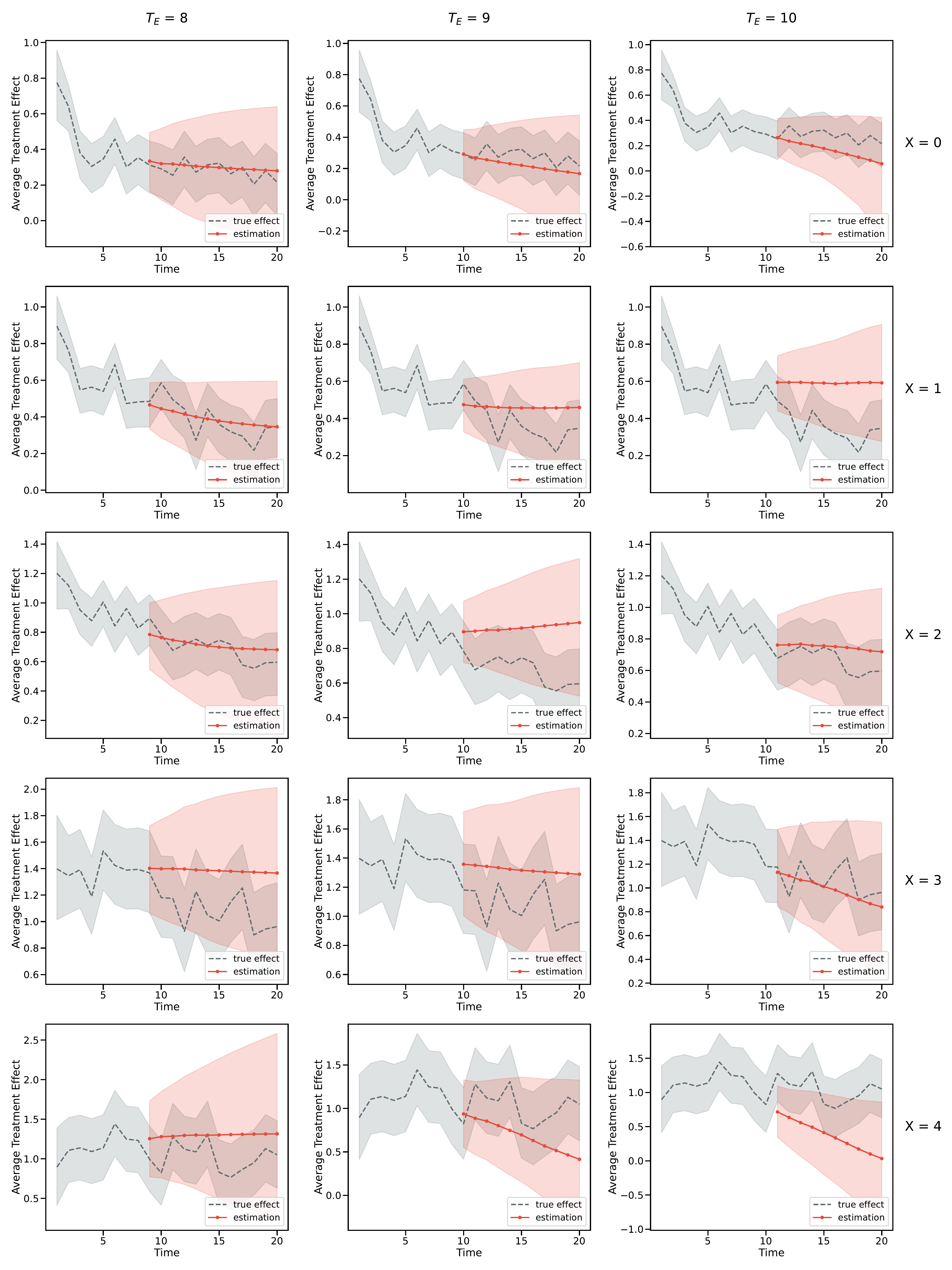}
\caption{Estimated effect using the linear surrogate model on heterogeneous groups for Experiment 2}
{\footnotesize \textit{Note}: Grey dashed curves represent the true average treatment effect on \textit{search\_uv} from week 1 to week 20. Solid red curves represent the estimated effects with the linear surrogate model. The five rows correspond to specific heterogeneous user groups, with a larger value indicating a more active user group. The three panels represent the scenarios when we use the first $T_E$ weeks as the experimental period and the last $T_F$ weeks as the future period.
}
\label{fig:searchDiscoveryHTE}
\end{figure}

\begin{table}[h!]
\centering
\caption{Comparison result between different heterogeneous groups for Bias and MSE for Experiment 1}\footnotesize\label{tb:synthetic2MSE1}

  \begin{tabular}{c c c ccc ccc}
    \toprule
    \multirow{2}{*}{\textbf{Expt.}}&\multirow{2}{*}{\textbf{Groups}}&\multirow{2}{*}{\textbf{Method}}&\multicolumn{3}{c}{\ \textbf{Bias}}&\multicolumn{3}{c}{\textbf{MSE}}\\
      & & & {$\mathbf{T_E=2}$} & {$\mathbf{T_E=3}$} & {$\mathbf{T_E=4}$} & {$\mathbf{T_E=2}$} & {$\mathbf{T_E=3}$} & {$\mathbf{T_E=4}$} \\
      \midrule
   \multirow{10}{*}{Exp 1} & \multirow{2}{*}{$X=0$} & Linear Surrogate Model & 0.286 & 0.096 & 0.188 & 0.253 & 0.192 & 0.220\\
   & & VAR Model (Baseline) & 0.351 & 0.065 & 0.126 & 0.210 & 0.579 & 0.348\\
   & \multirow{2}{*}{$X=1$} & Linear Surrogate Model & 0.222 & 0.177 & 0.305 & 0.332 & 0.390 & 0.481\\
   & & VAR Model (Baseline) & 0.418 & 0.182 & 0.220 & 0.308 & 1.113 & 0.534\\
   & \multirow{2}{*}{$X=2$} & Linear Surrogate Model & 0.252 & 0.314 & 0.500 & 0.566 & 0.605 & 0.911\\
   & & VAR Model (Baseline) & 0.438 & 0.432 & 0.387 & 0.415 & 2.195 & 0.919\\   
   & \multirow{2}{*}{$X=3$} & Linear Surrogate Model & 0.285 & 0.187 & 0.271 & 0.610 & 1.159 & 0.868\\
   & & VAR Model (Baseline) & 0.534 & 0.205 & 0.124 & 0.551 & 4.998 & 1.106\\   
   & \multirow{2}{*}{$X=4$} & Linear Surrogate Model & 0.347 & 0.283 & 0.080 & 2.516 & 3.002 & 2.206\\
   & & VAR Model (Baseline) & 0.458 & 0.171 & 0.273 & 0.984 & 1.763 & 9.570\\
    \bottomrule
  \end{tabular}
\end{table}

\begin{table}[h!]
\centering
\caption{Comparison result between different heterogeneous groups for Bias and MSE for Experiment 2}\label{tb:synthetic2MSE2}
\footnotesize
  \begin{tabular}{c c c ccc ccc}
    \toprule
    \multirow{2}{*}{\textbf{Expt.}}&\multirow{2}{*}{\textbf{Groups}}&\multirow{2}{*}{\textbf{Method}}&\multicolumn{3}{c}{\ \textbf{Bias}}&\multicolumn{3}{c}{\textbf{MSE}}\\
      & & & {$\mathbf{T_E=8}$} & {$\mathbf{T_E=9}$} & {$\mathbf{T_E=10}$} & {$\mathbf{T_E=8}$} & {$\mathbf{T_E=9}$} & {$\mathbf{T_E=10}$} \\
      \midrule
   \multirow{10}{*}{Exp 2} & \multirow{2}{*}{$X=0$} & Linear Surrogate Model & 0.035 & 0.060 & 0.117 & 0.123 & 0.121 & 0.136 \\
   & & VAR Model (Baseline) & 0.035 & 0.046 & 0.087 & 0.134 & 0.128 & 0.135 \\
   & \multirow{2}{*}{$X=1$} & Linear Surrogate Model & 0.060 & 0.112 & 0.239 & 0.133 & 0.154 & 0.225 \\
   & & VAR Model (Baseline) & 0.134 & 0.119 & 0.172 & 0.156 & 0.200 & 0.626 \\
   & \multirow{2}{*}{$X=2$} & Linear Surrogate Model & 0.062 & 0.245 & 0.084 & 0.302 & 0.311 & 0.273 \\
   & & VAR Model (Baseline) & 0.336 & 0.395 & 0.553 & 0.767 & 0.841 & 40.037 \\
   & \multirow{2}{*}{$X=3$} & Linear Surrogate Model & 0.290 & 0.250 & 0.108 & 0.470 & 0.495 & 0.379 \\
   & & VAR Model (Baseline) & 0.807 & 0.105 & 0.434 & 18.614 & 3.558 & 37.813 \\
   & \multirow{2}{*}{$X=4$} & Linear Surrogate Model & 0.283 & 0.356 & 0.668 & 1.523 & 0.957 & 1.522 \\
   & & VAR Model (Baseline) & 0.369 & 0.289 & 1.055 & 5.243 & 5.949 & 5.270 \\
    \bottomrule
  \end{tabular}
\end{table}

\subsubsection{Linear Surrogate Model with Regularization}
\label{sec:appendix:robust:regularization}

In the main context, we employ a linear surrogate model to generate the estimation of long-term effect in both two empirical studies. In reality, one potential challenge of this surrogate model is the curse of dimensionality. Given that in total $\#S$ surrogates are applied, there will be $\#S\times (T_E-1)$ features in the prediction models, and overfitting becomes a concern. To tackle this, we suggest employing the elastic net regularization \citep{zou2005regularization}. This method helps mitigate the effects of irrelevant dimensions and manage multicollinearity, common issues associated with the curse of dimensionality. In the following analysis, we would like to show that the estimation result is still robust given that the regularization method is applied. 

To effectively use elastic net regularization, we must tune two hyperparameters. To do this, we uniformly select 100 potential values within the range of $[0,1]$ for these hyperparameters. We then optimize the two hyperparameters using five-fold cross-validation to identify their most suitable values for the model. This approach allows us to find the best configuration for the linear surrogate model and improve its predictive performance. Figure~\ref{fig:searchHistoryRegularization} and Figure~\ref{fig:searchDiscoveryRegularization} present the result of the estimated long-term effect with the linear surrogate model with the regularization in the two empirical experiments. Comparing them to the estimation results shown in Figure~\ref{fig:searchHistoryNesting} and Figure~\ref{fig:searchDiscoveryNesting} respectively, we can observe that the estimates are very similar, which demonstrate the robustness of our method.

\begin{figure}[tb]
\centering
\includegraphics[width=0.9\linewidth]
{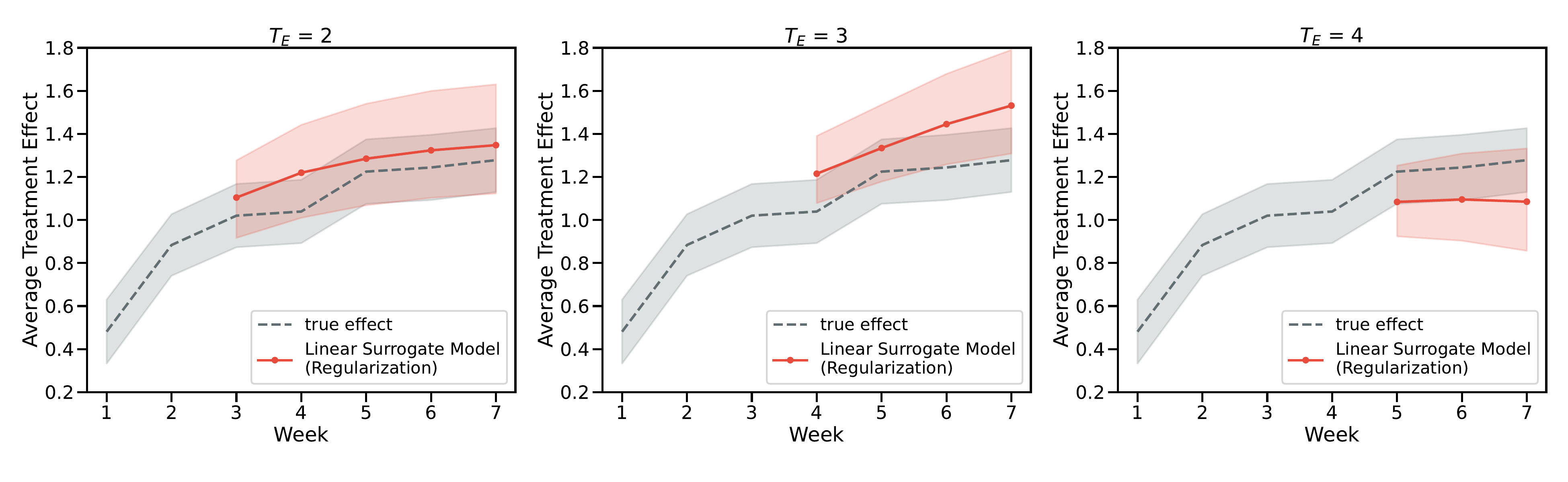}
\caption{Effects of Long-term Treatment using Pure Linear Surrogate Model for Experiment 1}
{\footnotesize \textit{Note}: Grey dashed curves represent the true average treatment effect on \textit{search\_uv} from week 1 to week 7. Solid red curves represent the estimated effects with the linear surrogate model without any regularization term. Shadows indicate 95\% confidence intervals. The three panels represent the scenarios when we use the first $T_E$ weeks as the experimental period and the last $T_F$ weeks as the future period.}
\label{fig:searchHistoryRegularization}
\end{figure}

\begin{figure}[tb]
\centering
\includegraphics[width=0.9\linewidth]
{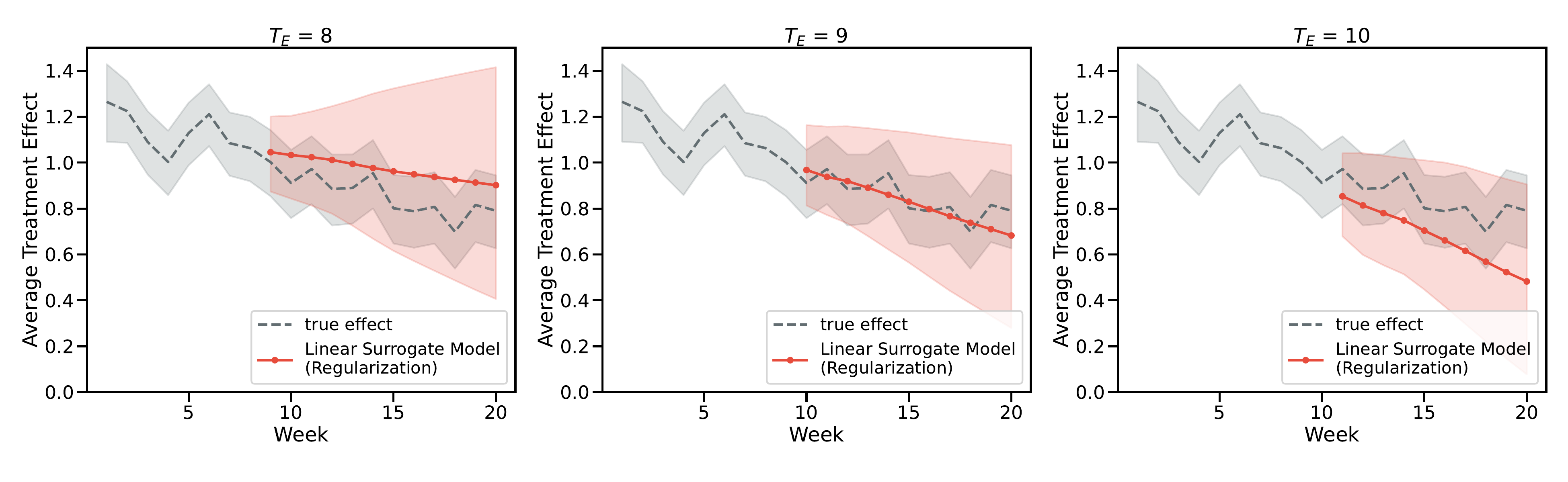}
\caption{Effects of Long-term Treatment using Pure Linear Surrogate Model for Experiment 2}
{\footnotesize \textit{Note}: Grey dashed curves represent the true average treatment effect on \textit{search\_uv} from week 1 to week 20. Solid red curves represent the estimated effects with the linear surrogate model without any regularization term. Shadows indicate 95\% confidence intervals. The three panels represent the scenarios when we use the first $T_E$ weeks as the experimental period and the last $T_F$ weeks as the future period.}
\label{fig:searchDiscoveryRegularization}
\end{figure}

\subsection{Choosing surrogates. }
\label{sec:guidelines:surrogates}

In practical applications, Assumption~\ref{asp:Surrogacy}—the longitudinal surrogacy assumption—is not testable. Past research indicates that bias can arise if this assumption is violated~\citep{athey2019surrogate}. Additionally, previous studies have provided guidelines for selecting surrogates~\citep{duan2021online}. In general, we recommend selecting intermediary metrics that are highly responsive to the treatment and meanwhile capture the diverse facets of user behavior leading to variations in primary outcomes during future periods. 

Both up-stream and down-stream behaviors during experimental periods can influence future primary outcomes. For instance, in the context where the click-through rates of products in an online marketplace serve as the primary outcome, up-stream behaviors might encompass actions such as users' search queries, their navigation patterns through categories, and the related content they peruse before ultimately clicking on a product. On the other hand, down-stream behaviors refer to the subsequent actions taken after clicking on a product, like reading reviews, adding the item to a cart, initiating a purchase, or even sharing the product with others. These up-stream behaviors are reflective of the users' intentions to click on products, and if a treatment can move these initial behaviors, it is plausible that it might impact down-stream outcomes in the future. Conversely, because down-stream behaviors represent deeper outcomes that indicate the realization of the primary outcome, they inherently have a strong predictive power for that primary outcome. In our real-world experiments, we mostly use down-stream behaviors as surrogates and demonstrate their effectiveness. Another potential surrogate to consider is simply the current primary outcome itself, which often exhibits a strong predictive power for the future primary outcome~\citep{deng2013improving}.

To improve the efficiency of selecting surrogates for A/B tests, experimenters in companies can create a library of surrogates for treatments and primary outcomes based on the relevant metrics used in previous experiments. By using similar surrogate groups across related experiments, we can streamline the surrogate selection process and obtain estimates easily. This approach can save time and effort in identifying appropriate surrogates for each new experiment, making the process more efficient.

Choosing either an excessive number of surrogates or too few can pose challenges. Choosing too few might not saturate the causal links between treatment and future primary outcomes and fulfill the longitudinal surrogacy assumption needed for reliable estimates, resulting in biased outcomes. Conversely, selecting an excessive number can lead to the ``curse of dimensionality," increasing the risk of overfitting. For practitioners, we advise starting with as many relevant metrics as feasible and then employing techniques to reduce dimensionality.
In our study, we utilized methods such as the elastic net \citep{zou2005regularization} to eliminate or compress irrelevant dimensions with low predictive power.\footnote{One may also use auto-encoders~\citep{rumelhart1986learning, vincent2008extracting}, a common neural network based dimensionality reduction approach. However, when there are only a moderate number of surrogates, auto-encoders may omit useful information and present less satisfactory performance than the elastic net. We thus only recommend using the elastic net for dimensionality reduction.
}

\subsection{Adopting Non-linear Models}
\label{sec:appendix:otherModelts}

In our main framework, the linear surrogate model is an extension of the longitudinal surrogate model with an additional linearity assumption.
Although it is intuitive to additionally assume  non-linear relationships between surrogates and future surrogates or outcomes (i.e., $\bm{G}$ or $h$ are non-linear functions), its empirical performance is not as satisfactory as linear models. 
Our exploration with practitioners reveals that common non-linear machine learning such as nearest neighbors, random forests, or neural networks, cannot well predict future treatment effects. In addition, due to the non-deterministic nature and the requirement of a large sample size of these models, the model variance is often unaffordably large and cannot produce reasonably small confidence intervals. 

As an illustration, we provide estimation results that are based on  the $k$-nearest neighbors ($k$NN) model.
Empirically, instead of assuming linear function forms for $\bm{g}_t$ and $h_t$, we obtain $\hat{\bm{{g}}}_t(\bm{s}, \bm{x}, \bm{w}_{1:t})$ and 
${\hat{h}}_t(\bm{s}, \bm{x}, \bm{w}_{1:t})$ 
as the averages of the $k$ nearest neighbors in terms of  $\bm{s}$ in the training set. We then take the averages of their primary outcomes ($Y_{it}$) and intermediate metrics ($\bm{S}_{it}$) to obtain $\hat{\bm{{g}}}_t$ and ${\hat{h}}_t$, respectively.

We apply the $k$-nearest neighbors model with $k=20$ to Experiment 1  as an illustrative example. 
As shown in Figure~\ref{fig:searchHistoryKnn}, while the $k$NN based model seems to have confidence intervals that cover the true effects, it has unaffordably wide confidence intervals. 
The explanation is that $k$NN-based estimation is more vulnerable than linear models to local random noises: that is, if there exists a large degree of variation and stochastic in the local nearest neighborhood, estimates of each future primary or intermediate outcomes would exhibit large variance consequently. This would lead to very wide confidence intervals in practice. For instance, the $k$NN-based estimator exhibits an average MSE value that is eight times greater than that of the constant extrapolation baseline (CEB) in Experiment 1.
This conclusion also holds when $k=10$ or $40$.
Note that this issue does not only persist in $k$NN, but also in many other machine learning models such as random forests or neural networks. Therefore, we recommend using linear models instead throughout this work.

\begin{figure}[tb]
\centering
\includegraphics[width=0.9\linewidth]
{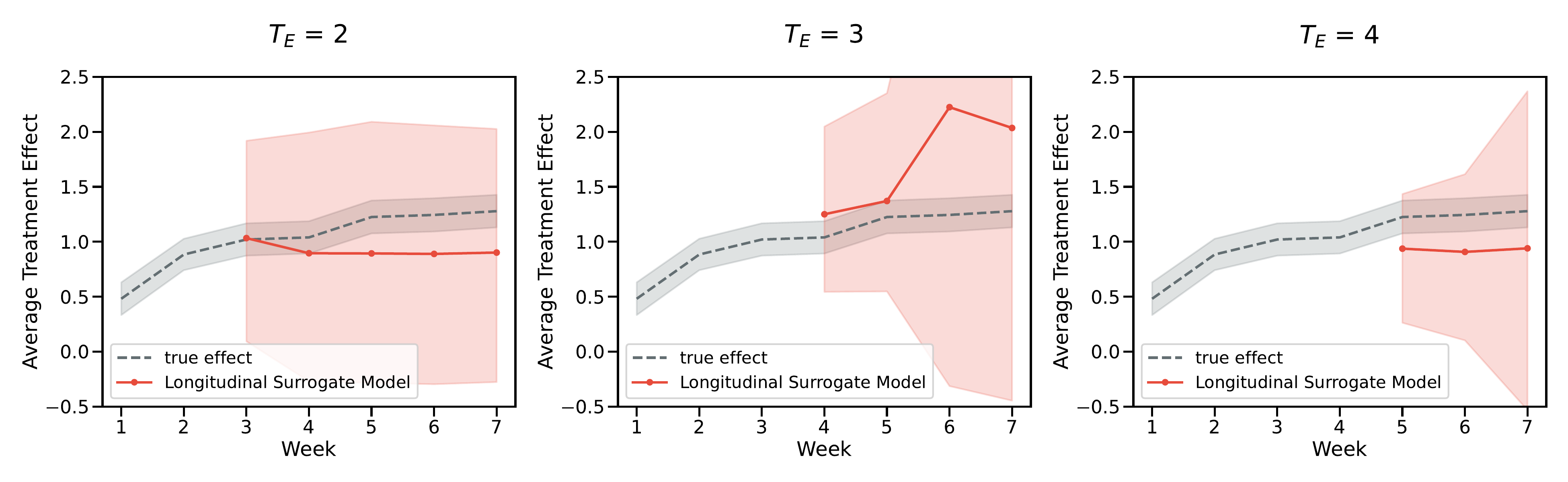}
\caption{Estimated effect for 7-week treatment with $T_E$ weeks observed data using $k$NN model for Experiment 1}
{\footnotesize \textit{Note}: Grey dashed curves represent the true average treatment effect on \textit{search\_uv} from week 1 to week 7. Solid red curves represent the estimated effects with the $k$NN Model. Shadows indicate 95\% confidence intervals. The three panels represent the scenarios when we use the first $T_E$ weeks as the experimental period and the last $T_F$ weeks as the future period.}
\label{fig:searchHistoryKnn}
\end{figure}

\subsection{Estimation with Pre-treatment Variables}
\label{sec:appendix:estimation:covariate}

In the main text, we present the estimation results using a linear surrogate model on two empirical experiments, highlighting our approach's performance in terms of bias and MSE. To ensure a fair comparison with baseline models, we only use surrogate variables to construct all models mentioned. Additionally, we discover that including pre-treatment variables can reduce estimation variance. 

We construct an enhanced linear surrogate model that includes both the surrogate variables used in the main text and a pre-treatment variable: the value of \textit{search\_uv} from a week before the experiment started, to estimate the long-term treatment effect. Tables~\ref{tb:searchHistoryMSEaddtional} and~\ref{tb:searchDiscoveryMSEaddtional} provide a comparison of the enhanced linear surrogate model with the previous linear surrogate model. We observe that incorporating pre-treatment variables effectively reduces variance and MSE compared to the results from not incorporating them. Bias is largely similar as we include pre-treatment variables in experimental data.

\begin{table}[h]
\centering
\caption{Comparison result between different methods in terms of Bias and MSE for Experiment 1}
\label{tb:searchHistoryMSEaddtional}\scriptsize
  \begin{tabular}{c ccc ccc}
    \toprule
    \multirow{2}{*}{\textbf{Method}}&\multicolumn{3}{c}{\textbf{Bias}}&\multicolumn{3}{c}{\textbf{MSE}}\\
      & {$\mathbf{T_E=2}$} & {$\mathbf{T_E=3}$} & {$\mathbf{T_E=4}$} & {$\mathbf{T_E=2}$} & {$\mathbf{T_E=3}$} & {$\mathbf{T_E=4}$} \\
      \midrule
    \makecell{Linear Surrogate Model\\ (with pre-treatment variables)} 	& 0.105 & 0.068	& 0.205
    & 0.228	& 0.216 & 0.294\\
    \makecell{Linear Surrogate Model\\ (without pre-treatment variables)}& 0.087 & 0.199 & 0.165 & 0.327 & 0.324 & 0.314 \\
    \bottomrule
  \end{tabular}
\end{table}

\begin{table}[h]
\centering
\caption{Comparison result between different methods in terms of Bias and MSE for Experiment 2}\scriptsize
  \begin{tabular}{c ccc ccc}
    \toprule
    \multirow{2}{*}{\textbf{Method}}&\multicolumn{3}{c}{\textbf{Bias}}&\multicolumn{3}{c}{\textbf{MSE}}\\
      & {$\mathbf{T_E=8}$} & {$\mathbf{T_E=9}$} & {$\mathbf{T_E=10}$} & {$\mathbf{T_E=8}$} & {$\mathbf{T_E=9}$} & {$\mathbf{T_E=10}$} \\
      \midrule
    \makecell{Linear Surrogate Model\\ (with pre-treatment variables)} & 0.110 & 0.047 & 0.146 & 0.212 & 0.127 & 0.186 \\
    \makecell{Linear Surrogate Model\\ (without pre-treatment variables)} & 0.098 & 0.048 & 0.158 & 0.233 & 0.136 & 0.201 \\
    \bottomrule
  \end{tabular}
\label{tb:searchDiscoveryMSEaddtional}
\end{table}

\subsection{Bootstrap with Full Sample}
\label{sec:appendix:bootstrap}

In the main analysis, we perform statistical inference with a bootstrap procedure that resamples 50\% of users (with replacement) in each replicate—striking a pragmatic balance between accuracy and computational cost. To verify robustness, we also apply a full‑sample bootstrap that resamples the entire user set with replacement. The corresponding treatment‑effect estimates for the two empirical experiments are shown in Figures~\ref{fig:searchHistoryFullSample} and~\ref{fig:searchDiscoveryFullSample}. The full‑sample results closely match those from the subsampling bootstrap, indicating that our findings are largely insensitive to the specific resampling scheme. Subsampling only half of the users therefore offers substantial computational savings without compromising inference reliability.

\begin{figure}[h]
\centering
\includegraphics[width=0.9\linewidth]
{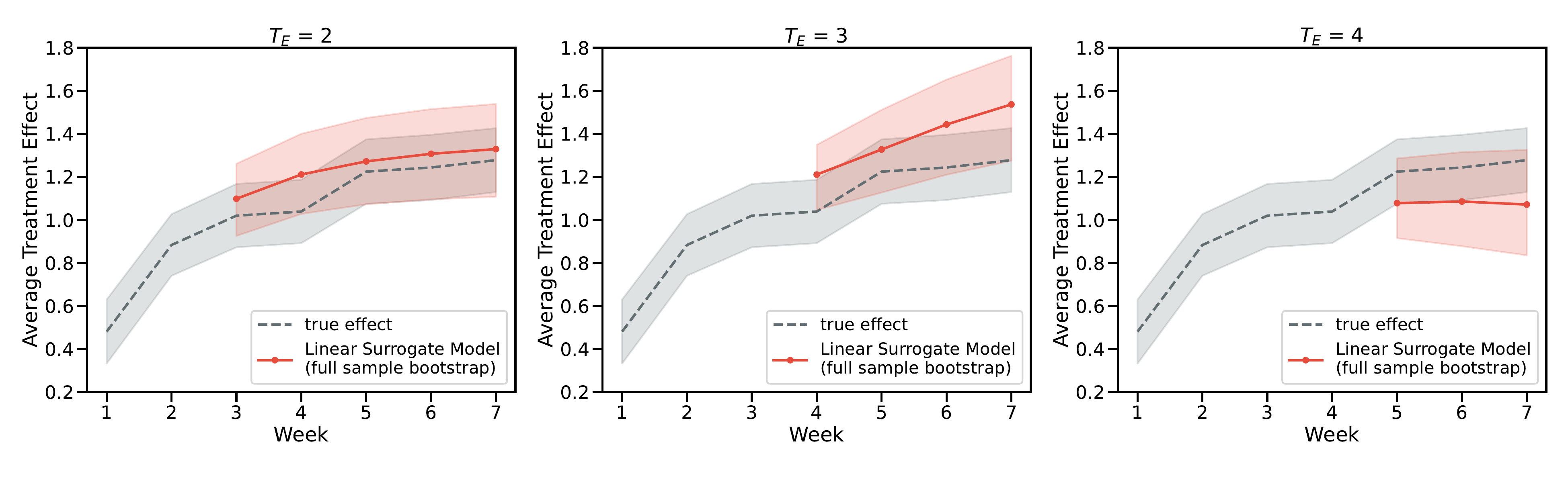}
\caption{Estimated Effects of Long-term Treatment with full sample bootstrap or Experiment 1}
{\footnotesize \textit{Note}: Grey dashed curves represent the true average treatment effect on \textit{search\_uv} from week 1 to week 7. Solid red curves represent the estimated effects with the linear surrogate model. Shadows indicate 95\% confidence intervals. The three panels represent the scenarios when we use the first $T_E$ weeks as the experimental period and the last $T_F$ weeks as the future period.}
\label{fig:searchHistoryFullSample}
\end{figure}

\begin{figure}[h]
\centering
\includegraphics[width=0.9\linewidth]
{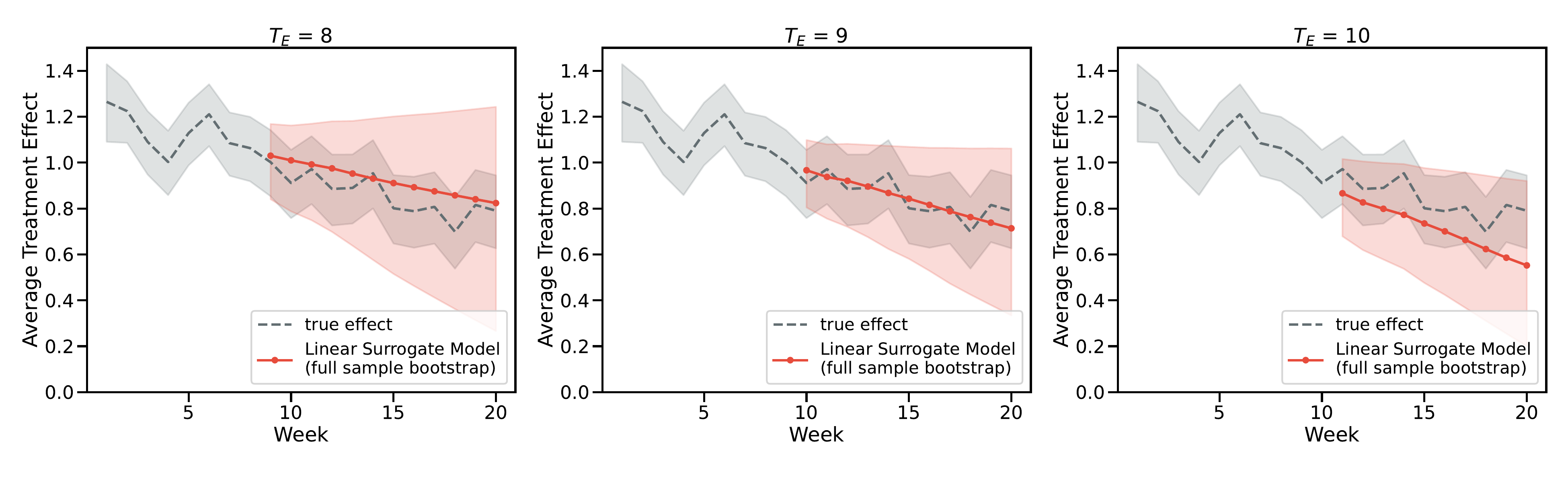}
\caption{Estimated Effects of Long-term Treatment with full sample bootstrap or Experiment 2}
{\footnotesize \textit{Note}: Grey dashed curves represent the true average treatment effect on \textit{search\_uv} from week 1 to week 20. Solid red curves represent the estimated effects with the linear surrogate model. Shadows indicate 95\% confidence intervals. The three panels represent the scenarios when we use the first $T_E$ weeks as the experimental period and the last $T_F$ weeks as the future period.}
\label{fig:searchDiscoveryFullSample}
\end{figure}

\section{Validation of Comparability Assumption}
\label{sec:appendix:comparability}

\subsection{Testing results for Assumption~\ref{asp:Comparability}}
\label{sec:appendix:comparabilityDirect}

We provide example results on directly testing Assumption~\ref{asp:Comparability} in both Empirical Experiment 1 and Empirical Experiment 2. The detailed approach is outlined in Section~\ref{sec:validation}. For each experiment, we use the same observable experimental time period \(T_E\) from the main text, testing on various combinations of \(t\) and \(t'\). We set \(\delta=1\), matching the criteria on five surrogate variables one period before \(t\) and \(t'\). The ``$\bm{\#}$Tests" column indicates the number of tests on the possible sets of values in $S_{i,t - \delta}$ ($S_{i,t' - \delta}$). In Experiment 1, the average number of tests conducted for each stratum is around six thousand, while in Experiment 2 it's over nine thousand. Among these tests, some of them only include a small number of matched pairs. Across tests under all conditions, about 7\% have a p-value less than 0.05, and about 11\% have a p-value less than 0.1 in both experiments. The result suggests that there is no strong evidence that Assumption~\ref{asp:Comparability} is violated, though stronger support for this claim is still needed. 

\begin{table}[h]
\centering
\caption{Testing results of the comparability assumption for two empirical experiments}\scriptsize
\begin{tabular}{cccccccccc}
\toprule
\textbf{Experiment} & \textbf{Groups} & $\bm{t}$ & $\bm{t'}$ & $\bm{\#}$\textbf{Tests} & $\bm{\# p<0.1}$ & $\bm{\# p<0.05}$ & $\bm{p<0.1 (\%)}$ & $\bm{p<0.05 (\%)}$ \\
\midrule
\multirow{6}{*}{Experiment 1} & \multirow{3}{*}{Control} & 2 & 3 & 6,115 & 693 & 456 & 11.33 & 7.46 \\
& & 2 & 4 & 5,994 & 673 & 440 & 11.23 & 7.34 \\
& & 3 & 4 & 6,015 & 663 & 413 & 11.02 & 6.87 \\
\cmidrule{3-9}
& \multirow{3}{*}{Treatment} & 2 & 3 & 6,219 & 710 & 454 & 11.42 & 7.30 \\
& & 2 & 4 & 6,253 & 753 & 472 & 12.04 & 7.55 \\
& & 3 & 4 & 6,215 & 724 & 466 & 11.65 & 7.50 \\
\midrule
\multirow{12}{*}{Experiment 2} & \multirow{6}{*}{Control} & 7 & 8 & 9,085 & 997 & 617 & 10.97 & 6.79 \\
& & 7 & 9 & 9,335 & 1,112 & 699 & 11.91 & 7.49 \\
& & 7 & 10 & 9,393 & 1,048 & 661 & 11.16 & 7.04 \\
& & 8 & 9 & 9,314 & 990 & 612 & 10.63 & 6.57 \\
& & 8 & 10 & 9,376 & 1,026 & 629 & 10.94 & 6.71 \\
& & 9 & 10 & 9,504 & 1,078 & 687 & 11.34 & 7.23 \\
\cmidrule{3-9}
& \multirow{6}{*}{Treatment} & 7 & 8 & 9,550 & 1,070 & 675 & 11.20 & 7.07 \\
& & 7 & 9 & 9,536 & 1,076 & 708 & 11.28 & 7.42 \\
& & 7 & 10 & 9,617 & 1,086 & 679 & 11.29 & 7.06 \\
& & 8 & 9 & 9,605 & 1,016 & 638 & 10.58 & 6.64 \\
& & 8 & 10 & 9,665 & 1,116 & 695 & 11.55 & 7.19 \\
& & 9 & 10 & 9,655 & 1,126 & 704 & 11.66 & 7.29 \\
\bottomrule
\end{tabular}
\end{table}

Considering the dynamic nature of real-world environments, including market fluctuations, economic shocks, and product iterations, maintaining the comparability assumption can sometimes be challenging. However, the robustness of our method may still render it applicable. In practice, instead of strictly testing the comparability assumption, we suggest using weaker and alternative criteria to determine the feasibility of our approach. We explore such an alternative in the subsequent sections.

\subsection{Parallel Trends Assumption}
\label{sec:ParallelTrend}

Given that we would like to generalize the comparability assumption, we need to define the surrogate index at different time periods.
For any positive integer $\delta \in \bN^+$, we denote a generalized version of the longitudinal surrogate index and the pivot index at time $t$ as follows.

\begin{definitionp}{\ref{defn:SurrogateIndex}'}[Generalized Longitudinal Surrogate Index]
\label{defn:GeneralizedSurrogateIndex:appendix}
For any $t \in [T]$, $\delta \in \bN^+$, $\bm{s} \in \bS$, $\bm{x} \in \bX$, $\bm{w}_{t-\delta+1:t} \in \{\bm{0}_{\delta}, \bm{1}_{\delta}\}$, the generalized surrogate index is the conditional expectation of the primary outcome at time $t$, given the surrogate outcomes at time $t-\delta$, the pre-treatment variables, and the treatment assignments, i.e., 
\begin{align*}
h^{\delta}_t(\bm{s}, \bm{x}, \bm{w}_{t-\delta+1:t}) = \bE_{\cF} \left[ Y_{it} \vert \bm{S}_{i(t-\delta)} = \bm{s}, \bm{X}_i = \bm{x}, \bm{W}_{i,t-\delta+1:t} = \bm{w}_{t-\delta+1:t}\right],
\end{align*}
where the expectation is taken over $Y_{it}$.
\end{definitionp}

\begin{definitionp}{\ref{defn:PivotIndex}'}[Generalized Pivot Index]
\label{defn:GeneralizedPivotIndex:appendix}
For any $t \in [T]$, $\delta \in \bN^+$, $\bm{s} \in \bS$, $\bm{x} \in \bX$, $\bm{w}_{t-\delta+1:t} \in \{\bm{0}_{\delta}, \bm{1}_{\delta}\}$, the generalized pivot index is a vector of the conditional expectations of the surrogate outcomes at time $t$, given the surrogate outcomes at time $t-\delta$, the pre-treatment variables, and the treatment assignments, i.e.,
\begin{align*}
\bm{g}^{\delta}_t(\bm{s}, \bm{x}, \bm{w}_{t-\delta+1:t}) = \bE_{\cF} \left[ \bm{S}_{it} \vert \bm{S}_{i(t-\delta)} = \bm{s}, \bm{X}_i = \bm{x}, \bm{W}_{i,t-\delta+1:t} = \bm{w}_{t-\delta+1:t}\right],
\end{align*}
where the expectation is taken over $\bm{S}_{it}$.
Moreover, we denote the conditional surrogate outcomes at time $t$, given the surrogate outcomes at time $t-\delta$, the pre-treatment variables, and the treatment assignments, to be
\begin{align*}
\bm{G}^{\delta}_t(\bm{s}, \bm{x}, \bm{w}_{1:t}) \sim \bm{S}_{it} \vert \bm{S}_{i(t-\delta)} = \bm{s}, \bm{X}_i = \bm{x}, \bm{W}_{i,t-\delta+1:t} = \bm{w}_{t-\delta+1:t}.
\end{align*}
\end{definitionp}

The generalized longitudinal surrogate index and the generalized pivot index represent a broader and more flexible range of surrogate index functions.
When $\delta = t$, the generalized longitudinal surrogate index function (Definition~\ref{defn:GeneralizedSurrogateIndex:appendix}) reduces to the surrogate index function (Definition~\ref{defn:SurrogateIndex}), and the generalized pivot index function (Definition~\ref{defn:GeneralizedPivotIndex:appendix}) reduces to the pivot index function (Definition~\ref{defn:PivotIndex}), that is, $h_t(\bm{s}, \bm{x}, \bm{w}_{1:t}) = h^{t}_t(\bm{s}, \bm{x}, \bm{w}_{1:t})$ and $\bm{g}_t(\bm{s}, \bm{x}, \bm{w}_{t-\delta+1:t}) = \bm{g}^{t}_t(\bm{s}, \bm{x}, \bm{w}_{t-\delta+1:t})$. 
Using Definitions~\ref{defn:GeneralizedSurrogateIndex:appendix} and~\ref{defn:GeneralizedPivotIndex:appendix} , we introduce an assumption that relaxes the comparability assumption (Assumption~\ref{asp:Comparability}).

\begin{assumptionp}{\ref{asp:Comparability}'}[Parallel Trends]
\label{asp:ParallelTrends:appendix}
The difference in the conditional expectation of the primary outcomes across time periods under the treatment condition is equal to that under the control condition. 
For any $t,t' \in [T]$ and any positive integer $\delta \in \bN^+$, 
\begin{align*}
h^{\delta}_t(\bm{s}_{1}, \bm{x}_i, \bm{1}_{\delta}) - h^{\delta}_{t'}(\bm{s}_{1}, \bm{x}_i, \bm{1}_{\delta}) = h^{\delta}_t(\bm{s}_{0}, \bm{x}_i, \bm{0}_{\delta}) - h^{\delta}_{t'}(\bm{s}_{0}, \bm{x}_i, \bm{0}_{\delta}).
\end{align*}
\end{assumptionp}

\begin{assumptionp}{\ref{asp:Comparability}''}[Extended Parallel Trends]
\label{asp:ParallelTrends:Technical}
For any $t,t', u_0, ..., u_{K-1}\in [T]$ and any positive integer $\delta \in \bN^+$, 
\begin{multline*}
h_{u_0}(\bm{g}^{\delta}_t(\bm{s}_{1}, \bm{x}_i, \bm{1}_{\delta}), \bm{x}_i, \bm{1}_{u_0}) - h_{u_0}(\bm{g}^{\delta}_{t'}(\bm{s}_{1}, \bm{x}_i, \bm{1}_{\delta}), \bm{x}_i, \bm{1}_{u_0}) \\
= h_{u_0}(\bm{g}^{\delta}_t(\bm{s}_{0}, \bm{x}_i, \bm{0}_{\delta}), \bm{x}_i, \bm{0}_{u_0}) - h_{u_0}(\bm{g}^{\delta}_{t'}(\bm{s}_{0}, \bm{x}_i, \bm{0}_{\delta}), \bm{x}_i, \bm{0}_{u_0}),
\end{multline*}
\begin{multline*}
h_{u_0}(\bm{g}_{u_1}(\bm{g}^{\delta}_t(\bm{s}_{1}, \bm{x}_i, \bm{1}_{\delta}), \bm{x}_i, \bm{1}_{u_1}), \bm{x}_i, \bm{1}_{u_0}) - h_{u_0}(\bm{g}_{u_1}(\bm{g}^{\delta}_{t'}(\bm{s}_{1}, \bm{x}_i, \bm{1}_{\delta}), \bm{x}_i, \bm{1}_{u_1}), \bm{x}_i, \bm{1}_{u_0}) \\
= h_{u_0}(\bm{g}_{u_1}(\bm{g}^{\delta}_t(\bm{s}_{0}, \bm{x}_i, \bm{0}_{\delta}), \bm{x}_i, \bm{0}_{u_1}), \bm{x}_i, \bm{0}_{u_0}) - h_{u_0}(\bm{g}_{u_1}(\bm{g}^{\delta}_{t'}(\bm{s}_{0}, \bm{x}_i, \bm{0}_{\delta}), \bm{x}_i, \bm{0}_{u_1}), \bm{x}_i, \bm{0}_{u_0}),
\end{multline*}
\begin{multline*}
h_{u_0}(\bm{g}_{u_1}( ... \bm{g}_{u_{K-1}}(\bm{g}^{\delta}_t(\bm{s}_{1}, \bm{x}_i, \bm{1}_{\delta}), \bm{x}_i, \bm{1}_{u_{K-1}}) ..., \bm{x}_i, \bm{1}_{u_1}), \bm{x}_i, \bm{1}_{u_0}) \\
\qquad \qquad - h_{u_0}(\bm{g}_{u_1}( ... \bm{g}_{u_{K-1}}(\bm{g}^{\delta}_{t'}(\bm{s}_{1}, \bm{x}_i, \bm{1}_{\delta}), \bm{x}_i, \bm{1}_{u_{K-1}}) ..., \bm{x}_i, \bm{1}_{u_1}), \bm{x}_i, \bm{1}_{u_0}) \\
= h_{u_0}(\bm{g}_{u_1}( ... \bm{g}_{u_{K-1}}(\bm{g}^{\delta}_t(\bm{s}_{0}, \bm{x}_i, \bm{0}_{\delta}), \bm{x}_i, \bm{0}_{u_{K-1}}) ..., \bm{x}_i, \bm{0}_{u_1}), \bm{x}_i, \bm{0}_{u_0}) \qquad \qquad \\
- h_{u_0}(\bm{g}_{u_1}( ... \bm{g}_{u_{K-1}}(\bm{g}^{\delta}_{t'}(\bm{s}_{0}, \bm{x}_i, \bm{0}_{\delta}), \bm{x}_i, \bm{0}_{u_{K-1}}) ..., \bm{x}_i, \bm{0}_{u_1}), \bm{x}_i, \bm{0}_{u_0}).
\end{multline*}
\end{assumptionp}

We first explore the relationship between the parallel trends assumption and the comparability assumption.

\begin{proposition}
\label{prop:Sufficient:appendix}
Assumption~\ref{asp:Comparability} is a sufficient condition of Assumption~\ref{asp:ParallelTrends:appendix}.
\end{proposition}

\proof{Proof of Proposition~\ref{prop:Sufficient:appendix}.}

For any $t,t' \in [T]$, $\delta \in \bN^+$, $\bm{s}_0,\bm{s}_1 \in \bS$, $\bm{x} \in \bX$, $\bm{w}_{t-\delta+1:t} \in \{\bm{0}_{\delta}, \bm{1}_{\delta}\}$, Assumption~\ref{asp:Comparability} implies that
\begin{align*}
h^{\delta}_t(\bm{s}_{1}, \bm{x}_i, \bm{1}_{\delta}) = h^{\delta}_{t'}(\bm{s}_{1}, \bm{x}_i, \bm{1}_{\delta}),\ 
h^{\delta}_t(\bm{s}_{0}, \bm{x}_i, \bm{0}_{\delta}) = h^{\delta}_{t'}(\bm{s}_{0}, \bm{x}_i, \bm{0}_{\delta}).
\end{align*}
Therefore,
\begin{align*}
h^{\delta}_t(\bm{s}_{1}, \bm{x}_i, \bm{1}_{\delta}) - h^{\delta}_{t'}(\bm{s}_{1}, \bm{x}_i, \bm{1}_{\delta}) = 0 = h^{\delta}_t(\bm{s}_{0}, \bm{x}_i, \bm{0}_{\delta}) - h^{\delta}_{t'}(\bm{s}_{0}, \bm{x}_i, \bm{0}_{\delta}),
\end{align*}
which leads to Assumption~\ref{asp:ParallelTrends:appendix}.
\hfill \halmos
\endproof

We have proved Theorem~\ref{thm:IdentificationL2} under Assumptions~\ref{asp:Surrogacy},~\ref{asp:Comparability}, and~\ref{asp:Linearity} in the previous section. Here we show that without the comparability assumption, the conclusion of the Theorem~\ref{thm:IdentificationL2} continue to hold given that the generalized surrogate index function is linear and the parallel trends assumption is met. We first rewrite the linearity of surrogates assumption with the generalized longitudinal surrogate index.

\begin{assumptionp}{\ref{asp:Linearity}'}[Extended Linearity of Surrogates]
\label{asp:Linearity2}
\begin{enumerate}
\item The generalized surrogate index function is linear with respect to the surrogates, i.e., $\forall t \in [T],\ \delta \in \bN^+,\ d\in\{0,1,...,D\},\ \bm{x}\in\bX,\ \bm{w}_{t-\delta+1:t}\in\{\bm{0}_{\delta}, \bm{1}_{\delta}\}$, there exists $\alpha^{\delta}_{d,t}(\bm{x}, \bm{w}_{t-\delta+1:t})$, such that
\begin{align*}
h^{\delta}_t(\bm{s}, \bm{x}, \bm{w}_{t-\delta+1:t}) = \alpha^{\delta}_{0,t}(\bm{x}, \bm{w}_{t-\delta+1:t}) + \sum_{d=1}^D s_d \cdot \alpha^{\delta}_{d,t}(\bm{x}, \bm{w}_{t-\delta+1:t}).
\end{align*}
\item The generalized pivot index function is linear with respect to the surrogates, i.e., $\forall t \in [T],\ \delta \in \bN^+,\ d\in\{0,1,...,D\},\ \bm{x}\in\bX,\ \bm{w}_{t-\delta+1:t}\in\{\bm{0}_{\delta}, \bm{1}_{\delta}\}$, there exists $\beta^{\delta}_{d,d',t}(\bm{x}, \bm{w}_{t-\delta+1:t})$, such that for each $d\in[D]$,
\begin{align*}
g^{\delta}_{t,d}(\bm{s}, \bm{x}, \bm{w}_{t-\delta+1:t}) = \beta^{\delta}_{d,0,t}(\bm{x}, \bm{w}_{t-\delta+1:t}) + \sum_{d'=1}^D s_d \cdot \beta^{\delta}_{d,d',t}(\bm{x}, \bm{w}_{t-\delta+1:t}),
\end{align*}
where $g^{\delta}_{t,d}(\bm{s}, \bm{x}, \bm{w}_{t-\delta+1:t})$ stands for the $d$-th component of $\bm{g}^{\delta}_t(\bm{s}, \bm{x}, \bm{w}_{t-\delta+1:t})$ the pivot index.
\end{enumerate}
\end{assumptionp}

With the above alternative assumptions, we can rewrite the Theorem~\ref{thm:IdentificationL2} as follows.

\begin{theoremp}{\ref{thm:IdentificationL2}'}[Linear Surrogate Model]
\label{thm:IdentificationL22}
When $T_E = T_F$, assume Assumption~\ref{asp:ParallelTrends:appendix}; when $T_E < T_F$, assume Assumptions~\ref{asp:ParallelTrends:appendix} and~\ref{asp:ParallelTrends:Technical}. In addition, assume Assumptions~\ref{asp:Surrogacy} and~\ref{asp:Linearity2}, where Assumption~\ref{asp:Surrogacy} holds for $\bT = \{t_1,t_2,...,t_K\}$. Then, the average effect of long-term treatments on the primary outcome is equal to the following expression, 
\begin{multline*}
\tau_T = \bE_{\cF} \Big[ h_{\Delta t_{K+1}} \big( \bm{g}_{\Delta t_K}( ... \bm{g}_{\Delta t_1}(\bm{S}_{i0}, \bm{X}_i, \bm{1}_{\Delta t_1}) ... ,\bm{X}_i, \bm{1}_{\Delta t_K}), \bm{X}_i, \bm{1}_{\Delta t_{K+1}} \big) \Big] \\
- \bE_{\cF} \Big[ h_{\Delta t_{K+1}}\left( \bm{g}_{\Delta t_K}( ... \bm{g}_{\Delta t_1}(\bm{S}_{i0}, \bm{X}_i, \bm{0}_{\Delta t_1}) ... ,\bm{X}_i, \bm{0}_{\Delta t_K}), \bm{X}_i, \bm{0}_{\Delta t_{K+1}} \right) \Big],
\end{multline*}
where the expectation is taken over $\bm{S}_{i0}, \bm{X}_i$.
\end{theoremp}

Prior to proving Theorem~\ref{thm:IdentificationL22}, we first re-state the first half of Theorem~\ref{thm:IdentificationL22} in the following Lemma~\ref{lem:SpecialCaseL2}.
Lemma~\ref{lem:SpecialCaseL2} illustrates the main idea of the proof for easier understanding.

\begin{lemma}
\label{lem:SpecialCaseL2}
Consider the special case when $T_E = T_F$.
Under Assumptions~\ref{asp:Surrogacy}, ~\ref{asp:ParallelTrends:appendix}, and~\ref{asp:Linearity2}-(i), where Assumption~\ref{asp:Surrogacy} holds for $\bT = \{T_E\}$, the average effect of long-term treatments on the primary outcome is equal to the following expression, 
\begin{align*}
\tau_T =  \bE_{\cF} \Big[h_{T_E}\left( \bm{g}_{T_E}(\bm{S}_{i0}, \bm{X}_i, \bm{1}_{T_E}), \bm{X}_i, \bm{1}_{T_E} \right)\Big] - \bE_{\cF} \Big[h_{T_E}\left( \bm{g}_{T_E}(\bm{S}_{i0}, \bm{X}_i, \bm{0}_{T_E}), \bm{X}_i, \bm{0}_{T_E} \right)\Big],
\end{align*}
where the expectation is taken over $\bm{S}_{i0}, \bm{X}_i$.
\end{lemma}

\subsection{Proof of Lemma~\ref{lem:SpecialCaseL2} with Parallel Trends Assumption}
\proof{Proof of Lemma~\ref{lem:SpecialCaseL2}.}

From the definition of the causal effect,
\begin{align*}
\tau_T = \bE_{\cF}\bigg[ Y_{iT}(\bm{1}_{T}) - Y_{iT}(\bm{0}_{T}) \bigg],
\end{align*}
we start with the first part $\bE_{\cF}\left[ Y_{iT}(\bm{1}_{T})\right]$.
\begin{align}
& \ \bE_{\cF}\left[ Y_{iT}(\bm{1}_{T}) \right] \nonumber\\
= & \sum_{\bm{x}_i \in \bX, \bm{s}_{i0} \in \bS} \bE_{\cF}\left[ Y_{iT} \bigg\vert \bm{W}_{i,1:T} = \bm{1}_T, \bm{x}_i, \bm{s}_{i0} \right] \Pr(\bm{x}_i, \bm{s}_{i0}) \nonumber\\
= & \sum_{\bm{x}_i \in \bX, \bm{s}_{i0} \in \bS} \sum_{\bm{s} \in \bS} \bE_{\cF}\left[ Y_{iT}  \bigg\vert  \bm{S}_{i T_E} = \bm{s}, \bm{W}_{i,1:T} = \bm{1}_{T}, \bm{x}_i, \bm{s}_{i0} \right] \cdot \Pr\left(\bm{S}_{i T_E} = \bm{s} \bigg\vert \bm{W}_{i,1:T} = \bm{1}_{T}, \bm{x}_i, \bm{s}_{i0}\right) \cdot \Pr(\bm{x}_i, \bm{s}_{i0})
\label{eqn:expansion1}
\end{align}

Note that, for any $\bm{s}, \bm{x}_i$,
\begin{align}
& \ \bE_{\cF}\left[ Y_{iT} \bigg\vert \bm{S}_{i T_E} = \bm{s}, \bm{W}_{i,1:T}
= \bm{1}_{T}, \bm{x}_i, \bm{s}_{i0} \right] \nonumber\\
= & \bE_{\cF}\left[ Y_{iT} \bigg\vert \bm{S}_{i T_E} = \bm{s}, \bm{W}_{i,T_E+1:T} = \bm{1}_{T_F}, \bm{x}_i \right] \nonumber\\
= & \bE_{\cF}\left[Y_{i T_E} \bigg\vert \bm{S}_{i 0} = \bm{s}, \bm{W}_{i,1:T_E} = \bm{1}_{T_E}, \bm{x}_i \right] \nonumber\\
& \hspace{3cm} + \left(\bE_{\cF}\left[ Y_{iT} \bigg\vert \bm{S}_{i T_E} = \bm{s}, \bm{W}_{i,T_E+1:T} = \bm{1}_{T_F}, \bm{x}_i \right] - \bE_{\cF}\left[Y_{i T_E} \bigg\vert \bm{S}_{i 0} = \bm{s}, \bm{W}_{i,1:T_E} = \bm{1}_{T_E}, \bm{x}_i \right]\right) \nonumber\\
= & h^{T_E}_{T_E}(\bm{s}, \bm{x}_i, \bm{1}_{T_E}) + \left( h^{T_F}_{T}\Big(\bm{s}, \bm{x}_i, \bm{1}_{T_F}\Big) - h^{T_E}_{T_E}\Big(\bm{s}, \bm{x}_i, \bm{1}_{T_E}\Big)\right),
\label{eqn:reform1}
\end{align}
where the first equality is due to Assumption~\ref{asp:Surrogacy} when $t=T_E, t'=T$; the second equality is because we add and subtract the same term; the third equality uses short-hand notation for the generalized longitudinal surrogate index.

Putting \eqref{eqn:reform1} back into \eqref{eqn:expansion1}, we have
\begin{align}
& \ \bE_{\cF}\left[ Y_{iT}(\bm{1}_{T}) \right] \nonumber\\
= & \sum_{\bm{x}_i \in \bX, \bm{s}_{i0} \in \bS} \sum_{\bm{s} \in \bS} h^{T_E}_{T_E}(\bm{s}, \bm{x}_i, \bm{1}_{T_E}) \cdot \Pr\left(\bm{S}_{i T_E} = \bm{s} \bigg\vert \bm{W}_{i,1:T} = \bm{1}_{T}, \bm{x}_i, \bm{s}_{i0}\right) \cdot \Pr(\bm{x}_i, \bm{s}_{i0}) \nonumber\\
& + \sum_{\bm{x}_i \in \bX, \bm{s}_{i0} \in \bS} \sum_{\bm{s} \in \bS} \left( h^{T_F}_{T}\Big(\bm{s}, \bm{x}_i, \bm{1}_{T_F}\Big) - h^{T_E}_{T_E}\Big(\bm{s}, \bm{x}_i, \bm{1}_{T_E}\Big)\right) \cdot \Pr\left(\bm{S}_{i T_E} = \bm{s} \bigg\vert \bm{W}_{i,1:T} = \bm{1}_{T}, \bm{x}_i, \bm{s}_{i0}\right) \cdot \Pr(\bm{x}_i, \bm{s}_{i0})
\label{eqn:combine1}
\end{align}

Note that, the treatment assignment remains the same throughout the entire horizon, which means that,
\begin{align*}
\Pr\left(\bm{S}_{i T_E} = \bm{s}  \bigg\vert \bm{W}_{i,1:T} = \bm{1}_{T}, \bm{x}_i, \bm{s}_{i0}\right) = \Pr\left(\bm{S}_{i T_E} = \bm{s}  \bigg\vert \bm{W}_{i,1:T_E} = \bm{1}_{T_E}, \bm{x}_i, \bm{s}_{i0}\right).
\end{align*}

Using the above expression, \eqref{eqn:combine1} can be rewritten as
\begin{align}
\bE_{\cF}\left[ Y_{iT}(\bm{1}_{T}) \right] = & \ \sum_{\bm{x}_i \in \bX, \bm{s}_{i0} \in \bS} \sum_{\bm{s} \in \bS} h^{T_E}_{T_E}\Big(\bm{s}, \bm{x}_i, \bm{1}_{T_E}\Big) \cdot \Pr\left(\bm{S}_{i T_E} = \bm{s}  \bigg\vert \bm{W}_{i,1:T_E} = \bm{1}_{T_E}, \bm{x}_i, \bm{s}_{i0}\right) \cdot \Pr(\bm{x}_i, \bm{s}_{i0}) \nonumber\\
& + \sum_{\bm{x}_i \in \bX, \bm{s}_{i0} \in \bS} \sum_{\bm{s} \in \bS} h^{T_F}_{T}\Big(\bm{s}, \bm{x}_i, \bm{1}_{T_F}\Big) \cdot \Pr\left(\bm{S}_{i T_E} = \bm{s}  \bigg\vert \bm{W}_{i,T_E+1:T} = \bm{1}_{T_F}, \bm{x}_i, \bm{s}_{i0}\right) \cdot \Pr(\bm{x}_i, \bm{s}_{i0}) \nonumber\\
& - \sum_{\bm{x}_i \in \bX, \bm{s}_{i0} \in \bS} \sum_{\bm{s} \in \bS} h^{T_E}_{T_E}\Big(\bm{s}, \bm{x}_i, \bm{1}_{T_E}\Big) \cdot \Pr\left(\bm{S}_{i T_E} = \bm{s}  \bigg\vert \bm{W}_{i,1:T_E} = \bm{1}_{T_E}, \bm{x}_i, \bm{s}_{i0}\right) \cdot \Pr(\bm{x}_i, \bm{s}_{i0}) \nonumber\\
= & \ \sum_{\bm{x}_i \in \bX, \bm{s}_{i0} \in \bS} \bE_{\cF}\left[h^{T_E}_{T_E}(\bm{G}_{T_E}(\bm{s}_{i0}, \bm{x}_i, \bm{1}_{T_E}), \bm{x}_i, \bm{1}_{T_E})\right] \cdot \Pr(\bm{x}_i, \bm{s}_{i0}) \nonumber\\
& + \sum_{\bm{x}_i \in \bX, \bm{s}_{i0} \in \bS} \bE_{\cF}\left[h^{T_F}_{T}(\bm{G}_{T_E}(\bm{s}_{i0}, \bm{x}_i, \bm{1}_{T_E}), \bm{x}_i, \bm{1}_{T_F})\right] \cdot \Pr(\bm{x}_i, \bm{s}_{i0}) \nonumber\\
& - \sum_{\bm{x}_i \in \bX, \bm{s}_{i0} \in \bS} \bE_{\cF}\left[h^{T_E}_{T_E}(\bm{G}_{T_E}(\bm{s}_{i0}, \bm{x}_i, \bm{1}_{T_E}), \bm{x}_i, \bm{1}_{T_E})\right] \cdot \Pr(\bm{x}_i, \bm{s}_{i0}) 
\label{eqn:treatment1}
\end{align}

Denote $\bm{G}_{T_E} = (G_{T_E, 1}, ..., G_{T_E, D})$ and $\bm{g}_{T_E} = (g_{T_E, 1}, ..., g_{T_E, D})$. 
Then for any pre-treatment variables and surrogates $\bm{x}_i, \bm{s}_{i0}$, we can write $\bE_{\cF}\left[h^{T_F}_{T}(\bm{G}_{T_E}(\bm{s}_{i0}, \bm{x}_i, \bm{1}_{T_E}), \bm{x}_i, \bm{1}_{T_F})\right]$ as
\begin{align*}
\bE_{\cF}\left[h^{T_F}_{T}(\bm{G}_{T_E}(\bm{s}_{i0}, \bm{x}_i, \bm{1}_{T_E}), \bm{x}_i, \bm{1}_{T_F})\right] = &\ \bE_{\cF}\left[ \alpha^{T_F}_{0,t}(\bm{x}_i, \bm{1}_{T_F}) + \sum_{d=1}^D G_{T_E, d}(\bm{s}_{i0}, \bm{x}_i, \bm{1}_{T_E}) \cdot \alpha^{T_F}_{d,t}(\bm{x}_i, \bm{1}_{T_F}) \right] \\
= &\ \alpha^{T_F}_{0,t}(\bm{x}_i, \bm{1}_{T_F}) + \sum_{d=1}^D \bE_{\cF}\left[G_{T_E, d}(\bm{s}_{i0}, \bm{x}_i, \bm{1}_{T_E})\right] \cdot \alpha^{T_F}_{d,t}(\bm{x}_i, \bm{1}_{T_F}) \\
= &\ \alpha^{T_F}_{0,t}(\bm{x}_i, \bm{1}_{T_F}) + \sum_{d=1}^D g_{T_E, d}(\bm{s}_{i0}, \bm{x}_i, \bm{1}_{T_E}) \cdot \alpha^{T_F}_{d,t}(\bm{x}_i, \bm{1}_{T_F}) \\
= &\ h^{T_F}_{T}(\bm{g}_{T_E}(\bm{s}_{i0}, \bm{x}_i, \bm{1}_{T_E}), \bm{x}_i, \bm{1}_{T_F}),
\end{align*}
where the first equality is due to Assumption~\ref{asp:Linearity2}; the second equality is due to linearity of expectation; the third equality is due to Definition~\ref{defn:GeneralizedPivotIndex:appendix}.

Similarly, we have
\begin{align*}
\bE_{\cF}\left[h^{T_E}_{T_E}(\bm{G}_{T_E}(\bm{s}_{i0}, \bm{x}_i, \bm{1}_{T_E}), \bm{x}_i, \bm{1}_{T_E})\right] = h^{T_E}_{T_E}(\bm{g}_{T_E}(\bm{s}_{i0}, \bm{x}_i, \bm{1}_{T_E}), \bm{x}_i, \bm{1}_{T_E})
\end{align*}

Putting the above expressions in~\eqref{eqn:treatment1}, and noting that $h^{T_E}_{T_E}(\cdot, \cdot, \cdot) = h_{T_E}(\cdot, \cdot, \cdot)$, we have
\begin{multline*}
\bE_{\cF}\left[ Y_{iT}(\bm{1}_{T}) \right] 
= \bE_{\cF} \Big[h_{T_E}\left( \bm{g}_{T_E}(\bm{S}_{i0}, \bm{X}_i, \bm{1}_{T_E}), \bm{X}_i, \bm{1}_{T_E} \right)\Big] \\
+ \sum_{\bm{x}_i \in \bX, \bm{s}_{i0} \in \bS} \Big(h^{T_F}_{T}(\bm{g}_{T_E}(\bm{s}_{i0}, \bm{x}_i, \bm{1}_{T_E}), \bm{x}_i, \bm{1}_{T_F}) - h^{T_E}_{T_E}(\bm{g}_{T_E}(\bm{s}_{i0}, \bm{x}_i, \bm{1}_{T_E}), \bm{x}_i, \bm{1}_{T_E})\Big) \cdot \Pr(\bm{x}_i, \bm{s}_{i0}) 
\end{multline*}

Similarly, 
\begin{multline*}
\bE_{\cF}\left[ Y_{iT}(\bm{0}_{T}) \right] 
= \bE_{\cF} \Big[h_{T_E}\left( \bm{g}_{T_E}(\bm{S}_{i0}, \bm{X}_i, \bm{0}_{T_E}), \bm{X}_i, \bm{0}_{T_E} \right)\Big] \\
+ \sum_{\bm{x}_i \in \bX, \bm{s}_{i0} \in \bS} \Big(h^{T_F}_{T}(\bm{g}_{T_E}(\bm{s}_{i0}, \bm{x}_i, \bm{0}_{T_E}), \bm{x}_i, \bm{0}_{T_F}) - h^{T_E}_{T_E}(\bm{g}_{T_E}(\bm{s}_{i0}, \bm{x}_i, \bm{0}_{T_E}), \bm{x}_i, \bm{0}_{T_E})\Big) \cdot \Pr(\bm{x}_i, \bm{s}_{i0}) 
\end{multline*}

Combining both parts,
\begin{align*}
\tau_T = &\ \bE_{\cF} \Big[h_{T_E}\left( \bm{g}_{T_E}(\bm{S}_{i0}, \bm{X}_i, \bm{1}_{T_E}), \bm{X}_i, \bm{1}_{T_E} \right)\Big] - \bE_{\cF} \Big[h_{T_E}\left( \bm{g}_{T_E}(\bm{S}_{i0}, \bm{X}_i, \bm{0}_{T_E}), \bm{X}_i, \bm{0}_{T_E} \right)\Big] \\
& + \sum_{\bm{x}_i \in \bX, \bm{s}_{i0} \in \bS} \Big(h^{T_F}_{T}(\bm{g}_{T_E}(\bm{s}_{i0}, \bm{x}_i, \bm{1}_{T_E}), \bm{x}_i, \bm{1}_{T_F}) - h^{T_E}_{T_E}(\bm{g}_{T_E}(\bm{s}_{i0}, \bm{x}_i, \bm{1}_{T_E}), \bm{x}_i, \bm{1}_{T_E})\Big) \cdot \Pr(\bm{x}_i, \bm{s}_{i0}) \\
& - \sum_{\bm{x}_i \in \bX, \bm{s}_{i0} \in \bS} \Big(h^{T_F}_{T}(\bm{g}_{T_E}(\bm{s}_{i0}, \bm{x}_i, \bm{0}_{T_E}), \bm{x}_i, \bm{0}_{T_F}) - h^{T_E}_{T_E}(\bm{g}_{T_E}(\bm{s}_{i0}, \bm{x}_i, \bm{0}_{T_E}), \bm{x}_i, \bm{0}_{T_E})\Big) \cdot \Pr(\bm{x}_i, \bm{s}_{i0}) \\
= &\ \bE_{\cF} \Big[h_{T_E}\left( \bm{g}_{T_E}(\bm{S}_{i0}, \bm{X}_i, \bm{1}_{T_E}), \bm{X}_i, \bm{1}_{T_E} \right)\Big] - \bE_{\cF} \Big[h_{T_E}\left( \bm{g}_{T_E}(\bm{S}_{i0}, \bm{X}_i, \bm{0}_{T_E}), \bm{X}_i, \bm{0}_{T_E} \right)\Big],
\end{align*}
where the equality is due to Assumption~\ref{asp:ParallelTrends:appendix} when $t=T,\ t'=T_E, \delta=T_E=T_F$, and we view $\bm{s}_1 = \bm{g}_{T_E}(\bm{s}_{i0}, \bm{x}_i, \bm{1}_{T_E})$ and $\bm{s}_0 = \bm{g}_{T_E}(\bm{s}_{i0}, \bm{x}_i, \bm{0}_{T_E})$.
\hfill \halmos
\endproof

\subsection{Proof of Theorem~\ref{thm:IdentificationL22} with Parallel Trends Assumption}

Lemma~\ref{lem:SpecialCaseL2} is a special case of Theorem~\ref{thm:IdentificationL2} when $T_E=T_F$. 
Below we iteratively employ the same approach to prove Theorem~\ref{thm:IdentificationL22}.

\proof{Proof of Theorem~\ref{thm:IdentificationL22}.}
From the definition,
\begin{align*}
\tau_T = \bE_{\cF}\bigg[ Y_{iT}(\bm{1}_{T}) - Y_{iT}(\bm{0}_{T}) \bigg],
\end{align*}
We start from the first part $\bE_{\cF}\left[ Y_{iT}(\bm{1}_{T})\right]$, which can be expressed as
\begin{align}
& \ \bE_{\cF}\left[ Y_{iT}(\bm{1}_{T}) \right] \nonumber\\
= & \sum_{\bm{x}_i \in \bX, \bm{s}_{i0} \in \bS} \bE_{\cF}\left[ Y_{iT} \bigg\vert \bm{W}_{i,1:T} = \bm{1}_T, \bm{x}_i, \bm{s}_{i0} \right] \Pr(\bm{x}_i, \bm{s}_{i0}) \nonumber\\
= & \sum_{\bm{x}_i \in \bX, \bm{s}_{i0} \in \bS} \sum_{\bm{s}_{t_K} \in \bS} \bE_{\cF}\left[ Y_{iT}  \bigg\vert  \bm{S}_{i t_K} = \bm{s}_{t_K}, \bm{W}_{i,1:T} = \bm{1}_{T}, \bm{x}_i, \bm{s}_{i0} \right] \cdot \Pr\left(\bm{S}_{i t_K} = \bm{s}_{t_K} \bigg\vert \bm{W}_{i,1:T} = \bm{1}_{T}, \bm{x}_i, \bm{s}_{i0}\right) \cdot \Pr(\bm{x}_i, \bm{s}_{i0})
\label{eqn:expansion}
\end{align}

We divide the following proof into two stages. 
First, we reformulate the term $\bE_{\cF}\left[ Y_{iT}(\bm{1}_{T})\right]$ by iteratively adding and subtracting the same term. 
Second, we use Assumption~\ref{asp:Linearity2} (i.e., the linearity assumption) to further simplify the expression of $\bE_{\cF}\left[ Y_{iT}(\bm{1}_{T})\right]$, and apply Assumptions~\ref{asp:ParallelTrends:appendix} and~\ref{asp:ParallelTrends:Technical} (i.e., the parallel trends assumption) to conclude the proof.

\vspace{0.2cm}
\noindent \textbf{\textit{Step 1.}} 

Note that,
\begin{align}
& \ \Pr\left(\bm{S}_{i t_K} = \bm{s}_{t_K} \bigg\vert \bm{W}_{i,1:T} = \bm{1}_{T}, \bm{x}_i, \bm{s}_{i0}\right) \nonumber \\
= & \sum_{\bm{s}_{t_{K-1}}, \bm{s}_{t_{K-2}}, ..., \bm{s}_{t_{2}}, \bm{s}_{t_1} \in \bS} 
\Pr\left(\bm{S}_{i t_K} = \bm{s}_{t_K} \bigg\vert \bm{W}_{i,1:t_K} = \bm{1}_{t_K}, \bm{S}_{i t_{K-1}} = \bm{s}_{t_{K-1}}, \bm{x}_i, \bm{s}_{i0}\right) \cdot \nonumber \\
& \hspace{3cm} \Pr\left(\bm{S}_{i t_{K-1}} = \bm{s}_{t_{K-1}} \bigg\vert \bm{W}_{i,1:t_{K-1}} = \bm{1}_{t_{K-1}}, \bm{S}_{i t_{K-2}} = \bm{s}_{t_{K-2}}, \bm{x}_i, \bm{s}_{i0}\right) \cdot \nonumber \\
& \hspace{3cm} \ldots \cdot \Pr\left(\bm{S}_{i t_2} = \bm{s}_{t_2} \bigg\vert \bm{W}_{i,1:t_2} = \bm{1}_{t_2}, \bm{S}_{i t_{1}} = \bm{s}_{t_{1}}, \bm{x}_i, \bm{s}_{i0}\right) \cdot \Pr\left(\bm{S}_{i t_{1}} = \bm{s}_{t_{1}} \bigg\vert \bm{W}_{i,1:t_{1}} = \bm{1}_{t_{1}}, \bm{x}_i, \bm{s}_{i0}\right) \nonumber \\
= & \sum_{\bm{s}_{t_{K-1}}, \bm{s}_{t_{K-2}}, ..., \bm{s}_{t_{2}}, \bm{s}_{t_1} \in \bS} 
\Pr\left(\bm{S}_{i t_K} = \bm{s}_{t_K} \bigg\vert \bm{W}_{i,t_{K-1}+1:t_K} = \bm{1}_{\Delta t_K}, \bm{S}_{i t_{K-1}} = \bm{s}_{t_{K-1}}, \bm{x}_i\right) \cdot \nonumber \\
& \hspace{3cm} \Pr\left(\bm{S}_{i t_{K-1}} = \bm{s}_{t_{K-1}} \bigg\vert \bm{W}_{i,t_{K-2}+1:t_{K-1}} = \bm{1}_{\Delta t_{K-1}}, \bm{S}_{i t_{K-2}} = \bm{s}_{t_{K-2}}, \bm{x}_i \right) \cdot \nonumber \\
& \hspace{3cm} \ldots \cdot \Pr\left(\bm{S}_{i t_2} = \bm{s}_{t_2} \bigg\vert \bm{W}_{i,t_1+1:t_2} = \bm{1}_{\Delta t_2}, \bm{S}_{i t_{1}} = \bm{s}_{t_{1}}, \bm{x}_i \right) \cdot \Pr\left(\bm{S}_{i t_{1}} = \bm{s}_{t_{1}} \bigg\vert \bm{W}_{i,1:t_{1}} = \bm{1}_{\Delta t_{1}}, \bm{x}_i, \bm{s}_{i0}\right), \label{eqn:proof:thm:Probabilities2_2}
\end{align}
where the first equality is using the law of total probability;
the second is because when any unit is assigned to the treatment group, it is always in the treatment group, that is,
\begin{align*}
\Pr\left(\bm{W}_{i,1:t_{k}} = \bm{1}_{t_{k}} \bigg\vert\bm{W}_{i,t_{k}+1:t_{k+1}} = \bm{1}_{\Delta t_{k+1}}\right) = 1.
\end{align*}

Putting the above expression \eqref{eqn:proof:thm:Probabilities2_2} into \eqref{eqn:expansion}, we have
\begin{align}
& \ \bE_{\cF}\left[ Y_{iT}(\bm{1}_{T}) \right] \nonumber \\
= & \ \sum_{\bm{x}_i \in \bX, \bm{s}_{i0} \in \bS} \sum_{\bm{s}_{t_K}, ..., \bm{s}_{t_1} \in \bS} h^{\Delta t_{K+1}}_{T}(\bm{s}_{t_K}, \bm{x}_i, \bm{1}_{\Delta t_{K+1}}) \cdot \Pr\left(\bm{S}_{i t_{K-1}} = \bm{s}_{t_{K-1}} \bigg\vert \bm{W}_{i,t_{K-1}+1:t_K} = \bm{1}_{\Delta t_K}, \bm{S}_{i t_{K-1}} = \bm{s}_{t_{K-1}}, \bm{x}_i\right) \nonumber \\
& \hspace{4cm} \cdot \Pr\left(\bm{S}_{i t_{K-2}} = \bm{s}_{t_{K-2}} \bigg\vert \bm{W}_{i,t_{K-2}+1:t_{K-1}} = \bm{1}_{\Delta t_{K-1}}, \bm{S}_{i t_{K-2}} = \bm{s}_{t_{K-2}}, \bm{x}_i \right) \nonumber \\
& \hspace{4.5cm} ... \cdot \Pr\left(\bm{S}_{i t_{1}} = \bm{s}_{t_{1}} \bigg\vert \bm{W}_{i,1:t_1} = \bm{1}_{\Delta t_{1}}, \bm{x}_i, \bm{s}_{i0}\right) \cdot \Pr(\bm{x}_i, \bm{s}_{i0}) \nonumber \\
= & \ \sum_{\bm{x}_i \in \bX, \bm{s}_{i0} \in \bS} \bE_{\cF} \left[ h^{\Delta t_{K+1}}_{T}\left( \bm{G}^{\Delta t_K}_{t_K}( ... \bm{G}^{\Delta t_2}_{t_2}(\bm{G}^{\Delta t_1}_{t_1}(\bm{s}_{i0}, \bm{x}_i, \bm{1}_{\Delta t_1}), \bm{x}_i, \bm{1}_{\Delta t_2}) ... ,\bm{x}_i, \bm{1}_{\Delta t_K}), \bm{x}_i, \bm{1}_{\Delta t_{K+1}} \right) \right] \cdot \Pr(\bm{x}_i, \bm{s}_{i0}) \nonumber \\
= & \ \sum_{\bm{x}_i \in \bX, \bm{s}_{i0} \in \bS} \bE_{\cF} \left[ h^{\Delta t_{K+1}}_{T}\left( \bm{G}^{\Delta t_K}_{t_K}( ... \bm{G}^{\Delta t_2}_{t_2}(\bm{G}^{\Delta t_1}_{t_1}(\bm{s}_{i0}, \bm{x}_i, \bm{1}_{\Delta t_1}), \bm{x}_i, \bm{1}_{\Delta t_2}) ... ,\bm{x}_i, \bm{1}_{\Delta t_K}), \bm{x}_i, \bm{1}_{\Delta t_{K+1}} \right) \right] \cdot \Pr(\bm{x}_i, \bm{s}_{i0}) \nonumber \\
& - \sum_{\bm{x}_i \in \bX, \bm{s}_{i0} \in \bS} \bE_{\cF} \left[ h^{\Delta t_{K+1}}_{\Delta t_{K+1}} \left( \bm{G}^{\Delta t_K}_{t_K}( ... \bm{G}^{\Delta t_2}_{t_2}(\bm{G}^{\Delta t_1}_{t_1}(\bm{s}_{i0}, \bm{x}_i, \bm{1}_{\Delta t_1}), \bm{x}_i, \bm{1}_{\Delta t_2}) ... ,\bm{x}_i, \bm{1}_{\Delta t_K}), \bm{x}_i, \bm{1}_{\Delta t_{K+1}} \right) \right] \cdot \Pr(\bm{x}_i, \bm{s}_{i0}) \nonumber \\
& + \sum_{\bm{x}_i \in \bX, \bm{s}_{i0} \in \bS} \bE_{\cF} \left[ h^{\Delta t_{K+1}}_{\Delta t_{K+1}} \left( \bm{G}^{\Delta t_K}_{t_K}( ... \bm{G}^{\Delta t_2}_{t_2}(\bm{G}^{\Delta t_1}_{t_1}(\bm{s}_{i0}, \bm{x}_i, \bm{1}_{\Delta t_1}), \bm{x}_i, \bm{1}_{\Delta t_2}) ... ,\bm{x}_i, \bm{1}_{\Delta t_K}), \bm{x}_i, \bm{1}_{\Delta t_{K+1}} \right) \right] \cdot \Pr(\bm{x}_i, \bm{s}_{i0}) \nonumber \\
& - \sum_{\bm{x}_i \in \bX, \bm{s}_{i0} \in \bS} \bE_{\cF} \left[ h^{\Delta t_{K+1}}_{\Delta t_{K+1}} \left( \bm{G}^{\Delta t_K}_{\Delta t_K}( \bm{G}^{\Delta t_{K-1}}_{ t_{K-1}}( ... \bm{G}^{\Delta t_1}_{t_1}(\bm{s}_{i0}, \bm{x}_i, \bm{1}_{\Delta t_1}) ... ,\bm{x}_i, \bm{1}_{\Delta t_{K-1}}),\bm{x}_i, \bm{1}_{\Delta t_K}), \bm{x}_i, \bm{1}_{\Delta t_{K+1}} \right) \right] \cdot \Pr(\bm{x}_i, \bm{s}_{i0}) \nonumber \\
& + \sum_{\bm{x}_i \in \bX, \bm{s}_{i0} \in \bS} \bE_{\cF} \left[ h^{\Delta t_{K+1}}_{\Delta t_{K+1}} \left( \bm{G}^{\Delta t_K}_{\Delta t_K}( \bm{G}^{\Delta t_{K-1}}_{ t_{K-1}}( ... \bm{G}^{\Delta t_1}_{t_1}(\bm{s}_{i0}, \bm{x}_i, \bm{1}_{\Delta t_1}) ... ,\bm{x}_i, \bm{1}_{\Delta t_{K-1}}),\bm{x}_i, \bm{1}_{\Delta t_K}), \bm{x}_i, \bm{1}_{\Delta t_{K+1}} \right) \right] \cdot \Pr(\bm{x}_i, \bm{s}_{i0}) \nonumber \\
& \ldots \nonumber \\
& - \sum_{\bm{x}_i \in \bX, \bm{s}_{i0} \in \bS} \bE_{\cF} \left[ h^{\Delta t_{K+1}}_{\Delta t_{K+1}} \left( \bm{G}^{\Delta t_K}_{\Delta t_K}( ... \bm{G}^{\Delta t_2}_{\Delta t_2}(\bm{G}^{\Delta t_1}_{t_1}(\bm{s}_{i0}, \bm{x}_i, \bm{1}_{\Delta t_1}), \bm{x}_i, \bm{1}_{\Delta t_2}) ... ,\bm{x}_i, \bm{1}_{\Delta t_K}), \bm{x}_i, \bm{1}_{\Delta t_{K+1}} \right) \right] \cdot \Pr(\bm{x}_i, \bm{s}_{i0}) \nonumber \\
& + \sum_{\bm{x}_i \in \bX, \bm{s}_{i0} \in \bS} \bE_{\cF} \left[ h^{\Delta t_{K+1}}_{\Delta t_{K+1}} \left( \bm{G}^{\Delta t_K}_{\Delta t_K}( ... \bm{G}^{\Delta t_2}_{\Delta t_2}(\bm{G}^{\Delta t_1}_{t_1}(\bm{s}_{i0}, \bm{x}_i, \bm{1}_{\Delta t_1}), \bm{x}_i, \bm{1}_{\Delta t_2}) ... ,\bm{x}_i, \bm{1}_{\Delta t_K}), \bm{x}_i, \bm{1}_{\Delta t_{K+1}} \right) \right] \cdot \Pr(\bm{x}_i, \bm{s}_{i0}) \nonumber \\
& - \sum_{\bm{x}_i \in \bX, \bm{s}_{i0} \in \bS} \bE_{\cF} \left[ h^{\Delta t_{K+1}}_{\Delta t_{K+1}} \left( \bm{G}^{\Delta t_K}_{\Delta t_K}( ... \bm{G}^{\Delta t_2}_{\Delta t_2}(\bm{G}^{\Delta t_1}_{\Delta t_1}(\bm{s}_{i0}, \bm{x}_i, \bm{1}_{\Delta t_1}), \bm{x}_i, \bm{1}_{\Delta t_2}) ... ,\bm{x}_i, \bm{1}_{\Delta t_K}), \bm{x}_i, \bm{1}_{\Delta t_{K+1}} \right) \right] \cdot \Pr(\bm{x}_i, \bm{s}_{i0}) \nonumber \\
& + \sum_{\bm{x}_i \in \bX, \bm{s}_{i0} \in \bS} \bE_{\cF} \left[ h^{\Delta t_{K+1}}_{\Delta t_{K+1}} \left( \bm{G}^{\Delta t_K}_{\Delta t_K}( ... \bm{G}^{\Delta t_2}_{\Delta t_2}(\bm{G}^{\Delta t_1}_{\Delta t_1}(\bm{s}_{i0}, \bm{x}_i, \bm{1}_{\Delta t_1}), \bm{x}_i, \bm{1}_{\Delta t_2}) ... ,\bm{x}_i, \bm{1}_{\Delta t_K}), \bm{x}_i, \bm{1}_{\Delta t_{K+1}} \right) \right] \cdot \Pr(\bm{x}_i, \bm{s}_{i0}), \label{eqn:combineFirstIteration1}
\end{align}
where the second equality is due to Definition~\ref{defn:GeneralizedPivotIndex:appendix};
the last equality is adding and subtracting the same $K$ quantities (there are $2K$ additional terms).
Similarly, we have
\begin{align}
& \ \bE_{\cF}\left[ Y_{iT}(\bm{0}_{T}) \right] \nonumber \\
= & \ \sum_{\bm{x}_i \in \bX, \bm{s}_{i0} \in \bS} \sum_{\bm{s}_{t_K}, ..., \bm{s}_{t_1} \in \bS} h^{\Delta t_{K+1}}_{T}(\bm{s}_{t_K}, \bm{x}_i, \bm{0}_{\Delta t_{K+1}}) \cdot \Pr\left(\bm{S}_{i t_{K-1}} = \bm{s}_{t_{K-1}} \bigg\vert \bm{W}_{i,t_{K-1}+1:t_K} = \bm{0}_{\Delta t_K}, \bm{S}_{i t_{K-1}} = \bm{s}_{t_{K-1}}, \bm{x}_i\right) \nonumber \\
& \hspace{4cm} \cdot \Pr\left(\bm{S}_{i t_{K-2}} = \bm{s}_{t_{K-2}} \bigg\vert \bm{W}_{i,t_{K-2}+1:t_{K-1}} = \bm{0}_{\Delta t_{K-1}}, \bm{S}_{i t_{K-2}} = \bm{s}_{t_{K-2}}, \bm{x}_i \right) \nonumber \\
& \hspace{4.5cm} ... \cdot \Pr\left(\bm{S}_{i t_{1}} = \bm{s}_{t_{1}} \bigg\vert \bm{W}_{i,1:t_1} = \bm{0}_{\Delta t_{1}}, \bm{x}_i, \bm{s}_{i0}\right) \cdot \Pr(\bm{x}_i, \bm{s}_{i0}) \nonumber \\
= & \ \sum_{\bm{x}_i \in \bX, \bm{s}_{i0} \in \bS} \bE_{\cF} \left[ h^{\Delta t_{K+1}}_{T}\left( \bm{G}^{\Delta t_K}_{t_K}( ... \bm{G}^{\Delta t_2}_{t_2}(\bm{G}^{\Delta t_1}_{t_1}(\bm{s}_{i0}, \bm{x}_i, \bm{0}_{\Delta t_1}), \bm{x}_i, \bm{0}_{\Delta t_2}) ... ,\bm{x}_i, \bm{0}_{\Delta t_K}), \bm{x}_i, \bm{0}_{\Delta t_{K+1}} \right) \right] \cdot \Pr(\bm{x}_i, \bm{s}_{i0}) \nonumber \\
= & \ \sum_{\bm{x}_i \in \bX, \bm{s}_{i0} \in \bS} \bE_{\cF} \left[ h^{\Delta t_{K+1}}_{T}\left( \bm{G}^{\Delta t_K}_{t_K}( ... \bm{G}^{\Delta t_2}_{t_2}(\bm{G}^{\Delta t_1}_{t_1}(\bm{s}_{i0}, \bm{x}_i, \bm{0}_{\Delta t_1}), \bm{x}_i, \bm{0}_{\Delta t_2}) ... ,\bm{x}_i, \bm{0}_{\Delta t_K}), \bm{x}_i, \bm{0}_{\Delta t_{K+1}} \right) \right] \cdot \Pr(\bm{x}_i, \bm{s}_{i0}) \nonumber \\
& - \sum_{\bm{x}_i \in \bX, \bm{s}_{i0} \in \bS} \bE_{\cF} \left[ h^{\Delta t_{K+1}}_{\Delta t_{K+1}} \left( \bm{G}^{\Delta t_K}_{t_K}( ... \bm{G}^{\Delta t_2}_{t_2}(\bm{G}^{\Delta t_1}_{t_1}(\bm{s}_{i0}, \bm{x}_i, \bm{0}_{\Delta t_1}), \bm{x}_i, \bm{0}_{\Delta t_2}) ... ,\bm{x}_i, \bm{0}_{\Delta t_K}), \bm{x}_i, \bm{0}_{\Delta t_{K+1}} \right) \right] \cdot \Pr(\bm{x}_i, \bm{s}_{i0}) \nonumber \\
& + \sum_{\bm{x}_i \in \bX, \bm{s}_{i0} \in \bS} \bE_{\cF} \left[ h^{\Delta t_{K+1}}_{\Delta t_{K+1}} \left( \bm{G}^{\Delta t_K}_{t_K}( ... \bm{G}^{\Delta t_2}_{t_2}(\bm{G}^{\Delta t_1}_{t_1}(\bm{s}_{i0}, \bm{x}_i, \bm{0}_{\Delta t_1}), \bm{x}_i, \bm{0}_{\Delta t_2}) ... ,\bm{x}_i, \bm{0}_{\Delta t_K}), \bm{x}_i, \bm{0}_{\Delta t_{K+1}} \right) \right] \cdot \Pr(\bm{x}_i, \bm{s}_{i0}) \nonumber \\
& - \sum_{\bm{x}_i \in \bX, \bm{s}_{i0} \in \bS} \bE_{\cF} \left[ h^{\Delta t_{K+1}}_{\Delta t_{K+1}} \left( \bm{G}^{\Delta t_K}_{\Delta t_K}( \bm{G}^{\Delta t_{K-1}}_{ t_{K-1}}( ... \bm{G}^{\Delta t_1}_{t_1}(\bm{s}_{i0}, \bm{x}_i, \bm{0}_{\Delta t_1}) ... ,\bm{x}_i, \bm{0}_{\Delta t_{K-1}}),\bm{x}_i, \bm{0}_{\Delta t_K}), \bm{x}_i, \bm{0}_{\Delta t_{K+1}} \right) \right] \cdot \Pr(\bm{x}_i, \bm{s}_{i0}) \nonumber \\
& + \sum_{\bm{x}_i \in \bX, \bm{s}_{i0} \in \bS} \bE_{\cF} \left[ h^{\Delta t_{K+1}}_{\Delta t_{K+1}} \left( \bm{G}^{\Delta t_K}_{\Delta t_K}( \bm{G}^{\Delta t_{K-1}}_{ t_{K-1}}( ... \bm{G}^{\Delta t_1}_{t_1}(\bm{s}_{i0}, \bm{x}_i, \bm{0}_{\Delta t_1}) ... ,\bm{x}_i, \bm{0}_{\Delta t_{K-1}}),\bm{x}_i, \bm{0}_{\Delta t_K}), \bm{x}_i, \bm{0}_{\Delta t_{K+1}} \right) \right] \cdot \Pr(\bm{x}_i, \bm{s}_{i0}) \nonumber \\
& \ldots \nonumber \\
& - \sum_{\bm{x}_i \in \bX, \bm{s}_{i0} \in \bS} \bE_{\cF} \left[ h^{\Delta t_{K+1}}_{\Delta t_{K+1}} \left( \bm{G}^{\Delta t_K}_{\Delta t_K}( ... \bm{G}^{\Delta t_2}_{\Delta t_2}(\bm{G}^{\Delta t_1}_{t_1}(\bm{s}_{i0}, \bm{x}_i, \bm{0}_{\Delta t_1}), \bm{x}_i, \bm{0}_{\Delta t_2}) ... ,\bm{x}_i, \bm{0}_{\Delta t_K}), \bm{x}_i, \bm{0}_{\Delta t_{K+1}} \right) \right] \cdot \Pr(\bm{x}_i, \bm{s}_{i0}) \nonumber \\
& + \sum_{\bm{x}_i \in \bX, \bm{s}_{i0} \in \bS} \bE_{\cF} \left[ h^{\Delta t_{K+1}}_{\Delta t_{K+1}} \left( \bm{G}^{\Delta t_K}_{\Delta t_K}( ... \bm{G}^{\Delta t_2}_{\Delta t_2}(\bm{G}^{\Delta t_1}_{t_1}(\bm{s}_{i0}, \bm{x}_i, \bm{0}_{\Delta t_1}), \bm{x}_i, \bm{0}_{\Delta t_2}) ... ,\bm{x}_i, \bm{0}_{\Delta t_K}), \bm{x}_i, \bm{0}_{\Delta t_{K+1}} \right) \right] \cdot \Pr(\bm{x}_i, \bm{s}_{i0}) \nonumber \\
& - \sum_{\bm{x}_i \in \bX, \bm{s}_{i0} \in \bS} \bE_{\cF} \left[ h^{\Delta t_{K+1}}_{\Delta t_{K+1}} \left( \bm{G}^{\Delta t_K}_{\Delta t_K}( ... \bm{G}^{\Delta t_2}_{\Delta t_2}(\bm{G}^{\Delta t_1}_{\Delta t_1}(\bm{s}_{i0}, \bm{x}_i, \bm{0}_{\Delta t_1}), \bm{x}_i, \bm{0}_{\Delta t_2}) ... ,\bm{x}_i, \bm{0}_{\Delta t_K}), \bm{x}_i, \bm{0}_{\Delta t_{K+1}} \right) \right] \cdot \Pr(\bm{x}_i, \bm{s}_{i0}) \nonumber \\
& + \sum_{\bm{x}_i \in \bX, \bm{s}_{i0} \in \bS} \bE_{\cF} \left[ h^{\Delta t_{K+1}}_{\Delta t_{K+1}} \left( \bm{G}^{\Delta t_K}_{\Delta t_K}( ... \bm{G}^{\Delta t_2}_{\Delta t_2}(\bm{G}^{\Delta t_1}_{\Delta t_1}(\bm{s}_{i0}, \bm{x}_i, \bm{0}_{\Delta t_1}), \bm{x}_i, \bm{0}_{\Delta t_2}) ... ,\bm{x}_i, \bm{0}_{\Delta t_K}), \bm{x}_i, \bm{0}_{\Delta t_{K+1}} \right) \right] \cdot \Pr(\bm{x}_i, \bm{s}_{i0}), \label{eqn:combineFirstIteration0}
\end{align}

\vspace{0.2cm}
\noindent \textbf{\textit{Step 2.}}

Next, similar to the proof of Lemma~\ref{lem:SpecialCaseL2}, we apply the extended linearity assumption to replace the random intermediate outcomes $\bm{G}$ by their expectations $\bm{g}$.
Denote, for any $t$ and $\delta$, $\bm{G}^\delta_t = (G^\delta_{t, 1}, ..., G^\delta_{t, D})$ and $\bm{g}^\delta_t = (g^\delta_{t, 1}, ..., g^\delta_{t, D})$.
We then have
\begin{align*}
& \ \bE_{\cF} \left[ h^{\Delta t_{K+1}}_{T}\left( \bm{G}^{\Delta t_K}_{t_K}( ... \bm{G}^{\Delta t_2}_{t_2}(\bm{G}^{\Delta t_1}_{t_1}(\bm{s}_{i0}, \bm{x}_i, \bm{1}_{\Delta t_1}), \bm{x}_i, \bm{1}_{\Delta t_2}) ... ,\bm{x}_i, \bm{1}_{\Delta t_K}), \bm{x}_i, \bm{1}_{\Delta t_{K+1}} \right) \right] \\
= & \bE_{\cF} \left[ \alpha^{\Delta t_{K+1}}_{0,T}(\bm{x}_i, \bm{1}_{\Delta t_{K+1}}) + \sum_{d=1}^D G^{\Delta t_K}_{t_K, d}( ... \bm{G}^{\Delta t_2}_{t_2}(\bm{G}^{\Delta t_1}_{t_1}(\bm{s}_{i0}, \bm{x}_i, \bm{1}_{\Delta t_1}), \bm{x}_i, \bm{1}_{\Delta t_2}) ... ,\bm{x}_i, \bm{1}_{\Delta t_K}) \cdot \alpha^{\Delta t_{K+1}}_{d,T}(\bm{x}_i, \bm{1}_{\Delta t_{K+1}})\right] \\
= & \alpha^{\Delta t_{K+1}}_{0,T}(\bm{x}_i, \bm{1}_{\Delta t_{K+1}}) \\
& + \sum_{d=1}^D \bE_{\cF} \left[g^{\Delta t_K}_{t_K, d}(\bm{G}^{\Delta t_{K-1}}_{t_{K-1}}( ... \bm{G}^{\Delta t_2}_{t_2}(\bm{G}^{\Delta t_1}_{t_1}(\bm{s}_{i0}, \bm{x}_i, \bm{1}_{\Delta t_1}), \bm{x}_i, \bm{1}_{\Delta t_2}) ... ,\bm{x}_i, \bm{1}_{\Delta t_{K-1}}), \bm{x}_i, \bm{1}_{\Delta t_K})\right] \cdot \alpha^{\Delta t_{K+1}}_{d,T}(\bm{x}_i, \bm{1}_{\Delta t_{K+1}}) \\
= & \alpha^{\Delta t_{K+1}}_{0,T}(\bm{x}_i, \bm{1}_{\Delta t_{K+1}}) \\
& + \sum_{d=1}^D \bE_{\cF} \left[ \beta^{\Delta t_K}_{d,0,t_K}(\bm{x}_i, \bm{1}_{\Delta t_{K}}) + \sum_{d'=1}^D G^{\Delta t_{K-1}}_{t_{K-1}, d'}( ... \bm{G}^{\Delta t_1}_{t_1}(\bm{s}_{i0}, \bm{x}_i, \bm{1}_{\Delta t_1}) ... ,\bm{x}_i, \bm{1}_{\Delta t_{K-1}}) \cdot \beta^{\Delta t_K}_{d,d',t_K}(\bm{x}_i, \bm{1}_{\Delta t_{K}})\right] \cdot \alpha^{\Delta t_{K+1}}_{d,T}(\bm{x}_i, \bm{1}_{\Delta t_{K+1}}) \\
= & \alpha^{\Delta t_{K+1}}_{0,T}(\bm{x}_i, \bm{1}_{\Delta t_{K+1}}) \\
& + \sum_{d=1}^D \left(\beta^{\Delta t_K}_{d,0,t_K}(\bm{x}_i, \bm{1}_{\Delta t_{K}}) + \sum_{d'=1}^D \bE_{\cF} \left[g^{\Delta t_{K-1}}_{t_{K-1}, d'}( ... \bm{G}^{\Delta t_1}_{t_1}(\bm{s}_{i0}, \bm{x}_i, \bm{1}_{\Delta t_1}) ... ,\bm{x}_i, \bm{1}_{\Delta t_{K-1}})\right] \cdot \beta^{\Delta t_K}_{d,d',t_K}(\bm{x}_i, \bm{1}_{\Delta t_{K}})\right) \cdot \alpha^{\Delta t_{K+1}}_{d,T}(\bm{x}_i, \bm{1}_{\Delta t_{K+1}}) \\
= & \ldots \\
= & h^{\Delta t_{K+1}}_{T}\left( \bm{g}^{\Delta t_K}_{t_K}( ... \bm{g}^{\Delta t_2}_{t_2}(\bm{g}^{\Delta t_1}_{t_1}(\bm{s}_{i0}, \bm{x}_i, \bm{1}_{\Delta t_1}),\bm{x}_i, \bm{1}_{\Delta t_2}) ... ,\bm{x}_i, \bm{1}_{\Delta t_K}), \bm{x}_i, \bm{1}_{\Delta t_{K+1}} \right)
\end{align*}

where the first equality is due to Assumption~\ref{asp:Linearity2}; the second and forth equality is due to linearity of expectation; the third equality is due to Assumption~\ref{asp:Linearity}, with which the equation can be iteratively expressed into a linear form, and the position of the expectation notion can be adjusted inside.

Similarly, 
\begin{align*}
& \ \bE_{\cF} \left[ h^{\Delta t_{K+1}}_{\Delta t_{K+1}}\left( \bm{G}^{\Delta t_K}_{t_K}( ... \bm{G}^{\Delta t_2}_{t_2}(\bm{G}^{\Delta t_1}_{t_1}(\bm{s}_{i0}, \bm{x}_i, \bm{1}_{\Delta t_1}), \bm{x}_i, \bm{1}_{\Delta t_2}) ... ,\bm{x}_i, \bm{1}_{\Delta t_K}), \bm{x}_i, \bm{1}_{\Delta t_{K+1}} \right) \right] \\
= & \alpha^{\Delta t_{K+1}}_{0,\Delta t_{K+1}}(\bm{x}_i, \bm{1}_{\Delta t_{K+1}}) \\
& + \sum_{d=1}^D \left(\beta^{\Delta t_K}_{d,0,t_K}(\bm{x}_i, \bm{1}_{\Delta t_{K}}) + \sum_{d'=1}^D \bE_{\cF} \left[g^{\Delta t_{K-1}}_{t_{K-1}, d'}( ... \bm{G}^{\Delta t_1}_{t_1}(\bm{s}_{i0}, \bm{x}_i, \bm{1}_{\Delta t_1}) ... ,\bm{x}_i, \bm{1}_{\Delta t_{K-1}})\right] \cdot \beta^{\Delta t_K}_{d,d',t_K}(\bm{x}_i, \bm{1}_{\Delta t_{K}})\right) \cdot \alpha^{\Delta t_{K+1}}_{d,\Delta t_{K+1}}(\bm{x}_i, \bm{1}_{\Delta t_{K+1}}) \\
= & \ldots \\
= & h^{\Delta t_{K+1}}_{\Delta t_{K+1}}\left( \bm{g}^{\Delta t_K}_{t_K}( ... \bm{g}^{\Delta t_2}_{t_2}(\bm{g}^{\Delta t_1}_{t_1}(\bm{s}_{i0}, \bm{x}_i, \bm{1}_{\Delta t_1}),\bm{x}_i, \bm{1}_{\Delta t_2}) ... ,\bm{x}_i, \bm{1}_{\Delta t_K}), \bm{x}_i, \bm{1}_{\Delta t_{K+1}} \right)
\end{align*}

Similarly,
\begin{align*}
& \ \bE_{\cF} \left[ h^{\Delta t_{K+1}}_{\Delta t_{K+1}} \left( \bm{G}^{\Delta t_K}_{\Delta t_K}( \bm{G}^{\Delta t_{K-1}}_{ t_{K-1}}( ... \bm{G}^{\Delta t_1}_{t_1}(\bm{s}_{i0}, \bm{x}_i, \bm{1}_{\Delta t_1}) ... ,\bm{x}_i, \bm{1}_{\Delta t_{K-1}}),\bm{x}_i, \bm{1}_{\Delta t_K}), \bm{x}_i, \bm{1}_{\Delta t_{K+1}} \right) \right] \\
= & \alpha^{\Delta t_{K+1}}_{0,\Delta t_{K+1}}(\bm{x}_i, \bm{1}_{\Delta t_{K+1}}) \\
& + \sum_{d=1}^D \Bigg(\beta^{\Delta t_K}_{d,0,\Delta t_K}(\bm{x}_i, \bm{1}_{\Delta t_{K}}) + \sum_{d'=1}^D \Big(\beta^{\Delta t_{K-1}}_{d',0,\Delta t_{K-1}}(\bm{x}_i, \bm{1}_{\Delta t_{K-1}}) \\
& \hspace{2cm} + \sum_{d''=1}^D \bE_{\cF} \left[ ... \bm{G}^{\Delta t_1}_{t_1}(\bm{s}_{i0}, \bm{x}_i, \bm{1}_{\Delta t_1}) ... \right] \cdot \beta^{\Delta t_{K-1}}_{d',d,''\Delta t_{K-1}}(\bm{x}_i, \bm{1}_{\Delta t_{K-1}})\Big) \cdot \beta^{\Delta t_K}_{d,d',\Delta t_K}(\bm{x}_i, \bm{1}_{\Delta t_{K}})\Bigg) \cdot \alpha^{\Delta t_{K+1}}_{d,\Delta t_{K+1}}(\bm{x}_i, \bm{1}_{\Delta t_{K+1}}) \\
= & \ldots \\
= & h^{\Delta t_{K+1}}_{\Delta t_{K+1}}\left( \bm{g}^{\Delta t_K}_{\Delta t_K}( \bm{g}^{\Delta t_{K-1}}_{ t_{K-1}}( ... \bm{g}^{\Delta t_1}_{t_1}(\bm{s}_{i0}, \bm{x}_i, \bm{1}_{\Delta t_1}) ... ,\bm{x}_i, \bm{1}_{\Delta t_{K-1}}),\bm{x}_i, \bm{1}_{\Delta t_K}), \bm{x}_i, \bm{1}_{\Delta t_{K+1}} \right)
\end{align*}

Iteratively applying the above action, we are able to rewrite equation~\eqref{eqn:combineFirstIteration1} into the following expression:
\begin{align*}
& \ \bE_{\cF}\left[ Y_{iT}(\bm{1}_{T}) \right] \\
= & \ \sum_{\bm{x}_i \in \bX, \bm{s}_{i0} \in \bS} h^{\Delta t_{K+1}}_{T}\left( \bm{g}^{\Delta t_K}_{t_K}( ... \bm{g}^{\Delta t_2}_{t_2}(\bm{g}^{\Delta t_1}_{t_1}(\bm{s}_{i0}, \bm{x}_i, \bm{1}_{\Delta t_1}), \bm{x}_i, \bm{1}_{\Delta t_2}) ... ,\bm{x}_i, \bm{1}_{\Delta t_K}), \bm{x}_i, \bm{1}_{\Delta t_{K+1}} \right) \cdot \Pr(\bm{x}_i, \bm{s}_{i0}) \\
& - \sum_{\bm{x}_i \in \bX, \bm{s}_{i0} \in \bS} h^{\Delta t_{K+1}}_{\Delta t_{K+1}} \left( \bm{g}^{\Delta t_K}_{t_K}( ... \bm{g}^{\Delta t_2}_{t_2}(\bm{g}^{\Delta t_1}_{t_1}(\bm{s}_{i0}, \bm{x}_i, \bm{1}_{\Delta t_1}), \bm{x}_i, \bm{1}_{\Delta t_2}) ... ,\bm{x}_i, \bm{1}_{\Delta t_K}), \bm{x}_i, \bm{1}_{\Delta t_{K+1}} \right) \cdot \Pr(\bm{x}_i, \bm{s}_{i0}) \\
& + \sum_{\bm{x}_i \in \bX, \bm{s}_{i0} \in \bS} h^{\Delta t_{K+1}}_{\Delta t_{K+1}} \left( \bm{g}^{\Delta t_K}_{t_K}( ... \bm{g}^{\Delta t_2}_{t_2}(\bm{g}^{\Delta t_1}_{t_1}(\bm{s}_{i0}, \bm{x}_i, \bm{1}_{\Delta t_1}), \bm{x}_i, \bm{1}_{\Delta t_2}) ... ,\bm{x}_i, \bm{1}_{\Delta t_K}), \bm{x}_i, \bm{1}_{\Delta t_{K+1}} \right) \cdot \Pr(\bm{x}_i, \bm{s}_{i0}) \\
& - \sum_{\bm{x}_i \in \bX, \bm{s}_{i0} \in \bS} h^{\Delta t_{K+1}}_{\Delta t_{K+1}} \left( \bm{g}^{\Delta t_K}_{\Delta t_K}( \bm{g}^{\Delta t_{K-1}}_{ t_{K-1}}( ... \bm{g}^{\Delta t_1}_{t_1}(\bm{s}_{i0}, \bm{x}_i, \bm{1}_{\Delta t_1}) ... ,\bm{x}_i, \bm{1}_{\Delta t_{K-1}}),\bm{x}_i, \bm{1}_{\Delta t_K}), \bm{x}_i, \bm{1}_{\Delta t_{K+1}} \right) \cdot \Pr(\bm{x}_i, \bm{s}_{i0}) \\
& + \sum_{\bm{x}_i \in \bX, \bm{s}_{i0} \in \bS}  h^{\Delta t_{K+1}}_{\Delta t_{K+1}} \left( \bm{g}^{\Delta t_K}_{\Delta t_K}( \bm{g}^{\Delta t_{K-1}}_{ t_{K-1}}( ... \bm{g}^{\Delta t_1}_{t_1}(\bm{s}_{i0}, \bm{x}_i, \bm{1}_{\Delta t_1}) ... ,\bm{x}_i, \bm{1}_{\Delta t_{K-1}}),\bm{x}_i, \bm{1}_{\Delta t_K}), \bm{x}_i, \bm{1}_{\Delta t_{K+1}} \right) \cdot \Pr(\bm{x}_i, \bm{s}_{i0}) \\
& \ldots \nonumber \\
& - \sum_{\bm{x}_i \in \bX, \bm{s}_{i0} \in \bS} h^{\Delta t_{K+1}}_{\Delta t_{K+1}} \left( \bm{g}^{\Delta t_K}_{\Delta t_K}( ... \bm{g}^{\Delta t_2}_{\Delta t_2}(\bm{g}^{\Delta t_1}_{t_1}(\bm{s}_{i0}, \bm{x}_i, \bm{1}_{\Delta t_1}), \bm{x}_i, \bm{1}_{\Delta t_2}) ... ,\bm{x}_i, \bm{1}_{\Delta t_K}), \bm{x}_i, \bm{1}_{\Delta t_{K+1}} \right) \cdot \Pr(\bm{x}_i, \bm{s}_{i0}) \\
& + \sum_{\bm{x}_i \in \bX, \bm{s}_{i0} \in \bS} h^{\Delta t_{K+1}}_{\Delta t_{K+1}} \left( \bm{g}^{\Delta t_K}_{\Delta t_K}( ... \bm{g}^{\Delta t_2}_{\Delta t_2}(\bm{g}^{\Delta t_1}_{t_1}(\bm{s}_{i0}, \bm{x}_i, \bm{1}_{\Delta t_1}), \bm{x}_i, \bm{1}_{\Delta t_2}) ... ,\bm{x}_i, \bm{1}_{\Delta t_K}), \bm{x}_i, \bm{1}_{\Delta t_{K+1}} \right) \cdot \Pr(\bm{x}_i, \bm{s}_{i0}) \\
& - \sum_{\bm{x}_i \in \bX, \bm{s}_{i0} \in \bS} h^{\Delta t_{K+1}}_{\Delta t_{K+1}} \left( \bm{g}^{\Delta t_K}_{\Delta t_K}( ... \bm{g}^{\Delta t_2}_{\Delta t_2}(\bm{g}^{\Delta t_1}_{\Delta t_1}(\bm{s}_{i0}, \bm{x}_i, \bm{1}_{\Delta t_1}), \bm{x}_i, \bm{1}_{\Delta t_2}) ... ,\bm{x}_i, \bm{1}_{\Delta t_K}), \bm{x}_i, \bm{1}_{\Delta t_{K+1}} \right) \cdot \Pr(\bm{x}_i, \bm{s}_{i0}) \\
& + \sum_{\bm{x}_i \in \bX, \bm{s}_{i0} \in \bS} h^{\Delta t_{K+1}}_{\Delta t_{K+1}} \left( \bm{g}^{\Delta t_K}_{\Delta t_K}( ... \bm{g}^{\Delta t_2}_{\Delta t_2}(\bm{g}^{\Delta t_1}_{\Delta t_1}(\bm{s}_{i0}, \bm{x}_i, \bm{1}_{\Delta t_1}), \bm{x}_i, \bm{1}_{\Delta t_2}) ... ,\bm{x}_i, \bm{1}_{\Delta t_K}), \bm{x}_i, \bm{1}_{\Delta t_{K+1}} \right) \cdot \Pr(\bm{x}_i, \bm{s}_{i0})
\end{align*}

Similarly, equation~\eqref{eqn:combineFirstIteration0} can be rewritten as
\begin{align*}
& \ \bE_{\cF}\left[ Y_{iT}(\bm{0}_{T}) \right] \\
= & \ \sum_{\bm{x}_i \in \bX, \bm{s}_{i0} \in \bS} h^{\Delta t_{K+1}}_{T}\left( \bm{g}^{\Delta t_K}_{t_K}( ... \bm{g}^{\Delta t_2}_{t_2}(\bm{g}^{\Delta t_1}_{t_1}(\bm{s}_{i0}, \bm{x}_i, \bm{0}_{\Delta t_1}), \bm{x}_i, \bm{0}_{\Delta t_2}) ... ,\bm{x}_i, \bm{0}_{\Delta t_K}), \bm{x}_i, \bm{0}_{\Delta t_{K+1}} \right) \cdot \Pr(\bm{x}_i, \bm{s}_{i0}) \\
& - \sum_{\bm{x}_i \in \bX, \bm{s}_{i0} \in \bS} h^{\Delta t_{K+1}}_{\Delta t_{K+1}} \left( \bm{g}^{\Delta t_K}_{t_K}( ... \bm{g}^{\Delta t_2}_{t_2}(\bm{g}^{\Delta t_1}_{t_1}(\bm{s}_{i0}, \bm{x}_i, \bm{0}_{\Delta t_1}), \bm{x}_i, \bm{0}_{\Delta t_2}) ... ,\bm{x}_i, \bm{0}_{\Delta t_K}), \bm{x}_i, \bm{0}_{\Delta t_{K+1}} \right) \cdot \Pr(\bm{x}_i, \bm{s}_{i0}) \\
& + \sum_{\bm{x}_i \in \bX, \bm{s}_{i0} \in \bS} h^{\Delta t_{K+1}}_{\Delta t_{K+1}} \left( \bm{g}^{\Delta t_K}_{t_K}( ... \bm{g}^{\Delta t_2}_{t_2}(\bm{g}^{\Delta t_1}_{t_1}(\bm{s}_{i0}, \bm{x}_i, \bm{0}_{\Delta t_1}), \bm{x}_i, \bm{0}_{\Delta t_2}) ... ,\bm{x}_i, \bm{0}_{\Delta t_K}), \bm{x}_i, \bm{0}_{\Delta t_{K+1}} \right) \cdot \Pr(\bm{x}_i, \bm{s}_{i0}) \\
& - \sum_{\bm{x}_i \in \bX, \bm{s}_{i0} \in \bS} h^{\Delta t_{K+1}}_{\Delta t_{K+1}} \left( \bm{g}^{\Delta t_K}_{\Delta t_K}( \bm{g}^{\Delta t_{K-1}}_{ t_{K-1}}( ... \bm{g}^{\Delta t_1}_{t_1}(\bm{s}_{i0}, \bm{x}_i, \bm{0}_{\Delta t_1}) ... ,\bm{x}_i, \bm{0}_{\Delta t_{K-1}}),\bm{x}_i, \bm{0}_{\Delta t_K}), \bm{x}_i, \bm{0}_{\Delta t_{K+1}} \right) \cdot \Pr(\bm{x}_i, \bm{s}_{i0}) \\
& + \sum_{\bm{x}_i \in \bX, \bm{s}_{i0} \in \bS} h^{\Delta t_{K+1}}_{\Delta t_{K+1}} \left( \bm{g}^{\Delta t_K}_{\Delta t_K}( \bm{g}^{\Delta t_{K-1}}_{ t_{K-1}}( ... \bm{g}^{\Delta t_1}_{t_1}(\bm{s}_{i0}, \bm{x}_i, \bm{0}_{\Delta t_1}) ... ,\bm{x}_i, \bm{0}_{\Delta t_{K-1}}),\bm{x}_i, \bm{0}_{\Delta t_K}), \bm{x}_i, \bm{0}_{\Delta t_{K+1}} \right) \cdot \Pr(\bm{x}_i, \bm{s}_{i0}) \\
& \ldots \nonumber \\
& - \sum_{\bm{x}_i \in \bX, \bm{s}_{i0} \in \bS} h^{\Delta t_{K+1}}_{\Delta t_{K+1}} \left( \bm{g}^{\Delta t_K}_{\Delta t_K}( ... \bm{g}^{\Delta t_2}_{\Delta t_2}(\bm{g}^{\Delta t_1}_{t_1}(\bm{s}_{i0}, \bm{x}_i, \bm{0}_{\Delta t_1}), \bm{x}_i, \bm{0}_{\Delta t_2}) ... ,\bm{x}_i, \bm{0}_{\Delta t_K}), \bm{x}_i, \bm{0}_{\Delta t_{K+1}} \right)\cdot \Pr(\bm{x}_i, \bm{s}_{i0}) \\
& + \sum_{\bm{x}_i \in \bX, \bm{s}_{i0} \in \bS} h^{\Delta t_{K+1}}_{\Delta t_{K+1}} \left( \bm{g}^{\Delta t_K}_{\Delta t_K}( ... \bm{g}^{\Delta t_2}_{\Delta t_2}(\bm{g}^{\Delta t_1}_{t_1}(\bm{s}_{i0}, \bm{x}_i, \bm{0}_{\Delta t_1}), \bm{x}_i, \bm{0}_{\Delta t_2}) ... ,\bm{x}_i, \bm{0}_{\Delta t_K}), \bm{x}_i, \bm{0}_{\Delta t_{K+1}} \right) \cdot \Pr(\bm{x}_i, \bm{s}_{i0}) \\
& - \sum_{\bm{x}_i \in \bX, \bm{s}_{i0} \in \bS} h^{\Delta t_{K+1}}_{\Delta t_{K+1}} \left( \bm{g}^{\Delta t_K}_{\Delta t_K}( ... \bm{g}^{\Delta t_2}_{\Delta t_2}(\bm{g}^{\Delta t_1}_{\Delta t_1}(\bm{s}_{i0}, \bm{x}_i, \bm{0}_{\Delta t_1}), \bm{x}_i, \bm{0}_{\Delta t_2}) ... ,\bm{x}_i, \bm{0}_{\Delta t_K}), \bm{x}_i, \bm{0}_{\Delta t_{K+1}} \right) \cdot \Pr(\bm{x}_i, \bm{s}_{i0}) \\
& + \sum_{\bm{x}_i \in \bX, \bm{s}_{i0} \in \bS}  h^{\Delta t_{K+1}}_{\Delta t_{K+1}} \left( \bm{g}^{\Delta t_K}_{\Delta t_K}( ... \bm{g}^{\Delta t_2}_{\Delta t_2}(\bm{g}^{\Delta t_1}_{\Delta t_1}(\bm{s}_{i0}, \bm{x}_i, \bm{0}_{\Delta t_1}), \bm{x}_i, \bm{0}_{\Delta t_2}) ... ,\bm{x}_i, \bm{0}_{\Delta t_K}), \bm{x}_i, \bm{0}_{\Delta t_{K+1}} \right) \cdot \Pr(\bm{x}_i, \bm{s}_{i0})
\end{align*}

Combining both parts,
\begin{align}
\tau_T = &\ \sum_{\bm{x}_i \in \bX, \bm{s}_{i0} \in \bS} \Bigg(h^{\Delta t_{K+1}}_{T}\left( \bm{g}^{\Delta t_K}_{t_K}( ... \bm{g}^{\Delta t_2}_{t_2}(\bm{g}^{\Delta t_1}_{t_1}(\bm{s}_{i0}, \bm{x}_i, \bm{1}_{\Delta t_1}), \bm{x}_i, \bm{1}_{\Delta t_2}) ... ,\bm{x}_i, \bm{1}_{\Delta t_K}), \bm{x}_i, \bm{1}_{\Delta t_{K+1}} \right) \nonumber\\
& \hspace{2cm} - h^{\Delta t_{K+1}}_{\Delta t_{K+1}} \left( \bm{g}^{\Delta t_K}_{t_K}( ... \bm{g}^{\Delta t_2}_{t_2}(\bm{g}^{\Delta t_1}_{t_1}(\bm{s}_{i0}, \bm{x}_i, \bm{1}_{\Delta t_1}), \bm{x}_i, \bm{1}_{\Delta t_2}) ... ,\bm{x}_i, \bm{1}_{\Delta t_K}), \bm{x}_i, \bm{1}_{\Delta t_{K+1}} \right)\Bigg) \cdot \Pr(\bm{x}_i, \bm{s}_{i0}) \nonumber \\
& - \sum_{\bm{x}_i \in \bX, \bm{s}_{i0} \in \bS} \Bigg(h^{\Delta t_{K+1}}_{T}\left( \bm{g}^{\Delta t_K}_{t_K}( ... \bm{g}^{\Delta t_2}_{t_2}(\bm{g}^{\Delta t_1}_{t_1}(\bm{s}_{i0}, \bm{x}_i, \bm{0}_{\Delta t_1}), \bm{x}_i, \bm{0}_{\Delta t_2}) ... ,\bm{x}_i, \bm{0}_{\Delta t_K}), \bm{x}_i, \bm{0}_{\Delta t_{K+1}} \right) \nonumber \\
& \hspace{2cm} - h^{\Delta t_{K+1}}_{\Delta t_{K+1}} \left( \bm{g}^{\Delta t_K}_{t_K}( ... \bm{g}^{\Delta t_2}_{t_2}(\bm{g}^{\Delta t_1}_{t_1}(\bm{s}_{i0}, \bm{x}_i, \bm{0}_{\Delta t_1}), \bm{x}_i, \bm{0}_{\Delta t_2}) ... ,\bm{x}_i, \bm{0}_{\Delta t_K}), \bm{x}_i, \bm{0}_{\Delta t_{K+1}} \right)\Bigg) \cdot \Pr(\bm{x}_i, \bm{s}_{i0}) \label{eq:1}\\
& + \sum_{\bm{x}_i \in \bX, \bm{s}_{i0} \in \bS} \Bigg(h^{\Delta t_{K+1}}_{\Delta t_{K+1}} \left( \bm{g}^{\Delta t_K}_{t_K}( ... \bm{g}^{\Delta t_2}_{t_2}(\bm{g}^{\Delta t_1}_{t_1}(\bm{s}_{i0}, \bm{x}_i, \bm{1}_{\Delta t_1}), \bm{x}_i, \bm{1}_{\Delta t_2}) ... ,\bm{x}_i, \bm{1}_{\Delta t_K}), \bm{x}_i, \bm{1}_{\Delta t_{K+1}} \right) \nonumber \\
& \hspace{2cm} - h^{\Delta t_{K+1}}_{\Delta t_{K+1}} \left( \bm{g}^{\Delta t_K}_{\Delta t_K}( \bm{g}^{\Delta t_{K-1}}_{ t_{K-1}}( ... \bm{g}^{\Delta t_1}_{t_1}(\bm{s}_{i0}, \bm{x}_i, \bm{1}_{\Delta t_1}) ... ,\bm{x}_i, \bm{1}_{\Delta t_{K-1}}),\bm{x}_i, \bm{1}_{\Delta t_K}), \bm{x}_i, \bm{1}_{\Delta t_{K+1}} \right)\Bigg) \cdot \Pr(\bm{x}_i, \bm{s}_{i0}) \nonumber \\
& - \sum_{\bm{x}_i \in \bX, \bm{s}_{i0} \in \bS} \Bigg(h^{\Delta t_{K+1}}_{\Delta t_{K+1}} \left( \bm{g}^{\Delta t_K}_{t_K}( ... \bm{g}^{\Delta t_2}_{t_2}(\bm{g}^{\Delta t_1}_{t_1}(\bm{s}_{i0}, \bm{x}_i, \bm{0}_{\Delta t_1}), \bm{x}_i, \bm{0}_{\Delta t_2}) ... ,\bm{x}_i, \bm{0}_{\Delta t_K}), \bm{x}_i, \bm{0}_{\Delta t_{K+1}} \right) \nonumber \\
& \hspace{2cm} - h^{\Delta t_{K+1}}_{\Delta t_{K+1}} \left( \bm{g}^{\Delta t_K}_{\Delta t_K}( \bm{g}^{\Delta t_{K-1}}_{ t_{K-1}}( ... \bm{g}^{\Delta t_1}_{t_1}(\bm{s}_{i0}, \bm{x}_i, \bm{0}_{\Delta t_1}) ... ,\bm{x}_i, \bm{0}_{\Delta t_{K-1}}),\bm{x}_i, \bm{0}_{\Delta t_K}), \bm{x}_i, \bm{0}_{\Delta t_{K+1}} \right)\Bigg) \cdot \Pr(\bm{x}_i, \bm{s}_{i0}) \label{eq:2}\\
& +\ \cdots\ -\ \cdots  \nonumber\\
& + \sum_{\bm{x}_i \in \bX, \bm{s}_{i0} \in \bS} \Bigg(h^{\Delta t_{K+1}}_{\Delta t_{K+1}} \left( \bm{g}^{\Delta t_K}_{\Delta t_K}( ... \bm{g}^{\Delta t_2}_{\Delta t_2}(\bm{g}^{\Delta t_1}_{t_1}(\bm{s}_{i0}, \bm{x}_i, \bm{1}_{\Delta t_1}), \bm{x}_i, \bm{1}_{\Delta t_2}) ... ,\bm{x}_i, \bm{1}_{\Delta t_K}), \bm{x}_i, \bm{1}_{\Delta t_{K+1}} \right) \nonumber \\
& \hspace{2cm} - h^{\Delta t_{K+1}}_{\Delta t_{K+1}} \left( \bm{g}^{\Delta t_K}_{\Delta t_K}( ... \bm{g}^{\Delta t_2}_{\Delta t_2}(\bm{g}^{\Delta t_1}_{\Delta t_1}(\bm{s}_{i0}, \bm{x}_i, \bm{1}_{\Delta t_1}), \bm{x}_i, \bm{1}_{\Delta t_2}) ... ,\bm{x}_i, \bm{1}_{\Delta t_K}), \bm{x}_i, \bm{1}_{\Delta t_{K+1}} \right)\Bigg) \cdot \Pr(\bm{x}_i, \bm{s}_{i0}) \nonumber \\
& - \sum_{\bm{x}_i \in \bX, \bm{s}_{i0} \in \bS} \Bigg(h^{\Delta t_{K+1}}_{\Delta t_{K+1}} \left( \bm{g}^{\Delta t_K}_{\Delta t_K}( ... \bm{g}^{\Delta t_2}_{\Delta t_2}(\bm{g}^{\Delta t_1}_{t_1}(\bm{s}_{i0}, \bm{x}_i, \bm{0}_{\Delta t_1}), \bm{x}_i, \bm{0}_{\Delta t_2}) ... ,\bm{x}_i, \bm{0}_{\Delta t_K}), \bm{x}_i, \bm{0}_{\Delta t_{K+1}} \right) \nonumber \\
& \hspace{2cm} - h^{\Delta t_{K+1}}_{\Delta t_{K+1}} \left( \bm{g}^{\Delta t_K}_{\Delta t_K}( ... \bm{g}^{\Delta t_2}_{\Delta t_2}(\bm{g}^{\Delta t_1}_{\Delta t_1}(\bm{s}_{i0}, \bm{x}_i, \bm{0}_{\Delta t_1}), \bm{x}_i, \bm{0}_{\Delta t_2}) ... ,\bm{x}_i, \bm{0}_{\Delta t_K}), \bm{x}_i, \bm{0}_{\Delta t_{K+1}} \right) \Bigg) \cdot \Pr(\bm{x}_i, \bm{s}_{i0}) \label{eq:3}\\
& + \sum_{\bm{x}_i \in \bX, \bm{s}_{i0} \in \bS} \Bigg(h^{\Delta t_{K+1}}_{\Delta t_{K+1}} \left( \bm{g}^{\Delta t_K}_{\Delta t_K}( ... \bm{g}^{\Delta t_2}_{\Delta t_2}(\bm{g}^{\Delta t_1}_{\Delta t_1}(\bm{s}_{i0}, \bm{x}_i, \bm{1}_{\Delta t_1}), \bm{x}_i, \bm{1}_{\Delta t_2}) ... ,\bm{x}_i, \bm{1}_{\Delta t_K}), \bm{x}_i, \bm{1}_{\Delta t_{K+1}} \right) \nonumber \\
& \hspace{2cm} - h^{\Delta t_{K+1}}_{\Delta t_{K+1}} \left( \bm{g}^{\Delta t_K}_{\Delta t_K}( ... \bm{g}^{\Delta t_2}_{\Delta t_2}(\bm{g}^{\Delta t_1}_{\Delta t_1}(\bm{s}_{i0}, \bm{x}_i, \bm{0}_{\Delta t_1}), \bm{x}_i, \bm{0}_{\Delta t_2}) ... ,\bm{x}_i, \bm{0}_{\Delta t_K}), \bm{x}_i, \bm{0}_{\Delta t_{K+1}} \right)\Bigg) \cdot \Pr(\bm{x}_i, \bm{s}_{i0}) \nonumber \\
= &\ \sum_{\bm{x}_i \in \bX, \bm{s}_{i0} \in \bS} \Bigg(h^{\Delta t_{K+1}}_{\Delta t_{K+1}} \left( \bm{g}^{\Delta t_K}_{\Delta t_K}( ... \bm{g}^{\Delta t_2}_{\Delta t_2}(\bm{g}^{\Delta t_1}_{\Delta t_1}(\bm{s}_{i0}, \bm{x}_i, \bm{1}_{\Delta t_1}), \bm{x}_i, \bm{1}_{\Delta t_2}) ... ,\bm{x}_i, \bm{1}_{\Delta t_K}), \bm{x}_i, \bm{1}_{\Delta t_{K+1}} \right) \nonumber \\
& \hspace{2cm} - h^{\Delta t_{K+1}}_{\Delta t_{K+1}} \left( \bm{g}^{\Delta t_K}_{\Delta t_K}( ... \bm{g}^{\Delta t_2}_{\Delta t_2}(\bm{g}^{\Delta t_1}_{\Delta t_1}(\bm{s}_{i0}, \bm{x}_i, \bm{0}_{\Delta t_1}), \bm{x}_i, \bm{0}_{\Delta t_2}) ... ,\bm{x}_i, \bm{0}_{\Delta t_K}), \bm{x}_i, \bm{0}_{\Delta t_{K+1}} \right)\Bigg) \cdot \Pr(\bm{x}_i, \bm{s}_{i0}) \nonumber \\
= &\ \bE_{\cF}\left[h_{\Delta t_{K+1}}\left( \bm{g}_{\Delta t_K}( ... \bm{g}_{\Delta t_1}(\bm{S}_{i0}, \bm{X}_i, \bm{1}_{\Delta t_1}) ... ,\bm{X}_i, \bm{1}_{\Delta t_K}), \bm{X}_i, \bm{1}_{\Delta t_{K+1}} \right)\right] \nonumber \\
& \hspace{5cm} - \bE_{\cF}\left[h_{\Delta t_{K+1}}\left( \bm{g}_{\Delta t_K}( ... \bm{g}_{\Delta t_1}(\bm{S}_{i0}, \bm{X}_i, \bm{0}_{\Delta t_1}) ... ,\bm{X}_i, \bm{0}_{\Delta t_K}), \bm{X}_i, \bm{0}_{\Delta t_{K+1}} \right)\right] \nonumber
\end{align}

where in the first equality, equation~\eqref{eq:1} is equal to zero due to Assumption~\ref{asp:ParallelTrends:appendix} when $t=T,\ t'=\Delta t_{K+1},\ \delta=\Delta t_{K+1}$;  equation~\eqref{eq:2} is equal to zero due to Assumption~\ref{asp:ParallelTrends:Technical} when $u_0=\Delta t_{K+1},\ t=t_{K},\ t'=\Delta t_{K},\ \delta=\Delta t_{K}$; ... ; equation~\eqref{eq:3} is equal to zero due to Assumption~\ref{asp:ParallelTrends:Technical} when $(u_0, ..., u_{K-1})=(\Delta t_{K+1},...,\Delta t_{2}),\ t=t_{1},\ t'=\Delta t_{1},\ \delta=\Delta t_{1}$. This concludes the proof.
\hfill \halmos

\endproof

\subsection{Testing Parallel Trends Assumption}
\label{sec:appendix:comparabilityParallel}

We further introduce a statistical test for whether the parallel trends assumption holds. This test focuses on two time periods, \(t\) and \(t'\), and a specified positive integer \(\delta\). For each unit $i$ characterized by the values of any observations of \(\bm{S}_{i,t-\delta}\) and \(\bm{X}_i\) at period $t$ in the treatment group (where \(\bm{W}_{i,t-\delta+1:t}=\bm{1}_{\delta}\)), we search for an exact match at time period \(t'\) based on the same values of \(\bm{S}_{i,t'-\delta}\), \(\bm{X}_i\) and \(\bm{W}_{i,t'-\delta+1:t'} = \bm{1}_{\delta}\). If an exact match is found, one match in period $t'$ is randomly selected for comparison; then we have a pair of ($Y_{it}$ vs $Y_{i't'}$ and $\bm{S}_{it}$ vs $\bm{S}_{i't'}$) for the treatment group. In cases where no match is found, the observation is not used in this evaluation. Similarly, we repeat the procedure for observations in the control group (where \(\bm{W}_{i,t-\delta+1:t}=\bm{0}_{\delta}\)), and obtain pairs of units at period $t$ and $t'$.

The above matching process ensures the paired observations in both treatment and control groups are conditioned on the same distribution of pre-period surrogates and pre-treatment variables.  By pooling these paired observations together, we are able to perform a test based on a difference-in-difference type regression to evaluate if the Assumption~\ref{asp:ParallelTrends:appendix} holds. Consider the following regression:
$$Y_i = \beta_0 + \beta_1\cdot \bI\{\bm{w}_{i,\delta}=\bm{1}_{\delta}\} + \beta_2\cdot \bI\{period = t\} + \beta_3\cdot \bI\{\bm{w}_{i,\delta}=\bm{1}_{\delta}\}\cdot\bI\{period = t\} + \epsilon_i$$

\noindent where variable \(period\) controls the time period of the observation, \(\bm{w}_{i,\delta}\) controls the treatment assignment condition. We conduct the regression, and apply a t-test for the null \(H_{0}: \beta_3=0\). We say that Assumption~\ref{asp:ParallelTrends:appendix} is not significantly violated if the null hypothesis can not be rejected.

Table~\ref{tb:comparabilityTest} presents the results of statistical testing for various combinations of $t$, $t'$, and $\delta$ across both Empirical Experiment 1 and Empirical Experiment 2. These tests are conducted on data collected during the observable experimental periods, which is week two to week four for Experiment 1, and week eight to week ten for Experiment 2. 
We observe that p-values for all of tests are greater than 0.05, and none of the tests were rejected, providing evidence that there is no significant violation of Assumption~\ref{asp:ParallelTrends:appendix} in either experiment.

\begin{table}[htbp]
    \centering
    \caption{Testing results of the parallel trends assumption for two empirical experiments}\scriptsize
    \begin{tabular}{cccccccc}
        \toprule
        \textbf{Experiment} &$\bm{t}$ & $\bm{t'}$ & $\bm{\delta}$ & $\widehat{\beta_3}$ & \textbf{t-statistic} & \textbf{p-value} & \textbf{Reject?} \\
        \midrule
        \multirow{4}{*}{Experiment 1} & 2 & 3 & 1 & -0.0009  & -1.057 & 0.290 & No \\
        & 2 & 4 & 1 & -0.0011  & -1.295 & 0.195 & No \\
        & 3 & 4 & 1 &  0.0001  &  0.064 & 0.949 & No \\
        & 3 & 4 & 2 & -0.0008  & -0.923 & 0.356 & No \\
        \midrule
        \multirow{16}{*}{Experiment 2} 
        & 8 & 9  & 6 &  0.00002  &  0.026 & 0.979 & No\\
& 8 & 9  & 7 &  0.0009  &  0.884 & 0.377 & No \\
&8 & 10 & 6 &  0.0003 &  0.284 & 0.776 & No \\
&8 & 10 & 7 &  0.0005  &  0.532 & 0.594 & No\\
&8 & 11 & 6 & -0.0010  & -1.122 & 0.261 & No\\
&8 & 11 & 7 & -0.0002 & -0.153 & 0.879 & No\\
&9 & 10 & 6 &  0.0003  &  0.333 & 0.739 & No\\
&9 & 10 & 7 & -0.0006 & -0.667 & 0.505 & No\\
&9 & 10 & 8 &  0.0005  &  0.452 & 0.651 & No\\
&9 & 11 & 6 &  0.0002 &  0.249 & 0.804 & No\\
&9 & 11 & 7 & -0.0006  & -0.663 & 0.507 & No\\
&9 & 11 & 8 &  0.0010  &  0.986 & 0.324 & No\\
&10 & 11 & 6 & 0.0008  &  0.827 & 0.408 & No\\
&10 & 11 & 7 & 0.0006 &  0.612 & 0.540 & No\\
&10 & 11 & 8 & 0.00001  &  0.006 & 0.996 & No\\
&10 & 11 & 9 & -0.0008  & -0.752 & 0.452 & No\\
        \bottomrule
    \end{tabular}
    \label{tb:comparabilityTest}
\end{table}

\section{Sensitivity Analysis on Longitudinal Surrogacy Assumption}
\label{sec:appendix:surrogacy}

The surrogacy assumption necessitates that the chosen surrogate variables encapsulate the entire causal path from the previous treatment to the future primary outcome. A illustrator of the typical violation of Assumption~\ref{asp:Surrogacy} is shown in Figure~\ref{fig:SurrogacyViolation}. Here we show that our method's estimation remains robust even when the longitudinal surrogacy assumption is moderately violated (i.e., there exists a causal path not blocked by surrogates over time) through both synthetic and empirical experiments.

\begin{figure}[h]
\includegraphics[width=0.35\textwidth]{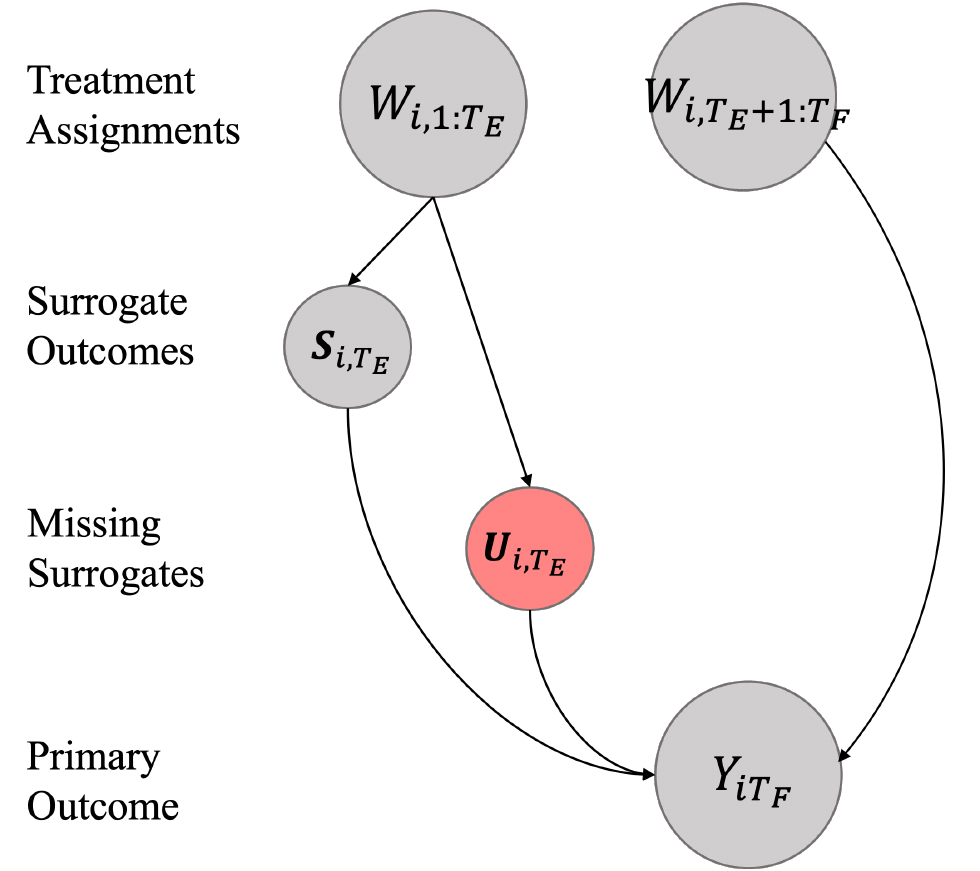}
\centering
\caption{An illustrator of a scenario where Assumption~\ref{asp:Surrogacy} is violated.}
{\footnotesize \textit{Note}: 
In this illustrator, each solid line represents a causal link. 
The treatment assignment at period $T_E+1:T_F$ impacts the primary outcomes at period $T_E+1:T_F$; The treatment assignment at period $T_E+1:T_F$ impact the primary outcome at period $T_E+1:T_F$ through both surrogate outcomes and omitted surrogates at period $1:T_E$. 
}
\label{fig:SurrogacyViolation}
\end{figure}

\subsection{Sensitivity Analysis with Subsets of Surrogate}
\label{sec:appendix:surrogateSubset}

\subsubsection{Evidence from the synthetic experiments}
We revisit the two synthetic experiments discussed in the main context. Both simulations initialize with four surrogates, denoted as \(\bm{S}_i = (S^1_{i0},S^2_{i0},S^3_{i0},S^4_{i0})\). Each surrogate follows a normal distribution: \(S^d_{i0}\sim\mathcal{N}(\mu_d,\sigma_d)\), where \(\mu_d \sim \mathcal{N}(2,1)\) and \(\sigma_d \sim \mathcal{N}(2,1)\) for \(d\in \{1,2,3,4\}\); if negative value is sampled we flip the sign. It is evident that these four surrogates are independent of each other. Throughout the experiment, the distribution of the surrogates remains unchanged in the control group, i.e. \(S^d_{it} \sim\mathcal{N}(\mu_d,\sigma_d)\) for \(d\in \{1,2,3,4\}\) and each period $t$. However, in the treatment group, the surrogate values decay over time governed by decay factors $\kappa_d$ of \([0.8,0.6,0.4,0.2]\) respectively, i.e. \(S^{d}_{i,t+1} = \kappa_d\cdot S^d_{t}\) for \(d\in \{1,2,3,4\}\) and each period $t$. The primary outcome, \(Y\), is designed as \(Y_{i,t+1} = -(0.1S^1_{it} + 0.1S^2_{it} + 0.4S^3_{it} + 0.4S^4_{it})\) in the first synthetic experiment, and \(Y_{i,t+1} = 0.1S^1_{it} + 0.1S^2_{it} + 0.4S^3_{it} + 0.4S^4_{it}\) in the second synthetic experiment. 

In the following analysis, we adhere to the same data generation process described above. However, we proceed as if we only observe a subset of the four underlying surrogates. Therefore, if we only use a subset as surrogates in our method, the longitudinal surrogacy assumption is violated.

Figures~\ref{fig:sim1NoSurrogacy} and~\ref{fig:sim2NoSurrogacy} showcase the results for the estimated long-term treatment effect, utilizing only subsets of the full surrogates in the two synthetic experiments. The specific set of surrogates used for prediction is indicated in the far-right column of the figures. It is evident that our linear surrogate model performs better with a more comprehensive set of surrogates. Furthermore, the influence of surrogacy violation varies based on the length of the experimental period data. Given the same set of surrogates, the estimation approach demonstrates greater robustness with a longer experimental period as opposed to a shorter one.

\begin{figure}[tb]
\centering
\includegraphics[width=0.9\linewidth]
{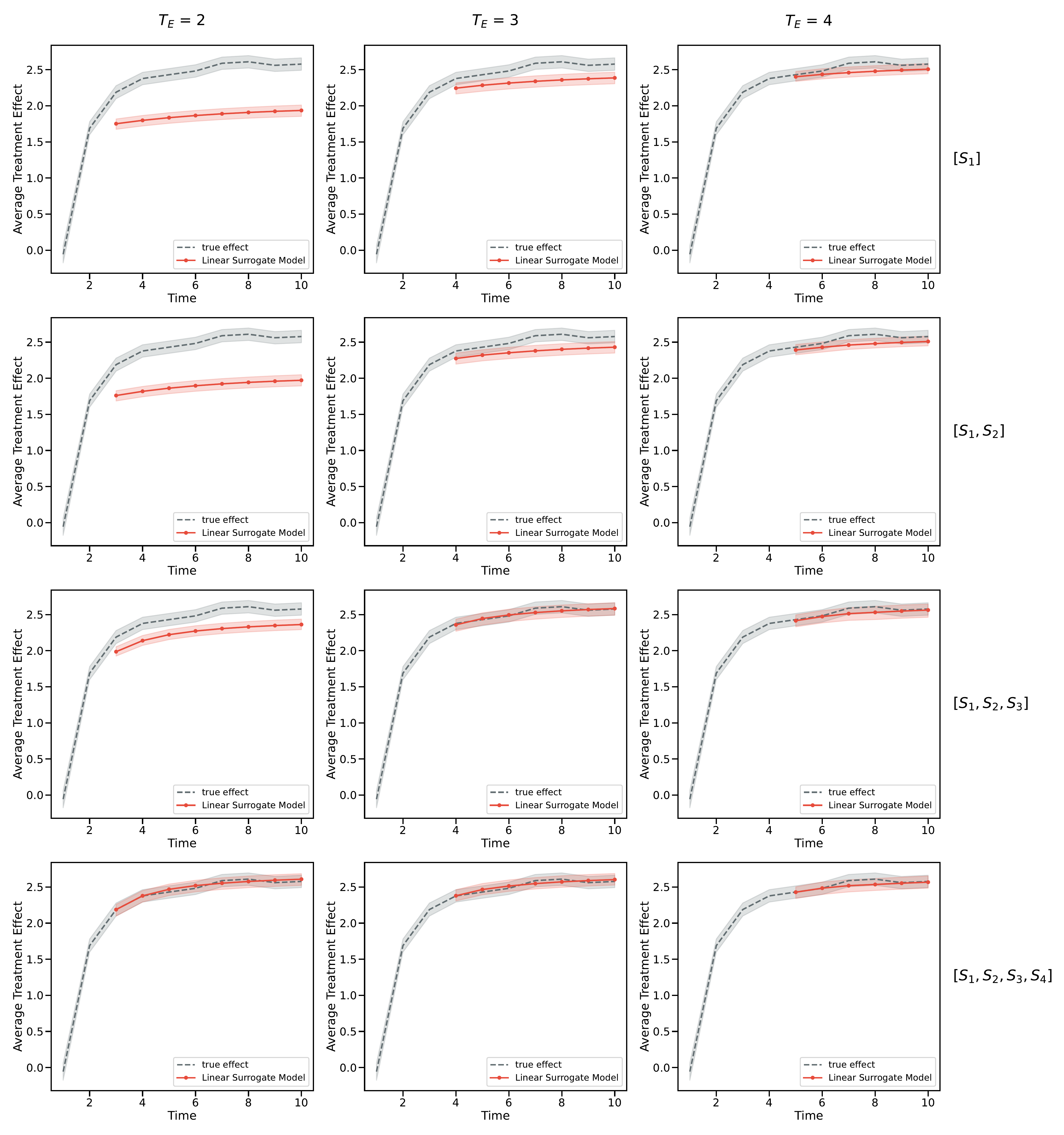}
\caption{Estimated effect with violated surrogacy assumption in the Synthetic Experiment 1}
{\footnotesize \textit{Note}: Grey dashed curves represent the true average treatment effect on $Y$ from periods 1 to periods 10. Solid red curves represent the estimated effects with the linear surrogate model. Shadows indicate 95\% confidence intervals. The four rows represent the specific set of surrogates used for prediction, where a smaller set signifies a more significant violation of the surrogacy assumption. The three panels represent the scenarios when we use the first $T_E$ weeks as the experimental period and the last $T_F$ weeks as the future period.}
\label{fig:sim1NoSurrogacy}
\end{figure}

\begin{figure}[tb]
\centering
\includegraphics[width=0.9\linewidth]
{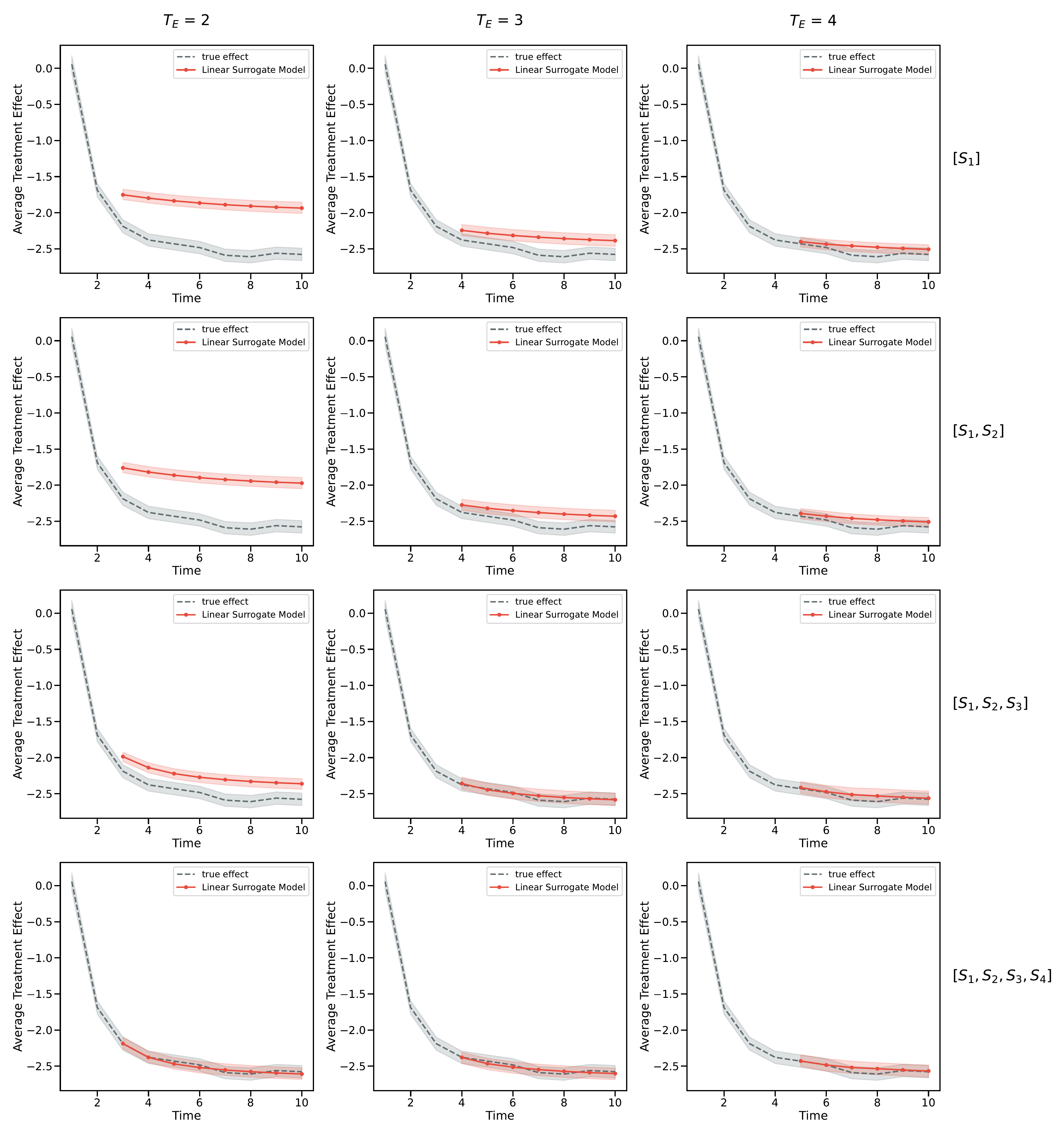}
\caption{Estimated effect with violated surrogacy assumption in the Synthetic Experiment 2}
{\footnotesize \textit{Note}: Grey dashed curves represent the true average treatment effect on $Y$ from periods 1 to periods 10. Solid red curves represent the estimated effects with the linear surrogate model. Shadows indicate 95\% confidence intervals. The four rows represent the specific set of surrogates used for prediction, where a smaller set signifies a more significant violation of the surrogacy assumption. The three panels represent the scenarios when we use the first $T_E$ weeks as the experimental period and the last $T_F$ weeks as the future period.}
\label{fig:sim2NoSurrogacy}
\end{figure}

\subsubsection{Evidence from the empirical experiments}

Similar to what has been done in the synthetic experiments, we attempt to use only a subset of surrogates to build the longitudinal surrogate model, and compare the estimation from which with the true effect in two empirical experiments. 

Figure~\ref{fig:searchHistoryLessSurrogate} and Figure~\ref{fig:searchDiscoveryLessSurrogate} present the estimation results obtained from the linear surrogate model using different subsets of the full surrogates for both empirical experiments. The specific surrogates employed for prediction are listed in the far-right column of the figures. In the real-world study, we can still observe an overall tendency that richer surrogates result in a more precise estimation.

\begin{figure}[h]
\centering
\includegraphics[width=0.9\linewidth]
{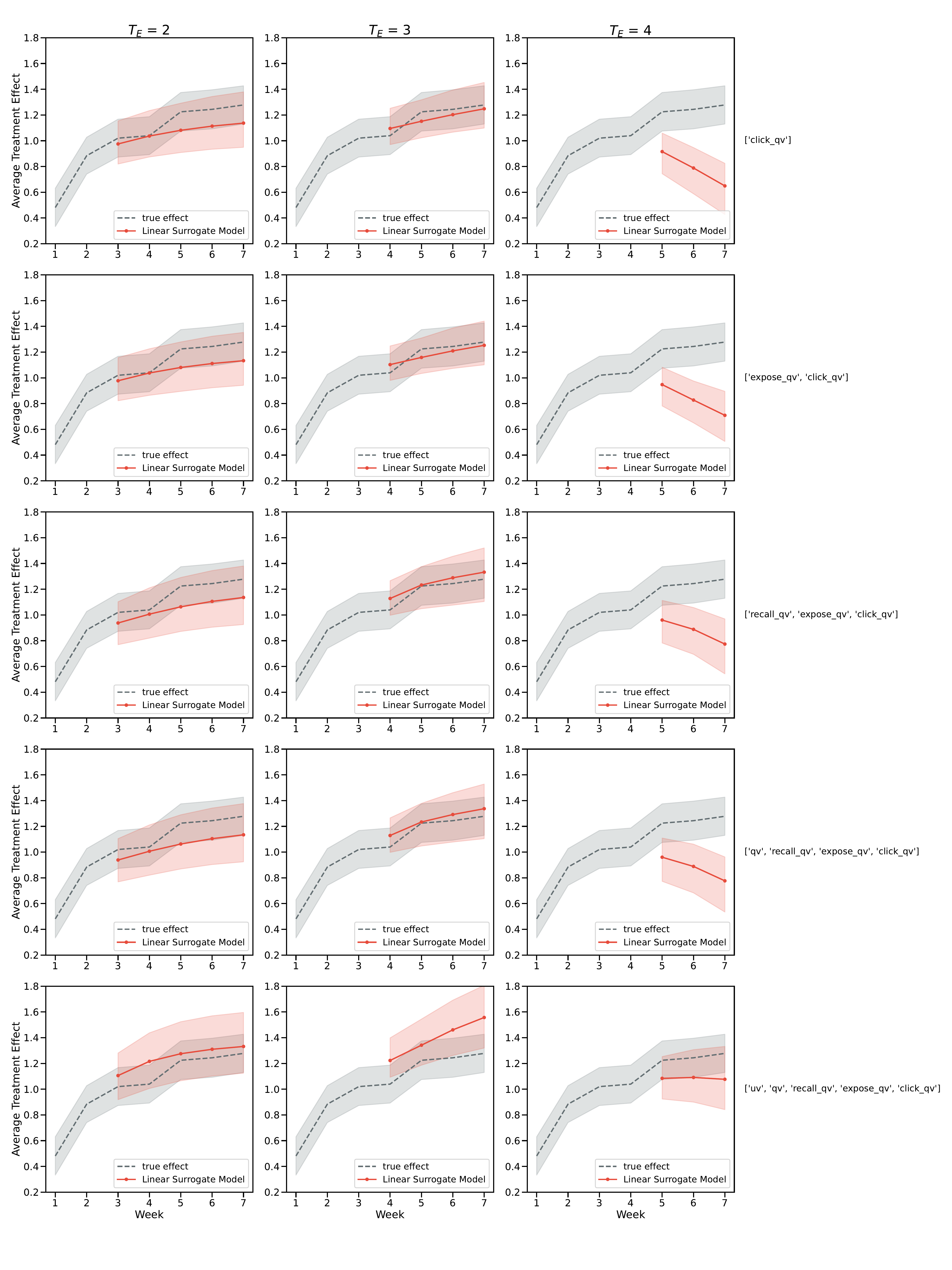}
\caption{Estimated effect with violated surrogacy assumption in Experiment 1}
{\footnotesize \textit{Note}: Grey dashed curves represent the true average treatment effect on \textit{search\_uv} from week 1 to week 7. Solid red curves represent the estimated effects with the linear surrogate model. Shadows indicate 95\% confidence intervals. The five rows represent the specific set of surrogates used for prediction, where a smaller set signifies a more significant violation of the surrogacy assumption. The three panels represent the scenarios when we use the first $T_E$ weeks as the experimental period and the last $T_F$ weeks as the future period.}
\label{fig:searchHistoryLessSurrogate}
\end{figure}

\begin{figure}[h]
\centering
\includegraphics[width=0.9\linewidth]
{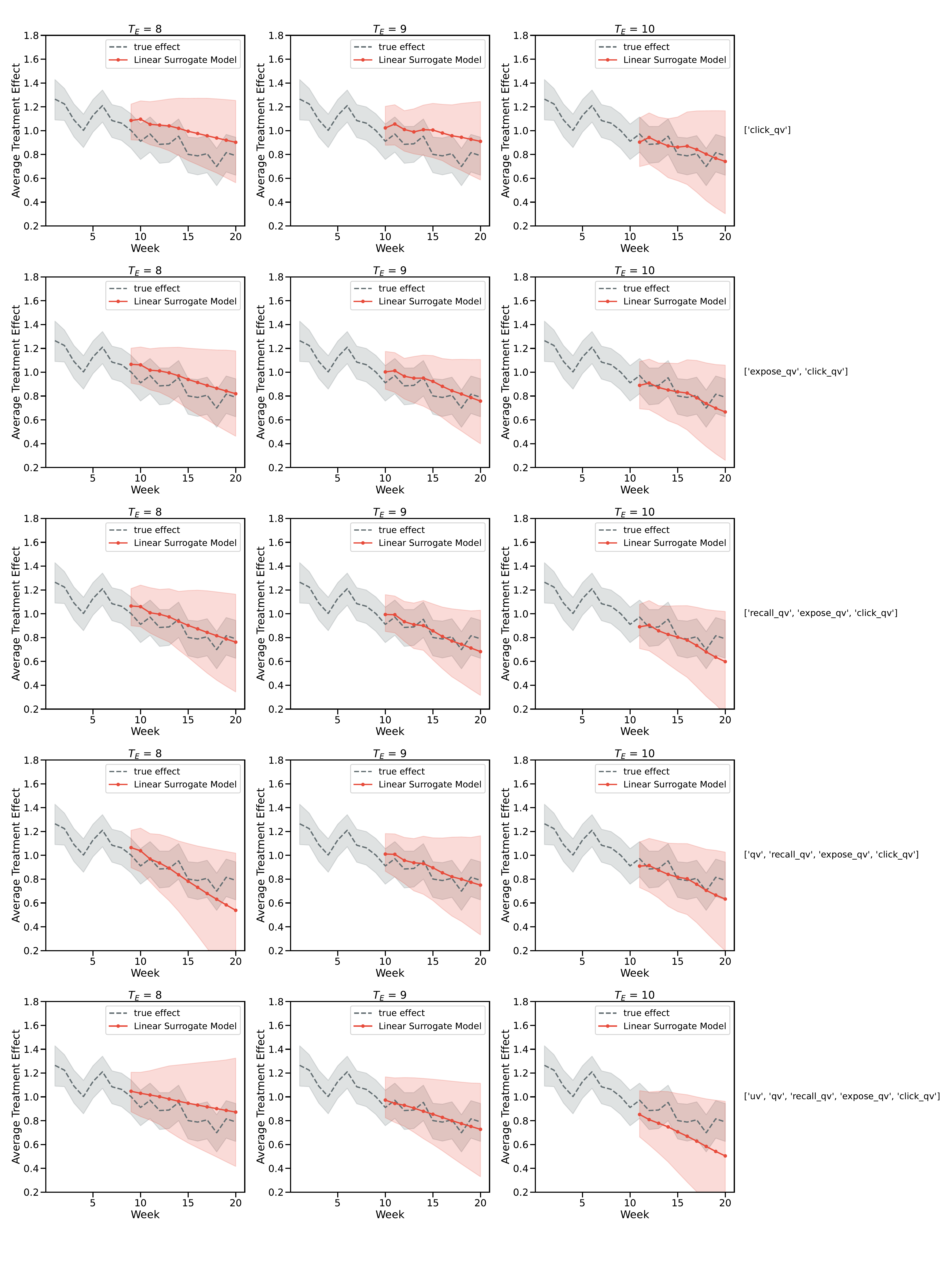}
\caption{Estimated effect with violated surrogacy assumption in Experiment 2}
{\footnotesize \textit{Note}: Grey dashed curves represent the true average treatment effect on \textit{search\_uv} from week 1 to week 20. Solid red curves represent the estimated effects with the linear surrogate model. Shadows indicate 95\% confidence intervals. The five rows represent the specific set of surrogates used for prediction, where a smaller set signifies a more significant violation of the surrogacy assumption. The three panels represent the scenarios when we use the first $T_E$ weeks as the experimental period and the last $T_F$ weeks as the future period.}
\label{fig:searchDiscoveryLessSurrogate}
\end{figure}

\subsection{Sensitivity Analysis of Omitted Surrogates}
\label{sec:appendix:missingSurrogates}

In real-world experiments, it is hard to quantify how much the surrogacy assumption is violated because we can't be certain if current surrogates fully satisfy this assumption. This challenge is similar to that faced in the literature regarding the validity of instrumental variables (IVs). 
Drawing on sensitivity analysis techniques used for IVs \citep{baiocchi2014instrumental}, we carry out sensitivity analysis on the both empirical experiments introduced in the main text to  demonstrate further the robustness of our method when the surrogacy assumption is relaxed in real-world scenarios. Suppose for any $i \in [N]$ and any $t \in \bT$, the treatment assignment affects the primary outcome not only through the identified surrogates, but also via an unobserved variable $\zeta_{it}$ which follows normal distribution with mean zero, and variance equals to the average variance of the primary outcome during the experimental period. We manually introduce an additional causal path between the treatment assignment and the primary outcome through variable $\zeta_{it}$ as follows:
\begin{align*}
\tilde{Y}_{it}(\bm{w}_{1:t}) = Y_{it}(\bm{w}_{1:t}) + \theta  \cdot \mathbbm{1}[\bm{w}_{1:t}=\bm{1}_{1:t}] \cdot \zeta_{it}
\end{align*}

In this way, if we use $\tilde{Y}$ instead of $Y$ as the outcome variable and consider only the existing surrogates, a larger $\theta$ indicates that our longitudinal surrogacy assumption is violated to a greater extent. Our objective is to illustrate that our model is relatively robust against this potential violation. Figure~\ref{fig:sensitivityTheta} presents a comparative analysis of bias and root mean squared error (RMSE) for Experiment 1 and Experiment 2, with varying values of $\theta$. We apply the same experimental periods from the main text, calculating the average bias and RMSE for each $\theta$. Specifically, for Experiment 1, this is averaged over $T_E = 2, 3, 4$, and for Experiment 2, over $T_E = 8, 9, 10$. The analysis indicates that, compared to our main model estimates (where $\theta = 0$), the bias and RMSE remain similar when $\theta$ is relatively small, demonstrating the robustness of the estimation.

\begin{figure}[h]
\centering
\includegraphics[width=0.7\linewidth]
{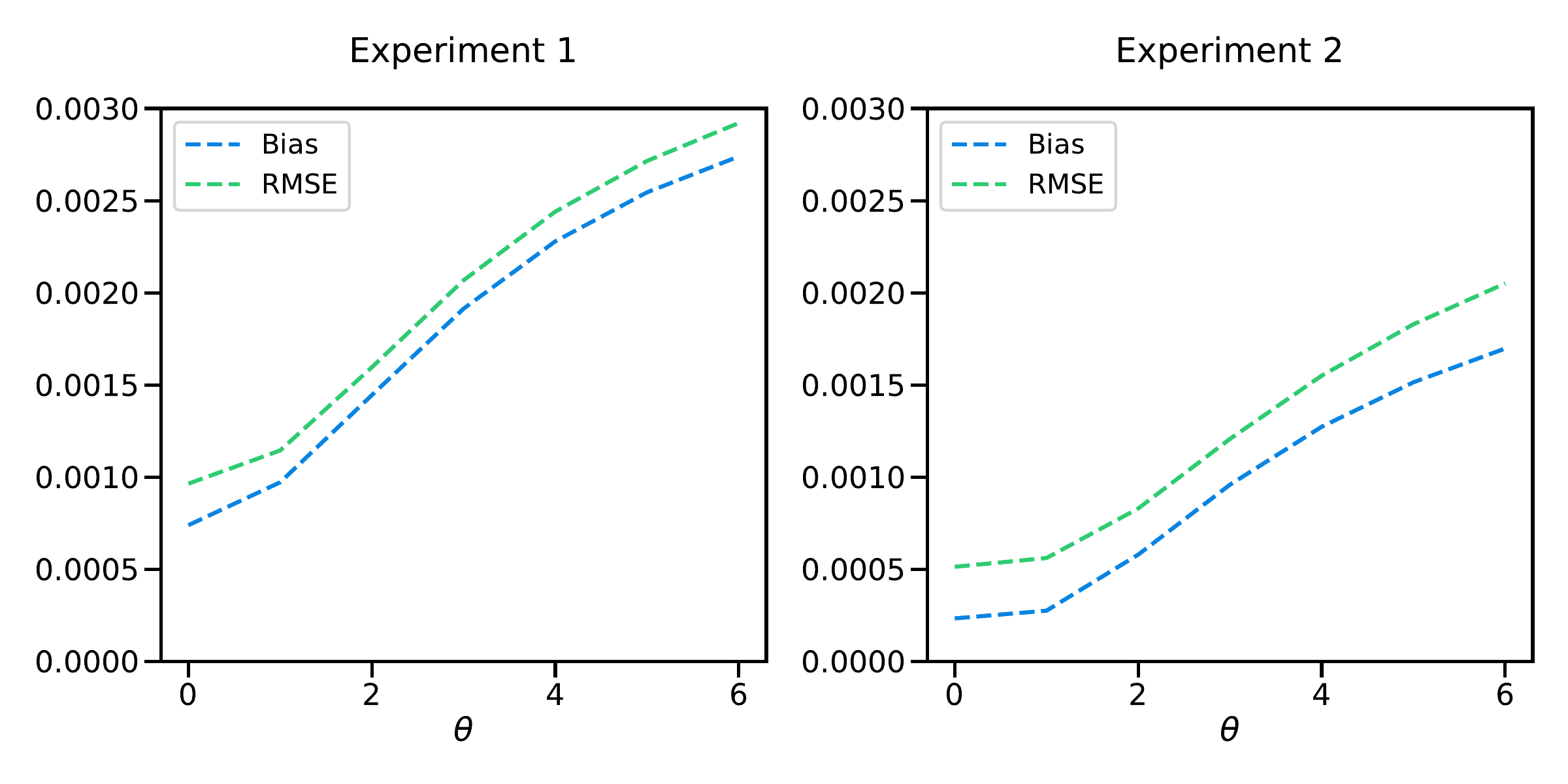}
\caption{Performance of Linear Surrogate Model in terms of Bias and RMSE under varying values of $\theta$ for Experiment 1 and Experiment 2}
\label{fig:sensitivityTheta}
\end{figure}

\section{Additional Results for Synthetic Experiments}
\label{sec:appendix:synthetic}

\subsection{Results for synthetic experiments with stabilized treatment effect}

Here we present the detailed results for the synthetic experiments discussed in Section~\ref{sec:Stabilized} of the main text. Figures~\ref{fig:sim1Baseline} and \ref{fig:sim2Baseline} graphically illustrate our model alongside baseline methods. As shown in both figures, our model effectively captures the decreasing and stabilizing trend of long-term treatment effects, a pattern that baseline models fail to capture. Numerical comparisons provided in Tables~\ref{tb:synthetic1MSE} and~\ref{tb:synthetic2MSE} further demonstrate that our approach outperforms baseline models in terms of both bias and MSE.

\begin{figure}[h]
\centering
\includegraphics[width=0.8\linewidth]
{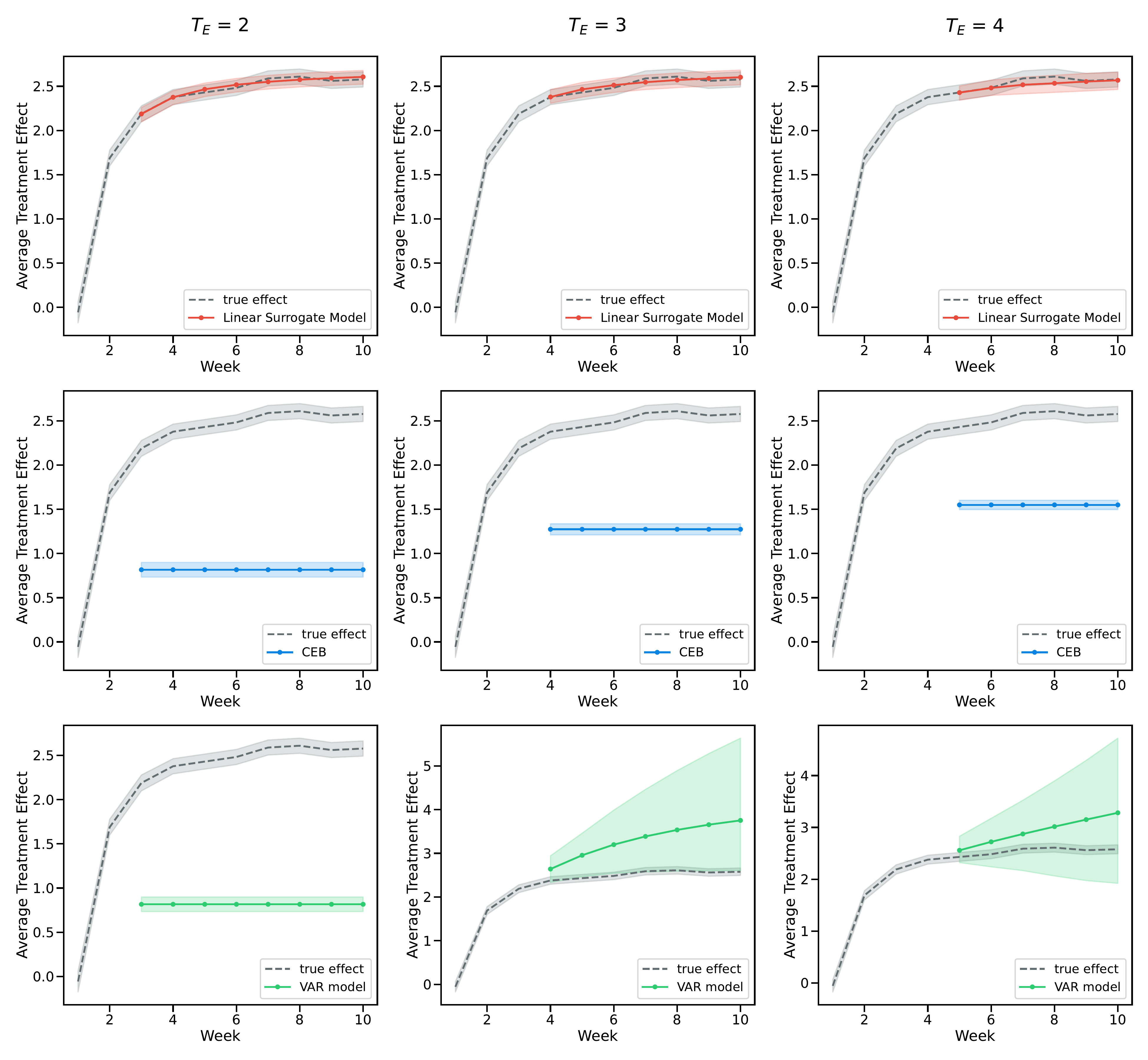}
\caption{Effects of Long-term Treatment using Linear Surrogate Model for the Synthetic Experiment 1}
{\footnotesize \textit{Note}: Grey dashed curves represent the true average treatment effect on $Y$ from periods 1 to periods 10. Solid red curves represent the estimated effects with the linear surrogate model. Shadows indicate 95\% confidence intervals. The three panels represent the scenarios when we use the first $T_E$ periods as the experimental period and the last $T_F$ periods as the future period.}
\label{fig:sim1Baseline}
\end{figure}

\begin{figure}[h]
\centering
\includegraphics[width=0.8\linewidth]
{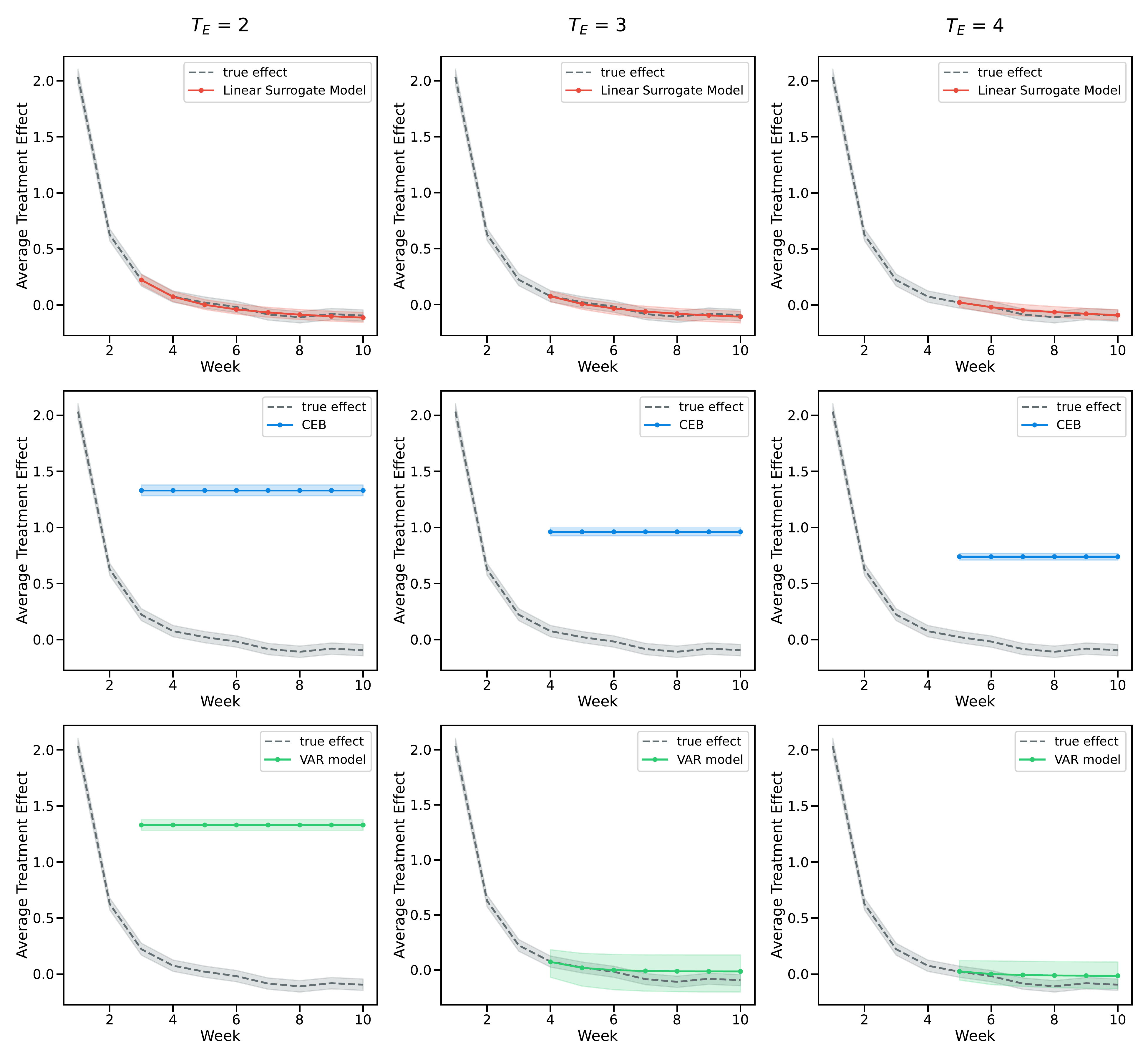}
\caption{Effects of Long-term Treatment using Linear Surrogate Model for the Synthetic Experiment 1}
{\footnotesize \textit{Note}: Grey dashed curves represent the true average treatment effect on $Y$ from periods 1 to periods 10. Solid red curves represent the estimated effects with the linear surrogate model. Shadows indicate 95\% confidence intervals. The three panels represent the scenarios when we use the first $T_E$ periods as the experimental period and the last $T_F$ periods as the future period.}
\label{fig:sim2Baseline}
\end{figure}

\begin{table}[h!]
\centering
\caption{Comparison result between different methods in terms of bias and MSE for synthetic experiment 1}\scriptsize
  \begin{tabular}{c ccc ccc}
    \toprule
    \multirow{2}{*}{\textbf{Method}}&\multicolumn{3}{c}{\textbf{Bias}}&\multicolumn{3}{c}{\textbf{MSE}}\\
      & {$\mathbf{T_E=2}$} & {$\mathbf{T_E=3}$} & {$\mathbf{T_E=4}$} & {$\mathbf{T_E=2}$} & {$\mathbf{T_E=3}$} & {$\mathbf{T_E=4}$} \\
      \midrule
   Linear surrogate model & $0.026$ & $0.029$ & $0.028$ & $0.003$ & $0.003$ & $0.004$ \\
    CEB & $1.661$ & $1.244$ & $0.992$ & $2.778$ & $1.567$ & $0.989$ \\
   VAR model & $1.661$ & $0.785$ & $0.392$ & $2.778$ & $1.073$ & $0.459$ \\
    \bottomrule
  \end{tabular}
\label{tb:synthetic1MSE}
\end{table}

\begin{table}[h!]
\centering
\caption{Comparison result between different methods in terms of Bias and MSE for synthetic experiment 2}\scriptsize  \begin{tabular}{c ccc ccc}
    \toprule
    \multirow{2}{*}{\textbf{Method}}&\multicolumn{3}{c}{\ \textbf{Bias}}&\multicolumn{3}{c}{\textbf{MSE}}\\
      & {$\mathbf{T_E=2}$} & {$\mathbf{T_E=3}$} & {$\mathbf{T_E=4}$} & {$\mathbf{T_E=2}$} & {$\mathbf{T_E=3}$} & {$\mathbf{T_E=4}$} \\
      \midrule
   Linear surrogate model & $0.015$ & $0.016$ & $0.015$ & $0.001$ & $0.001$ & $0.001$ \\
    CEB & $1.338$ & $1.002$ & $0.800$ & $1.802$ & $1.007$ & $0.642$ \\
   VAR model & $1.338$ & $0.048$ & $0.057$ & $1.802$ & $0.010$ & $0.007$ \\
    \bottomrule
  \end{tabular}
\label{tb:synthetic2MSE}
\end{table}

\subsection{Violation of Comparability}
\label{sec:appendix:additionalNonComparable} 

In order to better illustrate the necessity of satisfying Assumption~\ref{asp:Comparability} (Comparability Assumption), or alternatively, Assumption~\ref{asp:ParallelTrends:appendix} (Parallel Trends Assumption), we conduct following simulations where the primary outcome varies in how it breaches the comparability assumption.

In the initial simulations, we consider two surrogates, $S_{i0,1}, S_{i0,2}$, each following a normal distribution: \(S_{i0,1} \sim\mathcal{N}(\mu_1,\sigma_1^2)\) \(S_{i0,2} \sim\mathcal{N}(\mu_2,\sigma_2^2)\), where \(\mu_1,\mu_2,\sigma_1,\sigma_2 \sim \mathcal{N}(2,1)\)  before the experiment starts. For the control group, the distribution of surrogates remains unchanged throughout the experiment, i.e. \(S_{it,1}  \sim\mathcal{N}(\mu_1,\sigma_1^2)\) \(S_{it,2}\sim\mathcal{N}(\mu_2,\sigma_2^2)\) for each period $t$. In contrast, for the treatment group, the values of the surrogates decrease over time, influenced by decay factors \([0.8,0.6]\) respectively, which means that $S_{i,t+1,1} = 0.8\cdot S_{i,t,1}$ and $S_{i,t+1,2} = 0.6\cdot S_{i,t,1}$ for each period $t$. 

We define the primary outcome \(Y_{it}\) as a linear combination of these two surrogates, formulated as \(Y_{it} = -(0.1S_{it,1} + 0.4S_{it,2})\) at each period $t$, except for $t=2$. At $t=2$, we introduce an external shock for the treatment group, where the primary outcome \(Y_{i2}\) is formulated as \(Y_{i2} = -\gamma *(0.1S_{i2,1} + 0.4S_{i2,2})\) for subjects in the treatment group, to simulate the possible scenario where a festival amplifies the effect of the treatment. For the control group, the primary outcome remains unchanged. We set $\gamma$ to be \([1,1.5,2,2.5,3]\), with larger values indicating a more significant violation of the comparability assumption (no violation when $\gamma=1$). We focus on the scenario where \(t=2\), \(t'=3\) and \(\delta=1\), comparing the distribution of the primary outcome at time period two, conditional on surrogates from one period before, with the distribution at time period three under the same conditions. A violation of the comparability assumption will result in a discrepancy between these distributions. We expect that both the comparability assumption test and the parallel trends assumption test we proposed can detect this violation.

We first conduct the comparability assumption test for both the treatment and control groups under different values of \(\gamma\), as shown in Table~\ref{tb:comparabilityTestSim}.\footnote{Note that the total number of tests remains the same across different \(\gamma\) values in the treatment group, since the condition, i.e., the value of surrogates, is unchanged and unrelated to \(\gamma\).} We observe that Assumption~\ref{asp:Comparability} is most likely satisfied in the control group and in the treatment group when \(\gamma=1\). However, it is clearly violated in the treatment group as \(\gamma\) increases, with strong evidence. Table~\ref{tb:parallelTestSim}, which presents the results for the parallel trends assumption test, reinforces this conclusion, as the t-test is rejected for \(\gamma > 1.5\). For the test under \(\gamma=1.5\), it is not rejected but shows a relatively small p-value, indicating that a slight violation of the parallel trends assumption may not severely impact the estimation due to the robustness of our approach. Figure~\ref{fig:simNoComparability} displays the estimation results of our linear surrogate model under different \(\gamma\) values. We observe that as the degree of violation increases (\(\gamma\) becomes larger), the estimation becomes more biased.

\begin{table}[h]
\centering
\caption{Testing results of the comparability assumption with different values of $\gamma$}\scriptsize
\begin{tabular}{cccccccc}
\toprule
$\bm{\gamma}$ & \textbf{Groups} & $\bm{\#}$\textbf{Tests} & $\bm{\# p<0.1}$ & $\bm{\# p<0.05}$ & $\bm{p<0.1 (\%)}$ & $\bm{p<0.05 (\%)}$ \\
\midrule
1.0 & \multirow{5}{*}{Treatment} & 217 & 17 & 10 & 7.83 & 4.61 \\
1.5 & & 217 & 189 & 162 & 87.10 & 74.65 \\
2.0 & & 217 & 201 & 180 & 92.63 & 82.95 \\
2.5 & & 217 & 203 & 187 & 93.55 & 86.18 \\
3.0 & & 217 & 207 & 195 & 95.39 & 89.86 \\
\cmidrule{3-7}
/ & Control & 62 & 7 & 4 & 11.29 & 6.45 \\
\bottomrule
\end{tabular}
\label{tb:comparabilityTestSim}
\end{table}

\begin{table}[htbp]
    \centering
    \caption{Testing results of the parallel trends assumption with different values of $\gamma$}\scriptsize
    \begin{tabular}{ccccc}
        \toprule
        $\bm{\gamma}$ & $\widehat{\beta_3}$ & \textbf{t-statistic} & \textbf{p-value} & \textbf{Reject?} \\
        \midrule
        1.0 & -0.001 & -0.009 & 0.993 & No \\
        1.5 & -0.190 & -1.601 & 0.111 & No \\
        2.0 & 0.440 & -3.263 & 0.001 & Yes \\
        2.5 & -0.634 & -4.552 & 0.000 & Yes \\
        3.0 & -0.793 & -5.589 & 0.000  & Yes \\
        \bottomrule
    \end{tabular}
    \label{tb:parallelTestSim}
\end{table}

\begin{figure}[h]
\centering
\includegraphics[width=1\linewidth]
{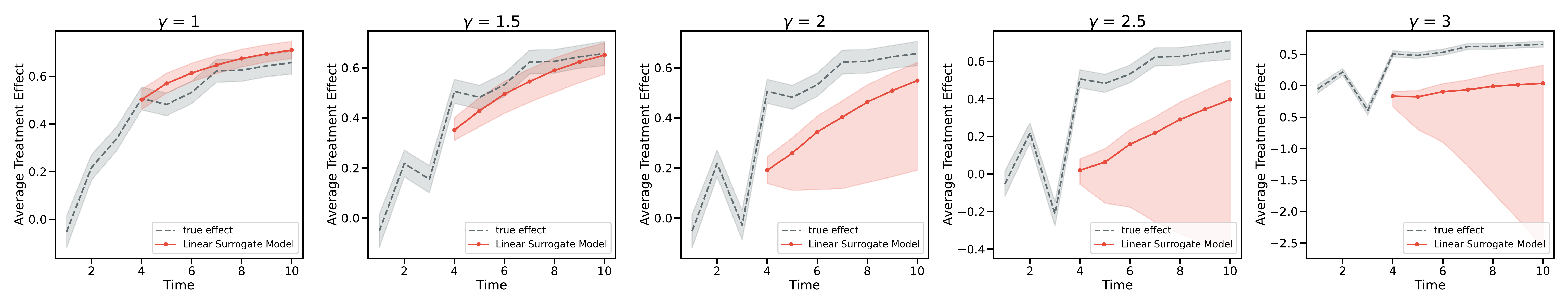}
\caption{Effect estimation in synthetic experiments when the comparability assumption is not satisfied.}
{\footnotesize \textit{Note}: Grey dashed curves represent the true average treatment effect on $Y$ from periods 1 to periods 10. Solid red curves represent the estimated effects with the linear surrogate model. Shadows indicate 95\% confidence intervals. The five columns represent a different degree of violation of the comparability assumption, with a larger $\gamma$ indicating a more severe violation (no violation when $\gamma=1$). We use the first three periods as the experimental period and the last seven periods as the future period.}
\label{fig:simNoComparability}
\end{figure}

\subsection{Non-linear} 
\label{sec:appendix:additionalNonlinear}
The previous synthetic experiments have had primary outcomes $Y_{it}$ as a linear combination of surrogates. As data distributions in real-world experiments may go beyond this linear formulation, we perturb the outcome functions in Section~\ref{sec:synthetic} from linear to non-linear to probe the sensitivity of our approach. This shows the robustness of our approach under non-linear scenarios when Assumption~\ref{asp:Linearity} is not satisfied. 

The simulations are initialized with two surrogate variables, denoted as $S_{i0,1}, S_{i0,2}$, each following a normal distribution: \(S_{i0,1} \sim\mathcal{N}(\mu_1,\sigma_1^2)\) \(S_{i0,2} \sim\mathcal{N}(\mu_2,\sigma_2^2)\), where \(\mu_1,\mu_2,\sigma_1,\sigma_2 \sim \mathcal{N}(2,1)\)  before the experiment starts.   Once the experiment begins, the surrogates in the control group continue to follow the same distribution as in the pre-treatment period, i.e. \(S_{it,1}  \sim\mathcal{N}(\mu_1,\sigma_1^2)\) \(S_{it,2}\sim\mathcal{N}(\mu_2,\sigma_2^2)\) for each period $t$. By contrast, the values of surrogates in the treatment group decrease over time, influenced by decay factors of \([0.8,0.6]\) respectively, which means that $S_{i,t+1,1} = 0.8\cdot S_{i,t,1}$ and $S_{i,t+1,2} = 0.6\cdot S_{i,t,1}$ for each period $t$.

In the first synthetic experiment, the primary outcome $Y$ is designed as a non-linear relationship, $Y_{i,t+1} = -(S_{it,1} + \theta\cdot e^{S_{it,2}})$ at each period t, where $\theta$ controls the magnitude of the exponential term. With this configuration, the average treatment effect of the long-term treatment initially increases and eventually converges to $\lim_{t\to\infty}\bE[Y_{i,t}(1)-Y_{i,t}(0)] = \mu_1+\theta\cdot e^{\mu_2+\sigma_2^2/2}-1$ at time $t$ approaches infinity. The exponential term introduces nonlinearity to the function, thereby violating the linearity assumption when estimating the treatment effect with a linear surrogate index function. We present the results of our estimation in Figure~\ref{fig:sim1Nonlinear}. We observe that the estimation is relatively stable and valid regardless of the magnitude  of $\theta$, which illustrates the robustness of our approach.

\begin{figure}[tb]
\centering
\includegraphics[width=0.9\linewidth]
{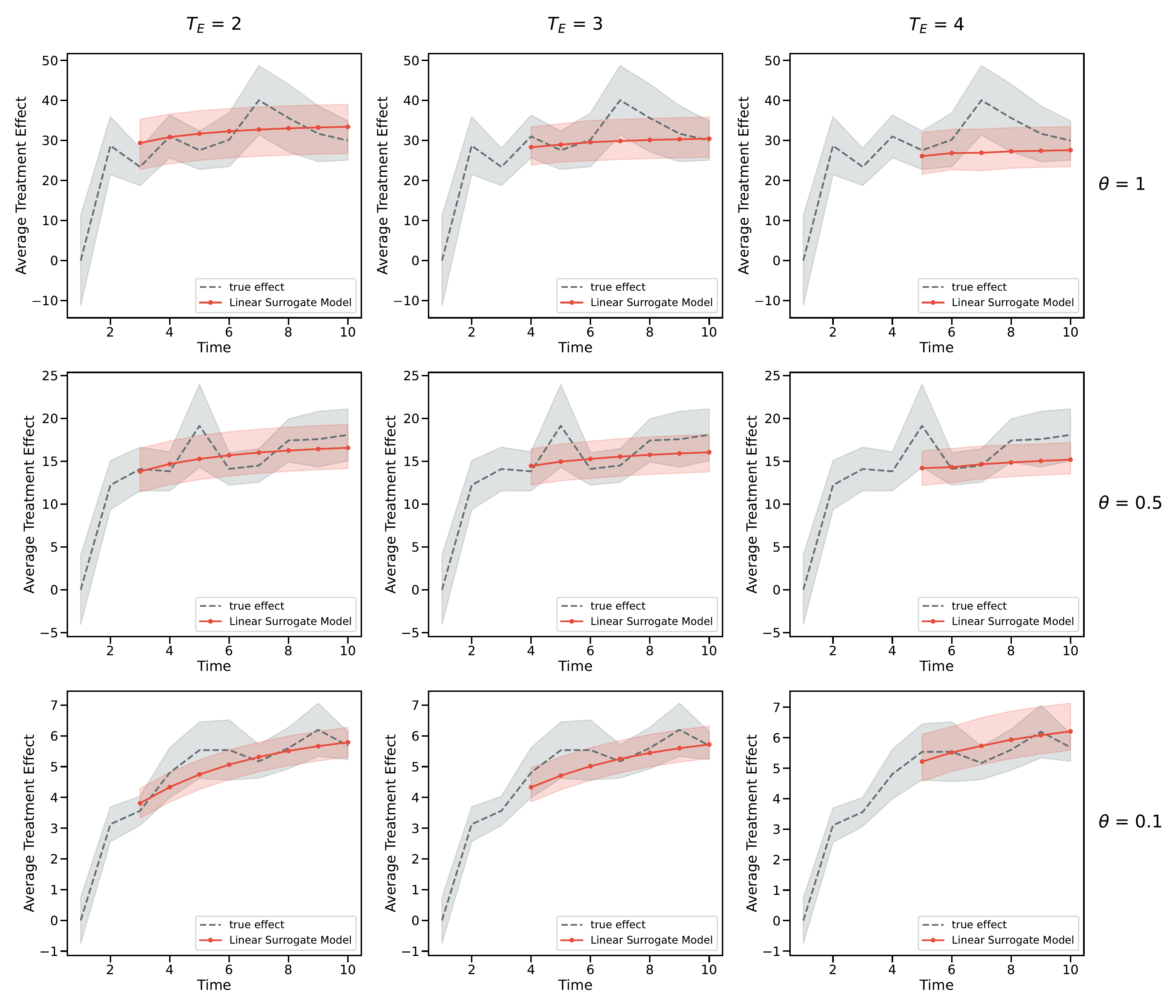}
\caption{Robust effect estimation in the first synthetic experiment when the linearity assumption is not satisfied.}
{\footnotesize \textit{Note}: Grey dashed curves represent the true average treatment effect on $Y$ from periods 1 to periods 10. Solid red curves represent the estimated effects with the linear surrogate model. Shadows indicate 95\% confidence intervals. The three rows represent a different degree of violation of the linearity assumption, with a larger $\theta$ indicating a more severe violation. The three panels represent the scenarios when we use the first $T_E$ periods as the experimental period and the last $T_F$ periods as the future period.}
\label{fig:sim1Nonlinear}
\end{figure}

In the second synthetic experiment, we similarly design the primary outcome $Y$ as $Y_{i,t+1} = S_{it,1} + \theta\cdot e^{S_{it,2}}$, where $\theta$ determines the magnitude of the exponential term. In this scenario, the average treatment effect of the long-term treatment initially decreases and then stabilizes around a certain level, which is $1-\mu_1-\theta\cdot e^{\mu_2+\sigma_2^2/2}$. Similarly, as shown in Figure~\ref{fig:sim2Nonlinear}, the estimation remains reasonably accurate under these conditions, further demonstrating the robustness of our method.

\begin{figure}[tb]
\centering
\includegraphics[width=0.9\linewidth]
{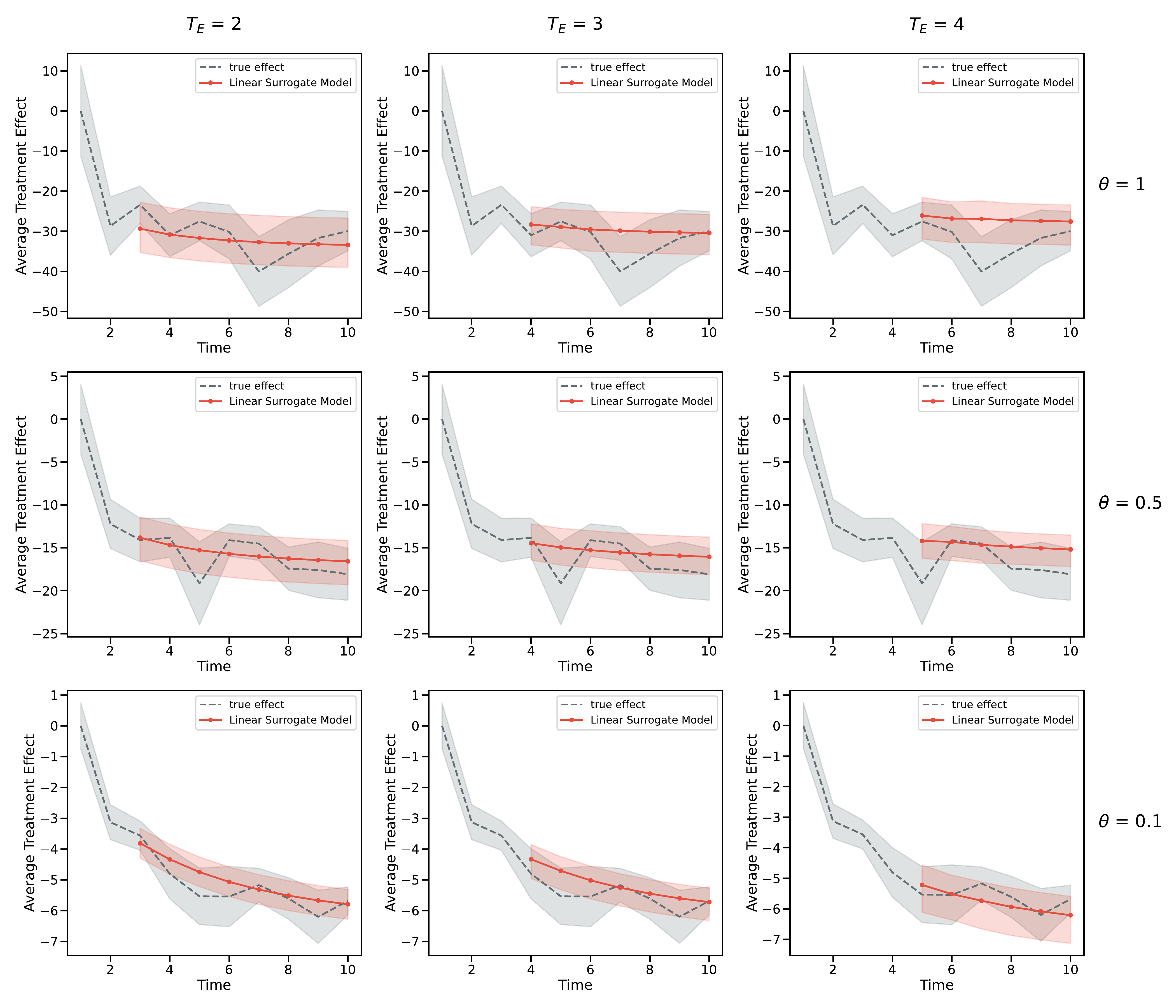}
\caption{Robust effect estimation  in the second synthetic experiment when the linearity assumption is not satisfied.}
{\footnotesize \textit{Note}: Grey dashed curves represent the true average treatment effect on $Y$ from periods 1 to periods 10. Solid red curves represent the estimated effects with the linear surrogate model. Shadows indicate 95\% confidence intervals. The three rows represent a different degree of violation of the linearity assumption, with a larger $\theta$ indicating a more severe violation. The three panels represent the scenarios when we use the first $T_E$ periods as the experimental period and the last $T_F$ periods as the future period.}
\label{fig:sim2Nonlinear}
\end{figure}

\subsection{No Long-Term Treatment Effect}
\label{sec:appendix:additionalNoeffect}

So far our synthetic experiments have focused on estimating the effect of a long-term treatment that has a significantly positive impact. We present an additional synthetic experiment to demonstrate that our methodology remains effective even when the long-term treatment effect eventually diminishes to zero. The initialization of the simulation is the same as in previous settings, which started with four surrogate variables, \(\bm{S}= [S_{i0,1},S_{i0,2},S_{i0,3},S_{i0,4}]\), each adhering to a normal distribution: \(S_{i0,d}\sim\mathcal{N}(\mu_d,\sigma_d^2)\), where \(\mu_d \sim \mathcal{N}(2,1)\) and \(\sigma_d \sim \mathcal{N}(2,1)\) for \(d\in \{1,2,3,4\}\) before the start of the experiment. Different from the previous settings, the surrogates in both the control group and the treatment group maintain the pre-treatment distribution throughout the experiment, i.e. \(S_{it,d} \sim\mathcal{N}(\mu_d,\sigma_d^2)\) for \(d\in \{1,2,3,4\}\) and each period $t$.
The primary outcome $Y$ in the control group is formulated as \(Y_{i,t+1} = -(0.1S_{it,1}+ 0.1S_{it,2} + 0.4S_{it,3} + 0.4S_{it,4})\), while the $Y$ in the treatment group is formulated as \(Y_{t+1} = -(0.1S_{it,1}+ 0.1S_{it,2} + 0.4S_{it,3} + 0.4S_{it,4}) + \frac{(-1)^t}{(t+2)^3}\).

The additional term for treatment group's $Y$ controls the volatility of the treatment effect. Given that $\frac{(-1)^t}{(t+2)^3}$ converges to zero as $t$ approaches infinity, the expectation of the average treatment effect in the long term is \textit{zero}. The estimated effect in the short term may exhibit significant fluctuations due to the disturbance term, which adds complexity to the prediction of the long-term treatment effect. However, as shown in Figure~\ref{fig:sim1NoEffect}, our approach can anticipate the eventual convergence level using only short-term experimental data, illustrating the capability of our method under various conditions. 

\begin{figure}[tb]
\centering
\includegraphics[width=0.9\linewidth]
{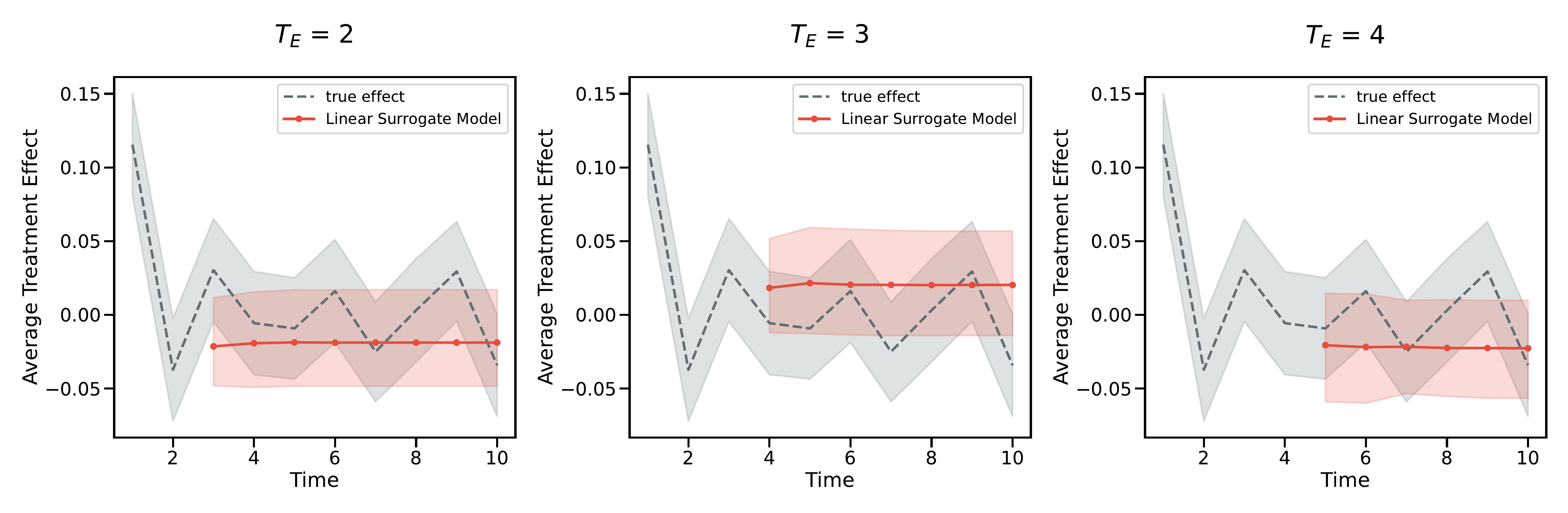}
\caption{Estimated effect using the linear surrogate model for a treatment with no long-term effect}
{\footnotesize \textit{Note}: Grey dashed curves represent the true average treatment effect on $Y$ from periods 1 to periods 10. Solid red curves represent the estimated effects with the linear surrogate model. Shadows indicate 95\% confidence intervals. The three panels represent the scenarios when we use the first $T_E$ periods as the experimental period and the last $T_F$ periods as the future period.}
\label{fig:sim1NoEffect}
\end{figure}

\subsection{Limitations of the Method}
\label{sec:appendix:limitation}
The additional synthetic experiments and sensitivity analyses demonstrate the robustness of our estimator from multiple angles.  
However, its predictive performance can deteriorate sharply when the underlying assumptions are grossly violated.  
To make these risks transparent, we examine representative failure modes associated with each of the three core assumptions—\emph{Surrogacy}, \emph{Comparability}, and \emph{Linearity}.

\subsubsection{Violation of Assumption~\ref{asp:Surrogacy}}

In Section~\ref{sec:appendix:surrogacy} we show that our estimator remains partially informative when Assumption~\ref{asp:Surrogacy} is only mildly violated---the resulting bias grows roughly in proportion to the severity of the violation.  
Here we explore two stark departures that can render the estimator unreliable.  
(i) \emph{Omitted surrogates}: the candidate set excludes key variables that carry most of the causal pathway from treatment to outcome.  
(ii) \emph{Irrelevant surrogates}: the set contains variables unrelated to the treatment effect, which inject noise and may aggravate any violation of the surrogacy condition.

Consider the following simulation setup: The simulations begin with two variables, \(S_{i0,1}\) and \(S_{i0,2}\), each drawn from a normal distribution  \(S_{i0,1} \sim\mathcal{N}(\mu_1,\sigma_1^2)\) \(S_{i0,2} \sim\mathcal{N}(\mu_2,\sigma_2^2)\), where \(\mu_1,\mu_2,\sigma_1,\sigma_2\) are themselves sampled from \(\mathcal{N}(2, 1)\) prior to the start of the experiment. For the control group, the variables remain unchanged throughout the experimental period, continuing to follow their initial distributions, i.e. \(S_{it,1}  \sim\mathcal{N}(\mu_1,\sigma_1^2)\) \(S_{it,2}\sim\mathcal{N}(\mu_2,\sigma_2^2)\) for each period $t$. In the treatment group, the variables fluctuate periodically, influenced by factors \([1-(-1)^t/5,1-(-1)^{t+1}/5]\) respectively. Specifically, for \(S_{i,t+1,1}\), \(S_{i,t+1,1} = 1.2 \cdot S_{i,t,1}\) when \(t\) is odd and \(S_{i,t+1,1} = 0.8 \cdot S_{i,t,1}\) when \(t\) is even; Similarly, \(S_{i,t+1,2} = 0.8 \cdot S_{i,t,2}\) when \(t\) is odd and \(S_{i,t+1,2} = 1.2 \cdot S_{i,t,2}\) when \(t\) is even for each period $t$. The primary outcome \(Y\) in period $t$ is formulated as \(Y_{it} = -(0.5\cdot S_{it,1} + 0.5\cdot S_{it,2})\). 

Based on the described data generation process, we apply our method using different surrogate variable sets to assess the impact of surrogate selection on estimation. First, we consider a complete set of surrogate variables, \([S_1, S_2, Y]\), which fully satisfies Assumption~\ref{asp:Surrogacy}, capturing all key variables necessary to explain the causal relationship between the treatment and outcomes. Second, we use only the primary outcome, \([Y]\), as the surrogate variable. This omission of \(S_1\) and \(S_2\), which are critical variables, reduces the model's ability to explain the causality. Finally, we introduce an irrelevant variable, \(S'\), which follows a normal distribution in the control group, \(S'_{it} \sim \mathcal{N}(\mu', \sigma'^2)\), where \(\mu'\) and \(\sigma'\) are sampled from \(\mathcal{N}(2, 1)\), and evolves dynamically in the treatment group as \(S'_{i,t+1} = 2 \cdot S'_{i,t}\). We apply our method with the surrogate set \([Y, S']\) to demonstrate the impact of including misleading variables to the estimation process. Figure~\ref{fig:simLimitSurrogacy} illustrates the estimation results of our method with these three surrogate sets given the experimental period \(T_E = 2\). The results show that only the first case produces an accurate estimation compared to the true effect, while the second and third cases fail to estimate the long-term treatment effect precisely. This analysis emphasizes the importance of selecting appropriate surrogate variables to ensure accurate and reliable results.

\begin{figure}[h]
\centering
\includegraphics[width=0.9\linewidth]
{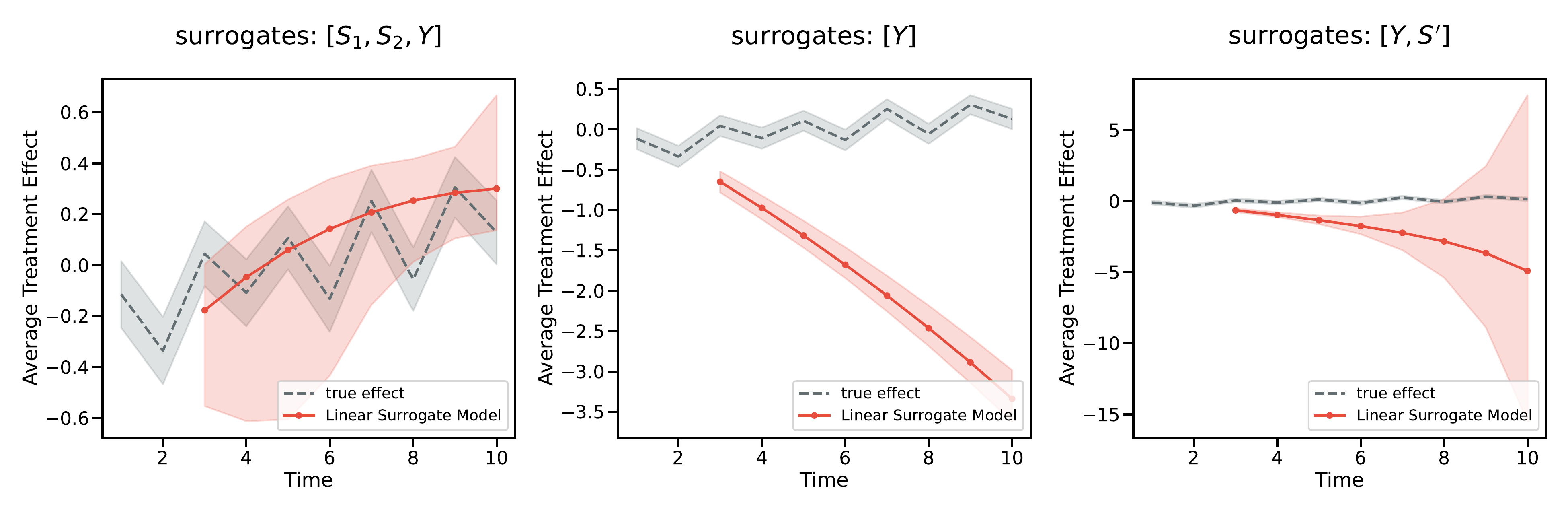}
\caption{Estimated effect using the linear surrogate model with different set of surrogates}
{\footnotesize \textit{Note}: Grey dashed curves represent the true average treatment effect on $Y$ from periods 1 to periods 10. Solid red curves represent the estimated effects with the linear surrogate model. Shadows indicate 95\% confidence intervals. The three panels represent the scenarios where we apply the linear surrogate model with different set of surrogates. We use the first 2 periods as the experimental period and the last 8 periods as the future period.}
\label{fig:simLimitSurrogacy}
\end{figure}

\subsubsection{Violation of Assumption~\ref{asp:Comparability}}

We now construct a scenario that explicitly violates the \emph{comparability} assumption to illustrate how our estimator can fail when Assumption~\ref{asp:Comparability} is ignored.

The simulated data is generated similarly to the synthetic experiment mentioned in Section~\ref{sec:Stabilized}, with one key difference: the distribution of surrogate variables in the treatment group shift before and after period \(t = 3\). For \(t \leq 3\), the decaying factors \(\gamma_d\) for each \(S_{it,d}\) are set as \(\gamma_d = 1 - (d + 1) \cdot (-1)^t / 10\), causing fluctuations in the primary outcome \(Y\) during the first three periods. For \(t > 3\), the decaying factors remain consistent with the settings, and all other parameters are identical to those described in the previous synthetic experiments. Apparently, the comparability assumption is violated in this scenario, as the conditional distribution of the primary outcome changes before and after \(t = 3\). We apply our method to this modified synthetic experiment, and the effect estimation results are presented in Figure~\ref{fig:simLimitComparability}. The poor performance of our method underscores the critical importance of verifying that the comparability assumption is satisfied before applying the method.

\begin{figure}[h]
\centering
\includegraphics[width=0.9\linewidth]
{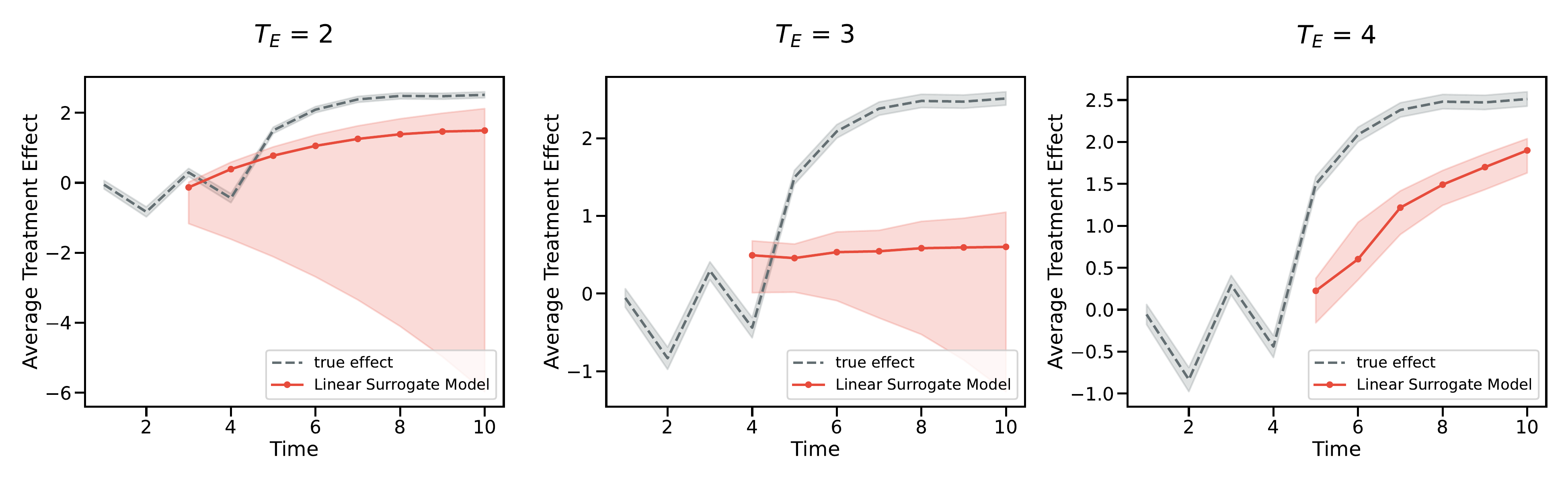}
\caption{Estimated effect using the linear surrogate model for a treatment violates the comparability assumption}
{\footnotesize \textit{Note}: Grey dashed curves represent the true average treatment effect on $Y$ from periods 1 to periods 10. Solid red curves represent the estimated effects with the linear surrogate model. Shadows indicate 95\% confidence intervals. The three panels represent the scenarios when we use the first $T_E$ periods as the experimental period and the last $T_F$ periods as the future period.}
\label{fig:simLimitComparability}
\end{figure}

    \subsubsection{Violation of Assumption~\ref{asp:Linearity}}
Section~\ref{sec:appendix:additionalNonlinear} shows that our estimator remains reliable when Assumption~\ref{asp:Linearity} is mildly relaxed: moderate nonlinearities can often be well approximated by linear functions once a rich set of surrogates is included.  
Nevertheless, when the data‐generating process is \emph{strongly} nonlinear, the linearity assumption becomes pivotal and our method can break down.

We illustrate this extreme scenario through a synthetic experiment, adopting a setup analogous to the nonlinear evaluation in Section~\ref{sec:appendix:additionalNonlinear} but with two critical modifications. First, the primary outcome function is defined as \(Y_{i,t+1} = \sin(\theta\cdot S_{it,1}) + \cos(\theta\cdot S_{it,2})\). Second, we vary \(\theta\) across \([2, 3, 4]\), which controls the minimum positive period of the sine and cosine functions. The results, illustrated in Figure~\ref{fig:simLimitLinearity}, reveal that our linear surrogate model struggles to accurately estimate the true treatment effects under this setting, particularly failing to capture the long-term trends of the true effect.

\begin{figure}[h]
\centering
\includegraphics[width=0.9\linewidth]
{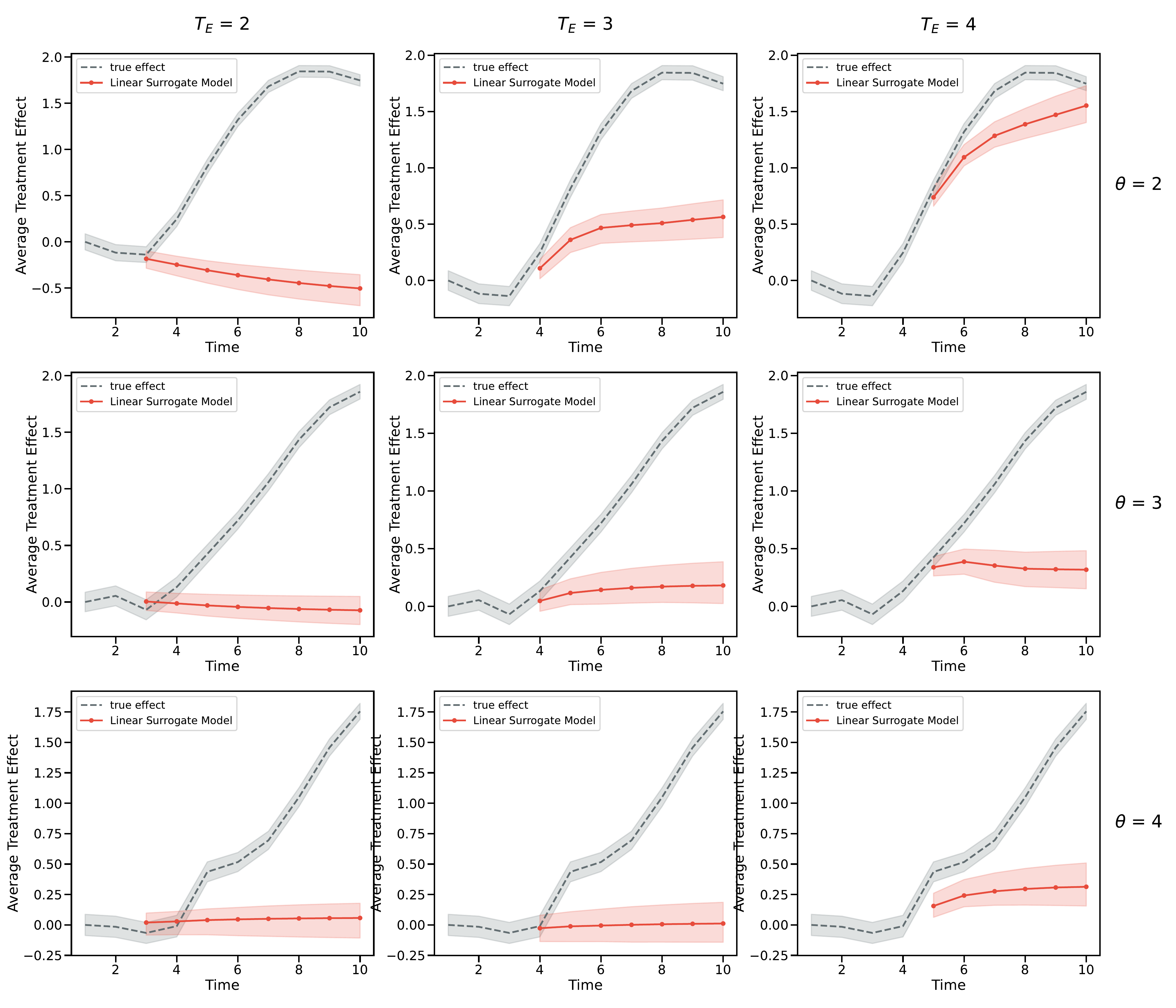}
\caption{Estimated effect using the linear surrogate model for a treatment violates the linearity assumption}
{\footnotesize \textit{Note}: Grey dashed curves represent the true average treatment effect on $Y$ from periods 1 to periods 10. Solid red curves represent the estimated effects with the linear surrogate model. Shadows indicate 95\% confidence intervals. The three rows represent a different type of violation of the linearity assumption. The three panels represent the scenarios when we use the first $T_E$ periods as the experimental period and the last $T_F$ periods as the future period.}
\label{fig:simLimitLinearity}
\end{figure}

\newpage

\end{APPENDICES}

%%%%%%%%%%%%%%%%%
\end{document}